\documentclass[11pt,a4paper]{article}

\usepackage{axodraw}
\usepackage{color}

\usepackage{epsfig,psfrag,graphicx}
\usepackage{amsfonts}
\usepackage{amssymb}
\usepackage{mathrsfs}
\usepackage{amsmath,amssymb,amscd}
\usepackage{subeqn}
\usepackage{bbm}
\usepackage[bookmarks=true]{hyperref}
\usepackage[nosort]{cite}

\ifx\hypersetup\sadfkjashdfkxja\else
\hypersetup{pdftitle={A First Course on Twistors, Integrability and Gluon 
Scattering Amplitudes}}
\hypersetup{pdfauthor={Martin Wolf}}
\hypersetup{pdfsubject={Mathematical Physics}}
\hypersetup{pdfkeywords={Twistors, Integrability, Scattering Theory}}
\hypersetup{plainpages=false}
\hypersetup{bookmarksnumbered=true}
\hypersetup{pdfstartview=FitH}
\hypersetup{pdfpagemode=None}
\hypersetup{colorlinks=false}
\hypersetup{citebordercolor={.5 1 .5}}
\hypersetup{urlbordercolor={.5 1 1}}
\hypersetup{linkbordercolor={1 .7 .7}}
\fi

\usepackage[british]{babel} 

\makeatletter\renewcommand{\section}{\@startsection
{section}{1}{\z@}{-2.5ex plus -1ex minus
    -.2ex}{2.3ex plus .2ex}{\centering\large\bf\mathversion{bold}}}

\makeatletter\renewcommand{\subsection}{\@startsection{subsection}{2}{\z@}{-3.25ex
plus -1ex minus
   -.2ex}{1.5ex plus .2ex}{\centering\bf\mathversion{bold}}}

\makeatletter\renewcommand{\subsubsection}{\@startsection{subsubsection}{3}{-2.45ex}{-3.25ex
plus -1ex minus -.2ex}{1.5ex plus .2ex}{\centering\bf\mathversion{bold}}}

\makeatletter\renewcommand{\paragraph}{\@startsection{paragraph}{4}{\z@}%
                                    {0.8ex \@plus1ex \@minus.2ex}%
                                    {-.5em}%
                                    {\normalfont\normalsize\bfseries\mathversion{bold}}}

\renewcommand{\thesection}{\arabic{section}.}

\numberwithin{paragraph}{section}
\renewcommand\theparagraph {\S\thesection\@arabic\c@paragraph.\kern-8pt}

\numberwithin{equation}{section}
\renewcommand{\theequation}{\thesection\arabic{equation}}

\numberwithin{table}{section}

\numberwithin{figure}{section}

\renewcommand*\l@section{\@dottedtocline{1}{0em}{1.5em}}
\renewcommand*\l@subsubsection{\@dottedtocline{4}{3.8em}{3.2em}}

\renewcommand\tableofcontents{%
    \section*{\large\contentsname
        \@mkboth{%
          \MakeUppercase\contentsname}{\MakeUppercase\contentsname}}%
       {\baselineskip=15pt plus 2pt minus 1pt
    \@starttoc{toc}}%
}

\renewenvironment{thebibliography}[1]
     {\section*{\large\centering{\refname}
        \@mkboth{\MakeUppercase\refname}{\MakeUppercase\refname}}%
     \list{\@biblabel{\@arabic\c@enumiv}}%
           {\settowidth\labelwidth{\@biblabel{#1}}%
            \leftmargin\labelwidth
            \advance\leftmargin\labelsep
            \@openbib@code
            \usecounter{enumiv}%
            \let\p@enumiv\@empty
            \renewcommand\theenumiv{\@arabic\c@enumiv}}%
      \sloppy
      \clubpenalty4000
      \@clubpenalty \clubpenalty
      \widowpenalty4000%
      \sfcode`\.\@m
 \catcode`\^^M=10%
}

\DeclareFontFamily{U}{rsf}{}
\DeclareFontShape{U}{rsf}{m}{n}{
  <5> <6> rsfs5 <7> <8> <9> rsfs7 <10-> rsfs10}{}
\DeclareMathAlphabet\Scr{U}{rsf}{m}{n}

\newcommand{\dbar}{\bar\partial}
\newcommand{\dd}{\mathrm{d}}
\newcommand{\di}{\mathrm{i}}
\newcommand{\de}{\mathrm{e}}

\newcommand{\ewith}{\quad\mbox{with}\quad}
\newcommand{\eand}{\quad\mbox{and}\quad}
\newcommand{\efor}{\quad\mbox{for}\quad}

\newcommand{\IC}{\mathbbm{C}}

\newcommand{\IR}{\mathbbm{R}}
\newcommand{\IZ}{\mathbbm{Z}}
\newcommand{\IP}{\mathbbm{P}}
\newcommand{\IE}{\mathbbm{E}}
\newcommand{\IM}{\mathbbm{M}}
\newcommand{\IN}{\mathbbm{N}}
\newcommand{\IK}{\mathbbm{K}}

\newcommand{\cC}{\mathscr{C}}
\newcommand{\cD}{\mathscr{D}}

\newcommand{\cP}{\mathscr{P}}

\newcommand{\cW}{\mathscr{W}}

\newcommand{\CA}{\mathcal{A}}
\newcommand{\CB}{\mathcal{B}}

\newcommand{\CG}{\mathcal{G}}

\newcommand{\CN}{\mathcal{N}}

\newcommand{\CO}{\mathcal{O}}

\newcommand{\CP}{\mathcal{P}}
\newcommand{\CF}{\mathcal{F}}
\newcommand{\CS}{\mathcal{S}}
\newcommand{\CR}{\mathcal{R}}
\newcommand{\CCD}{\mathcal{D}}
\newcommand{\CE}{\mathcal{E}}

\newcommand{\CI}{\mathcal{I}}
\newcommand{\CK}{\mathcal{K}}
\newcommand{\CZ}{\mathcal{Z}}

\newcommand{\sSU}{\mathsf{SU}}
\newcommand{\sSL}{\mathsf{SL}}
\newcommand{\sGL}{\mathsf{GL}}
\newcommand{\sSO}{\mathsf{SO}}

\newcommand{\sPSL}{\mathsf{PSL}}

\newcommand{\sSpin}{\mathsf{Spin}}

\newcommand{\fU}{\mathfrak{U}}
\newcommand{\fV}{\mathfrak{V}}

\newcommand{\ux}{\underline{x}}

\newcommand{\al}{{\alpha}}
\newcommand{\be}{{\beta}}
\newcommand{\ga}{{\gamma}}

\newcommand{\da}{{\dot\alpha}}
\newcommand{\db}{{\dot\beta}}
\newcommand{\dc}{{\dot\gamma}}

\newcommand{\Ac}{\overset{\circ}{A}}

\newcommand{\Wc}{\overset{\circ}{\phi}}
\newcommand{\cc}{\overset{\circ}{\chi}}

\newcommand{\Gc}{\overset{\circ}{G}}
\newcommand{\fc}{\overset{\circ}{f}}
\newcommand{\cnab}{\overset{\circ}{\nabla}}

\newtheorem{Exe}{Exercise}
\numberwithin{Exe}{section}

\newcommand{\Exercise}[1]{\begin{center}\hrule
\begin{minipage}{15.2cm}
 \vspace*{10pt}
  {\Exe #1}
 \vspace*{10pt}
\end{minipage}\hrule\end{center}}

\newtheorem{Rmk}{Remark}
\numberwithin{Rmk}{section}

\newcommand{\Remark}[1]{\begin{center}\hrule
\begin{minipage}{15.2cm}
 \vspace*{10pt}
  {\Rmk #1}
 \vspace*{10pt}
\end{minipage}\hrule\end{center}}

\newtheorem{Thm}{Theorem}
\numberwithin{Thm}{section}

\begin{document}

\begin{titlepage}

\setcounter{page}{0}
\renewcommand{\thefootnote}{\fnsymbol{footnote}}

\begin{flushright}
DAMTP 2010--05\\[.5cm]
\end{flushright}

\vspace*{1cm}

\begin{center}

{\LARGE\textbf{\mathversion{bold}A First Course on\\ Twistors, Integrability 
and Gluon Scattering Amplitudes}\par}

\vspace*{1cm}

{\large
 Martin Wolf\footnote{Also at the Wolfson College,
 Barton Road, Cambridge CB3 9BB, United Kingdom.} 
\footnote{{\it E-mail address:\/}
\href{mailto:m.wolf@damtp.cam.ac.uk}{\ttfamily m.wolf@damtp.cam.ac.uk}
}}

\vspace*{1cm}

{\it Department of Applied Mathematics and Theoretical Physics\\
University of Cambridge\\
Wilberforce Road, Cambridge CB3 0WA, United Kingdom}

\vspace*{1cm}

{\bf Abstract}
\end{center}

\vspace*{-.3cm}

\begin{quote}
These notes accompany an introductory lecture course on the twistor approach to
supersymmetric gauge theories aimed at early-stage PhD students. It was held by
the author at the University of Cambridge during the Michaelmas term in 2009.
The lectures assume a working knowledge of differential geometry and quantum
field theory. No prior knowledge of twistor theory is required. 
\vfill
\noindent
21st January 2010

\end{quote}

\setcounter{footnote}{0}\renewcommand{\thefootnote}{\arabic{thefootnote}}

\linespread{1.3}
\end{titlepage}

\section*{Preface}
The course is divided into two main parts: I) The re-formulation of gauge theory
on twistor space and II) the construction of tree-level gauge theory scattering
amplitudes. More specifically, the first few lectures deal with the basics of
twistor geometry and its application to free field theories. We then move on and
discuss the non-linear field equations of self-dual Yang--Mills theory. The
subsequent lectures deal with supersymmetric self-dual Yang--Mills theories
and the extension to the full non-self-dual supersymmetric Yang--Mills theory in
the case of maximal $\CN=4$ supersymmetry. Whilst studying the field equations
of these theories, we shall also discuss the associated action functionals on
twistor space. Having re-interpreted $\CN=4$ supersymmetric Yang--Mills theory
on twistor space, we discuss the construction of tree-level scattering
amplitudes. We first transform, to twistor space, the so-called
maximally-helicity-violating amplitudes. Afterwards we discuss the construction
of general tree-level amplitudes by means of the Cachazo--Svr\v cek--Witten
rules and the Britto--Cachazo--Feng--Witten recursion relations. Some
mathematical concepts underlying twistor geometry are summarised in several
appendices. The computation of scattering amplitudes beyond tree-level is not
covered here. 

My main motivation for writing these lecture notes was to provide an opportunity
for students and researchers in mathematical physics to get a grip of twistor
geometry and its application to perturbative gauge theory without having to go
through the wealth of text books and research papers but at the same time
providing as detailed derivations as possible. Since the present article should
be understood as notes accompanying an introductory lecture course rather than
as an exhaustive review article of the field, I emphasise that even though
I tried to refer to the original literature as accurately as possible, I had to
make certain choices for the clarity of presentation. As a result, the list of
references is by no means complete. Moreover, to keep the notes rather short in
length, I had to omit various interesting topics and recent developments.
Therefore, the reader is urged to consult  \htmladdnormallink{{Spires
HEP}}{http://www.slac.stanford.edu/spires/hep/search/} and
\htmladdnormallink{{\ttfamily arXiv.org}}{http://arxiv.org/} 
for the latest advancements and especially the
\htmladdnormallink{{citations}}{%
http://www-spires.dur.ac.uk/cgi-bin/spiface/hep?c=CMPHA,252,189} of Witten's
paper
on twistor string theory, published in \htmladdnormallink{{\ttfamily
Commun.~Math.~Phys.~252,~189~(2004)}}{%
http://www.springerlink.com/content/lxhrcf81x0j73b94/},
\htmladdnormallink{{\ttfamily
arXiv:hep-th/0312171}}{http://arxiv.org/abs/hep-th/0312171}.
 
Should you find any typos or mistakes in the text, please let me know by
sending an email to 
\href{mailto:m.wolf@damtp.cam.ac.uk}{\ttfamily m.wolf@damtp.cam.ac.uk}. 
For the most recent version of these lecture notes, please also check
\begin{center}
 \htmladdnormallink{{\ttfamily http://www.damtp.cam.ac.uk/user/wolf}}{http://www.damtp.cam.ac.uk/user/wolf}
\end{center}

\vspace*{.2cm}
\noindent
{\bf Acknowledgements.}
I am very grateful to  J.~Bedford, N.~Bouatta, D.~Correa, N.~Dorey, M.~Dunajski,
L.~Mason, R.~Ricci and C.~S{\"a}mann for many helpful discussions and
suggestions. Special thanks go to J.~Bedford for various
discussions and comments on the manuscript. I would also like to thank those
who attended the lectures for asking various interesting questions. This work
was
supported by an STFC Postdoctoral Fellowship and by a Senior Research Fellowship
at the Wolfson College, Cambridge, U.K.

\begin{flushright}
 Cambridge, 21st January 2010\\
 Martin Wolf
\end{flushright}


\newpage
\section*{Literature}

Amongst many others (see bibliography at the end of this article), the following
lecture notes and books have been used when compiling this article and are
recommended as references and for additional reading (chronologically ordered).

\vspace*{10pt}
\noindent{\bf Complex geometry:}
\vspace*{-5pt}
\begin{itemize}
\setlength{\itemsep}{-.4mm}
 \item[(i)] P.~Griffiths \& J.~Harris, {\it Principles of algebraic geometry}, 
            John Wiley \& Sons, New York, 1978
 \item[(ii)] R.~O.~Wells, {\it Differential analysis on complex manifolds}, 
             Springer Verlag, New York, 1980
 \item[(iii)] M.~Nakahara, {\it Geometry, topology and physics}, The
              Institute of Physics, Bristol--Philadel\-phia, 2002
 \item[(iv)] V. Bouchard, {\it  Lectures on complex geometry, Calabi--Yau 
             manifolds and toric geometry},
             \htmladdnormallink{{\ttfamily
arXiv:hep-th/0702063}}{http://arxiv.org/abs/hep-th/0702063}
\end{itemize}

\vspace*{10pt}
\noindent{\bf Supermanifolds and supersymmetry:}
\vspace*{-5pt}
\begin{itemize}
\setlength{\itemsep}{-.4mm}
 \item[(i)] Yu.~I.~Manin, {\it Gauge field theory and complex geometry}, 
            Springer Verlag, New York, 1988
 \item[(ii)] C.~Bartocci, U.~Bruzzo \& D.~Hernand\'ez-Ruip\'erez, {\it The 
             geometry of supermanifolds}, Kluwer, Dordrecht, 1991
 \item[(iii)] J.~Wess \& J.~Bagger, {\it Supersymmetry and supergravity},
Princeton University Press, Princeton, 1992
 \item[(iv)] C.~S\"amann, {\it Introduction to supersymmetry}, 
            \htmladdnormallink{{Lecture Notes}}{http://www.christiansaemann.de/files/LecturesOnSUSY.pdf}, 
             Trinity College Dublin, 2009 
\end{itemize}

\vspace*{10pt}
\noindent{\bf Twistor geometry:}
\vspace*{-5pt}
\begin{itemize}
\setlength{\itemsep}{-.4mm}
 \item[(i)] R.~S.~Ward \& R.~O.~Wells, {\it Twistor geometry and field theory}, 
            Cambridge University Press, Cambridge, 1989
 
 \item[(ii)] S.~A.~Huggett \& K.~P.~Tod, {\it  An introduction to twistor
theory},
            Cambridge University Press, Cambridge, 1994
 \item[(iii)] L.~J.~Mason \& N.~M.~J.~Woodhouse, {\it Integrability,
self-duality, and twistor theory}, Clarendon Press, Oxford, 1996 
 \item[(iv)] M.~Dunajski, {\it Solitons, instantons and twistors}, Oxford
University Press, Oxford, 2009
\end{itemize}

\vspace*{10pt}
\noindent{\bf Tree-level gauge theory scattering amplitudes and twistor theory:}
\vspace*{-5pt}
\begin{itemize}
\setlength{\itemsep}{-.4mm}
 \item[(i)] F.~Cachazo \& P.~Svr\v cek, {\it Lectures on twistor strings and 
 perturbative Yang--Mills theory},
            PoS {\bf RTN2005} (2005) 004, \htmladdnormallink{{\ttfamily
arXiv:hep-th/0504194}}{http://arxiv.org/abs/hep-th/0504194}
 \item[(ii)] J.~A.~P.~Bedford, {\it  On perturbative field theory and twistor
string theory},
             \htmladdnormallink{{\ttfamily arXiv:0709.3478}}{http://arxiv.org/abs/0709.3478}, 
              PhD thesis, Queen Mary, University of London (2007)
 \item[(iii)] C.~Vergu, {\it Twistors, strings and supersymmetric gauge theories
},
             \htmladdnormallink{{\ttfamily
arXiv:0809.1807}}{http://arxiv.org/abs/0809.1807}, 
              PhD thesis, Universit\'e Paris IV--Pierre et Marie Curie (2008)
\end{itemize}


\newpage
\tableofcontents

\bigskip
\bigskip
\hrule
\bigskip
\bigskip

\newpage
\thispagestyle{empty}
\vspace*{5cm}
\addtocontents{toc}{\hspace{-0cm}{\bf \centerline{Part I: Twistor
re-formulation of gauge theory}\\}} 
\begin{center}
 {\bf\Large Part I}

\vspace*{1cm}
 {\bf\large Twistor re-formulation of gauge theory}
\end{center}


\newpage
\section{Twistor space}

\subsection{Motivation}

Usually, the equations of motion of physically interesting theories are
complicated systems of coupled non-linear partial differential equations. This
thus makes it extremely hard to find explicit solutions. However, among the
theories of interest are some which are completely solvable in the sense of
allowing for the construction
(in principle) of all solutions to the corresponding equations of motion. We
shall refer to these systems as integrable systems. It should be noted at this
point that there are various distinct notions of integrability in the
literature and here we shall use the word `integrability' in the loose sense of
`complete solvability' without any concrete assumptions. The prime examples of
integrable theories are the self-dual Yang--Mills and gravity theories in four
dimensions including their various reductions to lower space-time dimensions.
See e.g.~\cite{Mason:1991rf,Dunajski:2009aa} for details.  

Twistor theory has turned out to be a very powerful tool in analysing integrable
systems. The key ingredient of twistor theory is the substitution of space-time
as a background for physical processes by an auxiliary space called
twistor space. The term `twistor space' is used collectively and refers to
different spaces being associated with different physical theories
under consideration. All these twistor spaces have one thing in common in that
they are (partially) complex manifolds, and
moreover, solutions to the field equations on space-time of the theory in
question are encoded in terms of differentially unconstrained (partially)
complex analytic data on twistor space. This way one may sometimes even classify
all solutions to a problem. The goal of the first part of these lecture notes is
the twistor re-formulation of $\CN=4$ supersymmetric Yang--Mills theory on
four-dimensional flat space-time.

\subsection{Preliminaries}\label{sec:PL}

Let us consider $M^4\cong\IR^{p,q}$ for $p+q=4$, where $\IR^{p,q}$ is $\IR^{p+q}$ equipped with a metric $g=(g_{\mu\nu})=\mbox{diag}(-\mathbbm{1}_p,\mathbbm{1}_q)$ of signature $(p,q)$. Here and in the following, $\mu,\nu,\ldots$ run from 0 to 3.
In particular, for $(p,q)=(0,4)$ we shall speak of Euclidean ($\IE$) space, for
$(p,q)=(1,3)$ of Minkowski ($\IM$) space and for $(p,q)=(2,2)$ of Kleinian ($\IK$) space. The rotation group is then given by $\sSO(p,q)$. Below we shall only be interested in the connected component of the identity of the rotation group $\sSO(p,q)$ which is is commonly denoted by $\sSO_0(p,q)$.

If we let $\alpha,\beta,\ldots=1,2$ and $\da,\db,\ldots=\dot 1,\dot 2$,
then we may represent any real four-vector $x=(x^\mu)\in M^4$ as a $2\times 2$-matrix $\ux=(x^{\al\db})\in\mbox{Mat}(2,\IC)\cong\IC^4$ 
subject to the following reality conditions:\footnote{Note that for the Kleinian case one may alternatively impose $\bar\ux=\sigma_1\,\ux\,\sigma_1^{\rm t}$.}
\begin{equation}\label{eq:RC}
 \begin{aligned}
   \IE\ &:\ \bar\ux\ =\ -\sigma_2\, \ux\, \sigma_2^{\rm t}~,\\
   \IM\ &:\ \bar\ux\ =\ -\ux^{\rm t}~,\\
   \IK\ &:\ \bar\ux\ =\ \ux~,
 \end{aligned}
\end{equation}
where bar denotes complex conjugation, `t' transposition and $\sigma_i$, for $i,j,\ldots=1,2,3$, are the Pauli matrices
\begin{equation}
 \sigma_1\ =\ \begin{pmatrix}
               0 & 1\\ 1 & 0
              \end{pmatrix},\quad
 \sigma_2\ =\ \begin{pmatrix}
               0 & -\di\\ \di & 0
              \end{pmatrix}\eand
 \sigma_3\ =\ \begin{pmatrix}
               1 & 0\\ 0 & -1
              \end{pmatrix}.
\end{equation}
Recall that they obey
\begin{equation}
 \sigma_i\sigma_j\ =\ \delta_{ij}+\di \sum_k\varepsilon_{ijk}\sigma_k~,
\end{equation}
where $\delta_{ij}$ is the Kronecker symbol and $\varepsilon_{ijk}$ is totally 
anti-symmetric in
its indices with $\varepsilon_{123}=1$.
To be more concrete, the isomorphism $\sigma\,:\,x\mapsto\ux=\sigma(x)$ can be written as
\begin{subequations}\label{eq:SigmaIso}
\begin{equation}
 x^{\al\db}\ =\ \sigma_\mu^{\al\db}x^\mu\qquad\Longleftrightarrow\qquad
 x^\mu\ =\ \tfrac12\varepsilon_{\al\be}\varepsilon_{\da\db}\sigma_\mu^{\al\da}x^{\be\db}~,
\end{equation}
where $\varepsilon_{\al\be}=\varepsilon_{[\al\be]}$ with $\varepsilon_{12}=-1$
and $\varepsilon_{\al\gamma}\varepsilon^{\gamma\be}=\delta_\al^\be$ (and similar
relations for $\varepsilon_{\da\db}$)\footnote{We have chosen
particle physics literature conventions which are somewhat different from the
twistor literature.}
\begin{equation}
 \begin{aligned}
   \IE\ &:\ (\sigma_\mu^{\al\db})\ :=\ (\mathbbm{1}_2,\di\sigma_3,-\di\sigma_2,-\di\sigma_1)~,\\
   \IM\ &:\ (\sigma_\mu^{\al\db})\ :=\ (-\di\mathbbm{1}_2,-\di\sigma_1,-\di\sigma_2,-\di\sigma_3)~,\\
   \IK\ &:\ (\sigma_\mu^{\al\db})\ :=\ (\sigma_3,\sigma_1,-\di\sigma_2,\mathbbm{1}_2)~.
 \end{aligned}
\end{equation}
\end{subequations}

The line element $\dd s^2=g_{\mu\nu}\dd x^\mu\dd x^\nu$ on $M^4\cong\IR^{p,q}$ 
is then given by 
\begin{equation}\label{eq:LineElement}
\dd s^2\ =\ \det \dd\ux\ =\ \tfrac{1}{2}\varepsilon_{\al\be}\varepsilon_{\da\db}\dd x^{\al\da}\dd x^{\be\db}~
\end{equation}
Rotations (respectively, Lorentz transformations) act on $x^\mu$ according to $x^\mu\mapsto x'^\mu={\Lambda^\mu}_\nu x^\nu$ with $\Lambda=({\Lambda^\mu}_\nu)\in\sSO_0(p,q)$. The induced action on $\ux$ reads as
\begin{equation}
 \ux\ \mapsto\ \ux'\ =\ g_1\,\ux\,g_2\efor g_{1,2}\ \in\ \sGL(2,\IC)~.
\end{equation}
The $g_{1,2}$ are not arbitrary for several reasons. Firstly, any two pairs
$(g_1,g_2)$ and $(g'_1,g'_2)$ with $(g'_1,g'_2)=(t g_1,t^{-1}g_2)$ for $t\in\IC\setminus\{0\}$ induce the same transformation on $\ux$, hence we may regard the equivalence classes $[(g_1,g_2)]=\{(g'_1,g'_2)|
(g'_1,g'_2)=(t g_1,t^{-1}g_2)\}$. Furthermore, rotations preserve the line element and
from $\det \dd\ux=\det \dd\ux'$ we conclude that $\det g_1\det g_2=1$. Altogether, we may take $g_{1,2}\in\sSL(2,\IC)$ without loss of generality. In addition, the $g_{1,2}$ have to preserve the reality conditions \eqref{eq:RC}. For instance, on $\IE$ we find that $\bar g_{1,2}=-\sigma_1\, g_{1,2}\, \sigma_1^{\rm t}$.
Explicitly, we have
\begin{equation}
 g_{1,2}\ =\ \begin{pmatrix}
               a_{1,2} & b_{1,2}\\ c_{1,2} & d_{1,2} 
             \end{pmatrix}
        \ =\ \begin{pmatrix}
               a_{1,2} & b_{1,2}\\ -\bar b_{1,2} & \bar a_{1,2} 
             \end{pmatrix}.
\end{equation}
Since $\det g_{1,2}=1=|a_{1,2}|^2+|b_{1,2}|^2$ (which topologically describes a
three-sphere) we conclude that $g_{1,2}\in\sSU(2)$,
i.e.~$g_{1,2}^{-1}=g_{1,2}^\dagger$. In addition, if $g_{1,2}\in\sSU(2)$ then
also $\pm g_{1,2}\in\sSU(2)$ and since $g_{1,2}$ and $\pm g_{1,2}$ induce the
same transformation on $\underline{x}$, we have therefore established
\begin{equation}
 \sSO(4)\ \cong\ (\sSU(2)\times\sSU(2))/\IZ_2~.
\end{equation}

One may proceed similarly for $\IM$ and $\IK$ but we leave this as an exercise. Eventually, we arrive at
\begin{equation}\label{eq:RG}
 \begin{aligned}
    \IE\ &:\ \sSO(4)\ \cong\ (\sSU(2)\times\sSU(2))/\IZ_2~,\ewith \ux\ \mapsto\ g_1\,\ux\,g_2
         \eand g_{1,2}\ \in\ \sSU(2)~,\\
    \IM\ &:\ \sSO_0(1,3)\ \cong\ \sSL(2,\IC)/\IZ_2~,\ewith \ux\ \mapsto\ g\,\ux\,g^\dagger
        \eand g\ \in\ \sSL(2,\IC)~,\\
    \IK\ &:\ \sSO_0(2,2)\ \cong\ (\sSL(2,\IR)\times\sSL(2,\IR))/\IZ_2~,\ewith \ux\ \mapsto\ g_1\,\ux\,g_2
        \eand g_{1,2}\ \in\ \sSL(2,\IR)~.
 \end{aligned}
\end{equation}
Notice that in general one may write
\begin{equation}
 \sSO_0(p,q)\ \cong\ \sSpin(p,q)/\IZ_2~,
\end{equation}
where $\sSpin(p,q)$ is known as the spin group of $\IR^{p,q}$. In a more
mathematical terminology, $\sSpin(p,q)$ is the double cover of $\sSO_0(p,q)$
(for the sum $p+q$ not necessarily restricted to 4). For $p=0,1$ and $q>2$, the
spin group is simply connected and thus coincides with the universal
cover. Since the fundamental group (or first homotopy group) of
$\sSpin(2,2)$ is
non-vanishing, $\pi_1(\sSpin(2,2))\cong\IZ\times\IZ$, the spin
group $\sSpin(2,2)$  is not simply connected. See,
e.g.~\cite{Lawson:1989,Ward:1990vs} for more details on the spin groups.

In summary, we may either work with $x^\mu$ or with $x^{\al\db}$ and
making this identification amounts to identifying $g_{\mu\nu}$ with
$\frac12\varepsilon_{\al\be}\varepsilon_{\da\db}$.  Different signatures are
encoded in different reality conditions \eqref{eq:RC} on $x^{\al\db}$. Hence, in
the following we shall work with the complexification $M^4\otimes\IC\cong\IC^4$
and $\ux=(x^{\al\db})\in\mbox{Mat}(2,\IC)$ and impose the reality conditions
whenever appropriate. Therefore, the different cases of \eqref{eq:RG} can be
understood as different real forms of the complex version
\begin{equation}\label{eq:ComplexIso}
 \sSO(4,\IC)\ \cong\ (\sSL(2,\IC)\times\sSL(2,\IC))/\IZ_2~.
\end{equation}
For brevity, we denote $\ux$ by $x$ and $M^4\otimes\IC$ by $M^4$.

\Exercise{Prove that the rotation groups on $\IM$ and $\IK$ are given by \eqref{eq:RG}.}

\subsection{Twistor space}\label{sec:BTS}

In this section, we shall introduce Penrose's twistor space
\cite{Penrose:1967wn} by starting from complex space-time $M^4\cong\IC^4$ and
the identification $x^\mu\leftrightarrow x^{\al\db}$. According to the
discussion of the previous section, we view the tangent bundle $T M^4$ of $M^4$
according to 
\begin{equation}\label{eq:ident}
\begin{aligned}
 T M^4\ &\cong\ S\otimes\tilde S~,\\
  &\kern-2.3cm\partial_\mu\ :=\ \frac{\partial}{\partial x^\mu}\ \overset{\sigma_*}{\longleftrightarrow}\ 
    \partial_{\al\db}\ :=\ \frac{\partial}{\partial x^{\al\db}}~,
\end{aligned}
\end{equation}
where $S$ and $\tilde S$ are the two complex rank-2 vector bundles called the
bundles of dotted and undotted spinors. See Appendix \ref{app:VB}~for the
definition of a vector bundle. The two copies of $\sSL(2,\IC)$ in
\eqref{eq:ComplexIso} act independently on $S$ and $\tilde S$. Let us denote
undotted spinors
by $\mu^\al$ and dotted ones by $\lambda^\da$.\footnote{Notice that it is also
common to denote undotted spinors by $\lambda^\al$ and dotted spinors by
$\tilde\lambda^\da$. However, we shall stick to our above conventions.}  On $S$
and $\tilde S$ we have the symplectic forms $\varepsilon_{\al\be}$ and
$\varepsilon_{\da\db}$ from before which can be used to raise and lower spinor
indices: 
\begin{equation}
 \mu_\al\ =\ \varepsilon_{\al\be}\mu^\be\eand\lambda_\da\ =\ \varepsilon_{\da\db}\lambda^\db~. 
\end{equation}

\Remark{\label{rmk:confstru} Let us comment on conformal structures since the identification
\eqref{eq:ident} amounts to choosing a (holomorphic) conformal structure. This
can be seen as follows:
 The standard definition of a conformal structure on a four-dimensional 
complex manifold $X$ states that a conformal structure is an equivalence
class $[g]$, the conformal class, of holomorphic metrics $g$ on $X$, 
where two given metrics $g$
and $g'$ are called equivalent if $g'=\gamma^2 g$ for some nowhere vanishing
holomorphic function $\gamma$. Put differently, a conformal
structure is a line subbundle $L$ in $T^*X\odot T^*X$.
Another, maybe less familiar definition assumes a
factorisation of the holomorphic tangent bundle $TX$ of $X$
as a tensor product of two rank-2 holomorphic vector 
bundles $S$ and $\tilde S$, that is, 
$TX\cong S\otimes\tilde S$. This isomorphism in turn gives 
(canonically)  
the line subbundle $\Lambda^2S^*\otimes\Lambda^2\tilde S^*$ in
$T^*X\odot T^*X$ which, in fact, can be identified with
$L$. The metric $g$ is then given by the tensor product of the two
symplectic forms on $S$ and $\tilde S$ (as done above) which are
sections of  $\Lambda^2S^*$ and $\Lambda^2\tilde S^*$.
}

Let us now consider the projectivisation of the dual spin bundle $\tilde S^*$.
Since $\tilde S$ is of rank two, the projectivisation $\IP(\tilde S^*)\to M^4$
is a $\IC P^1$-bundle over $M^4$. Hence, $\IP(\tilde S^*)$ is a five-dimensional
complex manifold bi-holomorphic to $\IC ^4\times\IC P^1$. In what follows, we
shall denote it by $F^5$ and call it correspondence space. The reason for this
name becomes transparent momentarily. We take $(x^{\al\db},\lambda_\da)$ as
coordinates on $F^5$, where $\lambda_\da$ are homogeneous
coordinates on $\IC P^1$.

\vspace*{5pt}

\Remark{\label{rmk:cover} Remember that $\IC P^1$ can be covered by two
coordinate patches, $U_\pm$, with $\IC P^1=U_+\cup U_-$. If we let
$\lambda_\da=(\lambda_{\dot1},\lambda_{\dot2})^t$ be homogeneous coordinates on
$\IC P^1$ with $\lambda_\da\sim t\lambda_\da$ for $t\in\IC\setminus\{0\}$,
$U_\pm$ and the corresponding affine coordinates $\lambda_\pm$ can be defined as
follows:
$$
\begin{aligned}
 U_+\ &:\ \lambda_{\dot1}\ \neq\ 0\eand\lambda_+\ :=\ \frac{\lambda_{\dot2}}{\lambda_{\dot1}}~,\\
 U_-\ &:\ \lambda_{\dot2}\ \neq\ 0\eand\lambda_-\ :=\ \frac{\lambda_{\dot1}}{\lambda_{\dot2}}~.
\end{aligned}
$$
On $U_+\cap U_-\cong\IC\setminus\{0\}$ we have $\lambda_+=\lambda_-^{-1}$.
}

\vspace*{5pt}

On $F^5$ we may consider the following vector fields:
\begin{equation}\label{eq:VF}
 V_\al\ =\ \lambda^\db\partial_{\al\db}\ =\ \lambda^\db\frac{\partial}{\partial x^{\al\db}}~.
\end{equation}
They define an integrable rank-2 distribution on $F^5$ (i.e.~a rank-2
subbundle in $TF^5$) which is called the twistor distribution. Therefore, we
have a foliation of $F^5$ by two-dimensional complex manifolds. The 
 resulting quotient
will be twistor space, a three-dimensional complex manifold denoted by $P^3$. We
have thus established the following double fibration:
\begin{equation}\label{eq:DoubleFibration}
 \begin{picture}(50,40)
  \put(0.0,0.0){\makebox(0,0)[c]{$P^3$}}
  \put(64.0,0.0){\makebox(0,0)[c]{$M^4$}}
  \put(34.0,33.0){\makebox(0,0)[c]{$F^5$}}
  \put(7.0,18.0){\makebox(0,0)[c]{$\pi_1$}}
  \put(55.0,18.0){\makebox(0,0)[c]{$\pi_2$}}
  \put(25.0,25.0){\vector(-1,-1){18}}
  \put(37.0,25.0){\vector(1,-1){18}}
 \end{picture}
\end{equation}
The projection $\pi_2$ is the trivial projection and
$\pi_1\,:\,(x^{\alpha\db},\lambda_\da)\mapsto
(z^\al,\lambda_\da)=(x^{\alpha\db}\lambda_\db,\lambda_\da)$, where
$(z^\al,\lambda_\da)$ are homogeneous coordinates on $P^3$. The relation 
\begin{equation}\label{eq:incidence}
 z^\al\ =\ x^{\alpha\db}\lambda_\db
\end{equation}
is known as the incidence relation.
Notice that \eqref{eq:DoubleFibration} makes clear why $F^5$ is called correspondence space: It is the space that `links' space-time with twistor space.

Also $P^3$ can be covered by two coordinate patches, which we (again) denote by $U_\pm$ (see also Remark \ref{rmk:cover}):
\begin{equation}\label{eq:tw1}
\begin{aligned}
 U_+\ &:\ \lambda_{\dot1}\ \neq\ 0
  \eand z_+^\al\ :=\ \frac{z^\al}{\lambda_{\dot1}}
  \eand\lambda_+\ :=\ \frac{\lambda_{\dot2}}{\lambda_{\dot1}}~,\\
 U_-\ &:\ \lambda_{\dot2}\ \neq\ 0
  \eand z_-^\al\ :=\ \frac{z^\al}{\lambda_{\dot2}}
  \eand\lambda_-\ :=\ \frac{\lambda_{\dot1}}{\lambda_{\dot2}}~.
\end{aligned}
\end{equation}
On $U_+\cap U_-$ we have $z_+^\al=\lambda_+ z_-^\al$ and $\lambda_+=\lambda_-^{-1}$. This shows that twistor space $P^3$ can be identified with the total space of the holomorphic fibration
\begin{equation}\label{eq:tw2}
 \CO(1)\oplus\CO(1)\ \to\ \IC P^1~,
\end{equation}
where $\CO(1)$ is the dual of the tautological line bundle $\CO(-1)$ over $\IC P^1$,
\begin{equation}\label{eq:TB}
 \CO(-1)\ :=\ \{(\lambda_\da,\rho_\da)\in\IC
P^1\times\IC^2\,|\,\rho_\da\propto\lambda_\da\}~,
\end{equation}
i.e.~$\CO(1)=\CO(-1)^*$. The bundle $\CO(1)$ is also referred to as the hyperplane bundle.
Other line bundles, which we will frequently encounter below, are:
\begin{equation}
 \CO(-m)\ =\ \CO(-1)^{\otimes m}\eand
 \CO(m)\ =\ \CO(-m)^*
 \efor m\ \in\ \IN~.
\end{equation}
The incidence relation $z^\al=x^{\al\db}\lambda_\db$ identifies $x\in M^4$ with
holomorphic sections of \eqref{eq:tw2}.
Note that $P^3$ can also be identified 
with $\IC P^3\setminus\IC P^1$, where the deleted projective line is given by
$z^\al\neq0$ and $\lambda_\da=0$. 
\vspace*{5pt}

\Exercise{Let $\lambda_\da$ be homogeneous coordinates on $\IC P^1$ and $z$ be the fibre coordinates of $\CO(m)\to\IC P^1$ for $m\in\IZ$. Furthermore, let $\{U_\pm\}$ be the canonical cover as in Remark \ref{rmk:cover} Show that the transition function of $\CO(m)$ is given by $\lambda_+^m=\lambda_-^{-m}$. Show further that while $\CO(1)$ has global holomorphic sections, $\CO(-1)$ does not.}

\vspace*{5pt}

Having established the double fibration \eqref{eq:DoubleFibration}, we may ask
about the geometric correspondence, also known as the Klein correspondence,
between
space-time $M^4$ and twistor space $P^3$. In fact, for any point $x\in M^4$, the
corresponding manifold $L_x:=\pi_1(\pi_2^{-1}(x))\hookrightarrow P^3$ is a curve
which is bi-holomorphic to $\IC P^1$.
Conversely, any point $p\in P^3$ corresponds to a totally null-plane in $M^4$,
which
can be seen as follows. For some fixed $p=(z,\lambda)\in P^3$, the incidence
relation \eqref{eq:incidence} tells us that
$x^{\al\db}=x_0^{\al\db}+\mu^\al\lambda^\db$ since
$\lambda_\da\lambda^\da=\varepsilon^{\da\db}\lambda_\da\lambda_\db=0$. Here,
$x_0$ is a particular solution to \eqref{eq:incidence}.
Hence, this describes a two-plane in $M^4$ which is totally null since any
null-vector $x^{\al\db}$ is of the form $x^{\al\db}=\mu^\al\lambda^\db$.
In addition, \eqref{eq:incidence} implies that the removed line $\IC P^1$ of
$P^3\cong\IC P^3\setminus\IC P^1$ corresponds to the point `infinity' of
space-time. Thus, $\IC P^3$ can be understood as the twistor space of
conformally compactified complexified space-time.

\vspace*{5pt}

\Remark{\label{rmk:lightlike} Recall that a four-vector $x^\mu$ in $M^4$ is said
to be null if it has zero norm, i.e. $g_{\mu\nu}x^\mu x^\nu=0$. This is
equivalent to saying that $\det x=0$. Hence, the two columns/rows of $x$ must be
linearly dependent. Thus,
$x^{\al\db}=\mu^\al\lambda^\db$.}

\vspace*{10pt}

\section{Massless fields and the Penrose transform}\label{sec:MFPT}

The subject of this section is to sketch how twistor space can be used to derive
all solutions to zero-rest-mass field equations.

\subsection{Integral formul{\ae} for massless fields}\label{sec:IF}

To begin with, let $P^3$ be twistor space (as before) and consider a function
$f$ that is holomorphic on the intersection $U_+\cap U_-\subset P^3$.
Furthermore, let us pull back $f$ to the correspondence space $F^5$. The
pull-back of $f(z^\al,\lambda_\da)$ is $f(x^{\al\db}\lambda_\db,\lambda_\da)$,
since the tangent spaces of the leaves of the fibration $\pi_1\,:\,F^5\to P^3$
are spanned by \eqref{eq:VF} and so the pull-backs have to be annihilated by the
vector fields
\eqref{eq:VF}. Then we may consider following contour integral:
\begin{equation}\label{eq:PW0}
 \phi(x)\ =\ -\frac{1}{2\pi\di}\oint_\cC\dd\lambda_\da\lambda^\da\,f(x^{\al\db}\lambda_\db,\lambda_\da)~,
\end{equation}
where $\cC$ is a closed curve in $U_+\cap U_-\subset\IC P^1$.\footnote{As
before, we shall not make any notational distinction between the coordinate
patches covering $\IC P^1$ and the ones covering twistor space.} Since the
measure $\dd\lambda_\da\lambda^\da$ is of homogeneity 2, the function $f$ should
be of homogeneity $-2$ as only then is the integral well-defined. Put
differently, only if $f$ is of homogeneity $-2$, $\phi$ is a function defined on
$M^4$. 

 Furthermore, one readily computes
\begin{equation}
 \square\phi\ =\ 0~,\ewith\square\
:=\ \tfrac12\partial_{\al\db}\partial^{\al\db}~
\end{equation}
by differentiating under the integral. Hence, the function $\phi$ satisfies the
Klein--Gordon equation. Therefore, any $f$ with the above properties will yield
a solution to the Klein--Gordon equation via the contour integral
\eqref{eq:PW0}. This is the essence of twistor theory: Differentially
constrained data on space-time (in the present situation the function $\phi$) is
encoded in differentially unconstrained complex analytic data on twistor space
(in the present situtation the function $f$). 

\vspace*{5pt}

\Exercise{Consider the following function $f=1/(z^1z^2)$ which is holomorphic on $U_+\cap U_-\subset P^3$. Clearly, it is of homogeneity $-2$. Show that the integral \eqref{eq:PW0} gives rise to $\phi=1/\det x$. Hence, this $f$ yields the elementary solution to the Klein--Gordon equation based at the origin $x=0$.}

\vspace*{5pt}

What about the other zero-rest-mass field equations?
Can we say something similar about them? Consider a zero-rest-mass field
$\phi_{\da_1\cdots\da_{2h}}$ of positive helicity $h$ (with $h>0$). Then
\begin{equation}
 \phi_{\da_1\cdots\da_{2h}}(x)\ =\ -\frac{1}{2\pi\di}\oint_\cC\dd
                 \lambda_\da\lambda^\da\,\lambda_{\da_1}\cdots\lambda_{\da_{2h}}
                   f(x^{\al\db}\lambda_\db,\lambda_\da)
\end{equation}
solves the equation
\begin{equation}\label{eq:EOMH>0}
 \partial^{\al\da_1}\phi_{\da_1\cdots\da_{2h}}\ =\ 0~.
\end{equation}
Again, in order to have a well-defined integral, the integrand should have
total homogeneity zero, which is equivalent to requiring $f$ to be of homogeneity
$-2h-2$. Likewise, we may also consider a zero-rest-mass field 
$\phi_{\al_1\cdots\al_{2h}}$ of negative helicity $-h$ (with $h>0$) for which we
take
\begin{equation}
 \phi_{\al_1\cdots\al_{2h}}(x)\ =\ -\frac{1}{2\pi\di}\oint_\cC\dd
                 \lambda_\da\lambda^\da\,\frac{\partial}{\partial z^{\al_1}}
                 \cdots \frac{\partial}{\partial z^{\al_{2h}}}
                   f(x^{\al\db}\lambda_\db,\lambda_\da)
\end{equation}
such that $f$ is of homogeneity $2h-2$. Hence,
\begin{equation}\label{eq:EOMH<0}
\partial^{\al_1\da}\phi_{\al_1\cdots\al_{2h}}\ =\ 0~.
\end{equation}
These contour integral formul{\ae} provide the advertised Penrose transform
\cite{Penrose:1969ae,Penrose:1977in}. Sometimes, one refers to this transform
as the Radon--Penrose transform to emphasise that it is a generalisation of the
Radon transform.\footnote{\label{foot:radon} The Radon transform, named after
Johann Radon \cite{Radon:1917a},  is an integral
transform in two dimensions consisting of the integral of a function  over
straight lines. It plays an important role in computer assisted tomography. 
The higher dimensional analog of the Radon transform is the X-ray
transform; see footnote \ref{foot:xray}.}

In summary, any function on twistor space, provided it is of 
appropriate homogeneity $m\in\IZ$, 
can be used to construct solutions to zero-rest-mass field 
equations. However, there are a lot of different functions leading to the 
same solution. For instance, we could simply change $f$ by adding a 
function which has singularities on one side of the contour but is holomorphic
on the other, since the contour integral does not feel such functions. How can we understand
what is going on? Furthermore,
are the integral formul{\ae} invertible? 
In addition, we made use of particular coverings, so do the results depend on
these choices? The tool which helps clarify all these issues is sheaf
cohomology.\footnote{In Section \ref{sec:WHFT}~we present a discussion for
Kleinian signature which by-passes sheaf cohomology.}  For a detailed discussion
about sheaf theory, see e.g.~\cite{Ward:1990vs,Griffiths:1978}.

\subsection{{\v C}ech cohomology groups and Penrose's theorem  -- a sketch}\label{sec:Cech}

Consider some Abelian sheaf $\CS$ over some 
manifold $X$, that is, for any open subset $U\subset X$ one has an Abelian
group $\CS(U)$ subject to certain `locality conditions'; Appendix
\ref{app:Sh}~collects useful definitions regarding sheaves including some
examples. Furthermore, let $\fU=\{U_i\}$ be an open cover of $X$. A $q$-cochain
of the covering $\fU$ with values in $\CS$ is a collection 
$f=\{f_{i_0\cdots i_q}\}$ of sections of the sheaf $\CS$ over non-empty 
intersections $U_{i_0}\cap\cdots\cap U_{i_q}$. 

The set of all $q$-cochains has an Abelian group structure  (with respect to addition) and is denoted by
$C^q(\fU,\CS)$. Then we define the coboundary map by
\begin{subequations}
\begin{equation}
 \begin{aligned}
  \delta_q\,:\,C^q(\fU,\CS)\ &\to\ C^{q+1}(\fU,\CS)~,\\
            (\delta_q f)_{i_0\cdots i_{q+1}}\ &:=\ \sum_{k=0}^{q+1}(-)^i
                 r^{i_0\cdots\hat{i}_k\cdots i_{q+1}}_{i_0\cdots i_{q+1}}
                 f_{i_0\cdots\hat{i}_k\cdots i_{q+1}}~,
 \end{aligned}
\end{equation}
where
\begin{equation}
 r^{i_0\cdots\hat{i}_k\cdots i_{q+1}}_{i_0\cdots i_{q+1}}\,:\,
        \CS(U_{i_0}\cap\cdots\cap U_{\hat{i}_k}\cap\cdots\cap U_{i_{q+1}})
        \ \to\ \CS(U_{i_0}\cap\cdots\cap U_{i_{q+1}})
\end{equation}
\end{subequations}
is the sheaf restriction morphism and $\hat{i}_k$ means omitting 
$i_k$. It is clear that $\delta_q$ is a morphism of groups, and one may
check that $\delta_q\circ\delta_{q-1}=0$. 

\Exercise{Show that $\delta_q\circ\delta_{q-1}=0$ for $\delta_q$ as defined above.}

\noindent
Furthermore, we see straight away that $\ker\delta_0=\CS(X)$. Next we define
\begin{equation}
Z^q(\fU,\CS)\ :=\ \ker\delta_q\eand
       B^q(\fU,\CS)\ :=\ \mbox{im}\,\delta_{q-1}~.
\end{equation}
We call elements of $Z^q(\fU,\CS)$ $q$-cocycles and elements of 
$B^q(\fU,\CS)$ $q$-coboundaries, respectively. Cocycles are anti-symmetric in their indices.
Both $Z^q(\fU,\CS)$ and
$B^q(\fU,\CS)$ are Abelian groups and since the coboundary map is 
nil-quadratic, $B^q(\fU,\CS)$ is a (normal) subgroup of $Z^q(\fU,\CS)$. The
$q$-th \v Cech cohomology group is the quotient
\begin{equation}
H^q(\fU,\CS)\ :=\ Z^q(\fU,\CS)/B^q(\fU,\CS)~.
\end{equation}

In order to get used to these definitions, let us consider a simple example and
take the (Abelian) sheaf of holomorphic sections of the line bundle 
$\CO(m)\to\IC P^1$. As before we choose the canonical cover $\fU=\{U_\pm\}$ of $\IC P^1$. Since there is only a double intersection, all cohomology groups $H^q$ with $q>1$ vanish automatically. The following table then summarises $H^0$ and $H^1$:

\vspace*{5pt}

\begin{table}[h]
\begin{center}
\begin{tabular}{|c||c|c|c|c|c|c|c|c|c|}
 \hline
 $m$ & $\cdots$ & $-4$ & $-3$ & $-2$ & $-1$ & $0$ & $1$ & $2$ & $\cdots$ \\
 \hline
 $H^0(\fU,\CO(m))$ & $\cdots$ & $0$ & $0$ & $0$ & $0$ & $\IC^1$ & $\IC^2$ & $\IC^3$ & $\cdots$\\
 \hline
 $H^1(\fU,\CO(m))$ & $\cdots$ & $\IC^3$ & $\IC^2$ & $\IC^1$ & $0$ & $0$ & $0$ & $0$ & $\cdots$\\
 \hline
\end{tabular}
\caption{\it \v Cech cohomology groups for $\CO(m)\to\IC P^1$ with respect to the cover $\fU=\{U_\pm\}$.}
\label{tab:CCGA}
\end{center}
\end{table}


\noindent
Note that when writing $H^q(X,E)$ for some vector bundle $E\to X$ over some manifold $X$, we actually mean the (Abelian) sheaf $\CE$ of sections (either smooth or holomorphic depending on the context) of $E$. By a slight abuse of notation, we shall often not make a notational distinction between $E$ and its sheaf of sections $\CE$ and simply write $E$ in both cases.

Let us now compute $H^1(\fU,\CO(-m))$ for $m\geq 0$. The rest is left as an exercise. To this end, consider some representative $f=\{f_{+-}\}$ defined on $U_+\cap U_-\subset\IC P^1$.\footnote{Notice that in the preceding sections, we have not made a notational distinction between $f$ and $f_{+-}$, but strictly speaking we should have.} Clearly, $\delta_1 f=0$ as there are no triple intersections. Without loss of generality, $f$ might be taken as
\begin{equation}
 f_{+-}\ =\ \frac{1}{(\lambda_{\dot1})^m}\,\sum_{n=-\infty}^\infty c_n \left(\frac{\lambda_{\dot2}}{\lambda_{\dot1}}\right)^n~.
\end{equation}
This can be re-written according to
\begin{eqnarray}
 f_{+-}\! &=&\! \frac{1}{(\lambda_{\dot1})^m}\,\sum_{n=-\infty}^\infty c_n \left(\frac{\lambda_{\dot2}}{\lambda_{\dot1}}\right)^n\notag\\
 &=&\! \frac{1}{(\lambda_{\dot1})^m}\left[\sum_{n=-\infty}^{-m}+\sum_{n=-m+1}^{-1}+\sum_{n=0}^\infty\right] c_n \left(\frac{\lambda_{\dot2}}{\lambda_{\dot1}}\right)^n\notag\\
 &=&\!  
\underbrace{\frac{1}{(\lambda_{\dot1})^m}\sum_{n=0}^\infty c_{n}\left(\frac{\lambda_{\dot2}}{\lambda_{\dot1}}\right)^n}_{=:\ r^+_{+-}f_+}
+
\underbrace{\sum_{n=1}^{m-1}\frac{c_{-n}}{(\lambda_{\dot2})^n(\lambda_{\dot1})^{m-n}}}_{=:\ f'_{+-}}+
\underbrace{\frac{1}{(\lambda_{\dot2})^m}\sum_{n=0}^\infty c_{-n-m}\left(\frac{\lambda_{\dot1}}{\lambda_{\dot2}}\right)^n}_{=:\ -r^-_{+-}f_-}\notag\\
&=&\! f'_{+-}+r^+_{+-}f_+-r^-_{+-}f_-~,
\end{eqnarray}
where $r^\pm_{+-}$ are the restriction mappings. Since the $f_\pm$ are
holomorphic on $U_\pm$, we conclude that $f=\{f_{+-}\}$ is cohomologous to
$f'=\{f'_{+-}\}$ with
\begin{equation}
 f'_{+-}\ =\ \sum_{n=1}^{m-1}\frac{c_{-n}}{(\lambda_{\dot2})^n(\lambda_{\dot1})^{m-n}}~.
\end{equation}
There are precisely $m-1$ independent complex parameters, $c_{-1},\ldots,c_{-m+1}$, which parametrise $f'$. Hence, we have established $H^1(\fU,\CO(-m))\cong \IC^{m-1}$ whenever $m>1$ and $H^1(\fU,\CO(-m))=0$ for $m=0,1$.

\vspace*{5pt}

\Exercise{Complete the Table \ref{tab:CCGA}. }

\vspace*{5pt}

Table \ref{tab:CCGA}. hints that there is some sort of duality. In fact,
\begin{equation}\label{eq:Serre}
 H^0(\fU,\CO(m))\ \cong\ H^1(\fU,\CO(-m-2))^*~, 
\end{equation}
which is a special instance of Serre duality (see also Remark \ref{rmk:serre}). Here, the star denotes the
vector space dual. To understand this relation better, consider ($m\geq0$)
\begin{equation}
 g\ \in\  H^0(\fU,\CO(m))~,\ewith
       g\ =\ g^{\da_1\cdots\da_m}\lambda_{\da_1}\cdots\lambda_{\da_m}
\end{equation}
and $f\in H^1(\fU,\CO(-m-2))$. Then define the pairing
\begin{equation}\label{eq:pairing}
 (f,g)\ :=\ -\frac{1}{2\pi\di}\oint_\cC\dd
                 \lambda_\da\lambda^\da\,f(\lambda_\da)\,g(\lambda_\da)~,
\end{equation}
where the contour is chosen as before. This expression is
complex linear and non-degenerate and depends only on the cohomology class of
$f$. Hence, it gives the duality \eqref{eq:Serre}.

A nice way of writing \eqref{eq:pairing} is as 
$(f,g)=f_{\da_1\cdots\da_m}g^{\da_1\cdots\da_m}$, where
\begin{equation}
 f_{\da_1\cdots\da_m}\ :=\ -\frac{1}{2\pi\di}\oint_\cC\dd
      \lambda_\da\lambda^\da\,\lambda_{\da_1}\cdots\lambda_{\da_m}\, f(\lambda_\da)~, 
\end{equation}
such that Penrose's contour integral formula \eqref{eq:PW0} can be recognised as an 
instance of Serre duality (the coordinate $x$ being interpreted as some parameter). 

\vspace*{5pt}

\Remark{\label{rmk:serre} If $\CS$ is some Abelian sheaf over some compact complex manifold $X$ with covering $\fU$ and $\CK$ the sheaf of sections of the canonical line bundle $K:=\det T^*X$, then there is the following isomorphism which is referred to as Serre duality (or sometimes to as Kodaira--Serre duality):
$$
 H^q(\fU,\CS)\ \cong\ H^{d-q}(\fU,\CS^*\otimes\CK)^*~.
$$
Here, $d=\dim_\IC X$. See e.g.~\cite{Griffiths:1978} for more details. In our present case, $X=\IC P^1$ and so $d=1$ and $K=\det T^*\IC P^1=T^*\IC P^1\cong\CO(-2)$ and furthermore $\CS=\CO(m)$. 
}

\vspace*{5pt}

One technical issue remains to be clarified. Apparently all of our above
calculations seem to depend on the chosen cover. But is this really the case?

Consider again some manifold $X$ with cover $\fU$ together with some
Abelian sheaf $\CS$. If another cover
$\fV$ is the refinement of $\fU$, that is, for $\fU=\{U_i\}_{i\in I}$ 
and $\fV=\{V_j\}_{j\in J}$ there is a map $\rho\,:\,J\to I$ of index sets,
such that for any $j\in J$, $V_j\subseteq U_{\rho(j)}$, then there is a natural
group homomorphism (induced by the restriction mappings of the sheaf $\CS$)
\begin{equation}
 h^\fU_\fV\,:\, H^q(\fU,\CS)\ \to\ H^q(\fV,\CS)~.
\end{equation}
We can then define the inductive limit of these cohomology groups with 
respect to the partially ordered set of all coverings (see also Remark \ref{rmk:inductive}),
\begin{equation}
 H^q(X,\CS)\ :=\ \underset{\fU}{\mbox{lim\,ind}}\,H^q(\fU,\CS)
\end{equation}
which we call the $q$-th \v Cech cohomology group of $X$ with coefficients
in $\CS$.

\Remark{\label{rmk:inductive} Let us recall the definition of the inductive limit. If we let $I$ be a partially ordered set (with respect to `$\geq$') and $S_i$ a family of modules indexed by $I$ with homomorphisms $f^i_j\,:\,S_i\to S_j$ with $i\geq j$ and $f^i_i={\rm id}$, $f^i_j\circ f^j_k=f^i_k$ for $i\geq j\geq k$, then the inductive limit, $$\underset{i\in I}{\rm lim\,ind}\, S_i~,$$ is defined by quotienting the disjoint union $\dot\bigcup_{i\in I} S_i=\bigcup_{i\in I}\{(i,S_i)\}$ by the following equivalence relation: Two elements $x_i$ and $x_j$ of $\dot\bigcup_{i\in I} S_i$ are said to be equivalent if there exists a $k\in I$ such that $f^i_k(x_i)=f^j_k(x_j)$. }

\noindent
By the properties of inductive limits, we have a homomorphism
$H^q(\fU,\CS)\to H^q(X,\CS)$. Now the question is: When does this becomes an
isomorphism?
The following theorem tells us when this is going to happen.

{\Thm (Leray) Let $\fU=\{U_i\}$ be a covering of $X$ with the property that 
for all tuples $(U_{i_0},\ldots,U_{i_p})$ of the cover, 
$H^q(U_{i_0}\cap\cdots\cap U_{i_p},\CS)=0$ for all $q\geq1$. Then
$$
H^q(\fU,\CS)\ \cong\ H^q(X,\CS)~.
$$
}

\vspace*{-15pt}

\noindent
For a proof, see e.g.~\cite{Hirzebruch:1966,Griffiths:1978}.

Such covers are called Leray or acyclic covers and in fact our two-set cover $\fU=\{U_\pm\}$ of 
$\IC P^1$ is of this form. Therefore, Table \ref{tab:CCGA}. translates into Table \ref{tab:CCGB}.

\vspace*{10pt}

\begin{table}[h]
\begin{center}
\begin{tabular}{|c||c|c|c|c|c|c|c|c|c|}
 \hline
 $m$ & $\cdots$ & $-4$ & $-3$ & $-2$ & $-1$ & $0$ & $1$ & $2$ & $\cdots$ \\
 \hline
 $H^0(\IC P^1,\CO(m))$ & $\cdots$ & $0$ & $0$ & $0$ & $0$ & $\IC^1$ & $\IC^2$ & $\IC^3$ & $\cdots$\\
 \hline
 $H^1(\IC P^1,\CO(m))$ & $\cdots$ & $\IC^3$ & $\IC^2$ & $\IC^1$ & $0$ & $0$ & $0$ & $0$ & $\cdots$\\
 \hline
\end{tabular}
\caption{\it \v Cech cohomology groups for $\CO(m)\to\IC P^1$.}
\label{tab:CCGB}
\end{center}
\end{table}

\vspace*{-5pt}

\Remark{\label{rmk:kodaira} We have seen that twistor space $P^3\cong\IC
P^3\setminus\IC P^1\cong\CO(1)\oplus\CO(1)$; see \eqref{eq:tw1} and
\eqref{eq:tw2}. There is yet another interpretation. The Riemann sphere $\IC
P^1$ can be embedded into $\IC P^3$. The normal bundle $N_{\IC P^1|\IC P^3}$ of
$\IC P^1$ inside $\IC P^3$ is $\CO(1)\otimes\CO(1)$ as follows from the short
exact sequence:
$$ 0\ \overset{\varphi_1}{\longrightarrow}\ T\IC P^1\
\overset{\varphi_2}{\longrightarrow}\ T\IC P^3|_{\IC P^1}\
\overset{\varphi_3}{\longrightarrow}\ N_{\IC P^1|\IC P^3}\
\overset{\varphi_4}{\longrightarrow}\ 0~.$$
Exactness of this sequence means that ${\rm im\,}\varphi_i=\ker\varphi_{i+1}$.
If we take $(z^\al,\lambda_\da)$ as homogeneous coordinates on $\IC P^3$ with
the embedded $\IC P^1$ corresponding to $z^\al=0$ and $\lambda_\da\neq0$, then
the non-trivial mappings $\varphi_{2,3}$ are given by
$\varphi_2\,:\,
\partial/\partial\lambda_\da\mapsto\ell_\da\partial/\partial\lambda_\da$ while
$\varphi_3\,:\,\ell^\al\partial/\partial
z^\al+\ell_\da\partial/\partial\lambda_\da\mapsto \ell^\al$, where
$\ell^\al,\ell_\da$ are linear in $z^\al,\lambda_\da$ and
the restriction to $\IC P^1$ is understood. This shows that indeed $N_{\IC
P^1|\IC P^3}\cong\CO(1)\otimes\CO(1)$, i.e.~twistor space $P^3$ can be
identified with the normal bundle of $\IC P^1\hookrightarrow \IC P^3$. Kodaira's
theorem on relative deformation states that if $Y$ is a compact complex
submanifold of a not necessarily compact complex manifold $X$, and if
$H^1(Y,N_{Y|X})=0$, where $N_{Y|X}$ is the normal bundle of $Y$ in $X$, then
there exists a $d$-dimensional family of deformations of $Y$ inside $X$, where
$d:=\dim_\IC H^0(Y,N_{Y|X})$. See e.g.~\cite{Burns:1979,Kodaira:1986} for more
details. In our example, $Y=\IC P^1$, $X=\IC P^3$ and $N_{\IC P^1|\IC
P^3}\cong\CO(1)\otimes\CO(1)$. Using Table \ref{tab:CCGB}.,
we conclude that $H^1(\IC P^1,\CO(1)\oplus\CO(1))=0$ and $d=4$. In fact, complex
space-time $M^4\cong \IC^4$ is precisely this family of deformations. To be more
concrete, any $L_x\cong \IC P^1$ has $\CO(1)\otimes\IC^2$ as normal bundle, and
the tangent space $T_xM^4$ at $x\in M^4$ arises as 
$T_xM^4\cong H^0(L_x,\CO(1)\otimes\IC ^2)\cong H^0(L_x,\IC^2)\otimes
H^0(L_x,\CO(1))\cong S_x\otimes\tilde S_x$,
where $S_x:=H^0(L_x,\IC^2)$ and $\tilde S_x:=H^0(L_x,\CO(1))$ which is the
factorisation \eqref{eq:ident}.
}

In summary, the functions $f$ on twistor space from Section
\ref{sec:IF}~leading to solutions of zero-rest-mass field equations should be
thought of as representatives of sheaf
cohomology classes in $H^1(P^3,\CO(\mp 2h-2))$. Then we can state the following
theorem:

{\Thm\label{thm:penrose} (Penrose \cite{Penrose:1977in}) If we let $\CZ_{\pm h}$
 be the sheaf of (sufficiently well-behaved) solutions to the helicity $\pm h$
(with $h\geq0$) zero-rest-mass field equations on $M^4$, then
 $$H^1(P^3,\CO(\mp 2h-2))\ \cong\ H^0(M^4,\CZ_{\pm h})~.$$
}

\vspace*{-15pt}

\noindent
The proof of this theorem requires more work including a weightier mathematical
machinery. It therefore lies somewhat far afield from the main thread of
development
and we refer the interested reader to e.g.~\cite{Ward:1990vs} for details.

\vspace*{5pt}

\section{Self-dual Yang--Mills theory and the Penrose--Ward transform}

So far, we have discussed free field equations. The subject of this section is a
generalisation of our above  discussion to the non-linear field equations of
self-dual Yang--Mills theory on four-dimensional space-time. Self--dual
Yang--Mills theory can be regarded as a subsector of Yang--Mills theory and in
fact, the self--dual Yang--Mills equations are the Bogomolnyi equations of
Yang--Mills theory. Solutions to the self-dual Yang--Mills equations
are always solutions to the Yang--Mills equations, while the converse may not be
true.

\subsection{Motivation}

To begin with, let $M^4$ be $\IE$ and $E\to M^4$ a (complex) vector bundle over
$M^4$ with structure group $G$. For the moment, we shall assume that $G$ is
semi-simple and compact. This allows us to normalise the generators $t_a$ of $G$
according to $\mbox{tr}(t_a^\dagger t_b)=-\mbox{tr}(t_a t_b)=C(r)\delta_{ab}$
with $C(r)>0$. Furthermore,
let $\nabla\,:\,\Omega^p(M^4,E)\to\Omega^{p+1}(M^4,E)$ be a connection on $E$
with curvature $F=\nabla^2\in H^0(M^4,\Omega^2(M^4,{\rm End}\,E))$. Here,
$\Omega^p(M^4)$ are the $p$-forms on $M^4$ and
$\Omega^p(M^4,E):=\Omega^p(M^4)\otimes E$. Then $\nabla=\dd+A$ and $F=\dd
A+A\wedge A$, where $A$ is the ${\rm End}\,E$-valued connection one-form. The
reader unfamiliar with these quantities may wish to consult Appendix
\ref{app:VB}~for their definitions. In the coordinates $x^\mu$ on $M^4$ we have
\begin{subequations}
\begin{equation}
 A\ =\ \dd x^\mu A_\mu\eand\nabla\ =\ \dd x^\mu\nabla_\mu~,\ewith\nabla_\mu\ =\ \partial_\mu+A_\mu
\end{equation}
and therefore
\begin{equation}
 F\ =\ \tfrac12\dd x^\mu\wedge\dd x^\nu  F_{\mu\nu}~,\ewith
 F_{\mu\nu}\ =\ [\nabla_\mu,\nabla_\nu]\ =\ \partial_\mu A_\nu-\partial_\nu A_\mu+[A_\mu,A_\nu]~.
\end{equation}
\end{subequations}

The Yang--Mills action functional is defined by
\begin{equation}\label{eq:YM}
 S\ =\ -\frac{1}{g_{\rm YM}^2}\int_{M^4}\mbox{tr}(F\wedge{*F})~,
\end{equation}
where $g_{\rm YM}$ is the Yang--Mills coupling constant and `$*$' denotes the Hodge star on $M^4$. The corresponding field equations read as
\begin{equation}\label{eq:YME}
 \nabla{*F}\ =\ 0\qquad\Longleftrightarrow\qquad\nabla^\mu F_{\mu\nu}\ =\ 0~.
\end{equation}

\Exercise{Derive \eqref{eq:YME} by varying \eqref{eq:YM}.}

\vspace*{5pt}

\noindent
Solutions to the Yang--Mills equations are critical points of the Yang--Mills action. The critical points may be local maxima of the action, local minima, or saddle points. To find the field configurations that truly minimise \eqref{eq:YM}, we consider the following inequality:
\begin{equation}
 \mp\int_{M^4}\mbox{tr}\big[(F\pm{*F})\wedge(F\pm{*F})\big]\ \geq\ 0~.
\end{equation}
A short calculation then shows that
\begin{equation}
 -\int_{M^4}\mbox{tr}(F\wedge{*F})\ \geq\ \pm\int_{M^4}\mbox{tr}(F\wedge F)
\end{equation}
and therefore
\begin{equation}
 S\ \geq\ \pm\frac{1}{g_{\rm YM}^2}\int_{M^4}\mbox{tr}(F\wedge F)\qquad\Longrightarrow\qquad
 S\ \geq\ \frac{8\pi^2}{g_{\rm YM}^2}|Q|~,
\end{equation}
where $Q\in\IZ$ is called topological charge or instanton number,
\begin{equation}\label{eq:charge}
 Q\ =\ -\frac{1}{8\pi^2}\int_{M^4}\mbox{tr}(F\wedge F)\ =\ -c_2(E)~.
\end{equation}
Here, $c_2(E)$ denotes the second Chern class of $E$; see Appendix
\ref{app:VB}~for the definition.

Equality is achieved for configurations that obey
\begin{equation}\label{eq:SDYM}
 F\ =\ \pm {*F}\qquad\Longleftrightarrow\qquad F_{\mu\nu}\ =\ \pm\tfrac12\varepsilon_{\mu\nu\lambda\sigma} F^{\lambda\sigma}~
\end{equation}
with $\varepsilon_{\mu\nu\lambda\sigma}=\varepsilon_{[\mu\nu\lambda\sigma]}$ and $\varepsilon_{0123}=1$.
These equations are called the self-dual and anti-self-dual Yang--Mills equations. Solutions to these equations with finite charge $Q$ are referred to as instantons and anti-instantons. The sign of $Q$ has been chosen such that $Q>0$ for instantons while $Q<0$ for anti-instantons. Furthermore, by virtue of the Bianchi identity, $\nabla F=0$ $\Longleftrightarrow$ $\nabla_{[\mu}F_{\nu\lambda]}=0$,  solutions to \eqref{eq:SDYM} automatically satisfy the second-order Yang--Mills equations \eqref{eq:YME}.

Remember from our discussion in Section \ref{sec:PL}~that the rotation group
$\sSO(4)$ is given by
\begin{equation}
 \sSO(4)\ \cong\ (\sSU(2)\times\sSU(2))/\IZ_2~.
\end{equation}
Therefore, the anti-symmetric tensor product of two vector representations ${\bf 4}\wedge{\bf 4}$ decomposes under this isomorphism as ${\bf 4}\wedge{\bf 4}\cong{\bf 3}\oplus{\bf 3}$. More concretely, by taking the explicit isomorphism 
\eqref{eq:SigmaIso}, we can write
\begin{equation}\label{eq:decomp}
 F_{\al\da\be\db}\ :=\ \tfrac14 \sigma^\mu_{\al\da}\sigma^\nu_{\be\db}F_{\mu\nu}\ =\ 
                    \varepsilon_{\al\be}f_{\da\db}+\varepsilon_{\da\db}f_{\al\be}~,
\end{equation}
with $f_{\al\be}=f_{\be\al}$ and $f_{\da\db}=f_{\db\da}$. Since each of these symmetric rank-2 tensors has three independent components, we have made the decomposition ${\bf 4}\wedge{\bf 4}\cong{\bf 3}\oplus{\bf 3}$ explicit.
Furthermore, if we write $F=F^++F^-$ with $F^\pm:=\frac12(F\pm{*F})$, i.e.~$F^\pm=\pm{*F^\pm}$, then
\begin{equation}\label{eq:decomp2}
 F^+\ \Longleftrightarrow\ f_{\al\be}\eand 
 F^-\ \Longleftrightarrow\ f_{\da\db}~.
\end{equation}
Therefore, the self-dual Yang--Mills equations correspond to
\begin{equation}\label{eq:SDYMSpin}
 F\ =\ {*F}\  \Longleftrightarrow\  F^-\ =\ 0\ \Longleftrightarrow\ f_{\da\db}\ =\ 0
\end{equation}
and similarly for the anti-self-dual Yang--Mills equations.

\Exercise{\label{exe:SDYM} Verify \eqref{eq:decomp} and \eqref{eq:decomp2} explicitly. Show further that $F\wedge{*F}$ corresponds to $f_{\al\be}f^{\al\be}+f_{\da\db}f^{\da\db}$ while $F\wedge F$ to $f_{\al\be}f^{\al\be}-f_{\da\db}f^{\da\db}$.}

Most surprisingly, even though they are non-linear, the (anti-)self-dual
Yang--Mills equations are integrable in the sense that one can give, at least in
principle, all solutions. We shall establish this by means of twistor geometry
shortly, but again we will not be too rigorous in our discussion. Furthermore,
$f_{\al\be}=0$ or $f_{\da\db}=0$ make perfect sense in the complex setting.
For convenience, we shall therefore work in the complex setting from now on and
impose reality conditions later on when necessary. Notice that contrary to the
Euclidean and Kleinian cases, the (anti-)self-dual Yang--Mills equations on
Minkowski space only make sense for complex Lie groups $G$. This is so because
$*^2=-1$ on two-forms in Minkowski space. 

\subsection{Penrose--Ward transform}\label{sec:PWtransform}

The starting point is the double fibration \eqref{eq:DoubleFibration}, which we
state again for the reader's convenience,
\begin{equation}\label{eq:DoubleFibration2}
 \begin{picture}(50,40)
  \put(0.0,0.0){\makebox(0,0)[c]{$P^3$}}
  \put(64.0,0.0){\makebox(0,0)[c]{$M^4$}}
  \put(34.0,33.0){\makebox(0,0)[c]{$F^5$}}
  \put(7.0,18.0){\makebox(0,0)[c]{$\pi_1$}}
  \put(55.0,18.0){\makebox(0,0)[c]{$\pi_2$}}
  \put(25.0,25.0){\vector(-1,-1){18}}
  \put(37.0,25.0){\vector(1,-1){18}}
 \end{picture}
\end{equation}
Consider now a rank-$r$ holomorphic vector bundle $E\to P^3$ together with its
pull-back $\pi_1^*E\to F^5$. Their structure groups are thus $\sGL(r,\IC)$. We
may impose the additional condition of having a trivial determinant line bundle,
$\det E$, which reduces $\sGL(r,\IC)$ to $\sSL(r,\IC)$. Furthermore, we again
choose the two-patch covering $\fU=\{U_\pm\}$ of $P^3$. Similarly, $F^5$ may be
covered by two coordinate patches which we denote by $\hat\fU=\{\hat U_\pm\}$.
Therefore, $E$ and $\pi_1^*E$ are characterised by the transition functions
$f=\{f_{+-}\}$ and $\pi_1^*f=\{\pi_1^*f_{+-}\}$. As before, the pull-back of
$f_{+-}(z^\al,\lambda_\da)$ is
$f_{+-}(x^{\al\db}\lambda_\db,\lambda_\da)$, i.e.~it is annihilated by the vector fields \eqref{eq:VF} and therefore constant along $\pi_1\,:\,F^5\to P^3$. In the following, we shall not make a notational distinction between $f$ and $\pi_1^*f$ and simply write $f$ for both bundles. Letting $\dbar_P$ and $\dbar_F$ be the anti-holomorphic parts of the exterior derivatives on $P^3$ and $F^5$, respectively, we have $\pi_1^*\dbar_P=\dbar_F\circ\pi_1^*$. Hence, the transition function $f_{+-}$ is also annihilated by $\dbar_F$.  

We shall also assume that $E$ is holomorphically trivial when restricted to any projective line $L_x=\pi_1(\pi_2^{-1}(x))\hookrightarrow P^3$ for $x\in M^4$. This then implies that there exist matrix-valued functions $\psi_\pm$ on $\hat U_\pm$, which define trivialisations of $\pi_1^*E$ over $\hat U$, such that $f_{+-}$ can be decomposed as (see also Remark \ref{rmk:BG})
\begin{equation}
 f_{+-}\ =\ \psi_+^{-1}\psi_-
\end{equation}
with $\dbar_F\psi_\pm=0$, i.e.~the $\psi_\pm=\psi_\pm(x,\lambda_\pm)$ are holomorphic on $\hat U_\pm$. 
Clearly, this splitting is not unique, since one can always perform the transformation 
\begin{equation}\label{eq:GTF}
 \psi_\pm\ \mapsto\ g^{-1}\,\psi_\pm~,
\end{equation}
where $g$ is some globally defined matrix-valued function holomorphic function
on $F^5$. Hence, it is constant on $\IC P^1$, i.e.~it depends on $x$ but not
on $\lambda_\pm$. We shall see momentarily, what the transformation
\eqref{eq:GTF} corresponds to on space-time $M^4$.

Since $V_\alpha^\pm f_{+-}=0$, 
where $V^\pm_\alpha$ are the restrictions of the vector fields $V_\alpha$ given in \eqref{eq:VF} to the coordinate patches $\hat U_\pm$, we find
\begin{equation}\label{eq:splitting}
 \psi_+ V_\alpha^+\psi_+^{-1}\ =\ \psi_- V_\alpha^+\psi_-^{-1}~
\end{equation}
on $\hat U_+\cap\hat U_-$. Explicitly, $V^\pm_\alpha=\lambda^\db_\pm\partial_{\alpha\db}$ with $\lambda^+_\da:=\lambda_\da/\lambda_{\dot 1}
=(1,\lambda_+)^{\rm t}$ and $\lambda^-_\da:=\lambda_\da/\lambda_{\dot 2}=(\lambda_-,1)^{\rm t}$. Therefore, by an
extension of Liouville's theorem, the expressions \eqref{eq:splitting} can be at most
linear in $\lambda_+$. This can also be understood by noting
\begin{equation}
 \psi_+ V_\alpha^+\psi_+^{-1}\ =\ \psi_- V_\alpha^+\psi_-^{-1}\ =\ \lambda_+ \psi_- V_\alpha^-\psi_-^{-1}
\end{equation}
and so it is of homogeneity 1. Thus, we may introduce a Lie algebra-valued one-form $A$
on $ F^5$ which has components only along $\pi_1\,:\,F^5\to P^3$,
\begin{equation}
 V_\alpha\lrcorner A|_{\hat U_\pm}\ :=\ A_\alpha^\pm\ =\ \psi_\pm V_\alpha^\pm \psi_\pm^{-1}\ =\ 
 \lambda^\db_\pm A_{\alpha\db}~,
\end{equation}
where $A_{\alpha\db}$ is $\lambda_\pm$-independent.
This can be re-written as
\begin{equation}\label{eq:LinSys}
 (V^\pm_\alpha+A^\pm_\alpha)\psi_\pm\ =\ \lambda^\db_\pm\nabla_{\alpha\db}\psi_\pm\ =\ 0~,
 \ewith
 \nabla_{\alpha\db}\ :=\ \partial_{\alpha\db}+A_{\alpha\db}~.
\end{equation}
The compatibility conditions for this linear system read as
\begin{equation}
 [\nabla_{\al\da},\nabla_{\be\db}]+[\nabla_{\al\db},\nabla_{\be\da}]\ =\ 0~,
\end{equation}
which is equivalent to saying that the $f_{\da\db}$-part of
\begin{equation}\label{eq:curvdec}
 [\nabla_{\alpha\da},\nabla_{\be\db}]\ =\ \varepsilon_{\alpha\be}f_{\da\db}
 +\varepsilon_{\da\db}f_{\al\be}~
\end{equation}
vanishes. However, $f_{\da\db}=0$ is nothing but the self-dual Yang--Mills equations \eqref{eq:SDYMSpin}
on $M^4$. 

Notice that the transformations of the form \eqref{eq:GTF} induce the transformations
\begin{equation}\label{eq:GTM}
 A_{\al\db}\ \mapsto\ g^{-1}\partial_{\al\db}g+g^{-1}A_{\al\db}g
\end{equation}
of $A_{\al\db}$ as can be seen directly from \eqref{eq:LinSys}. Hence, these transformations induce gauge transformations on space-time and so we may define gauge equivalence classes $[A_{\al\db}]$, where two gauge potentials are said to be equivalent if they differ by a transformation of the form \eqref{eq:GTM}.
On the other hand, transformations of the form\footnote{In Section \ref{sec:HCS}, we will formalise these transformations in the framework of non-Abelian sheaf cohomology.} 
\begin{equation}\label{eq:GTP}
 f_{+-}\ \mapsto\ h_+^{-1} f_{+-} h_-~,
\end{equation}
where $h_\pm$ are matrix-valued functions holomorphic on $\hat U_\pm$ with
$V_\al^\pm h_\pm=0$, leave the gauge potential $A_{\al\db}$ invariant. Since 
$V_\al^\pm h_\pm=0$, the functions $h_\pm$ descend down to twistor space $P^3$
and are
holomorphic on $U_\pm$ (remember that any function on twistor space that is
pulled back to the correspondence space must be annihilated by the vector fields
$V_\al$). Two transition functions that differ by a transformation of the form
\eqref{eq:GTP} are then said to be equivalent, as they define two
holomorphic vector bundles which are bi-holomorphic. Therefore, we may conclude
that an equivalence class $[f_{+-}]$ corresponds to an equivalence class
$[A_{\al\db}]$.

Altogether, we have seen that holomorphic vector bundles $E\to P^3$ over 
twistor space, which are holomorphically trivial on all projective lines
$L_x\hookrightarrow P^3$ yield solutions of the self-dual Yang--Mills equations
on $M^4$. In fact, the converse is also true: Any solution to the self-dual
Yang--Mills equations arises in this way. See e.g.~\cite{Ward:1990vs} for a
complete proof. Therefore, we have:

{\Thm\label{thm:ward} (Ward \cite{Ward:1977ta}) There is a one-to-one correspondence between gauge equivalence classes of solutions to the self-dual Yang--Mills equations on $M^4$ and equivalence classes of holomorphic vector bundles over the twistor space $P^3$ which are holomorphically trivial on any projective line $L_x=\pi_1(\pi_2^{-1}(x))\hookrightarrow P^3$. }

\vspace*{10pt}

\noindent
Hence, all solutions to the self-dual Yang--Mills equations are encoded in these
vector bundles and once more, 
differentially constrained data on space-time (the gauge potential $A_{\al\db}$)
is encoded in differentially unconstrained complex analytic data (the transition
function $f_{+-}$) on twistor space. The reader might be worried that our
constructions depend on the choice of coverings, but as in the case of
the Penrose transform, this is not the case as will become transparent
in Section \ref{sec:HCS}

As before, one may also write down certain integral formul{\ae} for the gauge potential $A_{\al\db}$. 
In addition, given a solution $A=\dd x^{\al\db} A_{\al\db}$ to the self-dual Yang--Mills equations, the matrix-valued functions $\psi_\pm$ are given by
\begin{equation}
 \psi_\pm\ =\ {\rm P}\exp\left(-\int_{\cC_\pm}A\right),
\end{equation}
where `P' denotes the path-ordering symbol and the contour $\cC_\pm$ is any real curve in the null-plane $\pi_2(\pi_1^{-1}(p))\hookrightarrow M^4$ for $p\in P^3$ running from some point $x_0$ to a point $x$ with $x^{\al\db}(s)=x_0^{\al\db}+s\mu^\al\lambda^\da_\pm$ for $s\in[0,1]$ and constant $\mu^\al$; the choice of contour plays no role, since the curvature is zero when restricted to the null-plane.

\vspace*{5pt}

\Exercise{\label{exe:PWT} Show that for a rank-1 holomorphic vector bundle $E\to
P^3$, the Ward theorem coincides with the Penrose transform for a helicity
$h=-1$ field. See also Appendix \ref{app:Sh} Thus, the Ward theorem gives a
non-Abelian generalisation of that case and one therefore often speaks of the
Penrose--Ward transform.}

\vspace*{5pt}

Before giving an explicit example of a real instanton solution, let us say a few
words about real structures. In Section \ref{sec:PL}, we introduced reality
conditions on $M^4$ leading to Euclidean, Minkowski and Kleinian spaces. In
fact, these conditions are induced from twistor space as we shall now explain.
For concreteness, let us restrict our attention to the Euclidean case. The
Kleinian case will be discussed in Section \ref{sec:WHFT} Remember that a
Minkowski signature does not allow for real (anti-)instantons.

A real structure on $P^3$ is an anti-linear involution $\tau\,:\,P^3\to P^3$. We
may choose it according to:
\begin{subequations}\label{eq:RS}
\begin{equation}\label{eq:RSa}
 \tau(z^\alpha,\lambda_\da)
 \ :=\ (\bar z^\beta{C_\beta}^\alpha, {C_{\da}}^\db \bar\lambda_{\db})~,
\end{equation}
where bar denotes complex conjugation as before and
\begin{equation}\label{eq:RSb}
 ({C_\alpha}^\beta)\ :=\ \begin{pmatrix}
                          0 & 1\\ -1 & 0
                         \end{pmatrix}\eand
 ({C_\da}^\db)\ :=\ \begin{pmatrix}
                          0 & -1\\ 1 & 0
                         \end{pmatrix}.
\end{equation}
\end{subequations}
By virtue of the incidence relation $z^\alpha=x^{\alpha\db}\lambda_\db$, we obtain an
induced involution on $ M^4$,\footnote{We shall use the same notation $\tau$ for the anti-holomorphic involutions induced on the different manifolds appearing in \eqref{eq:DoubleFibration2}.}
\begin{equation}
 \tau(x^{\alpha\db})\ =\ -\bar x^{\gamma\dot\delta} {C_{\gamma}}^\alpha {C_{\dot\delta}}^\db~.
\end{equation}
The set of fixed points $\tau(x)=x$ is given by $x^{1\dot1}=\bar x^{2\dot2}$ and $x^{1\dot2}=-\bar x^{1\dot2}$.
By inspecting \eqref{eq:RC}, we see that this corresponds to a Euclidean
signature real slice $\IE$ in $M^4$. Furthermore, $\tau$ can be extended to
$E\to P^3$ according to
$f_{+-}(z,\lambda)=(f_{+-}(\tau(z,\lambda))^\dagger$.\footnote{In fact, the
involution $\tau$ can be extended to any holomorphic function.} This will ensure
that the Yang--Mills gauge potential on space-time is real and in particular, we
find from \eqref{eq:LinSys} that $A_\mu=-A_\mu^\dagger$. Here, `$\dagger$'
denotes Hermitian conjugation.

\vspace*{5pt}

\Remark{\label{rmk:BG} Let us briefly comment on generic holomorphic vector
bundles over $\IC P^1$: So, let $E\to \IC P^1$ be a rank-$r$ holomorphic vector
bundle over $\IC P^1$. The Birkhoff--Grothendieck theorem (see
e.g.~\cite{Griffiths:1978} for details) then tells us that $E$ always decomposes
into a sum of holomorphic line bundles, $$\CO(k_1)\oplus\cdots\oplus\CO(k_r)\
\to\ \IC P^1~.$$ Therefore, if $\fU=\{U_\pm\}$ denotes the canonical cover of
$\IC P^1$, the transition function $f=\{f_{+-}\}$ of $E$ is always of the form
$$ f_{+-}\ =\ \psi_+^{-1} \Lambda_{+-}\psi_-~,\ewith \Lambda_{+-}\ :=\ {\rm
diag}(\lambda_+^{k_1},\ldots,\lambda_+^{k_r})~, $$ where the $\psi_\pm$ are
holomorphic on $U_\pm$. If $\det E$ is trivial then $\sum_i k_i=0$. If
furthermore $E$ is holomorpically trivial then $k_i=0$ and
$f_{+-}=\psi_+^{-1}\psi_-$. 

\hspace*{15pt}Notice that given some matrix-valued function $f_{+-}$ which is
holomorphic on $U_+\cap U_-\subset \IC P^1$, the problem of trying to split
$f_{+-}$ according to $f_{+-}=\psi_+^{-1}\psi_-$ with $\psi_\pm$ holomorphic on
$U_\pm$ is known as the Riemann--Hilbert problem and its solutions define
holomorphically trivial vector bundles on $\IC P^1$. 
If in addition $f_{+-}$ also depends on some parameter (in our above case
the
parameter is $x$), then one speaks of a parametric Riemann--Hilbert problem. A
solution to the parametric Riemann--Hilbert problem might not exist for all
values of the parameter, but if it exists at some point in parameter space, then
it exists in an open neighbourhood of that point. }

\vspace*{10pt}

\subsection{Example: Belavin--Polyakov--Schwarz--Tyupkin instanton}

Let us now present an explicit instanton solution on Euclidean space for the gauge group $\sSU(2)$. This amounts to considering a rank-$2$ holomorphic vector bundle $E\to P^3$ holomorphically trivial on any $L_x\hookrightarrow P^3$ with trivial determinant line bundle $\det E$ and to equipping twistor space with the real structure according to our previous discussion. 

Then let $E\to P^3$ and $\pi_1^*E\to F^5$ be defined by the following transition function $f=\{f_{+-}\}$ \cite{Crane:1987im}:
\begin{equation}
 f_{+-}\ =\ \frac{1}{\Lambda^2}
               \begin{pmatrix}
                 \Lambda^2-\frac{z^1z^2}{\lambda_{\dot1}\lambda_{\dot2}} &
                  \frac{(z^2)^2}{\lambda_{\dot1}\lambda_{\dot2}}\\
                 -\frac{(z^1)^2}{\lambda_{\dot1}\lambda_{\dot2}} &  
                 \Lambda^2+\frac{z^1z^2}{\lambda_{\dot1}\lambda_{\dot2}}
                               \end{pmatrix},
\end{equation}
where $\Lambda\in\IR\setminus\{0\}$. Evidently, $\det f_{+-}=1$ and so $\det E$ is trivial. Furthermore,
$f_{+-}(z,\lambda)=(f_{+-}(\tau(z,\lambda))^\dagger$, where $\tau$ is the involution \eqref{eq:RS} leading to Euclidean space. The main problem now is to find a solution to the
Riemann--Hilbert problem $f_{+-}=\psi_+^{-1}\psi_-$. Notice that if we succeed, we have automatically shown that $E\to P^3$ is holomorphically trivial on any projective line $L_x\hookrightarrow P^3$.

In terms of the coordinates on $U_+$, we have
\begin{equation}
 f_{+-}\ =\ \frac{1}{\Lambda^2}\begin{pmatrix}
                                \Lambda^2-\frac{z^1_+z^2_+}{\lambda_+} & \frac{(z_+^2)^2}{\lambda_+}\\
                                -\frac{(z_+^1)^2}{\lambda_+} &  \Lambda^2+\frac{z^1_+z^2_+}{\lambda_+}
                               \end{pmatrix}.
\end{equation}
As there is no generic algorithm, let us just present a solution \cite{Crane:1987im}:
\begin{equation}\label{eq:BPSTpsi}
 \psi_+\ =\ -\frac{1}{\Lambda}\frac{1}{\sqrt{x^2+\Lambda^2}}
            \begin{pmatrix}
             x^{2\dot2}z^1_++\Lambda^2 & -x^{2\dot2} z^2_+ \\
             x^{1\dot2}z^1_+ & -x^{1\dot2}z^2_++\Lambda^2
            \end{pmatrix}
 \eand
 \psi_-\ =\ \psi_+ f_{+-}~,
\end{equation}
where $x^2:=\det x$. 

It remains to determine the gauge potential and the curvature. We find
\begin{equation}\label{eq:BPSTA}
 A_{1\dot1}\ =\ \frac{1}{2(x^2+\Lambda^2)}\begin{pmatrix}
                                           x^{2\dot2} & 0\\
                                           2x^{1\dot2} & - x^{2\dot2}
                                          \end{pmatrix}~,\quad
 A_{2\dot1}\ =\ \frac{1}{2(x^2+\Lambda^2)}\begin{pmatrix}
                                           x^{1\dot2} & -2x^{2\dot2}\\
                                           0 & -x^{1\dot2}
                                          \end{pmatrix}
\end{equation}
and $A_{\al\dot2}=0$. Hence, our choice of gauge $\psi_\pm\mapsto g^{-1}\psi_\pm$ corresponds to gauging away $A_{\al\dot2}$. Furthermore, the only non-vanishing components of the curvature are
\begin{equation}\label{eq:BPSTF}
\begin{aligned}
 f_{11}\ =\ \frac{2\Lambda^2}{(x^2+\Lambda^2)^2}\begin{pmatrix}
                                                 0 & 0\\ 1 & 0
                                                \end{pmatrix}~,\quad
 f_{12}\ =\ \frac{\Lambda^2}{(x^2+\Lambda^2)^2}\begin{pmatrix}
                                                 1 & 0\\ 0 & -1
                                                \end{pmatrix}~,\\
 f_{22}\ =\ -\frac{2\Lambda^2}{(x^2+\Lambda^2)^2}\begin{pmatrix}
                                                 0 & 1\\ 0 & 0
                                                \end{pmatrix}~,\kern2.5cm
\end{aligned}
\end{equation}
which shows that we have indeed found a solution to the self-dual Yang--Mills equations. Finally, using 
\eqref{eq:charge}, we find that the instanton charge $Q=1$. We leave all the
details as an exercise. The above solution is the famous
Belavin--Polyakov--Schwarz--Tyupkin instanton \cite{Belavin:1975fg}. Notice that
$\Lambda$ is referred to as the `size modulus' as it determines the size of the
instanton. In addition, there are four translational moduli corresponding to
shifts of the form $x\mapsto x+c$ for constant $c$. Altogether, there are five
moduli characterising the charge one $\sSU(2)$ instanton. For details on
how to construct general instantons, see e.g.~\cite{Prasad:1980yy,Tong:2005un}.

\Exercise{Show that \eqref{eq:BPSTpsi} implies \eqref{eq:BPSTA} and
\eqref{eq:BPSTF} by using the linear system \eqref{eq:LinSys}. Furthermore, show
that $Q=1$. You might find the following integral useful:
$$\int_{\IE}\frac{\dd^4x}{(x^2+\Lambda^2)^4}\ =\ \frac{\pi^2}{6\Lambda^4}~,$$
where $x^2=x_\mu x^\mu$.}

\vspace*{10pt}

\section{Supertwistor space}

Up to now, we have discussed the purely bosonic setup. As our goal is the construction of amplitudes in supersymmetric gauge theories, we need to incorporate fermionic degrees of freedom. To this end, we start by
briefly discussing supermanifolds before we move on and introduce
supertwistor space and the supersymmetric generalisation of the self-dual
Yang--Mills equations.
For a detailed discussion about supermanifolds, we refer to \cite{Manin:1988ds,Bartocci:1991,DeWitt:1992cy}.

\subsection{A brief introduction to supermanifolds}

Let $R\cong R_0\oplus R_1$ be a $\IZ_2$-graded
ring, that is, $R_0R_0\subset R_0$, $R_1R_0\subset R_1$, $R_0R_1\subset R_1$
and $R_1R_1\subset R_0$. We call elements of $R_0$ 
Gra{\ss}mann even (or bosonic) and elements of $R_1$ Gra{\ss}mann odd (or fermionic). 
An element of $R$ is said to be homogeneous if it is either bosonic or fermionic. The 
degree (or Gra{\ss}mann parity) of a homogeneous element is defined to be $0$ if it 
is bosonic and $1$ if it is fermionic,
respectively. We denote the degree of a homogeneous element $r\in R$ by
$p_r$ ($p$ for parity). 

We define the supercommutator, $[\cdot,\cdot\}\,:\,R\times R\to R$, by
\begin{equation}\label{eq:supercommutator}
 [r_1,r_2\}\ :=\ r_1r_2-(-)^{p_{r_1}p_{r_2}}r_2r_1~,
\end{equation}
for all homogeneous elements $r_{1,2}\in R$. The $\IZ_2$-graded ring $R$ is
called supercommutative if the supercommutator vanishes for all of the ring's
elements. For our purposes, the most important example of such a
supercommutative ring is the Gra{\ss}mann or exterior algebra over $\IC^n$,
\begin{subequations}
\begin{equation}
 R\ =\ \Lambda^\bullet \IC^n\ :=\ \bigoplus_p\Lambda^p \IC^n~,
\end{equation}
with the $\IZ_2$-grading being
\begin{equation}
 R\ =\ \underbrace{\bigoplus_p\Lambda^{2p}\IC^n}_{=:\ R_0}\ \oplus\ 
 \underbrace{\bigoplus_p\Lambda^{2p+1}\IC^n}_{=:\ R_1}~.
\end{equation}
\end{subequations}

An $R$-module $M$ is a $\IZ_2$-graded bi-module which satisfies
\begin{equation}
 rm\ =\ (-)^{p_rp_m}mr~,
\end{equation}
for homogeneous $r\in R$, $m\in M$, with $M\cong M_0\oplus M_1$. Then there is a natural 
map\footnote{More precisely, it is a functor
from the category of $R$-modules to the category of $R$-modules. See Appendix
\ref{app:Ca}~for details.}
$\Pi$, called the parity operator, which is defined by
\begin{equation}\label{eq:paritymap}
 (\Pi M)_0\ :=\ M_1\qquad{\rm and}\qquad(\Pi M)_1\ :=\ M_0~.
\end{equation}
We should stress that $R$ is an $R$-module itself, and as such $\Pi R$ is an
$R$-module, as well. However, $\Pi R$ is no longer a $\IZ_2$-graded ring since
$(\Pi R)_1(\Pi R)_1\subset(\Pi R)_1$, for instance.

A free module of rank $m|n$ over $R$ is defined by
\begin{equation}
 R^{m|n}\ :=\ R^m\oplus(\Pi R)^n~,
\end{equation}
where $R^m:=R\oplus\cdots\oplus R$. This has a free system of generators, $m$ of which are bosonic and $n$ of
which are fermionic, respectively. We stress that the decomposition of $R^{m|n}$ into $R^{m|0}$ and 
$R^{0|n}$ has, in general, no invariant meaning and does not coincide with the
decomposition into bosonic and fermionic parts, $[R_0^m\oplus(\Pi
R_1)^n]\oplus[R_1^m\oplus(\Pi R_0)^n]$. Only when $R_1=0$, are these
decompositions the same. An example is $\IC^{m|n}$, where we
consider the complex numbers as a $\IZ_2$-graded ring (where $R=R_0$ with
$R_0=\IC$ and $R_1=0$). 

Let $R$ be a supercommutative ring and $R^{m|n}$ be a freely generated
$R$-module. Just as in the commutative case, morphisms between free
$R$-modules can be given by matrices. The standard matrix format
is 
\begin{equation}
A\ =\ \begin{pmatrix} A_1&A_2\\ A_3&A_4\end{pmatrix}, 
\end{equation}
where $A$ is said to be bosonic (respectively, fermionic) if $A_1$ and $A_4$ are
filled with bosonic (respectively, fermionic) elements of the ring while
$A_2$ and $A_3$ are filled with fermionic (respectively, bosonic) elements.
Furthermore, $A_1$ is a $p\times m$-, $A_2$ a $q\times m$-, $A_3$ 
a $p\times n$- and $A_4$ a $q\times n$-matrix. The set of matrices
in standard format with elements in $R$ is denoted by 
${\rm Mat}(m|n,p|q,R)$. It forms a $\IZ_2$-graded module which,
with the usual matrix multiplication, is naturally isomorphic
to ${\rm Hom}(R^{m|n},R^{p|q})$. We denote the endomorphisms of 
$R^{m|n}$ by ${\rm End}(m|n,R)$ and the automorphisms by 
${\rm Aut}(m|n,R)$, respectively. We use further the special symbols 
$\mathfrak{gl}(m|n,R)\subset{\rm End}(m|n,R)$ to denote the bosonic
endomorphisms  of $R^{m|n}$ and  $\sGL(m|n,R)\subset{\rm Aut}(m|n,R)$
to denote the bosonic automorphisms.

The supertranspose of $A\in{\rm Mat}(m|n,p|q,R)$ is defined according to
\begin{equation}
A^{\rm st}\ :=\    \begin{pmatrix}
                              A_1^{\rm t}& (-)^{p_A}\ A_3^{\rm t}\\
                              -(-)^{p_A} A_2^{\rm t} &A_4^{\rm t}
                            \end{pmatrix}, 
\end{equation}
where the superscript `t' denotes the usual transpose. The 
supertransposition satisfies 
$(A+B)^{\rm st}=\ A^{\rm st}+B^{\rm st}$ and
$(AB)^{\rm st}=(-)^{p_Ap_B} B^{\rm st} A^{\rm st}$.
We shall use the following definition of the supertrace of 
$A\in{\rm End}(m|n,R)$:
\begin{equation}\label{eq:str}
 {\rm str}A\ :=\  {\rm tr} A_1-(-)^{p_A}{\rm tr} A_4~.
\end{equation}
The supercommutator for matrices is defined analogously to \eqref{eq:supercommutator},
i.e.~$[A,B\}:=AB-(-)^{p_Ap_B}BA$ for $A,B\in{\rm End}(m|n,R)$. Then
${\rm str}[A,B\}=0$ and ${\rm str}A^{\rm st}={\rm str}A$.
Finally, let $A\in \sGL(m|n,R)$. The superdeterminant is given by
\begin{equation}\label{eq:sdet}
 {\rm sdet}\,A\ :=\ \det(A_1-A_2A_4^{-1}A_3)\det A_4^{-1}~,
\end{equation}
where the right-hand side is well-defined for $A_1\in \sGL(m|0,R_0)$
and $A_4\in \sGL(n|0,R_0)$. Furthermore, it belongs to $\sGL(1|0,R_0)$.
The superdeterminant satisfies the usual rules,
${\rm sdet}(AB)={\rm sdet}A\ {\rm sdetB}$ and ${\rm sdet} A^{\rm st}={\rm sdet}A$
for $A,B\in\sGL(m|n,R)$. Notice that sometimes ${\rm sdet}$ is referred to
as the Berezinian and also denoted by ${\rm Ber}$.

After this digression, we may now introduce the local model of a supermanifold.
Let $V$ be an open subset in $\IC^m$ and consider
$\CO_V(\Lambda^\bullet \IC^n):=\CO_V\otimes\Lambda^\bullet \IC^n~,$
where $\CO_V$ is the sheaf of holomorphic functions on $V\subset\IC^m$ 
which is also referred to as the structure sheaf of $V$. Thus,
$\CO_V(\Lambda^\bullet \IC^n)$ is a sheaf of supercommutative rings consisting of 
$\Lambda^\bullet \IC^n$-valued holomorphic functions on $V$. Let now $(x^1,\ldots,x^m)$
be coordinates on $V\subset\IC^m$ and $(\eta_1,\ldots,\eta_n)$ be a basis of the 
sections of
$\IC^n\cong\Lambda^1\IC^n$. Then $(x^1,\ldots,x^m,\eta_1,\ldots,\eta_n)$ are 
coordinates for the ringed space $V^{m|n}:=(V,\CO_V(\Lambda^\bullet \IC^n))$. Any
function $f$ can thus be Taylor-expanded as
\begin{equation}
 f(x,\eta)\ =\ \sum_I \eta_I f^I(x)~,
\end{equation}
where $I$ is a multi-index. These are the fundamental functions in 
supergeometry. 

To define a general supermanifold, let $X$ be some topological space of real dimension $2m$, and let $\CR_X$ 
be a sheaf of supercommutative rings on $X$. Furthermore, let $\CN$ be the ideal 
subsheaf in $\CR_X$ of all nil-potent elements in $\CR_X$, and define $\CO_X:=\CR_X/\CN.$\footnote{Instead of $\CR_X$, one often also writes $\CR$ and likewise for $\CO_X$.}
Then $X^{m|n}:=(X,\CR_X)$ is called a complex supermanifold of dimension $m|n$ if the following is fulfilled:
\begin{itemize}
\setlength{\itemsep}{-.8mm}
\item[(i)] $X^m:=(X,\CO_X)$ is an $m$-dimensional complex manifold which we call the
body of $X^{m|n}$.
\item[(ii)] For each point $x\in X$ there is a neighbourhood $U\ni x$ such that there is a local isomorphism
      $\CR_X|_U\cong \CO_X(\Lambda^\bullet(\CN/\CN^2))|_U$, where $\CN/\CN^2$ is
a rank-$n$ locally free sheaf of $\CO_X$-modules on $X$, i.e.~$\CN/\CN^2$ is
locally of the form $\CO_X\oplus\cdots\oplus\CO_X$ ($n$-times); $\CN/\CN^2$ is
called the characteristic sheaf of $X^{m|n}$.
\end{itemize}
Therefore, complex supermanifolds look locally like $V^{m|n}=(V,\CO_V(\Lambda^\bullet \IC^n))$. In view of this, we picture $\IC^{m|n}$ as $(\IC^m,\CO_{\IC^m}(\Lambda^\bullet\IC^n))$.
We shall refer to $\CR_X$ as the structure sheaf
of the supermanifold $X^{m|n}$ and to $\CO_X$ as the structure sheaf
of the body $X^m$ of $X^{m|n}$. Later on, we shall use a more common notation and re-denote $\CR_X$ by $\CO_X$ or simply by $\CO$ if there is no confusion with the structure sheaf of the body $X^m$ of $X^{m|n}$. In addition,
we sometimes write $X^{m|0}$ instead of $X^m$.
Furthermore, the tangent bundle $TX^{m|n}$ of a complex supermanifold $X^{m|n}$ is an example of a supervector bundle, where the transition functions are sections of the (non-Abelian) sheaf $\sGL(m|n,\CR_X)$ (see Section
\ref{sec:HCS}~for more details).

\vspace*{5pt}

\Remark{Recall that for a
ringed space $(X,\CO_X)$ with the property that for each $x\in X$ there is
a neighbourhood $U\ni x$ such that there is a ringed space isomorphism
$(U,\CO_X|_U)\cong(V,\CO_V)$, where $V\subset\IC^m$. Then
$X$ can be given the structure of a complex manifold and moreover, any 
complex manifold arises in this manner. By the usual abuse of notation,
$(X,\CO_X)$ is often denoted by $X$.}

\vspace*{5pt}

An important example of a supermanifold in the context of twistor geometry is the
complex projective superspace $\IC P^{m|n}$. It is given by
\begin{equation}\label{eq:PSS}
 \IC P^{m|n}\ :=\ (\IC P^m,\CO_{\IC P^m}(\Lambda^\bullet(\CO(-1)\otimes\IC^n)))~, 
\end{equation}
where $\CO(-1)$ is the tautological
line bundle over the complex projective space $\IC P^m$. It is defined analogously to $\IC P^1$ (see \eqref{eq:TB}). The reason for the 
appearance of $\CO(-1)$ is as follows. If we let 
$(z^0,\ldots,z^m,\eta_1,\ldots,\eta_n)$
be homogeneous coordinates\footnote{Note that they are
subject to the identification $(z^0,\ldots,z^m,\eta_1,\ldots,\eta_n)\sim 
(tz^0,\ldots,tz^m,t\eta_1,\ldots,t\eta_n)$, where $t\in\IC\setminus\{0\}$.}
 on $\IC P^{m|n}$, a holomorphic function $f$ on
$\IC P^{m|n}$ has the expansion
\begin{equation}
 f\ =\ \sum \eta_{i_1}\cdots\eta_{i_r}f^{i_1\cdots i_r}(z^0,\ldots,z^m)~. 
\end{equation}
Surely, for $f$ to be
well-defined the homogeneity of $f$ must be zero. Hence, 
$f^{i_1\cdots i_r}=f^{[i_1\cdots i_r]}$ must
be of homogeneity $-r$. This explains the above form of the structure
sheaf of the complex projective superspace.

\vspace*{5pt}

\Exercise{\label{exe:VB} Let $E\to X$ be a holomorphic vector bundle over a complex manifold
$X$. Show that $(X,\CO_X(\Lambda^\bullet E^*))$ is a supermanifold 
according to our definition given above.}

\vspace*{5pt}

Supermanifolds of the form as in the above exercise
are called globally split. We see that $\IC P^{m|n}$ is of the type $E\to\IC P^m$ with $E=\CO(1)\otimes\IC^n$. 
Due to a theorem of Batchelor \cite{Batchelor:1979} (see also
e.g.~\cite{Bartocci:1991}), any smooth supermanifold is globally split. This is
due to the existence of a (smooth) partition of unity. The reader should be
warned that, in general, complex supermanifolds are not of this type (basically
because of the lack of a holomorphic partition of unity).

\subsection{Supertwistor space}

Now we have all the necessary ingredients to generalise \eqref{eq:DoubleFibration} to the supersymmetric setting.
Supertwistors were first introduced by Ferber \cite{Ferber:1977qx}.

Consider $M^{4|2\CN}\cong\IC^{4|2\CN}$ together with the identification
\begin{equation}
 TM^{4|2\CN}\ \cong H\otimes\tilde S
\end{equation}
where the fibres $H_x$ of $H$ over $x\in M^{4|2\CN}$ are $\IC^{2|\CN}$ and
$\tilde S$ is again the dotted spin bundle. In this sense, $H$ is of rank
$2|\CN$ and $H\cong E\oplus S$, where $S$ is the undotted spin bundle and $E$ is
the rank-$0|\CN$ $R$-symmetry bundle. In analogy to $x^\mu\leftrightarrow
x^{\al\db}$, we now have $x^M\leftrightarrow x^{A\da}=(x^{\al\da},\eta^\da_i)$
for $A=(\al,i),B=(\be,j),\ldots$ and $i,j,\ldots=1,\ldots,\CN$. Notice that the
above factorisation of the tangent bundle can be understood as a generalisation
of a conformal structure (see Remark \ref{rmk:confstru}) known as
para-conformal structure (see e.g.~\cite{Bailey:1991xx}).

As in the bosonic setting, we may consider the projectivisation of $\tilde S^*$ to define the correspondence space $F^{5|2\CN}:=\IP(\tilde S^*)\cong\IC^{4|2\CN}\times\IC P^1$. Furthermore, we consider the vector fields
\begin{equation}\label{eq:TDsuper}
 V_A\ =\ \lambda^\da\partial_{A\da}\ =\ \lambda^\da\frac{\partial}{\partial x^{A\da}}~.
\end{equation}
They define an integrable rank-$2|\CN$ distribution on the correspondence space. The resulting quotient will be supertwistor $P^{3|\CN}$:
\begin{equation}\label{eq:DFsuper}
 \begin{picture}(50,40)
  \put(0.0,0.0){\makebox(0,0)[c]{$P^{3|\CN}$}}
  \put(64.0,0.0){\makebox(0,0)[c]{$M^{4|2\CN}$}}
  \put(34.0,33.0){\makebox(0,0)[c]{$F^{5|2\CN}$}}
  \put(7.0,18.0){\makebox(0,0)[c]{$\pi_1$}}
  \put(55.0,18.0){\makebox(0,0)[c]{$\pi_2$}}
  \put(25.0,25.0){\vector(-1,-1){18}}
  \put(37.0,25.0){\vector(1,-1){18}}
 \end{picture}
\end{equation}
The projection $\pi_2$ is the trivial projection and $\pi_1\,:\,(x^{A\da},\lambda_\da)\mapsto (z^A,\lambda_\da)=(x^{A\da}\lambda_\da,\lambda_\da)$, where $(z^A,\lambda_\da)=(z^\al,\eta_i,\lambda_\da)$ are homogeneous coordinates on $P^{3|\CN}$. 

As before, we may cover $P^{3|\CN}$ by two coordinate patches, which we (again) denote by $U_\pm$:
\begin{equation}
\begin{aligned}
 U_+\ &:\ \lambda_{\dot1}\ \neq\ 0
  \eand z_+^A\ :=\ \frac{z^A}{\lambda_{\dot1}}
  \eand\lambda_+\ :=\ \frac{\lambda_{\dot2}}{\lambda_{\dot1}}~,\\
 U_-\ &:\ \lambda_{\dot2}\ \neq\ 0
  \eand z_-^A\ :=\ \frac{z^A}{\lambda_{\dot2}}
  \eand\lambda_-\ :=\ \frac{\lambda_{\dot1}}{\lambda_{\dot2}}~.
\end{aligned}
\end{equation}
On $U_+\cap U_-$ we have $z_+^A=\lambda_+ z_-^A$ and
$\lambda_+=\lambda_-^{-1}$. This shows that $P^{3|\CN}$
can be identified with $\IC P^{3|\CN}\setminus\IC P^{1|\CN}$. It can also be
identified with the total space of the holomorphic fibration
\begin{equation}\label{eq:STF}
 \CO(1)\otimes\IC^{2|\CN}\ \to\ \IC P^1~.
\end{equation}
Another way of writing this is $\CO(1)\otimes\IC^2\oplus\Pi\CO(1)\otimes\IC^\CN\to\IC P^1$, where $\Pi$ is the
parity map given in \eqref{eq:paritymap}. In the following, we shall denote the
two patches covering the correspondence space $F^{5|2\CN}$ by $\hat U_\pm$.
Notice that Remark \ref{rmk:kodaira}~also applies to $P^{3|\CN}$.

Similarly, we may extend the geometric correspondence: A point $x\in M^{4|2\CN}$
corresponds to a projective line $L_x=\pi_1(\pi_2^{-1}(x))\hookrightarrow
P^{3|\CN}$, while a point $p=(z,\lambda)\in P^{3|\CN}$ corresponds to a
$2|\CN$-plane in superspace-time $M^{4|2\CN}$ that is parametrised by
$x^{A\da}=x^{A\da}_0+\mu^A\lambda^\da$, where $x_0^{A\da}$ is a particular
solution to the supersymmetric incidence relation
$z^A=x^{A\da}\lambda_\da$.\\[-25pt]

\subsection{Superconformal algebra}\label{sec:SCA}

Before we move on and talk about supersymmetric extensions of self-dual
Yang--Mills theory, let us digress a little and collect a few facts
about the superconformal algebra. The conformal algebra, $\mathfrak{conf}_4$, in
four dimensions is a real form of the complex Lie algebra
$\mathfrak{sl}(4,\IC)$. The concrete real form depends on the choice of
signature of space-time. For Euclidean signature we have
$\mathfrak{so}(1,5)\cong\mathfrak{su}^*(4)$ while for Minkowski and Kleinian
signatures we have $\mathfrak{so}(2,4)\cong\mathfrak{su}(2,2)$ and
$\mathfrak{so}(3,3)\cong\mathfrak{sl}(4,\IR)$, respectively. Likewise, the
$\CN$-extended conformal algebra---the superconformal algebra,
$\mathfrak{conf}_{4|\CN}$---is a real form of the complex Lie superalgebra
$\mathfrak{sl}(4|\CN,\IC)$ for $\CN<4$ and
$\mathfrak{psl}(4|4,\IC)$ for $\CN=4$. For a compendium of Lie superalgebras, see e.g.\cite{Frappat:1996pb}. In particular, for $\CN<4$ we have  $\mathfrak{su}^*(4|\CN)$, $\mathfrak{su}(2,2|\CN)$ and $\mathfrak{sl}(4|\CN,\IR)$  for Euclidean, Minkowski and Kleinian signatures while for $\CN=4$ the superconformal algebras are $\mathfrak{psu}^*(4|4)$, $\mathfrak{psu}(2,2|4)$ and $\mathfrak{psl}(4|4,\IR)$. Notice that for a Euclidean signature, the number $\CN$ of supersymmetries is restricted to be even. 

The generators of $\mathfrak{conf}_{4|\CN}$ are
\begin{equation}
 \mathfrak{conf}_{4|\CN}\ =\ \mbox{span}\big\{P_\mu,L_{\mu\nu},K^\mu,D,{R_i}^j,A\,|\,Q_{i\alpha},
 Q^i_\da,S^{i\alpha},S^{\da}_i\big\}~.
\end{equation}
Here, $P_\mu$ represents translations, $L_{\mu\nu}$ (Lorentz) rotations, $K^\mu$ special conformal transformations,
$D$ dilatations and ${R_i}^j$ the $R$-symmetry while $Q_{i\al}$, $Q^i_\da$ are the Poincar\'e
supercharges and $S^{i\al}$, $S^\da_i$ their superconformal partners. Furthermore, $A$ is the axial charge which
absent for $\CN=4$. Making use of the identification \eqref{eq:ident}, we may also write
\begin{equation}\label{eq:4dgen}
 \mathfrak{conf}_{4|\CN}\ =\ \mbox{span}\big\{P_{\al\db},L_{\al\be},L_{\da\db},K^{\al\db},D,{R_i}^j,A\,|\,Q_{i\alpha},
 Q^i_\da,S^{i\alpha},S^{\da}_i\big\}~,
\end{equation}
where the $L_{\al\be}$, $L_{\da\db}$
 are symmetric in their indices (see also \eqref{eq:decomp}). We may also
include a central
extension $\mathfrak{z}=\mbox{span}\{Z\}$ leading to
$\mathfrak{conf}_{4|\CN}\oplus\mathfrak{z}$, 
i.e.~$[\mathfrak{conf}_{4|\CN},\mathfrak{z}\}=0$ and
$[\mathfrak{z},\mathfrak{z}\}=0$.

The commutation relations for the centrally extended superconformal algebra
$\mathfrak{conf}_{4|\CN}\oplus\mathfrak{z}$ are 

\begin{equation}\label{eq:ComRelSuConAl}
\begin{aligned}
 \{Q_{i\alpha},Q_{\dot\beta}^j\}\ =\ -\delta_i^j P_{\al\dot\beta}~,\qquad
 \{S^{i\alpha},S_j^{\dot\beta}\}\ =\ -\delta^i_j K^{\al\dot\beta}~,\\
 \{Q_{i\alpha},S^{j\beta}\}\ =\ -\di\big[\delta_i^j{L_\al}^\be+\tfrac{1}{2}\delta_\al^\be\delta_i^j(D+Z)+
  2\delta_\al^\be{R_i}^j-\tfrac12\delta_\al^\be\delta_i^j(1-\tfrac{4}{\CN})A\big]~,\\
 \{Q^i_\da,S^\db_j\}\ =\ \di\big[\delta_j^i{L_\da}^\db+\tfrac{1}{2}\delta_\da^\db\delta_j^i(D-Z)-
  2\delta_\da^\db{R_j}^i+\tfrac12\delta_\da^\db\delta^i_j(1-\tfrac{4}{\CN})A\big]~,\\
 {[{R_i}^j,S_k^\da]}\ =\ -\tfrac{\di}{2}(\delta_k^j S_i^\da-\tfrac{1}{\CN}\delta_i^jS_k^\da)~,\qquad
 {[{R_i}^j,S^{k\alpha}]}\ =\ \tfrac{\di}{2}(\delta_i^k S^{j\alpha}-\tfrac{1}{\CN}\delta_i^jS^{k\alpha})~,\\
 {[{L_\al}^\be,S^{i\ga}]}\ =\ -\di(\delta^\ga_\al S^{i\be}-\tfrac12\delta_\al^\be S^{i\ga})~,\qquad
 {[{L_\da}^\db,S_i^\dc]}\ =\ -\di(\delta^\dc_\da S_i^\db-\tfrac12\delta_\da^\db S_i^\dc)~,\\
 {[S^{i\al},P_{\be\db}]}\ =\ -\delta_\be^\al Q^i_\db~,\qquad
 {[S_i^\da,P_{\be\db}]}\ =\ \delta_\db^\da Q_{i\be}~,\\
  {[D,S^{i\al}]}\ =\ -\tfrac{\di}{2}S^{i\al}~,\qquad
  {[D,S_i^\da]}\ =\ -\tfrac{\di}{2}S_i^\da~,\\
  {[A,S^{i\al}]}\ =\ \tfrac{\di}{2}S^{i\al}~,\qquad
  {[A,S_i^\da]}\ =\ -\tfrac{\di}{2}S_i^\da~,\\
 {[{R_i}^j,Q_{k\alpha}]}\ =\ -\tfrac{\di}{2}(\delta_k^j Q_{i\alpha}-\tfrac{1}{\CN}\delta^j_iQ_{k\alpha})~,\qquad
 {[{R_i}^j,Q^k_\da]}\ =\ \tfrac{\di}{2}(\delta_i^k Q^j_\da-\tfrac{1}{\CN}\delta^j_iQ^k_\da)~,\\
 {[{L_\al}^\be,Q_{i\gamma}]}\ =\ \di(\delta_\ga^\be Q_{i\al}-\tfrac12\delta_\al^\be Q_{i\ga})~,\qquad
 {[{L_\da}^\db,Q^i_\dc]}\ =\ \di(\delta_\dc^\db Q^i_\da-\tfrac12\delta_\da^\db Q^i_\dc)~,\\
 {[Q_{i\al},K^{\be\db}]}\ =\ \delta_\al^\be S^\db_i~,\qquad
 {[Q^i_\da,K^{\be\db}]}\ =\ -\delta_\da^\db S^{i\be}~,\\
 {[D,Q_{i\alpha}]}\ =\ \tfrac{\di}{2}Q_{i\alpha}~,\qquad
 {[D,Q^i_\da]}\ =\ \tfrac{\di}{2}Q^i_\da~,\\
 {[A,Q_{i\alpha}]}\ =\ -\tfrac{\di}{2}Q_{i\alpha}~,\qquad
 {[A,Q^i_\da]}\ =\ \tfrac{\di}{2}Q^i_\da~,\\
 {[{R_i}^j,{R_k}^l]}\ =\ \tfrac{\di}{2}(\delta^l_i{R_k}^j-\delta^j_k{R_i}^l)~,\\
 {[D,P_{\al\da}]}\ =\ \di P_{\al\da}~,\qquad
 {[D,K^{\al\da}]}\ =\ -\di K^{\al\da}~,\\
 {[{L_\al}^\be,P_{\ga\dc}]}\ =\ \di(\delta_\ga^\be P_{\al\dc}-\tfrac12\delta_\al^\be P_{\ga\dc})~,\qquad
 {[{L_\da}^\db,P_{\ga\dc}]}\ =\ \di(\delta_\dc^\db P_{\ga\da}-\tfrac12\delta_\da^\db P_{\ga\dc})~,\\
 {[{L_\al}^\be,K^{\ga\dc}]}\ =\ -\di(\delta^\ga_\al K^{\be\dc}-\tfrac12\delta_\al^\be K^{\ga\dc})~,\qquad
 {[{L_\da}^\db,K^{\ga\dc}]}\ =\ -\di(\delta^\dc_\da K^{\ga\db}-\tfrac12\delta_\da^\db K^{\ga\dc})~,\\ 
 {[{L_\al}^\be,{L_\ga^\delta}]}\ =\ \di(\delta_\ga^\be {L_\al}^\delta-\delta_\al^\delta {L_\ga}^\be)~,\qquad
 {[{L_\da}^\db,{L_\dc}^{\dot\delta}]}\ =\ \di(\delta_\dc^\db {L_\da}^{\dot\delta}-
      \delta_\da^{\dot\delta} {L_\dc}^\db)~,\\
 {[P_{\al\da},K^{\be\db}]}\ =\ -\di(\delta_\al^\be{L_\da}^\db+\delta_\da^\db{L_\al}^\be+
     \delta_\al^\be\delta_\da^\db D)~.
\end{aligned}
\end{equation}
Notice that for $\CN=4$, the axial charge $A$ decouples, as mentioned above.
Notice also that upon chosing a real structure, not all of the above commutation
relations are independent of each other. Some of them will be related via
conjugation.

If we let $(z^A,\lambda_\da)=(z^\al,\eta_i,\lambda_\da)$ be homogeneous coordinates on $P^{3|\CN}$, then 
$\mathfrak{conf}_{4|\CN}\oplus\mathfrak{z}$ can be realised in terms of the following vector fields:
\begin{equation}\label{eq:SCAP}
 \begin{aligned}
   P_{\al\da}\ =\ \lambda_\da\frac{\partial}{\partial z^\al}~,\qquad
   K^{\al\da}\ =\ z^\al\frac{\partial}{\partial \lambda_\da}~,\qquad 
   D\ =\ -\frac{\di}{2}
    \left(z^\al\frac{\partial}{\partial z^\al}-\lambda_\da\frac{\partial}{\partial \lambda_\da}\right)~,\\
   {L_\al}^\be\ =\ -\di\left(z^\be\frac{\partial}{\partial z^\al}-\frac12\delta_\al^\be z^\ga
                \frac{\partial}{\partial z^\ga}\right),\qquad
   {L_\da}^\db\ =\ \di\left(\lambda_\da\frac{\partial}{\partial \lambda_\db}-\frac12\delta_\da^\db 
                \lambda_\dc\frac{\partial}{\partial \lambda_\dc}\right),\\
   {R_i}^j\ =\ -\frac{\di}{2}\left(\eta_i\frac{\partial}{\partial \eta_j}-\frac{1}{\CN}
                  \eta_k\frac{\partial}{\partial \eta_k}\right),\qquad
   A\ =\ -\frac{\di}{2}\eta_i\frac{\partial}{\partial \eta_i}~,\\
   Z\ =\ -\frac{\di}{2}\left(z^\al\frac{\partial}{\partial z^\al}+\lambda_\da
            \frac{\partial}{\partial \lambda_\da}+\eta_i\frac{\partial}{\partial \eta_i}\right),\\
  Q_{i\al}\ =\ \di\eta_i\frac{\partial}{\partial z^\al}~,\qquad
  Q^i_\da\ =\ \di\lambda_\da\frac{\partial}{\partial \eta_i}~,\qquad
  S^{i\al}\ =\ \di z^\al\frac{\partial}{\partial \eta_i}~,\qquad
  S_i^\da\ =\ \di\eta_i\frac{\partial}{\partial \lambda_\da}~.
 \end{aligned}
\end{equation}
Using $\partial_\al z^\be =\delta_\al^\be$, $\partial^\da\lambda_\db=\delta^\da_\db$ and
$\partial^i\eta_j=\delta^i_j$ for $\partial_\al:=\partial/\partial z^\al$, 
$\partial^\da:=\partial/\partial\lambda_\da$ and
$\partial^i:=\partial/\partial\eta_i$, one can straightforwardly check that the
above commutations relations are satisfied. Furthermore, we emphasise that we
work non-projectively. Working projectively, the
central charge $Z$ is absent (when acting on holomorphic functions), as is
explained in Remark \ref{rmk:pojection} The
non-projective version will turn out to be more useful in our discussion of
scattering amplitudes.

\vspace*{5pt}

\Remark{\label{rmk:pojection} Consider complex projective superspace $\IC P^{m|n}$. Then we have the canonical projection $\pi\,:\,\IC^{m+1|0}\setminus\{0\}\times \IC^{0|m}\to\IC P^{m|n}$. Let now $(z^a,\eta_i)=(z^0,\ldots,z^m,\eta_1,\ldots,\eta_n)$ be linear coordinates on $\IC^{m+1|n}$ (or equivalently,
homogeneous coordinates on $\IC P^{m|n}$) for $a=0,\ldots,m$ and $i=1,\ldots,n$. Then
$$
 \pi_*\left(z^a\frac{\partial}{\partial z^a}+\eta_i\frac{\partial}{\partial \eta_i}\right)\ =\ 0
$$
as follows from a direct calculation in affine coordinates which are defined by 
$$\IC P^{m|n}\ \supset\ 
U_a\ :\ z^a\ \neq\ 0\eand (z^{\hat a}_{(a)},\eta_{i(a)})\ :=\ \Big(\frac{z^{\hat a}}{z^a},\frac{\eta_i}{z^a}\Big) $$ 
for $a=0,\ldots, m$ and $\hat a\neq a$, i.e. $\IC P^{m|n}=\bigcup_a U_a$. 
}

\vspace*{5pt}

Likewise, we have a realisation of $\mathfrak{conf}_{4|\CN}\oplus\mathfrak{z}$ in terms of vector fields on the correspondence space $F^{5|2\CN}$ compatible with the projection $\pi_1\,:\,F^{5|2\CN}\to P^{3|2\CN}$, i.e.~the vector fields \eqref{eq:SCAP} are the push-forward via $\pi_{1\,*}$ of the vector fields on $F^{5|2\CN}$. In particular, if we take $(x^{A\da},\lambda_\da)=(x^{\al\da},\eta^\da_i,\lambda_\da)$ as coordinates on $F^{5|2\CN}$, where $\lambda_\da$ are homogeneous coordinates on $\IC P^1$, we have
\begin{equation}\label{eq:SCAF}
 \begin{aligned}
  P_{\al\da}\ =\ \frac{\partial}{\partial x^{\al\da}}~,\qquad
  K^{\al\da}\ =\ -x^{\al\db}x^{\be\da}\frac{\partial}{\partial x^{\be\db}}-
                  x^{\al\db}\eta^\da_i\frac{\partial}{\partial \eta^\db_i}+x^{\al\db}\lambda_\db    
                  \frac{\partial}{\partial \lambda_\da}~,\\
  D\ =\ -\di\left(x^{\al\da}\frac{\partial}{\partial x^{\al\da}}+\frac12\eta^\da_i
                      \frac{\partial}{\partial \eta^\da_i}-
                 \frac12\lambda_\da\frac{\partial}{\partial \lambda_\da}\right),\\
  {L_\al}^\be\ =\ -\di\left(x^{\be\db}\frac{\partial}{\partial x^{\al\db}}-\frac12\delta_\al^\be    
                     x^{\ga\dc}\frac{\partial}{\partial x^{\ga\dc}}\right),\\
 {L_\da}^\db\ =\ -\di\left(x^{\be\db}\frac{\partial}{\partial x^{\be\da}}-\frac12\delta_\da^\db    
                     x^{\ga\dc}\frac{\partial}{\partial x^{\ga\dc}}\right)
                   -\di\left(\eta^\db_i\frac{\partial}{\partial \eta^\da_i}-\frac12\delta_\da^\db    
                     \eta^\dc_k\frac{\partial}{\partial \eta^\dc_k}\right)\\
                    +\ \di\left(\lambda_\da\frac{\partial}{\partial \lambda_\da}-\frac12\delta_\da^\db    
                     \lambda_\dc\frac{\partial}{\partial \lambda_\dc}\right),\kern3cm\\
 {R_i}^j\ =\ -\frac{\di}{2}\left(\eta^\da_i\frac{\partial}{\partial \eta^\da_j}
            -\frac{1}{\CN}\eta^\da_k\frac{\partial}{\partial \eta^\da_k}\right),\qquad
  A\ =\ -\frac{\di}{2}\eta^\da_i\frac{\partial}{\partial \eta^\da_i}~,\qquad
  Z\ =\ -\frac{\di}{2}\lambda_\da\frac{\partial}{\partial\lambda_\da}~,\\
  Q_{i\al}\ =\ \di\eta^\da_i\frac{\partial}{\partial x^{\al\da}}~,\qquad
  Q^i_\da\ =\ \di\frac{\partial}{\partial \eta^\da_i}~,\\
  S^{i\al}\ =\ \di x^{\al\da}\frac{\partial}{\partial \eta^\da_i}~,\qquad
  S^\da_i\ =\ -\di\eta^\db_i x^{\be\da}\frac{\partial}{\partial x^{\be\db}}
            -\di\eta^\db_i\eta^\da_j\frac{\partial}{\partial \eta_j^\db}+
            \di\eta^\db_i\lambda_\db\frac{\partial}{\partial\lambda_\da}~.
 \end{aligned}
\end{equation}
In order to understand these expressions, let us consider a holomorphic function
$f$ on $F^{5|2\CN}$ which descends down to $P^{3|\CN}$. Recall that such a
function is of the form $f=f(x^{A\da}\lambda_\da,\lambda_\da)=
f(x^{\al\da}\lambda_\da,\eta^\da_i\lambda_\da,\lambda_\da)$ since then $V_A f=0$. Then
\begin{equation}\label{eq:useful}
\begin{aligned}
 \left.\frac{\partial}{\partial x^{A\da}}\right|_{\lambda_\da}f\ &=\
  \left.\lambda_\da\frac{\partial}{\partial z^{A}}\right|_{\lambda_\da}f~,\\
 \left.\frac{\partial}{\partial\lambda_\da}\right|_{x^{A\da}}f\ &=\
  \left(x^{A\da}\left.\frac{\partial}{\partial z^A}\right|_{\lambda_\da}+
       \left.\frac{\partial}{\partial\lambda_\da}\right|_{z^A}\right)f~.
\end{aligned}
\end{equation}
Next let us exemplify the calculation for the generator ${L_\da}^\db$. The rest
is left as an exercise.
Using the relations \eqref{eq:useful}, we find
\begin{equation}
\begin{aligned}
 \left[-\di\left.\left(x^{A\db}\frac{\partial}{\partial x^{A\da}}-\frac12\delta_\da^\db x^{C\dc}
        \frac{\partial}{\partial x^{C\dc}}\right)\right|_{\lambda_\da}+
       \di\left.\left(\lambda_\da\frac{\partial}{\partial\lambda_\db}-\frac12\delta_\da^\db\lambda_\dc
        \frac{\partial}{\partial\lambda_\dc}\right)\right|_{x^{A\da}}\right]f\ =\ \\
  =\ \di\left.\left(\lambda_\da\frac{\partial}{\partial\lambda_\db}-\frac12\delta_\da^\db\lambda_\dc
        \frac{\partial}{\partial\lambda_\dc}\right)\right|_{z^A}f~.\kern3cm
\end{aligned}
\end{equation}
Therefore,
\begin{equation}
\begin{aligned}
 \pi_{1\,*}\left[-\di\left(x^{A\db}\frac{\partial}{\partial x^{A\da}}-\frac12\delta_\da^\db x^{C\dc}
        \frac{\partial}{\partial x^{C\dc}}\right)+
       \di\left(\lambda_\da\frac{\partial}{\partial\lambda_\db}-\frac12\delta_\da^\db\lambda_\dc
        \frac{\partial}{\partial\lambda_\dc}\right)\right]\ =\ \\
  =\ \di\left(\lambda_\da\frac{\partial}{\partial\lambda_\db}-\frac12\delta_\da^\db\lambda_\dc
        \frac{\partial}{\partial\lambda_\dc}\right),\kern3cm
\end{aligned}
\end{equation}
what is precisely the relation between the realisations of the ${L_\da}^\db$-generator on $F^{5|2\CN}$ and
$P^{3|\CN}$ as displayed in \eqref{eq:SCAP} and \eqref{eq:SCAF}.

\vspace*{5pt}

\Exercise{Show that all the generators \eqref{eq:SCAP} are the push-forward under $\pi_{1\,*}$ of the generators
        \eqref{eq:SCAF}.}

\vspace*{5pt}

It remains to give the vector field realisation of the superconformal algebra on space-time $M^{4|2\CN}$. This is rather trivial, however, since $\pi_2\,:\,F^{5|2\CN}\to M^{4|2\CN}$ is the trivial projection. We find
\begin{equation}\label{eq:SCAM}
 \begin{aligned}
  P_{\al\da}\ =\ \frac{\partial}{\partial x^{\al\da}}~,\qquad
  K^{\al\da}\ =\ -x^{\al\db}x^{\be\da}\frac{\partial}{\partial x^{\be\db}}-
                  x^{\al\db}\eta^\da_i\frac{\partial}{\partial \eta^\db_i}~,\\
  D\ =\ -\di\left(x^{\al\da}\frac{\partial}{\partial x^{\al\da}}+\frac12\eta^\da_i
                      \frac{\partial}{\partial \eta^\da_i}\right),\\
  {L_\al}^\be\ =\ -\di\left(x^{\be\db}\frac{\partial}{\partial x^{\al\db}}-\frac12\delta_\al^\be    
                     x^{\ga\dc}\frac{\partial}{\partial x^{\ga\dc}}\right),\\
 {L_\da}^\db\ =\ -\di\left(x^{\be\db}\frac{\partial}{\partial x^{\be\da}}-\frac12\delta_\da^\db    
                     x^{\ga\dc}\frac{\partial}{\partial x^{\ga\dc}}\right)
                   -\di\left(\eta^\db_i\frac{\partial}{\partial \eta^\da_i}-\frac12\delta_\da^\db    
                     \eta^\dc_k\frac{\partial}{\partial \eta^\dc_k}\right),\\
 {R_i}^j\ =\ -\frac{\di}{2}\left(\eta^\da_i\frac{\partial}{\partial \eta^\da_j}
            -\frac{1}{\CN}\eta^\da_k\frac{\partial}{\partial \eta^\da_k}\right),\qquad
  A\ =\ -\frac{\di}{2}\eta^\da_i\frac{\partial}{\partial \eta^\da_i}~,\\
  Q_{i\al}\ =\ \di\eta^\da_i\frac{\partial}{\partial x^{\al\da}}~,\qquad
  Q^i_\da\ =\ \di\frac{\partial}{\partial \eta^\da_i}~,\\
  S^{i\al}\ =\ \di x^{\al\da}\frac{\partial}{\partial \eta^\da_i}~,\qquad
  S^\da_i\ =\ -\di\eta^\db_i x^{\be\da}\frac{\partial}{\partial x^{\be\db}}
            -\di\eta^\db_i\eta^\da_j\frac{\partial}{\partial \eta_j^\db}~.
 \end{aligned}
\end{equation}

\vspace*{10pt}

\section{Supersymmetric self-dual Yang--Mills theory and the Penrose--Ward transform}\label{sec:SSDYM}

\subsection{Penrose--Ward transform}\label{sec:PWsuper}

By analogy with self-dual Yang--Mills theory, we may now proceed to construct
supersymmetrised versions of this theory within the twistor framework. The
construction is very similar to the bosonic setting, so we can be rather brief.

Take a holomorphic vector bundle  $E\to P^{3|\CN}$ and pull it back to
$F^{5|2\CN}$.  Note that although we restrict our discussion to ordinary vector
bundles, everything goes through for supervector bundles as well. Then the
transition function is constant along $\pi_1\,:\,F^{5|2\CN}\to P^{3|2\CN}$,
i.e.~$V_A^\pm f_{+-}=0$ where the $V_A^\pm$ are the restrictions of $V_A$ to the
patches $\hat U_\pm$ with $F^{5|2\CN}=\hat U_+\cup\hat U_-$.
Under the assumption that $E$ is holomorphically trivial on any $L_x=\pi_1(\pi_2^{-1}(x))\hookrightarrow P^{3¦\CN}$, we again split $f_{+-}$ according to $f_{+-}=\psi_+^{-1}\psi_-$ and hence $\psi_+^{-1}V_A^\pm\psi_+=\psi_-^{-1}V_A^\pm \psi_-$ on $\hat U_+\cap\hat U_-$. Therefore, we may again introduce a Lie algebra-valued one-form that has components only along $\pi_1\,:\,F^{5|2\CN}\to P^{3|2\CN}$:
\begin{equation}
 A_A^\pm\ =\ \lambda^\da_\pm A_{A\da} \ =\ \psi_\pm^{-1} V_A^\pm \psi_\pm~,
\end{equation}
where $A_{A\da}$ is $\lambda_\pm$-independent. Thus, we find
\begin{equation}
 \lambda^\da_\pm\nabla_{A\da}\psi_\pm\ =\ 0~,\ewith \nabla_{A\da}\ :=\ \partial_{A\da}+A_{A\da}~
\end{equation}
together with the compatibility conditions,
\begin{equation}\label{eq:constraint1}
 [\nabla_{A\da},\nabla_{B\db}\}+[\nabla_{A\db},\nabla_{B\da}\}\ =\ 0~.
\end{equation}
These equations are known as the constraint equations of $\CN$-extended supersymmetric self-dual Yang--Mills theory (see e.g.~\cite{Semikhatov:1982ig,Volovich:1983ii}).

Let us analyse these equations a bit more for $\CN=4$. Cases with $\CN<4$ can 
be obtained from the $\CN=4$ equations by suitable truncations. We may write 
the above constraint equations as
\begin{equation}\label{eq:constraint2}
 [\nabla_{A\da},\nabla_{B\db}\}\ =\ \varepsilon_{\da\db}F_{AB}~,\ewith
 F_{AB}\ =\ (-)^{p_Ap_B}F_{BA}~.
\end{equation}
We may then parametrise $F_{AB}$ as
\begin{equation}
 F_{AB}\ =\ (F_{\al\be},F^i_\al,F^{ij})\ :=\
(f_{\al\be},\tfrac{1}{\sqrt{2}}\chi^i_\al,-\phi^{ij})~.
\end{equation}
Furthermore, upon using Bianchi identities
\begin{equation}\label{eq:bianchi}
\begin{aligned}
 &{[\nabla_{A\da},\nabla_{B\db}\},\nabla_{C\dot\gamma}\}}
 +(-)^{p_A(p_B+p_C)}[\nabla_{B\db},\nabla_{C\dot\gamma}\},\nabla_{A\da}\}\\
 &\kern4cm+\,(-)^{p_C(p_A+p_B)}[\nabla_{C\dot\gamma},\nabla_{A\da}\},\nabla_{B\db}\}\ =\ 0~,
\end{aligned}
\end{equation}
we find two additional fields,
\begin{equation}
 \chi_{i\da}\ :=\ -\tfrac{\sqrt{2}}{3}\nabla^j_\da \phi_{ij}
       \eand
       G_{\da\db}\ :=\ \tfrac{1}{2\sqrt{2}}\nabla^i_{(\da}\chi_{i\db)}~,
\end{equation}
where we have introduced the common abbreviation
$\phi_{ij}:=\frac{1}{2!}\varepsilon_{ijkl}\phi^{kl}$ and parentheses mean
normalised symmetrisation. Altogether, we have obtained the fields displayed in Table \ref{tab:fields} Note that all these fields are superfields, i.e.~they live on $M^{4|8}\cong\IC^{4|8}$. 

\begin{table}[h]
\label{tab:fields}
\begin{center}
\begin{tabular}{|c||c|c|c|c|c|}
 \hline
 field & $f_{\al\be}$ & $\chi^i_\al$ & $\phi^{ij}$ & $\chi_{i\da}$ & $G_{\da\db}$ \\
 \hline
 helicity & $-1$ & $-\tfrac12$ & $0$ & $\tfrac12$ & $1$ \\
 \hline
 multiplicity &$1$ & $4$ & $6$ & $4$ & $1$ \\
 \hline
\end{tabular}
\caption{\it Field content of $\CN=4$ supersymmetry self-dual Yang--Mills theory.}
\end{center}
\vspace*{-10pt}
\end{table}

The question is, how can we construct fields and their corresponding equations of motion on $M^4$, since that is what we are actually after. 
The key idea is to impose the so-called transversal gauge condition \cite{Harnad:1984vk,Harnad:1985bc,Devchand:1996gv}:
\begin{equation}\label{eq:transversal}
 \eta^\da_i A^i_\da\ =\ 0~.
\end{equation}
This reduces supergauge transformations to ordinary ones as
follows. Generic infinitesimal supergauge transformations are of the form
$\delta
A_{A\da}=\nabla_{A\da}\varepsilon=\partial_{A\da}\varepsilon+[A_{A\da},
\varepsilon]$, where $\varepsilon$ is a bosonic Lie algebra-valued function on
$M^{4|8}$. Residual gauge transformations that preserve \eqref{eq:transversal}
are then given by
\begin{equation}
 \eta^\da_i\delta A^i_\da\ =\ 0\qquad\Longrightarrow\qquad
 \eta^\da_i\partial^i_\da\varepsilon\ =\ 0\qquad\Longleftrightarrow\qquad
 \varepsilon\ =\ \varepsilon(x^{\al\db})~,
\end{equation}
i.e.~we are left with gauge transformations on space-time $M^4$.
Then, by defining the recursion operator
$\cD:=\eta^\da_i\nabla^i_\da=\eta^\da_i\partial^i_\da$ and by using the Bianchi
identities \eqref{eq:bianchi}, after a somewhat lengthy calculation we obtain
the following set of recursion relations:
\begin{equation}\label{eq:recursion}
 \begin{aligned}
  \cD A_{\al\da}\ &=\ -\tfrac{1}{\sqrt{2}}\varepsilon_{\da\db}\eta^\db_i\chi^i_\al~,\\
        (1+\cD) A^i_\da\ &=\ -\varepsilon_{\da\db}\eta^\db_j\phi^{ij}~,\\
        \cD \phi_{ij}\ &=\ \sqrt{2}\eta^\da_{[i}\chi_{j]\da}~,\\
        \cD\chi^i_\al\ &=\ \sqrt{2}\eta^\da_j\nabla_{\al\da}\phi^{ij}~,\\
        \cD\chi_{i\da}\ &=\ -\tfrac{1}{\sqrt{2}}\eta^\db_i G_{\da\db}+\tfrac{1}{\sqrt{2}}\varepsilon_{\da\db}\eta^\db_j
       [\phi^{jk},\phi_{ki}]~,\\
        \cD G_{\da\db}\ &=\ \sqrt{2}\eta^{\dot\gamma}_i\varepsilon_{\dot\gamma(\da}[\chi_{j\db)},\phi^{ij}]~,
 \end{aligned}
\end{equation}
where, as before, parentheses mean normalised symmetrisation while the brackets
denote normalised anti-symmetrisation of the enclosed indices. These equations determine all 
superfields to order $n+1$, provided one knows them to $n$-th
order in the fermionic coordinates. 

At this point, it is helpful  
to present some formul{\ae} which simplify this argument  a great deal.
Consider some generic superfield $f$. Its explicit $\eta$-expansion has the form
\begin{equation}
 f\ =\ \fc+\sum_{k\geq1}\eta^{\dc_1}_{j_1}\cdots
\eta^{\dc_k}_{j_k}\,  f_{\dc_1\cdots\dc_k}^{j_1\cdots j_k}~.
\end{equation}
Here and in the following, the circle refers to the zeroth-order term in the
superfield expansion of some superfield $f$.
Furthermore, we have $\cD f=\eta^{\dc_1}_{j_1}[\ \
]_{\dc_1}^{j_1}$, where the bracket $[\ \ ]_{\dc_1}^{j_1}$ is a
composite expression of some superfields. For example, we have
$\cD A_{\al\da}=\frac{1}{\sqrt{2}}\eta^{\dc_1}_{j_1}
[\al\da]_{\dc_1}^{j_1}$,
with $[\al\da]_{\dc_1}^{j_1}=-
\varepsilon_{\da\dc_1}\chi^{j_1}_\da$.  Now let
\begin{equation}
 \cD[\ \ ]_{\dc_1\cdots\dc_k}^{j_1\cdots j_k}\ =\
\eta^{\dc_{k+1}}_{j_{k+1}}[\ \
]_{\dc_1\cdots\dc_{k+1}}^{j_1\cdots
       j_{k+1}}~.
\end{equation}
Then we find after a successive application of $\cD$
\begin{equation}
 f\ =\ \fc+\sum_{k\geq1}\frac{1}{k!}\,
\eta^{\dc_1}_{j_1}\cdots\eta^{\dc_k}_{j_k}\, \overset{\circ}{[\ \ ]}\
\!\!_{\dc_1\cdots\dc_k}^{j_1\cdots j_k}~.
\end{equation}
If the recursion relation of
$f$ is of the form $(1+\cD)f=\eta^{\dc_1}_{j_1}[\ \
]_{\dc_1}^{j_1}$ as it happens to be for $A^i_\da$, then
$\fc=0$ and the superfield expansion is of the form
\begin{equation}
 f\ =\ \sum_{k\geq1}\frac{k}{(k+1)!}\,
\eta^{\dc_1}_{j_1}\cdots\eta^{\dc_k}_{j_k}\, \overset{\circ}{[\ \ ]}\
\!\!_{\dc_1\cdots\dc_k}^{j_1\cdots j_k}~.
\end{equation}
Using these expressions, one obtains the following
results for the superfields $A_{\al\da}$ and $A^i_\da$:
\begin{equation}
 \begin{aligned}
   A_{\al\da}\ &=\ \Ac~\!\!_{\al\da}+\tfrac{1}{\sqrt{2}}\varepsilon_{\da\db}\cc\ 
                      \!\!^i_\al\eta^\db_i+\cdots,\\
     A^i_\da  \ &=\ -\tfrac{1}{2!}\varepsilon_{\da\db}\Wc~\!^{ij}
              \eta^\db_j-\tfrac{\sqrt{2}}{3!}
 \varepsilon^{ijkl}\varepsilon_{\da\db}\cc_{k\dc}\eta^\dc_l
                      \eta^\db_j\ +\\
                  &~~~~~~~\ +\tfrac{3}{2\cdot 4!}\varepsilon^{ijkl}\epsilon_{\da\db}
                      (-\Gc_{\dc\dot\delta}\delta^m_{l}+\varepsilon_{\dc\dot\delta}[\Wc~\!^{mn},
                      \Wc_{nl}])\eta^\dc_k\eta^{\dot\delta}_m\eta^\db_j+\cdots~.
 \end{aligned}
\end{equation}

Upon substituting these superfield expansions 
into the constraint equations \eqref{eq:constraint1}, \eqref{eq:constraint2}, we obtain
\begin{equation}\label{eq:fieldeq}
 \begin{aligned}
  \fc_{\da\db}\ &=\ 0~,\\
  \varepsilon^{\al\be}\cnab_{\al\da}\cc\ \!\!^i_\be\ &=\ 0~,\\
  \overset{\circ}{\square}\Wc~\!^{ij}\ &=\ \tfrac12
  \varepsilon^{\al\be}\{\cc\ \!\!^i_\al,\cc\ \!\!^j_\be\}~,\\
   \varepsilon^{\da\db}\cnab_{\al\da}\cc_{i\db}\ &=\ 
  -[\Wc_{ij},\cc\ \!\!^j_\al]~,\\
  \varepsilon^{\da\db}\cnab_{\al\da}\Gc_{\db\dc}\ &=\ 
  -\{\cc\ \!\!^i_\al,\cc_{i\dc}\}
  + \tfrac{1}{2}[\Wc_{ij},\cnab_{\al\dc}\Wc~\!^{ij}]~.
 \end{aligned}
\end{equation}
These are the equations of motion of $\CN=4$ supersymmetric self-dual Yang--Mills theory.
The equations for less
supersymmetry are obtained from these by suitable truncations.
We have also introduced the abbreviation
$
 \overset{\circ}{\square}:= 
    \tfrac{1}{2}\varepsilon^{\al\be}\varepsilon^{\da\db}
    \cnab_{\al\da}\cnab_{\be\db}.
$
We stress that \eqref{eq:fieldeq} represent the field equations
to lowest order in the superfield expansions. With the help of the
recursion operator $\cD$, one may verify that they are in one-to-one
correspondence with the constraint equations \eqref{eq:constraint1}.
For details, see e.g.~\cite{Harnad:1984vk,Harnad:1985bc,Devchand:1996gv}.

Altogether, we have a supersymmetric extension of Ward's theorem
\ref{thm:ward}:

{\Thm\label{thm:wardsusy} There is a one-to-one correspondence between gauge equivalence classes of solutions to the $\CN$-extended supersymmetric self-dual Yang--Mills equations on space-time $M^4$ and equivalence classes of holomorphic vector bundles over supertwistor space $P^{3|\CN}$ which are holomorphically trivial on any projective line $L_x=\pi_1(\pi_2^{-1}(x))\hookrightarrow P^{3|\CN}$. }

\vspace*{10pt}

\Exercise{Verify all equations from \eqref{eq:recursion} to \eqref{eq:fieldeq}.}

Finally, let us emphasise that the field equations \eqref{eq:fieldeq} also
follow from an action principle. Indeed, upon varying
\begin{equation}\label{eq:siegelaction}
 S\ =\ \int\dd^4x \,{\rm tr}
 \left\{\Gc\ \!\!^{\da\db}\fc_{\da\db}+
         \cc\ \!\!^{i\al}\cnab_{\al\da}\cc\ 
         \!\!^\da_i-\tfrac{1}{2}\Wc_{ij}
                \overset{\circ}{\square}\Wc\ \!\!^{ij}
             +\tfrac12\Wc_{ij}\{\cc\ \!\!_\al^i,\cc\ \!\!^{j\al}\} \right\},
\end{equation}
we find \eqref{eq:fieldeq}.
In writing this, we have implicitly assumed that a reality condition corresponding either to Euclidean or Kleinian signature has been chosen; see below for more details. This action functional is known as the Siegel action \cite{Siegel:1992za}.

\pagebreak

\Remark{Let us briefly comment on hidden symmetry structurs of self-dual
Yang--Mills theories. Since Pohlmeyer's
work \cite{Pohlmeyer:1979ya}, it has been known that self-dual 
Yang--Mills theory possess infinitely many hidden non-local symmetries.
Such symmetries are accompanied by conserved non-local charges. 
As was shown
in \cite{Chau:1981gi,Chau:1982mn,Chau:1982mj,Ueno:1982av,Dolan:1982dc,
Crane:1987im},
these symmetries are affine extensions of internal symmetries
with an underlying Kac--Moody structure. See
\cite{Dolan:1983bp} for a review. Subsequently, Popov \&
Preitschopf \cite{Popov:1995qb} found
affine extensions of conformal symmetries of 
Kac--Moody/Virasoro-type. 
A systematic investigation of symmetries based on 
twistor and cohomology theory
was performed in \cite{Popov:1998pc} (see also 
\cite{Popov:1998fb,Ivanova:1997cu} and the text book \cite{Mason:1991rf}),
where
all symmetries of the self-dual Yang--Mills equations were derived.
In \cite{Wolf:2004hp,Wolf:2005sd} (see \cite{Wolf:2006me} for a review), these
ideas were extended to $\CN$-extended self-dual Yang--Mills theory.
For some extensions to the full
$\CN=4$ supersymmetric Yang--Mills theory, see \cite{Popov:2006qu}.
Notice that the symmetries of the self-dual Yang--Mills equations are
intimitately connected with one-loop maximally-helicity-violating scattering
amplitudes \cite{Bardeen:1995gk,Cangemi:1996rx,Cangemi:1996pf,Rosly:1996vr}.
See also Part II of these lecture notes.
}

\vspace*{5pt}

\Exercise{Verify that the action functional \eqref{eq:siegelaction} is invariant under the following supersymmetry transformations ($\varrho^\da_i$ is some constant anti-commuting spinor):
$$
 \begin{aligned}
  \delta \Ac_{\al\da}\ &=\ -\tfrac{1}{\sqrt{2}}
\varepsilon_{\da\db}\varrho^\db_i\cc\ \!\!^i_\al~,\\
        \delta \Wc_{ij}\ &=\ \sqrt{2}\varrho^\da_{[i}\cc_{j]\da}~,\\
        \delta\cc\ \!\!^i_\al\ &=\
\sqrt{2}\varrho^\da_j\cnab_{\al\da}\Wc\ \!\!^{ij}~,\\
        \delta\cc_{i\da}\ &=\ -\tfrac{1}{\sqrt{2}}
       \varrho^\db_i \Gc_{\da\db}
            +\tfrac{1}{\sqrt{2}}\varepsilon_{\da\db}\varrho^\db_j
       [\Wc\ \!\!^{jk},\Wc_{ki}]~,\\
        \delta \Gc_{\da\db}\ &=\ \sqrt{2}
  \varrho^{\dot\gamma}_i\varepsilon_{\dot\gamma(\da}[\cc_{j\db)},\Wc\
\!\!^{ij}]~.
 \end{aligned}
$$
}

\vspace*{10pt}

\subsection{Holomorphic Chern--Simons theory}\label{sec:HCS}

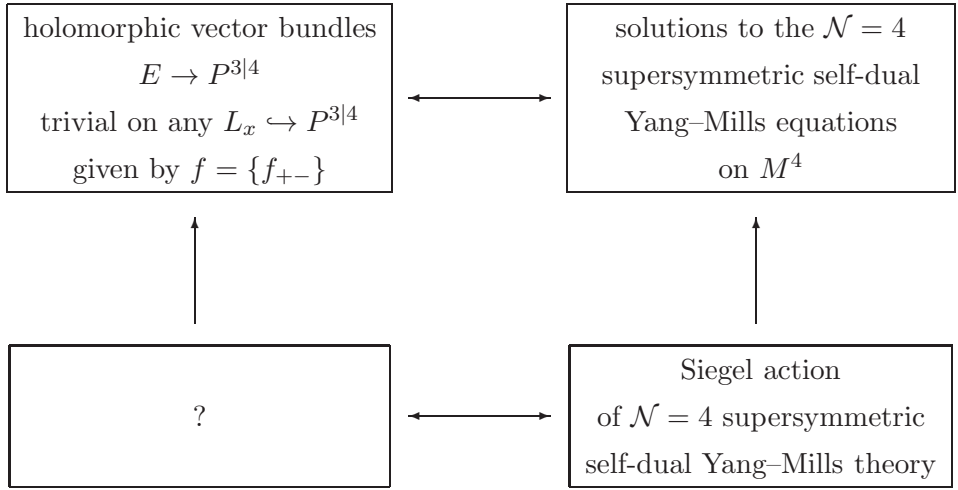
\begin{figure}[h] 
\begin{equation}\notag
\begin{aligned}
\begin{picture}(200,100)(0,0)
   \put(-10.0,-50.0){\makebox(0,0)[c]{
     \begin{tabular}{|c|}\hline { }\cr
                                {\kern2.2cm ?\kern2.2cm}\cr
                                { }\\ \hline \end{tabular}}}
  \put(200.0,-50.0){\makebox(0,0)[c]{
     \begin{tabular}{|c|}\hline {Siegel action}\cr
                                {of $\CN=4$ supersymmetric}\cr
                                {self-dual Yang--Mills theory}\\ \hline
\end{tabular}}}
  \put(-10.0,70.0){\makebox(0,0)[c]{
     \begin{tabular}{|c|}\hline {holomorphic vector bundles}\cr
                                {$E\to P^{3|4}$}\cr
                                {trivial on any $L_x\hookrightarrow P^{3|4}$}\cr
                                {given by $f=\{f_{+-}\}$}\\ 
                          \hline \end{tabular}}}
  \put(200.0,70.0){\makebox(0,0)[c]{
     \begin{tabular}{|c|}\hline {solutions to the $\CN=4$}\cr
                                {\kern.25cm supersymmetric
self-dual\kern.25cm}\cr 
                                {Yang--Mills equations}\cr
                                {on $M^4$}\\ 
                          \hline \end{tabular}}}
  \put(200.0,-15.0){\vector(0,1){40}}
  \put(-10.0,-15.0){\vector(0,1){40}}
  \put(70,-50.0){\vector(1,0){53}}
  \put(123,-50.0){\vector(-1,0){53}}
  \put(70,70.0){\vector(1,0){53}}
  \put(123,70.0){\vector(-1,0){53}}
\end{picture}
\end{aligned}
\end{equation}
\vspace*{2.2cm}
\caption{Correspondences between supertwistor space and space-time.}
\label{fig:summary}
\end{figure}

Let us pause for a moment and summarise what we have achieved so far. In the
preceding sections, we have discussed $\CN=4$ supersymmetric self-dual
Yang--Mills theory by means of holomorphic vector bundles $E\to P^{3|4}$ over
the supertwistor space $P^{3|4}$ that are holomorphically trivial on all
projective lines $L_x=\pi_1(\pi_2^{-1}(x))\hookrightarrow P^{3|4}$. These
bundles are given by holomorphic transition functions $f=\{f_{+-}\}$. We have
further shown that the field equations of 
$\CN=4$ supersymmetric self-dual Yang--Mills theory arise upon varying a certain action functional on space-time, the Siegel action. 
Figure \ref{fig:summary}.~summarises pictorially our previous discussion.

The question that now arises and which is depicted in Figure
\ref{fig:summary}. concerns the formulation of a corresponding action principle
on the supertwistor
space. Certainly, such an action, if it exists, should correspond to the Siegel
action on space-time. However, in constructing such a twistor space action, we
immediately face a difficulty. Our above approach to the twistor re-formulation
of field theories, either linear or non-linear, is intrinsically on-shell:
Holomorphic functions on twistor space correspond to solutions to field
equations on space-time and vice versa. In particular, holomorphic transition
functions of certain holomorphic vector bundles $E\to P^{3|4}$ correspond to
solutions to the $\CN=4$ supersymmetric self-dual Yang--Mills equations.
Therefore, we somehow need an `off-shell approach' to holomorphic vector
bundles, that is, we need a theory on supertwistor space that describes complex
vector bundles such that the  on-shell condition is the holomorphicity of these
bundles.

\vspace*{10pt}

Before we delve into this issue, 
let us formalise our above approach to holomorphic vector bundles (which is also
known as the \v Cech approach). Consider a complex (super)manifold $(X,\CO)$
with an open covering $\fU=\{U_i\}$. 
We are interested in holomorphic maps from open subsets of 
$X$ into $\sGL(r,\IC)$ as well as in the sheaf $\sGL(r,\CO)$ of such matrix-valued functions.\footnote{Basically everything we shall say below will also apply to $\sGL(r|s,\CO)$ and hence to supervector bundles. As we are only concerned with ordinary vector bundles (after all we are interested in $\sSU(r)$ gauge theory), we will stick to $\sGL(r,\CO)$ for concreteness. See e.g.~\cite{Wolf:2006me} for the following treatment in the context of supervector bundles.} Notice that $\sGL(r,\CO)$ is a non-Abelian sheaf contrary to the Abelian sheaves considered so far.
A $q$-cochain of the covering $\fU$ with values in $\sGL(r,\CO)$ is a collection $f=\{f_{i_0\cdots i_q}\}$ of sections of the sheaf $\sGL(r,\CO)$ over non-empty intersections $U_{i_0}\cap\cdots\cap U_{i_q}$. We will 
denote the set of such $q$-cochains by $C^q(\fU,\sGL(r,\CO))$. We stress that it has a
group structure, where the multiplication is just pointwise multiplication.

We may define the subsets of cocycles $Z^q(\fU,\sGL(r,\CO))\subset C^q(\fU,\sGL(r,\CO))$. 
For example, for $q=0,1$ they are given by
\begin{equation}
 \begin{aligned}
  Z^0(\fU,\sGL(r,\CO))\ &:=\ \{f\in C^0(\fU,\sGL(r,\CO))\ |\ f_i=f_j
           ~~{\rm on}~~U_i\cap U_j\neq\emptyset\}~,\\
           Z^1(\fU,\sGL(r,\CO))\ &:=\ \{f\in C^1(\fU,\sGL(r,\CO))\ |\ f_{ij}=
           f_{ji}^{-1}~~{\rm on}~~U_i\cap U_j\neq\emptyset\\
           &\kern3.5cm{\rm and}~~f_{ij}f_{jk}f_{ki}=1~~
           {\rm on}~~U_i\cap U_j\cap U_k\neq\emptyset\}~.
 \end{aligned}
\end{equation}
These sets will be of particular interest. We remark that from the 
first of these two definitions it follows that $Z^0(\fU,\sGL(r,\CO))$ coincides with 
the group of global sections, $H^0(\fU,\sGL(r,\CO))$, of the sheaf $\sGL(r,\CO)$. Note that in general
the subset $Z^1(\fU,\sGL(r,\CO))\subset C^1(\fU,\sGL(r,\CO))$ is not a subgroup of the group 
$C^1(\fU,\sGL(r,\CO))$. For notational reasons, we shall denote elements of
$C^0(\fU,\sGL(r,\CO))$ also by $h=\{h_i\}$.

We say that two cocycles $f,f'\in Z^1(\fU,\sGL(r,\CO))$ are equivalent if 
$f'_{ij}=h_i^{-1}f_{ij}h_j$ for some $h\in C^0(\fU,\sGL(r,\CO))$, since one can
always absorb the $h=\{h_i\}$ in a re-definition of the frame fields. Notice
that this is precisely the
transformation we already encountered in \eqref{eq:GTP}. The set
of equivalence classes induced by this equivalence relation is the first 
\v Cech cohomology set and denoted by $H^1(\fU,\sGL(r,\CO))$. If the $U_i$ are all Stein (see Remark \ref{rmk:stein})---in the case of supermanifolds $X$ we need the body to be covered by Stein manifolds---we have the bijection
\begin{equation}
 H^1(\fU,\sGL(r,\CO))\ \cong\ H^1(X,\sGL(r,\CO))~,
\end{equation}
otherwise one takes the inductive limit (see Remark \ref{rmk:inductive}). 

\vspace*{5pt}

\Remark{\label{rmk:stein} We call an ordinary complex manifold $(X,\CO)$ Stein if $X$ is 
holomorphically convex (that is, the holomorphically convex hull of any compact
subset of $X$ is again compact in $X$) and for any $x,y\in X$ with $x\neq y$ 
there is some $f\in\CO$ such that $f(x)\neq f(y)$.}

\vspace*{5pt}

\noindent
To sum up, we see that within the \v Cech approach, rank-$r$ holomorphic vector bundles over some complex (super)manifold $X$ are parametrised by $H^1(X,\sGL(r,\CO))$. Notice that our cover $\{U_\pm\}$ of the (super)twistor space is Stein and so $H^1(\{U_\pm\},\sGL(r,\CO))\cong H^1(P^{3|\CN},\sGL(r,\CO))$. This 
in turn explains that all of our above constructions are independent of the
choice of cover.

Another approach to holomorphic vector bundles is the so-called Dolbeault approach. Let $X$ be a complex (super)manifold and consider a rank-$r$ complex
vector bundle $E\to X$. Furthermore, we let $\Omega^{p,q}(X)$ be the smooth
differential $(p,q)$-forms on $X$ and
$\dbar\,:\,\Omega^{p,q}(X)\to\Omega^{p,q+1}(X)$ be 
the anti-holomorphic exterior derivative. A $(0,1)$-connection on
$E$ is defined by a covariant differential $\nabla^{0,1}\,:\,\Omega^{p,q}(X,E)\
\to\ \Omega^{p,q+1}(X,E)$ which satisfies the Leibniz formula.
Here, $\Omega^{p,q}(X,E):=\Omega^{p,q}(X)\otimes E$. Locally, it is of the form $\nabla^{0,1}=\dbar+A^{0,1}$,
where $A^{0,1}$ is a differential $(0,1)$-form with values in ${\rm End}\,E$
which we shall refer to as the connection $(0,1)$-form. The complex vector
bundle
$E$ is said to be holomorphic if the $(0,1)$-connection is flat, that is, if the
corresponding curvature vanishes,
\begin{equation}\label{eq:hCS1}
 F^{0,2}\ =\ (\nabla^{0,1})^2\ =\ \dbar A^{0,1}+
      A^{0,1}\wedge A^{0,1}\ =\ 0~.
\end{equation}
In other words, $\nabla^{0,1}$ defines a holomorphic structure on $E$.
 The group $H^0(X,\sGL(r,\CS))$, where $\CS$ is the sheaf of smooth functions on $X$, acts on $A^{0,1}$ by gauge transformations,
\begin{equation}\label{eq:HCSgauge}
 A^{0,1}\ \mapsto\ g^{-1} A^{0,1}g+g^{-1}\dbar g~,\ewith g\ \in\ H^0(X,\sGL(r,\CS))~.
\end{equation}
Clearly, such transformations leave \eqref{eq:hCS1} invariant, hence they do not
change the holomorphic structure on $E$. Therefore, two solutions to $F^{0,2}=0$
are regarded as equivalent if they differ by such a gauge transformation. We
shall then denote the space of equivalence classes by
$H^{0,1}_{\nabla^{0,1}}(X,E)$.

In summary, we have two apparently different approaches: 
In the \v Cech approach, holomorphic vector bundles are parametrised by 
$H^1(X,\sGL(r,\CO))$ while in the Dolbeault approach by $H^{0,1}_{\nabla^{0,1}}(X,E)$.
However, are these approaches really different? In fact, they turn out to be
equivalent by virtue of the following theorem:

{\Thm\label{thm:CD} Let $X$ be a complex
(super)manifold with an open Stein covering $\fU=\{U_i\}$
      and $E\to X$ be a rank-$r$ 
      complex vector bundle over $X$. Then there is a map $\rho\,:\,H^1(X,\sGL(r,\CO))\to H^1(X,\sGL(r,\CS))$ 
      of cohomology sets, such that $H^{0,1}_{\nabla^{0,1}}(X,E)\cong\ker\rho$. 
}

\vspace*{15pt}

\noindent
This means that given some holomorphic vector bundle $E\to X$ in the Dolbeault picture,
we can always find a holomorphic vector bundle $\check E\to X$ in the
\v Cech
picture and vice versa,
such that $E$ and $\check E$ are equivalent as complex vector bundles:
\begin{equation}
(E,f=\{f_{ij}\},\nabla^{0,1})\ \sim\ (\check E,\check f=\{\check
f_{ij}\},\dbar)~, 
\end{equation}
with $\check f_{ij}=\psi_i^{-1}f_{ij}\psi_j$ for some
$\psi=\{\psi_i\}\in C^0(\fU,\sGL(r,\CS))$. This theorem might be regarded as a
non-Abelian generalisation of the famous Dolbeault theorem (see Remark
\ref{rmk:HodgeDolbeault}) 

We shall not prove this theorem here but instead only make the following
observation: Given $(E,f=\{f_{ij}\},\nabla^{0,1})$, then any solution $A^{0,1}$
of $F^{0,2}=0$ is of the form $A^{0,1}|_{U_i}=A_i=\psi_i\dbar\psi_i^{-1}$ 
for some $\psi=\{\psi_i\}\in C^0(\fU,\sGL(r,\CS))$ with
\begin{equation}
 A_j\ =\ f_{ij}^{-1}\dbar f_{ij}+f_{ij}^{-1}A_i f_{ij}~,
\end{equation}
as patching conditions (see also Appendix \ref{app:VB}). Upon substituting
$A_i=\psi_i\dbar\psi_i^{-1}$ into these equations, we obtain
an $\check f_{ij}=\psi_i^{-1}f_{ij}\psi_j$ with $\dbar \check f_{ij}=0$.
Conversely,
starting from $\check f_{ij}=\psi_i^{-1}f_{ij}\psi_j$ with $\dbar \check
f_{ij}=0$ and
$\dbar f_{ij}\neq0$, one can recover an
$A^{0,1}|_{U_i}=A_i=\psi_i\dbar\psi_i^{-1}$ which obeys the patching conditions.
For more details on the proof of the theorem, see
e.g.~\cite{Popov:1998fb,Wolf:2006me}.

\vspace*{5pt}

\Remark{In fact, we already encountered a \v Cech--Dolbeault correspondence
before, but we did not allude to it as such. In Sections
\ref{sec:PWtransform}~and \ref{sec:PWsuper} we related the transition function
of a holomorphic vector bundle on (super)twistor space to those of the pull-back
on the correspondence space. By the definition of the pull-back, the transition
function is annihilated by the vector fields that span the tangent spaces of the
leaves of $\pi_1$. We then related the transition function to a Lie
algebra-valued (differential) one-form with components only along $\pi_1$. More
precisely, this one-form is the connection one-form of the so-called relative
connection along $\pi_1$ which by our very construction turned out to be flat.
In
this picture, the transition function is the \v Cech representative while the
Lie algebra-valued one-form is the Dolbeault representative of a relatively flat
bundle on the correspondence space.}

\vspace*{5pt}

The above theorem is good news in that it yields an `off-shell' formulation of holomorphic vector bundles: Given some complex vector bundle $E\to X$ with a $(0,1)$-connection $\nabla^{0,1}$ that is represented by $A^{0,1}$, we can turn it into a holomorphic vector bundle provided $A^{0,1}$ satisfies the `on-shell condition' \eqref{eq:hCS1}. The latter equation is also known as the equation of motion of holomorphic Chern--Simons theory. What about an action? Is it possible to write down an action functional which yields $F^{0,2}=0$. The answer is yes, but not always. The action that gives $F^{0,2}=0$ is \cite{Witten:1992fb}
\begin{equation}\label{eq:HCSA}
 S\ =\ -\frac{1}{2\pi\di}\int_X\Omega\wedge\underbrace{{\rm tr}\left\{A^{0,1}\wedge\dbar A^{0,1}+\tfrac23 A^{0,1}\wedge A^{0,1}\wedge A^{0,1}\right\}}_{=:\ \omega}\ =\ -\frac{1}{2\pi\di}\int_X\Omega\wedge\omega~,
\end{equation}
where the pre-factor has been chosen for later convenience.
A few words are in order. For a moment, let us assume that $X$ is an ordinary complex manifold. 
The holomorphic Chern--Simons form 
\begin{equation}
 \omega\ =\ {\rm tr}\left\{A^{0,1}\wedge\dbar A^{0,1}+\tfrac23 A^{0,1}\wedge A^{0,1}\wedge A^{0,1}\right\}
\end{equation}
is a $(0,3)$-form on $X$.  Furthermore, the action \eqref{eq:HCSA} is invariant
under \eqref{eq:HCSgauge} provided $\Omega$ is holomorphic (and $g$ is homotopic
to the identity and $g\to 1$ asymptotically).  Thus, $\dim_\IC X=3$ and so
$\Omega$ is a holomorphic $(3,0)$-form. In addition, $\Omega$ should be globally
defined, since the Chern--Simons form is. This in turn puts severe restrictions
on the geometric properties of $X$. Complex manifolds that admit such forms are
called formal Calabi--Yau manifolds. The name is chosen to distinguish them from
`honest' Calabi--Yau manifolds which are (compact) complex manifolds that are
K\"ahler and that admit globally defined holomorphic top-forms.\footnote{For a
beautiful recent exposition on Calabi--Yau manifolds, see e.g.~the lecture notes
\cite{Bouchard:2007ik}.}  Equivalently, we could say that the canonical bundle
$K:=\det T^{*\,1,0}X$ is trivial, since $\Omega$ is a section of $K$. Notice
that sections of $K$ are also called holomorphic volume forms. 
Therefore, while the equation $F^{0,2}=0$ makes sense on any complex manifold $X$, the corresponding action functional is defined only for three-dimensional formal Calabi--Yau manifolds.

Let us take a closer look at the twistor space $P^3$. Certainly, it is a
three-dimensional complex manifold. But does it admit a globally defined
holomorphic volume form $\Omega_0$? In fact it does not, since
\begin{equation}
 \Omega_0\ =\ \tfrac14\dd z^\al\wedge\dd z_\al\wedge\dd\lambda_\db\lambda^\db~,
\end{equation}
which shows that $K\cong\CO(-4)$ which is not a trivial bundle. Therefore, it is
not possible to write down an action principle for holomorphic Chern--Simons
theory on $P^3$. However, somehow we could have expected that as we do not have
an action on space-time either (apart from the Lagrange multiplier type
action $\int\dd^4 x\, G^{\da\db}f_{\da\db}$ with the
additional field $G_{\da\db}$; see Section \ref{eq:MotN4SYM} for more
details). 

But what about $\CN=4$ supersymmetric self-dual Yang--Mills theory? We know that there is an action principle.
The holomorphic volume form $\Omega$ on supertwistor space $P^{3|4}$ is given by
\begin{equation}\label{eq:holovol}
 \Omega\ =\ \Omega_0\otimes\Omega_1~,\ewith\Omega_1\ :=\ 
 \tfrac{1}{4!}\varepsilon^{ijkl}\dd\eta_i\dd\eta_j\dd\eta_k\dd\eta_l~,
\end{equation}
where $\Omega_0$ is as above. To determine whether this is globally defined or not, we have to take into account the definition of the Berezin integral over fermionic coordinates: 
\begin{equation}
 \int \dd\eta_i\,\eta_j\ =\ \delta_{ij}~.
\end{equation}
If one re-scales $\eta_i\mapsto t\,\eta_i$, with $t\in\IC\setminus\{0\}$, then
$\dd\eta_i\mapsto t^{-1}\dd\eta_i$. Therefore, the $\dd\eta_i$ transform
oppositely to $\eta_i$ and thus they are not differential forms. In fact,
they are so-called integral forms.  This shows that \eqref{eq:holovol} is indeed
globally defined, since $\Omega_0$ is a basis section of $\CO(-4)$ while
$\Omega_1$ is a basis section of $\CO(4)$ and so $\Omega$ is of homogeneity zero
since $\CO(-4)\otimes\CO(4)\cong\CO$. 
Hence, the Berezinian line bundle ${\rm Ber}$ (which is the supersymmetric
generalisation of the canonical bundle) of the supertwistor space $P^{3|4}$ is
trivial. In this sense, $P^{3|4}$ is a
formal Calabi--Yau supermanifold.\footnote{In fact, $P^{3|4}$ is also an
`honest' Calabi--Yau space in the sense of being K{\"a}hler and admitting a
super
Ricci-flat metric  \cite{Sethi:1994ch,Witten:2003nn}. Note that for ordinary
manifolds, the K{\"a}hler condition together with the existence of a globally
defined holomorphic volume form
always implies the existence of a Ricci-flat metric. This is the famous Yau
theorem. However, an analog of Yau's theorem in the context of supermanifolds
does not exist. See
e.g.~\cite{Sethi:1994ch,Rocek:2004bi,Zhou:2004su,Rocek:2004ha,Saemann:2004tt,
Lindstrom:2005uh,Ricci:2005cp}. } 
In Remark \ref{rmk:jet}, we show that any three-dimensional complex spin
manifold can be associated with a formal Calabi--Yau supermanifold via the
LeBrun construction \cite{LeBrun:2004}. Notice that if we had instead           
considered $P^{3|\CN}$, then $\Omega$ would have been of homogeneity $\CN-4$.
Altogether, we see that the holomorphic Chern--Simons action \eqref{eq:HCSA} can
only be written down for $\CN=4$. In this case, the symmetry group that
preserves the holomorphic measure $\Omega$ is $\sPSL(4|4,\IC)$. This is the
complexification of the $\CN=4$ superconformal group in four dimensions (which
is the global symmetry group of $\CN=4$ supersymmetric Yang--Mills theory). See
also Section \ref{sec:SCA}

We are almost done. We just need to clarify one issue. The connection one-form
$A^{0,1}$ is an `honest' differential form, i.e.~it is not an integral form like
$\Omega_1$. However, in general $A^{0,1}$ also depends on $\bar\eta_i$ and
furthermore on the differential $(0,1)$-forms $\dd\bar\eta_i$. As there is no
way of integrating fermionic differential forms, the integral \eqref{eq:HCSA}
seems not to make sense. However, this is not the case since the supertwistor
space $P^{3|4}$ can be regarded as a holomorphic vector bundle over the twistor
space $P^3$; see Exercise \ref{exe:VB}~and Equation \eqref{eq:PSS}.
Hence, there always exists a gauge in which
\begin{equation}\label{eq:wittengauge}
 \frac{\partial}{\partial\bar\eta_i}\lrcorner A^{0,1}\ =\ 0\eand
 \frac{\partial}{\partial\bar\eta_i}\lrcorner\, \dbar A^{0,1}\ =\ 0~,
\end{equation}
since the fibres $P^{3|4}\to P^3$ do not have sufficient non-trivial topology.
Thus, in this gauge $A^{0,1}$ only has a holomorphic dependence on
$\eta_i$. We shall refer to this gauge as Witten gauge \cite{Witten:2003nn}.

Summarising our above discussion, we have found an action principle on
supertwistor space for holomorphic vector bundles  in the case of $\CN=4$
supersymmetry.\footnote{An analogous action has been found by Sokatchev
\cite{Sokatchev:1995nj} in the harmonic superspace approach. Recently, Popov
gave alternative twistor space action principles in \cite{Popov:2009nx}.
Interestingly, similar actions also exist for maximal $\CN=8$ self-dual
supergravity in the harmonic space approach \cite{Karnas:1997it} as well as in
the twistor approach \cite{Mason:2007ct}.}
 Therefore, the missing box in Figure \ref{fig:summary}.~can be filled in with
the help of
\eqref{eq:HCSA}.  In the remainder of this section, we demonstrate
explicitly that 
\eqref{eq:HCSA} indeed implies \eqref{eq:siegelaction} on space-time. Our discussion below is based on 
\cite{Witten:2003nn,Popov:2004rb,Mason:2005zm,Boels:2006ir}.

For concreteness, let us choose reality conditions that lead to a Euclidean
signature real slice in $M^4$. At the end of Section \ref{sec:PWtransform}, we
saw that they arise from an anti-holomorphic involution on $P^3$. Similarly, we
may introduce an anti-holomorphic involution $\tau\,:\,P^{3|4}\to P^{3|4}$
leading to Euclidean superspace. In particular,
\begin{equation}\label{eq:RCESS}
 \tau(z^A,\lambda_\da)\ :=\ (\hat z^A,\hat\lambda_\da)\ :=\ (\bar z^B {C_B}^A,{C_\da}^\db\lambda_\db)~,
\end{equation}
where $({C_A}^B)={\rm diag}(({C_\al}^\be),({C_i}^j))$ with ${C_\al}^\be$ and ${C_\da}^\db$ given by \eqref{eq:RSb} and 
\begin{equation}
 ({C_i}^j)\ :=\ \begin{pmatrix}
                  0 & 1 & 0 & 0\\
                  -1 & 0 & 0 & 0\\
                  0 & 0 & 0 & 1\\
                  0 & 0 & -1 & 0
                \end{pmatrix}.
\end{equation}
By virtue of the incidence relation $z^A=x^{A\da}\lambda_\da$, we find 
\begin{equation}
 \tau(x^{A\da})\ =\ -\bar x^{B\db}{C_B}^A{C_\db}^\da~.
\end{equation}
The set of fixed points $\tau(x)=x$ corresponds to Euclidean superspace $\IR^{4|8}$ in $M^{4|8}$. 

\vspace*{5pt}

\Remark{\label{rmk:jet} Let $X$ be an $m$-dimensional complex manifold and $E\to
X$ be a rank-$n$ holomorphic vector bundle over $X$. In Exercise \ref{exe:VB}~we
saw that $X^{m|n}=(X,\CO_X(\Lambda^\bullet E^*))$ is a complex supermanifold.
The Berezinian line bundle is then given by ${\rm Ber}\cong K\otimes\Lambda^nE$,
where $K$ is the canonical bundle of $X$. Holomorphic volume forms on $X^{m|n}$
are sections of ${\rm Ber}$ and the existence of a globally defined holomorphic
volume form is equivalent to the triviality of ${\rm Ber}$.
Next one can show that there is a short exact sequence
$$ 0\ \overset{\varphi_1}{\longrightarrow}\ T^{*\,1,0}X\otimes V\
\overset{\varphi_2}{\longrightarrow}\ {\rm Jet}^1 V\
\overset{\varphi_3}{\longrightarrow}\ V\ \overset{\varphi_4}{\longrightarrow}\
0~,$$
where ${\rm Jet}^1V$ is the bundle of first-order jets of some other holomorphic
vector bundle  $V\to X$ (i.e.~sections of ${\rm Jet}^1V$ are sections of $V$
together with their first-order derivatives). Exactness of this sequence means
${\rm im\,}\varphi_i=\ker\varphi_{i+1}$.
Now let $m=3$ and furthermore, assume that $X$ is a spin manifold. In
particular, this means that $K$ has a square root denoted by $K^{1/2}$. Then $X$
can
be extended to a formal Calabi--Yau supermanifold of dimension $3|4$ by setting
$V:=K^{-1/2}$ and
$E:={\rm Jet}^1K^{-1/2}$, where $K^{-1/2}$ is the dual of $K^{1/2}$ (note that
for line bundles one often denotes the dual by the inverse since the tensor
product
of a line bundle with its dual is always trivial). 
To see that ${\rm Ber}$ is trivial, we use the above sequence, which implies
$\det{\rm Jet}^1V\cong\det V\otimes\det( T^{*\,1,0}X\otimes V)$, i.e.
$$ 
\begin{aligned}
 \Lambda^4 E\ &\cong\ K^{-1/2}\otimes\Lambda^3(T^{*\,1,0}X\otimes K^{-1/2})\\
             &\cong\ K^{-1/2}\otimes\Lambda^3T^{*\,1,0}X\otimes K^{-3/2}\
              \cong\ K^{-1/2}\otimes K\otimes K^{-3/2}\ \cong\ K^{-1}
\end{aligned}
$$
and hence ${\rm Ber}\cong K\otimes\Lambda^4E\cong K\otimes K^{-1}$ is trivial.
For instance, the supertwistor space $P^{3|4}$ fits into this construction
scheme, since $K\cong\CO(-4)$ and
${\rm Jet}^1K^{-1/2}\cong\CO(1)\otimes\IC^4$. The latter statement follows from
the Euler sequence (see also Appendix \ref{app:VB}):
$$
 0\ \longrightarrow\ \IC\ \longrightarrow\ \CO(1)\otimes\IC^4\ \longrightarrow\
TP^3\ \longrightarrow\ 0~.
$$
\vspace*{-20pt}
}

\vspace*{10pt}

Upon chosing this reality condition, we may invert the incidence relation. In Exercise \ref{exe:inc}, you
will derive the following result:
\begin{equation}\label{eq:invinc}
 x^{A\da}\ =\ \frac{z^A\hat\lambda^\da-\hat z^A\lambda^\da}{\lambda_\da\hat\lambda^\da}~.
\end{equation}
This shows that as real manifolds, we have the following diffeomorphisms:
\begin{equation}
 P^{3|4}\ \cong\ \IR^{4|8}\times\IC P^1\ \cong\ \IR^{4|8}\times S^2~,
\end{equation}
and hence we have a non-holomorphic fibration
\begin{equation}\label{eq:SFsuper}
\begin{aligned}
 &\pi\,:\,P^{3|4}\ \to\ \IR^{4|8}~,\\
 &\kern-1.5cm      (z^A,\lambda_\da)\ \mapsto\ x^{A\da}\ =\ \frac{z^A\hat\lambda^\da-\hat z^A\lambda^\da}{\lambda_\da\hat\lambda^\da}~.
\end{aligned}
\end{equation}
The same holds true for $P^{3|\CN}\cong\IR^{4|2\CN}\times\IC P^1\cong\IR^{4|2\CN}\times S^2$.
Therefore, in the real setting we may alternatively work with this single
fibration instead of the double fibration \eqref{eq:DFsuper} and we shall do so
in the following. This will be most convenient for our purposes.

\vspace*{5pt}

\Exercise{\label{exe:inc} Verify \eqref{eq:invinc}.}

\vspace*{5pt}

In order to write down the $\dbar$-operator on $P^{3|4}$, we need a basis for differential $(0,1)$-forms and $(0,1)$-vector fields in the non-holomorphic coordinates $(x^{A\da},\lambda_\da)$. Using the conventions
\begin{subequations}
\begin{equation}
 \frac{\partial}{\partial\lambda_\da}\lambda_\db\ =\ \delta^\da_\db~,\quad
 \frac{\partial}{\partial\lambda_\da}\lrcorner\dd\lambda_\db\ =\ \delta^\da_\db~,~\quad
 \frac{\partial}{\partial\hat\lambda_\da}\hat\lambda_\db\ =\ \delta^\da_\db~,\quad
 \frac{\partial}{\partial\hat\lambda_\da}\lrcorner\dd\hat\lambda_\db\ =\ \delta^\da_\db~
\end{equation}
and
\begin{equation}
 \frac{\partial}{\partial x^{A\da}} x^{B\db}\ =\ \delta_A^B\delta_\da^\db~,\quad
 \frac{\partial}{\partial x^{A\da}}\lrcorner\dd x^{B\db}\ =\ \delta_A^B\delta_\da^\db~,
\end{equation}
\end{subequations}
we have
\begin{subequations}\label{eq:formsvect}
\begin{equation}\label{eq:01forms}
 \bar e^0\ =\ \frac{[\dd\hat\lambda\hat\lambda]}{[\lambda\hat\lambda]^2}\eand
 \bar e^A\ =\ -\frac{\dd x^{A\da}\hat\lambda_\da}{[\lambda\hat\lambda]}~,
\end{equation}
where $[\rho\lambda]:=\varepsilon^{\da\db}\rho_\da\lambda_\db=\rho_\da\lambda^\da$, e.g.~$[\dd\hat\lambda\hat\lambda]=\dd\hat\lambda_\da\hat\lambda^\da$, together with
\begin{equation}
 \bar V_0\ =\ [\lambda\hat\lambda]\lambda_\da\frac{\partial}{\partial\hat\lambda_\da}\eand
 \bar V_A\ =\ \lambda^\da\partial_{A\da}~.
\end{equation}
\end{subequations}
In this sense, $\bar V_0\lrcorner\bar e^0=1$, $\bar V_A\lrcorner\bar e^B=\delta_A^B$, $\bar V_0\lrcorner\bar e^A=0$ and $\bar V_A\lrcorner\bar e^0=0$ and the $\dbar$-operator is $\dbar=\bar e^0\bar V_0+\bar e^A\bar V_A$.
Notice that the vector fields \eqref{eq:TDsuper} are $(0,1)$-vector fields in this real setting. The holomorphic volume form \eqref{eq:holovol} takes the form
\begin{equation}
\begin{aligned}
 \Omega\ &=\ \Omega_0\otimes\Omega_1\\
         &=\ \tfrac14[\lambda\hat\lambda]^4 e^0\wedge e^\al\wedge e_\al\otimes\Omega_1
 \ =\ \tfrac14[\dd\lambda\lambda]\wedge\varepsilon_{\al\be}\lambda_\da\lambda_\db\dd x^{\al\da}\wedge\dd x^{\be\db}
  \otimes\Omega_1~,
\end{aligned} 
\end{equation}
where $\Omega_1$ is given in \eqref{eq:holovol}. Here, we do not re-write the
$\dd\eta_i$ (or rather the corresponding integral forms) in terms of the $e_i$.
The reason for this becomes transparent in our subsequent discussion. Note also that the $e^0$, $e^A$ are obtained from \eqref{eq:01forms} via complex conjugation (or equivalently, via the involution $\tau$).

Working in Witten gauge \eqref{eq:wittengauge}, we have
\begin{equation}\label{eq:AinW}
 A^{0,1}\ =\ \bar e^0 A_0 + \bar e^\al A_\al~,
\end{equation}
where $A_0$ and $A_\al$ have a holomorphic dependence on $\eta_i$. In particular, we may expand $A^{0,1}$ according to
\begin{equation}\label{eq:etaexp}
 A^{0,1}\ =\ \underline{\Ac}+\eta_i\underline{\cc}\ \!\!^i+\tfrac{1}{2!}\eta_i\eta_j\underline{\Wc}\ \!\!^{ij}+
 \tfrac{1}{3!}\eta_i\eta_j\eta_k\varepsilon^{ijkl}\underline{\cc}_l+
 \tfrac{1}{4!}\eta_i\eta_j\eta_k\eta_l\varepsilon^{ijkl}\underline{\Gc}~,
\end{equation}
where the coefficients $\underline{\Ac}$, $\underline{\cc}\ \!\!^i$, $\underline{\Wc}\ \!\!^{ij}$, $\underline{\cc}_l$ and $\underline{\Gc}$
are differential $(0,1)$-forms of homogeneity $0$, $-1$, $-2$, $-3$ and $-4$ 
that depend only on $x,\lambda$ and $\hat\lambda$. 

At this stage, it is useful to digress a bit by discussing the linearised equations of motion.
At the linearised level, 
\eqref{eq:hCS1} reads as $\dbar A^{0,1}=0$ and the gauge transformations
\eqref{eq:HCSgauge} reduce to $A^{0,1}\mapsto \dbar\varepsilon$.
Therefore, the coefficient fields of 
\eqref{eq:etaexp} represent Lie algebra-valued elements of the Dolbeault
cohomology group 
$H^{0,1}_{\dbar}(P^3,\CO(-2h-2))$ for $h=-1,\ldots,1$. By the
Dolbeault theorem, we have $H^{0,1}_{\dbar}(P^3,\CO(-2h-2))\cong
H^1(P^3,\CO(-2h-2))$; see Remark \ref{rmk:HodgeDolbeault}~for more details.
Therefore, we can apply Theorem \ref{thm:penrose}~to conclude that on
space-time, these
fields correspond to $\fc_{\al\be}$, $\cc\ \!\!^i_\al$, $\Wc\ \!\!^{ij}$,
$\cc_{i\da}$ and $\Gc_{\da\db}$ which is precisely the field content displayed
in Table \ref{tab:fields}
Hence, one superfield $A^{0,1}$ on supertwistor space $P^{3|4}$ encodes the whole particle spectrum of $\CN=4$ self-dual supersymmetric Yang--Mills theory. 

As a next step, we need to remove the extra gauge
symmetry beyond the space-time gauge transformations (after all we want to
derive the Siegel action). To achieve this, we have to
further gauge-fix $A^{0,1}$ (on top of the already imposed Witten gauge). This
procedure is in
spirit of our discussion in
Section \ref{sec:PWsuper}, where we have shown how to go from the constraint
equations of supersymmetric self-dual Yang--Mills theory to its actual field
equations. To this end, we impose a gauge called space-time gauge
\cite{Popov:2004rb,Boels:2006ir,Boels:2007qn} 
\begin{equation}\label{eq:BMSgauge}
 \dbar^\dagger_L (\bar e^0A_0)\ =\ 0~,
\end{equation}
on all fibres of \eqref{eq:SFsuper}. Here, $\bar e^0A_0$ is understood as the
restriction of $A^{0,1}$ to the projective line $L_x\hookrightarrow P^{3|4}$ and
$\dbar^\dagger_L=-{*\dbar_L*}$ is the adjoint of $\dbar_L$ on $L_x$. This
amounts to choosing a metric on $L_x$ which we take to be the Fubini--Study
metric (see also Remark \ref{rmk:fubini}).  Put differently, we require $\bar
e^0A_0$ to be fibrewise co-closed with respect to the Fubini--Study metric on
each $L_x\cong \IC P^1$. Residual gauge transformations that preserve
\eqref{eq:BMSgauge} are then given by those $g\in H^0(P^{3|4},\sGL(r,\CS))$ that
obey
\begin{equation}
 \dbar^\dagger_L\dbar_L g\ =\ 0~,
\end{equation}
as one may straightforwardly check by inspecting \eqref{eq:HCSgauge}. If we let $(\cdot,\cdot)$ be a metric on the space of matrix-valued differential forms on each $L_x$, then $0=(g,\dbar^\dagger_L\dbar_L g)=(\dbar_L g,\dbar_Lg)\geq0$. Hence, $\dbar_L g=0$ and so $g$ is holomorphic on each fibre $L_x$ and thus it must be constant (since the $L_x$ are compact). Altogether, $g=g(x^{\al\db})$ and the residual gauge freedom is precisely space-time gauge transformations. 

\vspace*{5pt}

\Remark{\label{rmk:fubini} Consider $\IC P^1$ with the canonical cover $\{U_\pm\}$ as in Remark \ref{rmk:cover} In homogeneous coordinates $\lambda_\da$, the Fubini--Study metric is given by
$$\dd s^2\ =\ \frac{[\dd\lambda\lambda][\dd\hat\lambda\hat\lambda]}{[\lambda\hat\lambda]^2}~ $$
while in affine coordinates $\lambda_\pm$ we have
$$ \dd s^2|_{U_\pm}\ =\ \frac{\dd\lambda_\pm\dd\bar\lambda_\pm}{(1+\lambda_\pm\bar\lambda_\pm)^2}~,\ewith 
    \dd s^2|_{U_+}\ =\ \dd s^2|_{U_-}~.
$$
Here, $\hat\lambda_\da$ was defined in \eqref{eq:RCESS}. Notice that the Fubini--Study metric is the K\"ahler metric on $\IC P^1$ with the K\"ahler form
$$
 K\ =\ K(\lambda)\ =\ \frac{[\dd\lambda\lambda]\wedge[\dd\hat\lambda\hat\lambda]}{[\lambda\hat\lambda]^2}~.
$$
\vspace*{-15pt}
}

\vspace*{5pt}

Since $A^{0,1}$ is holomorphic in $\eta_i$, the gauge fixing condition $\dbar^\dagger_L (\bar e^0A_0)=0$ has to hold for each component, 
\begin{equation}
 \bar e^0 A_0\ =\ \bar e^0\underline{\Ac}_0+\eta_i\bar e^0\underline{\cc}\ \!\!^i_0+\tfrac{1}{2!}\eta_i\eta_j\bar e^0\underline{\Wc}\ \!\!^{ij}_0+
 \tfrac{1}{3!}\eta_i\eta_j\eta_k\varepsilon^{ijkl}\bar e^0\underline{\cc}_{l0}+
 \tfrac{1}{4!}\eta_i\eta_j\eta_k\eta_l\varepsilon^{ijkl}\bar e^0\underline{\Gc}_0~,
\end{equation}
i.e.
\begin{equation}
 \dbar^\dagger_L(\bar e^0\underline{\Ac}_0)\ =\ 
 \dbar^\dagger_L(\bar e^0\underline{\cc}\ \!\!^i_0)\ =\ 
 \dbar^\dagger_L(\bar e^0\underline{\Wc}\ \!\!^{ij}_0)\ =\ 
 \dbar^\dagger_L(\bar e^0\underline{\cc}_{i0})\ =\ 
 \dbar^\dagger_L(\bar e^0\underline{\Gc}_0)\ =\ 0~.
\end{equation}

\Remark{\label{rmk:HodgeDolbeault} Let $X$ be a complex manifold and $E\to X$ be
a holomorphic vector bundle. Furthermore, we denote the $\dbar$-harmonic
$(p,q)$-forms on $X$ with values in $E$ by ${\rm Harm}^{p,q}_{\dbar}(X,E)$,
where $\dbar$ is the anti-holomorphic exterior derivative. These are forms
$\omega\in{\rm Harm}^{p,q}_{\dbar}(X,E)$ that obey
$\dbar\omega=0=\dbar^\dagger\omega$. In addition, let $H^{p,q}_{\dbar} (X,E)$ be
the $\dbar$-cohomology groups with coefficients in $E$, i.e.~elements of 
$H^{p,q}_{\dbar} (X,E)$ are $\dbar$-closed $E$-valued $(p,q)$-forms which are
not $\dbar$-exact. If $X$ is compact, the Hodge theorem establishes the
following isomorphism:
$$ {\rm Harm}^{p,q}_{\dbar}(X,E)\ \cong\ H^{p,q}_{\dbar} (X,E)~.$$
The Dolbeault theorem establishes another isomorphism for $X$ not necessarily
compact
$$H^{p,q}_{\dbar} (X,E)\ \cong H^q(X,\Lambda^p T^*X\otimes E)~,$$
where $H^q(X,\Lambda^p T^*X\otimes E)$ is the $q$-th \v Cech cohomology group
with coefficients in
$\Lambda^p T^*X\otimes E$. For details, see
e.g.~\cite{Griffiths:1978,Nakahara:1990th}.
}

\noindent
Notice that $\bar e^0\underline{\Ac}_0,\ \bar e^0\underline{\cc}\ \!\!^i_0,\
\bar e^0\underline{\Wc}\ \!\!^{ij}_0,\ \bar e^0\underline{\cc}_{i0}$, and $\bar
e^0\underline{\Gc}_0$ take values in $\CO(m)\otimes{\rm End}\,E$ with $m$ being
$0,\ldots,-4$. Furthermore, $\bar e^0A_0$ is also $\dbar_L$-closed since
$\dim_\IC L_x=1$ and therefore it is harmonic along the fibres. Likewise, all
the
component fields of $\bar e^0A_0$ are fibrewise harmonic. Therefore, the Hodge
and
Dolbeault theorems (see Remark \ref{rmk:HodgeDolbeault}) then tell us that the
restriction of these components to the fibres are ${\rm End}\,E$-valued elements
of $H^1(\IC P^1,\CO(-2h-2))$ since $L_x\cong\IC P^1$. From Table \ref{tab:CCGB},
we conclude that $H^1(\IC P^1,\CO(-2h-2))=0$ for $h=-1,-\frac12$ and hence
$\underline{\Ac}_0=\underline{\cc}\ \!\!^i_0=0$. Furthermore, $H^1(\IC
P^1,\CO(-2h-2))\cong\IC^{2h+1}$ for $h\geq0$ and therefore, the other 
components may be taken as
(see also \cite{Woodhouse:id}):
\begin{equation}\label{eq:TF}
 \begin{aligned}
   \underline{\Ac}\ &=\  \bar e^0\underline{\Ac}_0+\bar e^\al \underline{\Ac}_\al\ =\ \bar e^\al \underline{\Ac}_\al~,\\
   \underline{\cc}\ \!\!^i\ &=\ \bar e^0\underline{\cc}\ \!\!^i_0+ \bar e^\al \underline{\cc}\ \!\!^i_\al\ =\ \bar e^\al \underline{\cc}\ \!\!^i_\al~,\\
   \underline{\Wc}\ \!\!^{ij}\ &=\ \bar e^0\underline{\Wc}\ \!\!^{ij}_0+\bar
e^\al\underline{\Wc}\ \!\!^{ij}_\al\ =\ \bar e^0\Wc\ \!\!^{ij}+\bar
e^\al\underline{\Wc}\ \!\!^{ij}_\al~,\\
   \underline{\cc}_{i}\ &=\ \bar e^0\underline{\cc}_{i0}+ \bar e^\al \underline{\cc}_{i\al}\ =\ 2\bar e^0\frac{\cc_{i\da}\hat\lambda^\da}{[\lambda\hat\lambda]}+
                           \bar e^\al \underline{\cc}_{i\al}~,\\
   \underline{\Gc}\ &=\ \bar e^0\underline{\Gc}_0+\bar e^\al
\underline{\Gc}_\al\ =\  3\bar
e^0\frac{\Gc_{\da\db}\hat\lambda^\da\hat\lambda^\db}{[\lambda\hat\lambda]^2}
           + \bar e^\al \underline{\Gc}_\al~.
 \end{aligned}
\end{equation}
Here, the fields $\Wc\ \!\!^{ij}$, $\cc_{i\da}$ and $\Gc_{\da\db}$ depend only on the space-time coordinates $x^{\al\db}$ while the remaining fields still may depend on $\lambda_\da$ and $\hat\lambda_\da$.
Notice that we use $\underline{\Wc}\ \!\!^{ij}=\frac12\varepsilon^{ijkl}\underline{\Wc}_{kl}$ and similarly for its components.

The non-trivial step in writing \eqref{eq:TF} is in setting $\underline\Ac_0=0$
(and $\underline{\cc}\ \!\!^i_0=0$ by supersymmetry). This implies that $E$ is
holomorphically trivial on any $L_x\hookrightarrow P^{3|4}$, which is one of the
assumptions in Ward's theorem; see Theorems \ref{thm:ward} and
\ref{thm:wardsusy} Notice that this is not guaranteed in general, but will
follow from the smallness assumption of $\underline\Ac_0$ in a perturbative
context.

In terms of the expansions \eqref{eq:TF}, the action \eqref{eq:HCSA} reads 
\begin{equation}\label{eq:eq3}
 \begin{aligned}
  S\ &=\ -\frac{1}{2\pi\di}\int\frac{\Omega_0\wedge\bar\Omega_0}{[\lambda\hat\lambda]^4}\,{\rm tr}
  \left\{\frac{\Gc_{\da\db}\hat\lambda^\da\hat\lambda^\db}{[\lambda\hat\lambda]^2}\left(
       -\lambda^\dc\partial_{\al\dc}\underline{\Ac}\ \!\!^\al+\tfrac12\Big[\underline{\Ac}\ \!\!^\al,
       \underline{\Ac}_\al\Big]\right)\right.\\
   &\kern2cm+\,\frac{\cc_{i\da}\hat\lambda^\da}{[\lambda\hat\lambda]}
     \left(\lambda^\db\partial_{\al\db}\underline{\cc}\ \!\!^{i\al}+\Big[\underline{\Ac}\
      \!\!_\al,\underline{\cc}\ \!\!^{i\al}\Big]\right)
     -\tfrac12\Wc_{ij}\left(\lambda^\db\partial_{\al\db}\underline{\Wc}\ \!\!^{ij\al}+\Big[
       \underline{\Ac}\ \!\!_\al,
        \underline{\Wc}\ \!\!^{ij\al}\Big]\right)\\
    &\kern1.5cm\left.\phantom{\frac{\Gc_{\da\db}\hat\lambda^\da\hat\lambda^\db}{(\lambda\cdot\hat\lambda)^2}}
        +\,\tfrac12\Wc_{ij}\{\underline{\cc}\ \!\!^{i}_\al,\underline{\cc}\ \!\!^{j\al}\}+
      \left(\underline{\Gc}\ \!\!^\al\bar V_0\underline{\Ac}_\al
          -\underline{\cc}\ \!\!_{i\al}\bar V_0\underline{\cc}\ \!\!^{i\al}+
           \tfrac14\underline{\Wc}\ \!\!_{ij\al}\bar V_0\underline{\Wc}\ \!\!^{ij\al}\right)\right\}.
 \end{aligned}
\end{equation}
Note that in actual fact one should also add a gauge-fixing term and a ghost
term. However, the ghost-matter interaction is such that the Fadeev--Popov
determinant is independent of $A^{0,1}$. See \cite{Boels:2006ir} for more
details. For the time being, we therefore omit these pieces.

The fields $\underline{\Gc}_\al$ and $\underline{\cc}_{i\al}$ appear only
linearly, so they are in fact Lagrange multiplier fields. Integrating them out
enforces
\begin{equation}
 \bar V_0 \underline{\Ac}_\al\ =\ 0\eand
  \bar V_0 \underline{\cc}\ \!\!^i_\al\ =\ 0~
\end{equation}

\pagebreak
\noindent
and so $\underline{\Ac}_\al$ and $\underline{\cc}\ \!\!^i_\al$ must be
holomorphic in $\lambda_\da$. Hence,
\begin{equation}\label{eq:eq1}
 \underline{\Ac}_\al\ =\ \lambda^\db\Ac_{\al\db}\eand
 \underline{\cc}\ \!\!^i_\al\ =\ \cc\ \!\!^i_\al
\end{equation}
since they are of homogeneity $1$ and $0$, respectively. Here, $\Ac_{\al\db}$ and $\cc\ \!\!^i_\al$
depend only on $x^{\al\db}$. Likewise, $\underline{\Wc}\ \!\!^{ij}_\al$ can be
eliminated as it appears only quadratically with a constant
coefficient.\footnote{Notice that if one integrates out generic quadratic pieces
in the path integral, then one picks up field-dependent functional determinants.
However, in our case the `$\phi^2$-piece' has a constant coefficient and
hence, the
determinant is constant and thus can be absorbed in the normalisation of the
path integral measure.} We have
\begin{equation}
 \bar V_0 \underline{\Wc}\ \!\!^{ij}_\al\ =\ \lambda^\db
 \left(
       \partial_{\al\db}\Wc\ \!\!^{ij}+\Big[\Ac_{\al\db},
       \Wc\ \!\!^{ij}\Big]\right)
\end{equation}
which yields
\begin{equation}\label{eq:eq2}
  \underline{\Wc}\ \!\!^{ij}_\al\ =\ \frac{1}{[\lambda\hat\lambda]}\hat\lambda^\db\cnab_{\al\db}
   \Wc\ \!\!^{ij}~,\ewith
   \cnab_{\al\db}\ :=\ \partial_{\al\db}+\Ac_{\al\db}~.
\end{equation}
Inserting \eqref{eq:eq1} and \eqref{eq:eq2} into \eqref{eq:eq3}, we find
\begin{equation}
 \begin{aligned}
  S\ &=\ -\frac{1}{2\pi\di}\int\frac{\Omega_0\wedge\bar\Omega_0}{[\lambda\hat\lambda]^4}\,{\rm tr}
  \left\{\Gc_{\da\db} \fc_{\dc\dot\delta}\frac{\hat\lambda^\da\hat\lambda^\db\lambda^\dc\lambda^{\dot\delta}}{[\lambda\hat\lambda]^2}+\cc_{i\da}\cnab\ \!\!_{\be\dc}\cc\ \!\!^{i\be}\frac{\hat\lambda^\da\hat\lambda^\dc}{[\lambda\hat\lambda]}\right.\\
   &\kern5cm\left.+\,\tfrac14\Wc_{ij}\cnab_{\al\db}\cnab\ \!\!^\al_{\ \dc}\Wc\ \!\!^{ij}
 \frac{\hat\lambda^\da\lambda^\dc}{[\lambda\hat\lambda]}+\tfrac12\Wc_{ij}\{\cc\ \!\!^{i}_\al,\cc\ \!\!^{j\al}\}\right\},
 \end{aligned}
\end{equation}
where $\fc_{\da\db}$ is the anti-self-dual part of the curvature of $\Ac_{\al\db}$ (see also \eqref{eq:curvdec}).
Using
\begin{equation}\label{eq:measure}
 \frac{\Omega_0\wedge\bar\Omega_0}{[\lambda\hat\lambda]^4}\ =\ 
 \dd^4 x\frac{[\dd\lambda\lambda][\dd\hat\lambda\hat\lambda]}{[\lambda\hat\lambda]^2}~
\end{equation}
together with the formula derived in Exercise \ref{exe:formula},
we can integrate out the fibres to finally arrive at
\begin{equation}\label{eq:siegelaction2}
 S\ =\ \int\dd^4x \,{\rm tr}
 \left\{\Gc\ \!\!^{\da\db}\fc_{\da\db}+
         \cc\ \!\!^{i\al}\cnab_{\al\da}\cc\ 
         \!\!^\da_i-\tfrac{1}{2}\Wc_{ij}
                \overset{\circ}{\square}\Wc\ \!\!^{ij}
             +\tfrac12\Wc_{ij}\{\cc\ \!\!_\al^i,\cc\ \!\!^{j\al}\} \right\}.
\end{equation}
This is nothing but the Siegel action \eqref{eq:siegelaction}.

Altogether, the holomorphic Chern--Simons action \eqref{eq:HCSA} on supertwistor
space $P^{3|4}$ corresponds to the Siegel action \eqref{eq:siegelaction} on
space-time. This concludes Figure \ref{fig:summary}.~and our discussion about
$\CN=4$ supersymmetric self-dual Yang--Mills theory.\footnote{For twistor
constructions including action functionals of other gauge field theories, see
\cite{Popov:2004nk,Saemann:2004tt,Giombi:2004xv,Chiou:2005jn,Popov:2005uv,
Lechtenfeld:2005xi,
Mason:2005kn,Boels:2006ir,Boels:2007qn,Bedford:2007qj}.}

\pagebreak

\Exercise{\label{exe:formula} Verify \eqref{eq:measure}. Furthermore, prove that
          $$
           -\frac{1}{2\pi\di}\int K\,
                f_{\da_1\cdots\da_m}g^{\db_1\cdots\db_m}\frac{\lambda_{\db_1}\cdots\lambda_{\db_m}
             \hat\lambda^{\da_1}\cdots\hat\lambda^{\da_m}}{[\lambda\hat\lambda]^m}\ =\ 
              \frac{1}{m+1}f_{\da_1\cdots\da_m}g^{\da_1\cdots\da_m}~,
          $$
          where $K$ is the K\"ahler form given in Remark \ref{rmk:fubini}~and
$g^{\da_1\cdots\da_m}$ and $f_{\da_1\cdots\da_m}$ do not depend on $\lambda$ and
$\hat\lambda$. In fact, this is the Dolbeault version of Serre duality discussed
in Remark \ref{rmk:serre}~(see also Remark \ref{rmk:HodgeDolbeault}): 
            $$ H^{0,0}_{\dbar}(\IC P^1,\CO(m))\ \cong\ H^{0,1}_{\dbar}(\IC P^1,\CO(-m-2))^*~.$$
          More concretely, $g=g^{\da_1\cdots\da_m}\lambda_{\da_1}\dots\lambda_{\da_m}$ represents an element of  
          $H^{0,0}_{\dbar}(\IC P^1,\CO(m))\cong H^0(\IC P^1,\CO(m))$ while 
          $$ f\ =\ (m+1)f_{\da_1\cdots\da_m}
          \frac{\hat\lambda^{\da_1}\cdots\hat\lambda^{\da_m}}{{[\lambda\hat\lambda]^m}}\,
                   \frac{[\dd\hat\lambda\hat\lambda]}{[\lambda\hat\lambda]^2}$$ 
          is an element of $H^{0,1}_{\dbar}(\IC P^1,\CO(-m-2))\cong H^1(\IC
P^1,\CO(-m-2))$. Here, we have used \eqref{eq:01forms}. Then the Dolbeault
version of the pairing \eqref{eq:pairing} is given by
          $$ (f,g)\ =\ -\frac{1}{2\pi\di}\int[\dd\lambda\lambda]\wedge fg \ =\ 
              f_{\da_1\cdots\da_m}g^{\da_1\cdots\da_m}~,$$
          which is nothing but the above formula.}

\vspace*{10pt}

\section{$\CN=4$ supersymmetric Yang--Mills theory from supertwistor space}\label{sec:SYMTP}

\subsection{Motivation}\label{eq:MotN4SYM}

So far, we have shown how $\CN=4$ supersymmetric self-dual Yang--Mills theory can be understood in terms of twistor geometry. The field content of this theory is displayed in Table \ref{tab:fields} and the corresponding action in \eqref{eq:siegelaction}. In fact, the full $\CN=4$ supersymmetric Yang--Mills theory has precisely the same field content but it differs in the interaction terms. The action is given by
\begin{equation}\label{eq:SYMAct}
 \begin{aligned}
   S\ &=\ \int\dd^4x \,{\rm tr}
 \left\{\fc\ \!\!^{\da\db}\fc_{\da\db}+\fc\ \!\!^{\al\be}\fc_{\al\be}
         +\Wc_{ij}\overset{\circ}{\square}\Wc\ \!\!^{ij}
         -2\cc\ \!\!^{i\al}\cnab_{\al\da}\cc\ 
         \!\!^\da_i \right.\\
  &\kern3cm\left.-\,\Wc_{ij}\{\cc\ \!\!_\al^i,\cc\ \!\!^{j\al}\}
         -\Wc\ \!\!^{ij}\{\cc\ \!\!_{i\da},\cc\ \!\!_j^\da\}
         +\tfrac18[\Wc_{ij},\Wc_{kl}][\Wc\ \!\!^{ij},\Wc\ \!\!^{kl}]\right\}.
 \end{aligned}
\end{equation}

By comparing the Siegel action \eqref{eq:siegelaction2} with the action \eqref{eq:SYMAct}, we realise that the Siegel action provides only parts of the interaction terms. What about the other terms? Can they also be understood in terms of twistors? 

Before delving into this, let us first consider purely bosonic self-dual Yang--Mills theory. This theory may be described by the action
\begin{equation}
 S\ =\ \int{\rm tr}(G\wedge F^-)~,
\end{equation}
since the corresponding field equations are
\begin{equation}
 F^-\ =\ 0\eand \nabla{*G}\ =\ 0~.
\end{equation}
Hence, we obtain the self-dual Yang--Mills equations plus an equation for an
anti-self-dual field $G$ propagating in the self-dual background. Notice that
$G$ is nothing but the field $G_{\da\db}$ we already encountered. Let us
modify the above action by adding the term \cite{Witten:2003nn}
\begin{equation}\label{eq:modYM}
 S_\varepsilon\ =\ -\tfrac12\varepsilon\,\int{\rm tr}(G\wedge G)~,\ewith
 S_{\rm tot}\ =\ S+S_\varepsilon
\end{equation}
where $\varepsilon$ is some small parameter. Upon
integrating out the field $G$, we find
\begin{equation}
 S_{\rm tot}\ =\ \tfrac{1}{2\varepsilon}\int{\rm tr}(F^-\wedge F^-)\ =\ 
                 -\tfrac{1}{4\varepsilon}\int{\rm tr}(F\wedge {*F})
                 +\tfrac{1}{4\varepsilon}\int{\rm tr}(F\wedge {F})~,
\end{equation}
where we have used $F^-=\frac12(F-{*F})$. Hence, we obtain the Yang--Mills
action \eqref{eq:YM} plus a term proportional to the topological charge
\eqref{eq:charge}, provided we identify $\varepsilon$ with the Yang--Mills
coupling constant $g_{\rm YM}^2$. Therefore, small $\varepsilon$ corresponds to
small $g_{\rm YM}$. As we are about to study perturbation theory, the
topological term will not play a role and we may therefore work with
\eqref{eq:modYM}. Consequently, we may re-phrase our question: Can we
derive $\int{\rm tr}(G\wedge G)$ or rather its $\CN=4$ supersymmetric extension
from twistor space? This will be answered in the next section.

\vspace*{5pt}

\Exercise{Show that the action \eqref{eq:SYMAct} is invariant under the
following supersymmetry transformations ($\varrho^{i\al}$ and $\varrho_i^\da$
are constant anti-commuting spinors):
$$
\begin{aligned}
 \delta\Ac_{\al\da}\ &=\
-\varepsilon_{\al\be}\varrho^{i\be}\cc_{i\da}+\varepsilon_{\da\db}
\varrho^\db_i\cc\ \!\!^i_\al~,\\
 \delta\Wc_{ij}\ &=\ \varepsilon_{ijkl}\varrho^{k\al}\cc\
\!\!^l_\al-2\varrho^\da_{[i}\cc_{j]\da}~,\\
 \delta\cc\ \!\!^i_\al\ &=\
-2\varrho^{i\be}\fc_{\al\be}+\varepsilon_{\al\be}\varrho^{j\be}[\Wc\ \!\!^{ik},
 \Wc_{jk}]-2\varrho^\da_j\cnab_{\al\da}\Wc\ \!\!^{ij}~,\\
 \delta\cc_{i\da}\ &=\
2\varrho^{j\al}\cnab_{\al\da}\Wc_{ij}+2\varrho^\db_i\fc_{\da\db}+\varepsilon_{
\da\db}
 \varrho^\db_j[\Wc\ \!\!^{jk},\Wc_{ik}]~.
\end{aligned}
$$
\vspace*{-10pt}
}

\vspace*{10pt}

\subsection{$\CN=4$ supersymmetric Yang--Mills theory from supertwistor
space}\label{sec:SYMFromTwistor}

In parts, the following discussion follows the lines of
\cite{Mason:2005zm,Boels:2006ir}.
To obtain the full $\CN=4$ supersymmetric Yang--Mills action, one considers the
following modification:
\begin{subequations}
\begin{equation}\label{eq:full}
 S_{\rm tot}\ =\ S+S_\varepsilon~,
\end{equation}
where $S$ is the holomorphic Chern--Simons action \eqref{eq:HCSA} and
$S_\varepsilon$ is given by \cite{Boels:2006ir}
\begin{equation}\label{eq:HCSmod}
 S_\varepsilon\ =\ -\varepsilon \int\dd^{4|8}x\, \log\det\nabla^{0,1}|_{L_x}~.
\end{equation}
\end{subequations}
One takes the $(0,1)$-connection $\nabla^{0,1}$, restricts it to the
fibre
$L_x\hookrightarrow P^{3|4}$, constructs the determinant of this operator and
finally integrates the logarithm of this determinant over $x\in M^{4|8}$
parametrising $L_x\cong\IC P^1$. Here, we have used the common abbreviation
$\dd^{4|8}x:=\dd^4x\dd^8\eta$.
Note that the action \eqref{eq:HCSmod} is invariant under the gauge transformations \eqref{eq:HCSgauge}.
This will be verified in Exercise \ref{exe:anomaly} Before showing that $S_\varepsilon$ really gives the missing interactions, let us emphasise that all the constructions presented in the preceding section also apply to 
the full twistor space action \eqref{eq:full} without alteration. We may
therefore stick to $S_\varepsilon$ for the remainder of this section.

\Exercise{\label{exe:anomaly} Argue that the action \eqref{eq:HCSmod} is invariant under the gauge transformations
\eqref{eq:HCSgauge}. You might find it useful to know that the
determinant $\det\nabla^{0,1}|_{L_x}$ behaves under infinitesimal
gauge transformations $\delta A^{0,1}=\nabla^{0,1}\varepsilon$, with
$\varepsilon$ being an infinitesimal gauge parameter, as
$$ \delta\det\nabla^{0,1}|_{L_x}\ =\ \left(\frac{1}{2\pi}\int_{L_x}{\rm
tr}(A^{0,1}\wedge\partial\varepsilon)\right)\det\nabla^{0,1}|_{L_x}~.
$$
This formula is, in fact, the so-called chiral anomaly with $\nabla^{0,1}$ on $L_x$ being the chiral Dirac operator. From a physical point of view, it can best be understood in a path integral language. If we consider two fermionic fields $\alpha$ and $\beta$ of homogeneity $-1$ on $L_x$ taking values in $E|_{L_x}$ and $E^*|_{L_x}$ respectively with the action
$$S_{\beta\alpha}\ =\ \int_{L_x}[\dd\lambda\lambda]\wedge\beta\,\nabla^{0,1}\alpha~,$$
then
$$ \det\nabla^{0,1}|_{L_x}\ =\ \de^{-W[A^{0,1}]}\ =\ \int \CCD\alpha\CCD\beta\,\exp\left(-\int_{L_x}[\dd\lambda\lambda]\wedge\beta\,\nabla^{0,1}\alpha\right),$$
where $W[A^{0,1}]$ is the effective action.
The above formula of the gauge variation of the determinant arises by studying
variations $A^{0,1}\mapsto A^{0,1}+\delta
A^{0,1}=A^{0,1}+\nabla^{0,1}\varepsilon$ of the path integral. See
e.g.~\cite{Nakahara:1990th,Bertlmann:1996} for more details.
}

\vspace*{10pt}

In what follows, we shall need the Green function of the $\dbar$-operator on
$\IC P^1$ and so it will be useful to recapitulate some of its properties first.
To this end, notice that any differential $(0,1)$-form on $\IC P^1$ is automatically $\dbar$-closed since $\dim_\IC\IC P^1=1$. Furthermore, the Dolbeault cohomology groups $H^{0,1}_{\dbar}(\IC P^1,\CO(m))$ vanish for $m\geq-1$ since $H^{0,1}_{\dbar}(\IC P^1,\CO(m))\cong H^1(\IC P^1,\CO(m))$ and the \v Cech cohomology groups $H^1(\IC P^1,\CO(m))$ are given in Table \ref{tab:CCGB}. Therefore, any differential $(0,1)$-form $\omega$ of homogeneity $m\geq-1$ is necessarily $\dbar$-exact, i.e.~$\omega=\dbar\rho$ for some $\rho$ of homogeneity $m$.  We shall denote the element $\rho$ by $\dbar^{-1}\omega$, in the following.
Again from Table \ref{tab:CCGB}., we conclude that $\dbar^{-1}\omega$ is not unique for $m\geq0$ as one can always add a global holomorphic function $\alpha$ of homogeneity $m$: $\dbar(\dbar^{-1}\omega+\alpha)=\omega$.
For $m=-1$, however, there are no global holomorphic functions and so $\dbar^{-1}\omega$ is unique. Explicitly, we have
\begin{equation}
 \dbar^{-1}\omega(\lambda)\ =\ -\frac{1}{2\pi\di}\int\frac{[\dd\lambda'\lambda']}{[\lambda\lambda']}\wedge\omega(\lambda')~.
\end{equation}
In deriving this result, one makes use of the basis \eqref{eq:formsvect}
\begin{equation}
 \omega(\lambda)\ =\ \bar e^0 \omega_0(\lambda_\da,\hat\lambda_\da)\ =\ \frac{[\dd\hat\lambda\hat\lambda]}{[\lambda\hat\lambda]^2}\,\omega_0(\lambda_\da,\hat\lambda_\da)~
\end{equation}
together with
\begin{subequations}
\begin{equation}\label{eq:delta}
 \dbar(\lambda)\,\frac{1}{[\lambda\lambda']}\ =\
2\pi[\dd\hat\lambda\hat\lambda]\,\frac{[\xi\hat\lambda']}{[\xi\hat\lambda]}\,
\delta^{(2)}([\lambda\lambda'],[\hat\lambda\hat\lambda'])~.
\end{equation}
Here, $\xi^\da$ is some
constant spinor. The complex delta function is given by
\begin{equation}
\delta^{(2)}([\lambda\lambda'],[\hat\lambda\hat\lambda'])\ :=\ 
\tfrac12\delta(\mathfrak{Re}[\lambda\lambda'])\delta(\mathfrak{Im}[\lambda\lambda'])~
\end{equation}
\end{subequations}
and obeys $\delta^{(2)}(t[\lambda\lambda'],\bar t[\hat\lambda\hat\lambda'])=(t\bar t)^{-1}\delta^{(2)}([\lambda\lambda'],[\hat\lambda\hat\lambda'])$ for $t\in\IC\setminus\{0\}$.
On support of the delta function, $\lambda_\da\propto\lambda_\da'$ and therefore the expression \eqref{eq:delta} is independent of the choice of $\xi^\da$. Notice that
$\omega_0(t\lambda_\da,\bar t\hat\lambda_\da)=t\omega_0(\lambda_\da,\hat\lambda_\da)$ for $t\in\IC\setminus\{0\}$
since $\omega$ is of homogeneity $-1$. Notice also that one often writes
\begin{equation}\label{eq:Greens}
 G(\lambda,\lambda')\ :=\ -\frac{1}{2\pi\di\,[\lambda\lambda']}\qquad\Longrightarrow\qquad
 \dbar^{-1}\omega(\lambda)\ =\ \int
K(\lambda')\,G(\lambda,\lambda')\,\omega_0(\lambda')~,
\end{equation}
where $G(\lambda,\lambda')$ is referred to as the Green function of the
$\dbar$-operator or equivalently, as the integral kernel of
$\dbar^{-1}$-operator. Furthermore, $K(\lambda)$ is the K\"ahler form
introduced in Remark \ref{rmk:fubini}~and
$\omega_0(\lambda)$ is short-hand notation for
$\omega_0(\lambda_\da,\hat\lambda_\da)$. 

\vspace*{5pt}

\Exercise{Verify \eqref{eq:delta}. You may find it useful to consider
($\varepsilon\in\IR$)
$$\dbar(\lambda)\,\frac{[\hat\lambda\hat\lambda']}{[\lambda\lambda']
[\hat\lambda\hat\lambda']+\varepsilon^2}
$$
in the limit $\varepsilon\to0$.
}

\vspace*{5pt}

After this digression, let us come back to the action $S_\varepsilon$. To compute its space-time form,
we recall the identity 
\begin{equation}
 \log\det\nabla^{0,1}|_{L_x}\ =\ {\rm tr}\log \nabla^{0,1}|_{L_x}
\end{equation}
and consider the following power series expansion:
\begin{eqnarray}\label{eq:power}
  {\rm tr}\log \nabla^{0,1}|_{L_x} \!&=&\! {\rm tr}\log(\dbar+A^{0,1})|_{L_x}\notag\\
    \! &=&\! {\rm tr}\,\big[\log\dbar|_{L_x}+\log(1+\dbar^{-1}A^{0,1})|_{L_x}\big]\notag\\
   \!&=&\! {\rm tr}\left[\log\dbar|_{L_x}+\sum_{r=1}^\infty\frac{(-1)^r}{r}
      \int_{L_x}\prod_{s=1}^r K(\lambda_s)G(\lambda_{s-1},\lambda_s)A_0(\lambda_s)\right],
\end{eqnarray}
with the identification $\lambda_0\equiv\lambda_r$. Here, we have inserted the
Green function \eqref{eq:Greens}. In addition, we used \eqref{eq:AinW},
i.e.~each $A_0$ is the $\bar e^0$-coefficient of $A^{0,1}=\bar e^0 A_0+\bar
e^\al A_\al$ and restricted to lie on a copy of the fibre over 
the same point $x\in M^{4|8}$, 
i.e.~$\lambda_s\in L_x\hookrightarrow P^{3|4}$ for all $s$.
Notice that $A_0$ is of homogeneity $2$ as follows from the scaling behaviour of $\bar e^0$.

Adopting the gauge \eqref{eq:BMSgauge}, 
we see that the series \eqref{eq:power} terminates after the fourth term, since
in this gauge $A_0$ starts only at second-order in the $\eta_i$-expansion
(compare \eqref{eq:etaexp} with
\eqref{eq:TF}). In addition, we eventually integrate over $\dd^{4|8}x$, so we only need to keep terms that contain all $\eta_i^\da$. In this respect, recall that on $L_x$ we have $\eta_i=\eta_i^\da\lambda_\da$. Then there are only three types of relevant terms: Firstly, we have the $\Gc_{\da\db}\Gc\ \!\!^{\da\db}$-term which is
\begin{equation}
\begin{aligned}
 &-\varepsilon\int\dd^{4|8}x\,\frac12\left(\frac{3}{2\pi\di}\right)^2\int_{L_x}
 \prod_{s=1}^2
 \frac{K(\lambda_s)}{[\lambda_{s-1}\lambda_{s}]}\\
 &\kern2cm\times\,{\rm tr}\left\{
 \frac{1}{4!}\varepsilon^{ijkl}\eta_{1i}\eta_{1j}\eta_{1k}\eta_{1l}
 \frac{\Gc_{\da\db}\hat\lambda_1^\da\hat\lambda_1^\db}{[\lambda_1\hat\lambda_1]^2}\,
 \frac{1}{4!}\varepsilon^{ijkl}\eta_{2i}\eta_{2j}\eta_{2k}\eta_{l\,2}
 \frac{\Gc_{\da\db}\hat\lambda_2^\da\hat\lambda_2^\db}{[\lambda_2\hat\lambda_2]^2}
 \right\}.
\end{aligned}
\end{equation}
The integration over the fermionic coordinates can be performed using Nair's lemma
\begin{equation}\label{eq:Nairlemma}
 \int\dd^8\eta\, (\eta_1)^4(\eta_2)^4\ =\ [\lambda_1\lambda_2]^4~,\ewith
(\eta_s)^4\ :=\ 
 \tfrac{1}{4!}\varepsilon^{ijkl}\eta_{si}\eta_{sj}\eta_{sk}\eta_{sl}~
\end{equation}
and $\eta_{si}=\eta^\da_i\lambda_{s\da}$,
while the integration over $L_x$ is performed with the help of the formula derived in Exercise \ref{exe:formula} We find
\begin{equation}
 -\tfrac{\varepsilon}{2}\int\dd^4x\,{\rm tr}\big\{\Gc_{\da\db}\Gc\ \!\!^{\da\db}\big\}~.
\end{equation}

Next let us look at terms of the form $\Wc\ \!\!^{ij}\cc_{i\da}\cc_{j\db}$ and
terms that are quartic in the scalars $\Wc_{ij}$. They come from the expressions
\begin{equation}\label{eq:term1}
\begin{aligned}
 &-\varepsilon\int\dd^{4|8}x\,\frac{4}{(2\pi\di)^3}\int_{L_x}\prod_{s=1}^3
 \frac{K(\lambda_s)}{[\lambda_{s-1}\lambda_{s}]}\,\\
 &\kern2cm\times\,{\rm tr}
 \left\{\frac{1}{2!}
 \eta_{1i}\eta_{1j}\Wc\ \!\!^{ij}\,
 \frac{1}{3!}\varepsilon^{ijkl}\eta_{2i}\eta_{2j}\eta_{2k}
 \frac{\cc_{l\da}\hat\lambda^\da_2}{[\lambda_2\hat\lambda_2]}\,
 \frac{1}{3!}\varepsilon^{ijkl}\eta_{3i}\eta_{3j}\eta_{3k}
 \frac{\cc_{l\da}\hat\lambda^\da_3}{[\lambda_3\hat\lambda_3]}
\right\}
\end{aligned}
\end{equation}
and
\begin{equation}\label{eq:term2}
\begin{aligned}
 &-\varepsilon\int\dd^{4|8}x\,\frac{1}{4(2\pi\di)^4}\int_{L_x}\prod_{s=1}^4
 \frac{K(\lambda_s)}{[\lambda_{s-1}\lambda_{s}]}\\
 &\kern2cm\times\,{\rm tr}\left\{\frac{1}{2!}\eta_{1i}\eta_{1j}\Wc\ \!\!^{ij}\, 
         \frac{1}{2!}\eta_{2i}\eta_{2j}\Wc\ \!\!^{ij}\,
         \frac{1}{2!}\eta_{3i}\eta_{3j}\Wc\ \!\!^{ij}\, 
         \frac{1}{2!}\eta_{4i}\eta_{4j}\Wc\ \!\!^{ij}\right\},
\end{aligned}
\end{equation}
respectively. In similarity with the $\Gc_{\da\db}\Gc\ \!\!^{\da\db}$-term, we
may now integrate out the fermionic coordinates and the fibres. We leave this as
an exercise and only state the final result. 
Collecting all the terms, we find that the action \eqref{eq:HCSmod} reduces to
\begin{equation}\label{eq:term3}
 S_\varepsilon\ =\ -\varepsilon\int\dd^4x\,{\rm tr}\left\{\tfrac12 \Gc_{\da\db}\Gc\ \!\!^{\da\db}+
  \Wc\ \!\!^{ij}\{\cc_{i\da},\cc\ \!\!_j^\da\}-\tfrac{1}{8}[\Wc\ \!\!^{ij},\Wc\ \!\!^{kl}]
  [\Wc_{ij},\Wc_{kl}]\right\}.
\end{equation}
Combining this with the Siegel action \eqref{eq:siegelaction2} and integrating out $\Gc_{\da\db}$, we arrive at the full $\CN=4$ supersymmetric Yang--Mills action \eqref{eq:SYMAct}, modulo the topological term
(see also Exercise \ref{exe:SDYM}). Therefore, the twistor action
\eqref{eq:full} is indeed perturbatively equivalent to the $\CN=4$
supersymmetric  Yang--Mills action on space-time.

\Exercise{Complete the calculation by integrating in \eqref{eq:term1}, \eqref{eq:term2} over the fermionic coordinates and the fibres to arrive at \eqref{eq:term3}. Verify Nair's lemma \eqref{eq:Nairlemma}.}

This concludes our discussion of the twistor re-formulation of gauge theory.
The next part of these lecture notes is devoted to the construction of
tree-level gauge theory scattering amplitudes by means of twistors.

\vspace*{5pt}

\Remark{Apart from the approach discussed here, there exist alternative
twistor constructions of (full) supersymmetric Yang--Mills theory. See
\cite{Witten:1978xx,Isenberg:1978kk,Manin:1988ds,Eastwood:1987,Harnad:1988rs,
Howe:1995md}. In \cite{Mason:2005kn}, a twistor action for $\CN=4$
supersymmetric Yang--Mills theory was proposed using the so-called ambidextrous
approach that uses both twistors and dual twistors. For an ambidextrous approach
to maximally supersymmetric Yang--Mills theory in three dimensions, see
\cite{Saemann:2005ji}.}

\vspace*{5pt}

\Remark{Since most of the things said above in the supersymmetric setting were
sparked off by Witten's twistor string theory \cite{Witten:2003nn}, let us make
a few comments on that. His work is based on three facts: I) Holomorphic vector
bundles over $P^{3|4}$ are related to $\CN=4$ supersymmetric self-dual
Yang--Mills theory on $M^4$, II) the supertwistor space $P^{3|4}$ is a
Calabi--Yau supermanifold and III) the existence of a string theory, the open
topological B model, whose effective action is the holomorphic Chern--Simons
action \cite{Witten:1992fb}. Roughly speaking, topological string theories are
simplified versions of string theories obtained by giving different spins to the
worldsheet description of ordinary string theories. Topological string theories
come in two main versions, the A and B models, which are related by a duality
called mirror symmetry. 
See e.g.~\cite{Hori:2003ic} for more details.
Unlike ordinary string theories, topological string theories are exactly solvable in the sense that the computation of correlation functions of physical observables is reduced to classical questions in geometry.

\hspace*{15pt} Witten then interprets the perturbative expansion of the full $\CN=4$ supersymmetric Yang--Mills theory as an instanton expansion of the open topological B model on $P^{3|4}$ (see also
\cite{Berkovits:2004hg,Mason:2007zv} for alternative formulations). At the end
of the day, this corresponds to complementing the holomorphic Chern--Simons
action by an additional term, similarly to what we have done above. Witten's
string approach leads to $\int\det\nabla^{0,1}$. Witten \& Berkovits
\cite{Berkovits:2004jj} soon realised, however, that $\CN=4$ supersymmetric
Yang--Mills theory can be described this way only at tree-level in perturbation
theory, since at loop-level conformal supergravity is inextricably mixed in
with the
gauge theory. The reason for that is the non-gauge invariance of the string
theory formula $\int\det\nabla^{0,1}$. Gauge invariance of
$\int\det\nabla^{0,1}$ may be restored by compensating gauge tranformations of
fields from the closed string sector \cite{Berkovits:2004jj}. These observations
led Boels, Mason \& Skinner \cite{Boels:2006ir} to suggest the twistor action
\eqref{eq:full} with \eqref{eq:HCSmod} which is gauge invariant and thus free of
conformal supergravity (roughly speaking, the logarithm cancels out the
multi-trace contributions responsible for conformal supergravity). 
However, the derivation of \eqref{eq:HCSmod} from string theory remains
unclear.}

\newpage
\thispagestyle{empty}
\vspace*{5cm}
\addtocontents{toc}{\hspace{-0.61cm}{\bf\\ \centerline{Part II: Tree-level
gauge
theory scattering amplitudes}\\}} 
\begin{center}
 {\bf\Large Part II}

\vspace*{1cm}
 {\bf\large Tree-level gauge theory scattering amplitudes}
\end{center}


\newpage
\section{Scattering amplitudes in Yang--Mills theories}

\subsection{Motivation and preliminaries}\label{sec:MotAmp}

In order to compute scattering amplitudes in a quantum field theory, one usually
takes the (local) Lagrangian, derives the corresponding set of Feynman rules and
constructs the amplitudes order by order in perturbation theory. However, gauge
theories---with or without matter---present many technical challenges as the
calculational complexity grows rapidly with the number of external states
(i.e.~the number of particles one is scattering) and the number of loops. For
instance, even at tree-level where there are no loops to consider, the number of
Feyman diagrams describing $n$-particle scattering of gluons in pure Yang--Mills
theory grows faster than factorially with $n$
\cite{Kleiss:1988ne,Mangano:1990by}:

\vspace*{10pt}

\begin{table}[h]
\begin{center}
\begin{tabular}{|c||c|c|c|c|c|c|c|}
 \hline
 $n$ & $4$ & $5$ & $6$ & $7$ & $8$ & $9$ & $10$  \\
 \hline
 number of diagrams & $4$ & $25$ & $220$ & $2,485$ & $34,300$ & $559,405$ & $10,525,900$ \\
 \hline
\end{tabular}
\caption{\it The number of Feynman diagrams relevant for tree-level $n$-gluon scattering.}
\label{tab:gluons}
\end{center}
\end{table}

\vspace*{-5pt}

\noindent
We should stress that these numbers are relevant for the case where one is considering a single colour structure only (see below). The total number of diagrams after summing over all possible colour structures is much larger.

In contrast to the complexity of the calculation, the final result is often
surprisingly simple and elegant. To jump ahead of our story a bit, the prime
example is the so-called maximally-helicity-violating (MHV) amplitude describing
the scattering of two gluons (say $r$ and $s$) of positive helicity with $n-2$
gluons of negative helicity. At tree-level, the (momentum space) amplitude can
be recasted as:
\begin{equation}
 A^{\rm MHV}_{0,n}(r^+,s^+)\ =\ g^{n-2}_{\rm
YM}(2\pi)^4\delta^{(4)}\left(\sum_{r'=1}^n
p_{r'\al\db}\right)\frac{[rs]^4}{[12][23]\cdots[n1]}~.
\end{equation}
Our notation will be explained shortly. This is the famous Parke--Taylor
formula, which was first conjectured by Parke \& Taylor in \cite{Parke:1986gb}
and later proved by Berends \& Giele in \cite{Berends:1987me}. Equivalently, one
could consider scattering of two gluons of negative helicity and of $n-2$ gluons
of positive helicity. This leads to the so-called $\overline{\mbox{MHV}}$ or
`googly' amplitude\footnote{The term `googly' is borrowed from cricket and
refers to a ball thrown with the opposite of the natural spin.}, which in a
Minkowski signature space-time is obtained by complex conjugating the MHV
amplitude. Notice that our conventions are somewhat opposite
from the scattering theory literature, where $n$-gluon amplitudes with two
positive helicity and $n-2$ gluons of negative helicity are actually called
$\overline{\mbox{MHV}}$ amplitudes.
The maximally-helicity-violating amplitudes
are of phenomenological importance: For instance,
the $n=4$ and $n=5$ MHV amplitudes dominate the two-jet and three-jet production
in hadron colliders at very high energies. The tree-level $n$-gluon amplitudes
with all gluons of the same helicity or all but one gluon of the same helicity
are even simpler: They vanish.\footnote{This is true for $n\geq4$. For complex
momenta or for signatures other than Minkowski signature, the three-gluon
amplitude with all but one gluon of the same helicity is not necessarily zero
but just a special case of the MHV or $\overline{\mbox{MHV}}$ amplitude
\cite{Witten:2003nn}. See also below.} This follows, for instance, from
supersymmetric Ward identities
\cite{Grisaru:1976vm,Grisaru:1977px,Parke:1985pn,Kunszt:1985mg,Bern:1996ja} (see
also \cite{Dixon:1996wi,Mangano:1990by} for reviews). 

It is natural to wonder why the final form of these amplitudes (and others not
mentioned here) is so simple despite that their derivation is extremely
complicated. One of the main reasons for this is that the Feynman prescription
involves (gauge-dependent) off-shell states. The question then arises: Are
there
alternative methods which do not suffer these issues therefore leading to
simpler derivations and thus explaining the simplicity of the results? The
remainder of these lecture notes is devoted to precisely this question in the
context of $\CN=4$ supersymmetric Yang--Mills theory. But before delving into
this, 
let us give some justification for why it is actually very useful to consider
the
construction of `physical' quantities like scattering amplitudes in such a
 `non-physical' theory. 

It is a crucial observation that 
$\CN=4$ amplitudes are identical to or at least part of the physical amplitudes.
For instance, gluon scattering amplitudes at tree-level are the same in both
pure Yang--Mills theory and $\CN=4$ supersymmetric Yang--Mills theory as follows
from inspecting the interaction terms in the action  \eqref{eq:SYMAct}. Hence,
pure Yang--Mills gluon scattering amplitudes at tree-level have a `hidden'
$\CN=4$ supersymmetry,
\begin{equation}
 \CA_{0,n}^{\rm YM}\ =\ \CA_{0,n}^{\CN=4}~,
\end{equation}
and it does not matter which one of the two theories we use to compute them.
The same can of course be said about any supersymmetric gauge theory with adjoint matter fields when one is concerned with scattering of external gluons at tree-level. 
If there is no confusion, we will occasionally omit either one or even both of
the subscripts appearing on the symbol $\CA$ which denotes a scattering
amplitude.

Likewise, we can find a supersymmetric decomposition for gluon scattering at
one-loop. It is
\begin{equation}\label{eq:SUSYDEC1L}
 \CA_{1,n}^{\rm YM}\ =\ \CA_{1,n}^{\CN=4}-4\CA_{1,n}^{\CN=1,{\rm chiral}}+2\CA_{1,n}^{\CN=0,{\rm scalar}}~.
\end{equation}
In words this says that the $n$-gluon amplitude in pure Yang--Mills theory at one-loop can be decomposed into three terms: Firstly, a term where the whole $\CN=4$ multiplet propagates in the loop. This is represented by
$\CA_{1,n}^{\CN=4}$. Secondly, there is the term  $\CA_{1,n}^{\CN=1,{\rm
chiral}}$ where an $\CN=1$ chiral multiplet propagates in the loop. Lastly,
there is the term $\CA_{1,n}^{\CN=0,{\rm scalar}}$ where a scalar propagates
in the loop. The reason for this becomes most transparent when one
considers the multiplicities of the various particle multiplets in question:
Recall that the $\CN=4$ multiplet is 
$h_m^{\CN=4}=(-1_1,-\frac12\ \!\!_4,0_6,\frac12\ \!\!_4,1_1)$; the subscript $m$ denotes the multiplicity of the respective helicity-$h$ field. The $\CN=1$ chiral multiplet is 
$h_m^{\CN=1,{\rm chiral}}=(-1_0,-\frac12\ \!\!_1,0_2,\frac12\ \!\!_1,1_0)$ while the scalar multiplet is just
$h_m^{\rm scalar}=(-1_0,-\frac12\ \!\!_0,0_1,\frac12\ \!\!_0,1_0)$. The pure Yang--Mills multiplet contains of course just a vector field, i.e.~$h_m^{\rm YM}=(-1_1,-\frac12\ \!\!_0,0_0,\frac12\ \!\!_0,1_1)$. Altogether,
\begin{equation}
\begin{aligned}
 &(-1_1,-\tfrac12\ \!\!_0,0_0,\tfrac12\ \!\!_0,1_1)\ =\ 
 (-1_1,-\tfrac12\ \!\!_4,0_6,\tfrac12\ \!\!_4,1_1)\\
 &\kern5cm-\,4(-1_0,-\tfrac12\ \!\!_1,0_2,\tfrac12\ \!\!_1,1_0)\\
 &\kern6cm+\,2(-1_0,-\tfrac12\ \!\!_0,0_1,\tfrac12\ \!\!_0,1_0)~.
\end{aligned}
\end{equation}
The left-hand-side of \eqref{eq:SUSYDEC1L} is extremely complicated to evaluate.
However, the three pieces on the right-hand-side are easier to deal with. The first two 
pieces are contributions coming from supersymmetric field theories and these
extra supersymmetries greatly help to reduce the complexity of the calculation 
there. Much of the difficulty is thus pushed into the last term which is the most
complex of the three, but still far easier than the left-hand-side, mainly because
a scalar instead of a gluon is propagating in the loop and thus one does
not have to deal
with polarisations. 

The upshot is then that supersymmetric field theories are not
only simpler toy models 
which one can use to try to understand
the gauge theories of the 
Standard Model of Particle Physics, but relevant theories themselves which
contribute parts and sometimes the entirety of the answer to calculations in
physically relevant theories. 

The next two sections will set up some notation and conventions used in later
discussions.

\subsection{Colour ordering}

One important simplification in computing amplitudes comes from the concept of colour ordering.
For concreteness, let us consider $\sSU(N)$ gauge theory. Let $t_a$ be the generators of $\sSU(N)$ and
${f_{ab}}^c$ be the structure constants; $a,b,\ldots=1,\ldots,N^2-1$. Then $[t_a,t_b]={f_{ab}}^ct_c$. We shall assume that the $t_a$ are anti-Hermitian, i.e.~$t_a^\dagger=-t_a$. In a given matrix representation, we will write ${(t_a)_m}^n$ for $m,n,\ldots=1,\ldots,d(r)$, where $d(r)$ is the dimension of the representation. For instance, $d(r)=N$ for the fundamental  representation while $d(r)\equiv d(G)=N^2-1$ for the adjoint representation. Furthermore,
$g_{ab}=\mbox{tr}\,(t_a^\dagger t_b)=-\mbox{tr}\,(t_at_b)=C(r)\delta_{ab}$, with $C(r)=\frac12$ for the fundamental representation and
$C(r)=N$ for the adjoint representation. Using $g_{ab}$, we may re-write the structure constants $f_{abc}:={f_{ab}}^d g_{dc}$ as
\begin{equation}
 f_{abc}\ =\ -\mbox{tr}\,([t_a,t_b]t_c)~.
\end{equation}
In re-writing $f_{abc}$ this way, all the colour factors appearing in the
Feynman rules can be replaced by strings of the $t_a$ and their traces, e.g.
\begin{equation}
 \sum g^{bc}\mbox{tr}\,(\cdots t_at_b\cdots)\mbox{tr}\,(\cdots t_ct_d\cdots)\mbox{tr}\,(\cdots t_e\cdots)~
\end{equation}
if we only have external gluons (or adjoint matter as in $\CN=4$ supersymmetric
Yang--Mills theory). Here, $g^{ab}$ is the inverse of $g_{ab}$,
i.e.~$g_{ac}g^{cb}=\delta_a^b$. Likewise, if external matter in a different
representation is present, we have sums of the type
\begin{equation}
 \sum \cdots g^{bc} g^{de}{(t_a\cdots t_b)_m}^n\mbox{tr}\,(t_c\cdots t_d){(t_e\cdots t_f)_k}^l\cdots~,
\end{equation}
but this case will not be of further interest in the following, as we will
solely be dealing with scattering amplitudes in pure Yang--Mills theory
or in its $\CN=4$  supersymmetric extension. 

In order to simplify the number of traces, let us recall the following
identity in the fundamental representation of $\mathfrak{su}(N)$ ($d(r)=N$):
\begin{equation}\label{eq:Fierz}
 g^{ab}{(t_a)_m}^n{(t_b)_k}^l\ =\
\delta_m^l\delta_k^n-\frac{1}{N}\delta_m^n\delta_k^l
\end{equation}
This is nothing but the completeness relation for the generators $t_a$ in the
fundamental representation.

\vspace*{5pt}

\Exercise{Check \eqref{eq:Fierz}.}

\vspace*{5pt}

\noindent
As an immediate consequence of \eqref{eq:Fierz}, we have
\begin{equation}\label{eq:Fierzs}
 g^{ab}\mbox{tr}\,(Xt_a)\mbox{tr}\,(t_bY)\ =\ \mbox{tr}\,(XY)-\frac{1}{N}\mbox{tr}\,X\,\mbox{tr}\,Y~.
\end{equation}

The term in \eqref{eq:Fierz}, \eqref{eq:Fierzs} proportional to $1/N$
corresponds to the subtraction of the trace part in $\mathfrak{u}(N)$ in which
$\mathfrak{su}(N)$ is embedded. As such, terms involving it disappear at
tree-level after one sums over all the permutations present.  Hence, if one
considers gluon scattering at tree-level, the amplitude contains only
single-trace terms,
\begin{equation}\label{eq:TreeAmp}
 \CA_{0,n}\ =\ g^{n-2}_{\rm YM}\sum_{\sigma\in S_n/Z_n}\mbox{tr}\,\big(t_{a_{\sigma(1)}}\cdots t_{a_{\sigma(n)}}\big)
                     A_{0,n}(\sigma(1),\ldots,\sigma(n))~,
\end{equation}
where $S_n$ is the permutation group of degree $n$ and $Z_n$ is the group of
cyclic permutations of order $n$. Note that $A_{0,n}$ is assumed to contain the
momentum conserving delta function (we shall take all momenta as
incoming). Furthermore, we have re-scaled the gauge potential $A_\mu$ according
to $A_\mu\mapsto g_{\rm YM}A_\mu$ to bring the Yang--Mills action \eqref{eq:YM}
into a form suitable for perturbation theory. Therefore, the three-gluon vertex
is
of order $g_{\rm YM}$ while the four-gluon vertex is of order $g_{YM}^2$
meaning that tree-level $n$-gluon amplitudes scale like $g^{n-2}_{\rm YM}$. The
amplitude $A_{0,n}$ is called  the colour-stripped or partial amplitude and it
is this object which we will be interested in our subsequent discussion.

Let us note in passing that there is a similar colour decomposition at one-loop.
In pure Yang--Mills theory at one-loop, we have the following expression for the
$n$-gluon amplitude  \cite{Bern:1990ux}:
\begin{equation}\label{eq:1LoopColour}
\begin{aligned}
 &\CA_{1,n}\ =\ g^{n}_{\rm YM}
              \left[N \sum_{\sigma\in S_n/Z_n}\mbox{tr}\,\big(t_{a_{\sigma(1)}}\cdots t_{a_{\sigma(n)}}\big)
                     A_{1,n}^{(1)}(\sigma(1),\ldots,\sigma(n))\right.\\
 &\kern1cm \left.+\sum_{c=1}^{\lfloor\frac{n}{2}\rfloor}\sum_{\sigma\in S_n/(Z_c\times Z_{n-c})}
                   \mbox{tr}\,\big(t_{a_{\sigma(1)}}\cdots t_{a_{\sigma(c)}}\big)
                   \mbox{tr}\,\big(t_{a_{\sigma(c+1)}}\cdots t_{a_{\sigma(n)}} \big)
                     A_{1,n}^{(c)}(\sigma(1),\ldots,\sigma(n))\right].
\end{aligned}
\end{equation}
Hence, if one is only interested in the so-called large $N$-limit which is also
referred to as the planar limit (see \cite{'tHooft:1973jz}), only the
single-trace term survives. In fact, it is a general feature that in the planar
limit the amplitude to $\ell$-loop order is given by a single-trace expression.
Of course, for finite $N$ this is not the case and the colour structure of the
amplitudes will look much more complicated. Note that due to a remarkable
result of Bern, Dixon, Dunbar \& Kosower \cite{Bern:1994zx}, the double-trace
expressions $A_{1,n}^{(c)}$ in \eqref{eq:1LoopColour} are obtained as sums over
permutations of the single-trace term $A_{1,n}^{(1)}$. This also applies to
gauge theories with external particles and those running in the loop both in the
adjoint representation in general and to $\CN=4$ supersymmetric Yang--Mills
theory in particular.

\subsection{Spinor-helicity formalism re-visited}\label{sec:Revisted}

In the conventional description as given in most standard textbooks, scattering amplitudes are  considered as a function
of the external momenta of the particles (in fact the Mandelstam invariants)
together with the spin information such as polarisation vectors if one considers
photons or gluons (leaving aside colour degrees of freedom). Furthermore, see
e.g.~Weinberg's book \cite{Weinberg:1995mt} for the details. However, as
we shall see momentarily, there is a more convenient way to encode both the
momentum and spin information. In the following, we  continue to follow the
philosophy of working in a complex setting for most of the time. Reality
conditions are imposed (explicitly or implicitly) whenever needed.

Consider a null-momentum $p_{\al\da}$. From Remark \ref{rmk:lightlike}~in
Section \ref{sec:BTS}~we know that $p_{\al\da}$
can be written as $p_{\al\da}=\tilde k_{\al}k_{\da}$ for two co-spinors $\tilde
k_{\al}$ and $k_{\da}$. Clearly, this decomposition is not unique since one can
always perform the following transformation:
\begin{equation}\label{eq:MomSca}
 (\tilde k_\al,k_\da)\ \mapsto\ (t^{-1}\tilde k_\al,tk_\da)~,\efor t\ \in\
\IC\setminus\{0\}~.
\end{equation}
This is in fact the action of the little group of $\sSO(4,\IC)$ on the
co-spinors. Therefore, one associates a helicity of $-1/2$ with $\tilde k_\al$
and a helicity of $1/2$ with $k_\da$ respectively, so that $p_{\al\db}$ has
helicity zero.
In addition, we say that a quantity carries (a
generalised)
helicity $h\in\frac12\IZ$ 
whenever it scales as $t^{2h}$ under \eqref{eq:MomSca}.
Later on, we shall need the derivatives with respect to $\tilde k_\al$ and
$k_\da$. Our conventions are
\begin{equation}\label{eq:SomeConventions}
 \frac{\partial}{\partial\tilde k_\al}\tilde k_\be\ =\ \delta^\al_\be\eand
 \frac{\partial}{\partial k_\da}k_\db\ =\ \delta^\da_\db~.
\end{equation}

Let us now consider a collection of null-momenta $p_{r\al\da}$ labelled by
$r,s,\ldots\in\{1,\ldots,n\}$. 
Each of these can then be decomposed as $p_{r\al\da}=\tilde k_{r\al}k_{r\da}$. Next we define the spinor brackets
\begin{equation}
 \langle\tilde k_r\tilde k_s\rangle\ :=\ \langle rs\rangle\ :=\ \varepsilon_{\al\be}\tilde k^\al_r
   \tilde k^\be_s\eand
 [k_rk_s]\ :=\ [rs]\ :=\ \varepsilon^{\da\db}k_{r\da}k_{s\db}~.
\end{equation}
Recall that the second of these products was already introduced in \eqref{eq:formsvect}. The inner product of two null-momenta, $p_{r\al\da}$ and $p_{s\al\da}$, is then\footnote{\label{foot:Metric}Note that since the metric is $g_{\al\da\be\db}=\frac12\varepsilon_{\al\be}\varepsilon_{\da\db}$ the inverse metric is
$g^{\al\da\be\db}=2\varepsilon^{\al\be}\varepsilon^{\da\db}$.}
\begin{equation}\label{eq:inner}
 p_r\cdot p_s\ =\ 2 \varepsilon^{\al\be}\varepsilon^{\da\db}p_{r\al\da}p_{s\be\db}\ =\ -2
    \langle rs\rangle[rs]~.
\end{equation}

Once we are given a null-momentum and its spinor decomposition, we have enough
information to decribe the different helicity-$h$ wavefunctions (plane wave
solutions) modulo gauge equivalence  \cite{Berends:1987cv}. Since we
are interested in gauge theories, we restrict our attention to $|h|\leq1$.
Let us start with helicity $h=0$. The wavefunction  of a scalar field is just
\begin{equation}\label{eq:H0PW}
 \phi(x)\ =\ \de^{\di x^{\al\db}\tilde k_\al k_\db}~,
\end{equation}
i.e.~$\partial_{\al\da}\partial^{\al\da}\phi=0$, for some fixed null-momentum
$p_{\al\db}=\tilde k_\al k_\db$.

Likewise, the helicity $h=\pm\frac12$ plane waves are
\begin{equation}\label{eq:H1/2PW}
 \chi_\da(x)\ =\ k_\da\,\de^{\di x^{\al\db}\tilde k_\al k_\db}\eand
 \chi_\al(x)\ =\ \tilde k_\al\, \de^{\di x^{\al\db}\tilde k_\al k_\db}~,
\end{equation}
i.e.~$\partial_{\al\da}\chi^\da=0=\partial_{\al\da}\chi^\al$.

For a massless particle of helicity $h=\pm1$, the usual method is to specify the
polarisation co-vector $\epsilon_{\al\db}$ in addition to its momentum
$p_{\al\db}$ together with the constraint $p_{\al\db}\epsilon^{\al\db}=0$.
This constraint is equivalent to the Lorenz gauge condition and deals with
fixing the gauge invariance inherent in gauge theories. It is clear that if we
add any multiple of  
$p_{\al\db}$ to $\epsilon_{\al\db}$ then this condition is still satisfied and we have the gauge invariance
\begin{equation}\label{eq:PolGT}
 \epsilon_{\al\db}\ \mapsto\ \epsilon'_{\al\db}\ =\ \epsilon_{\al\db}+\rho\, p_{\al\db}~,\efor\rho\ \in\ \IC~.
\end{equation}
Since we have the decomposition of the momentum co-vector into two co-spinors,
we can take the $h=\pm1$ polarisation co-vectors to be
\cite{Siegel:1999ew,Witten:2003nn}
\begin{equation}\label{eq:PolarisationCoVectors}
 \epsilon^+_{\al\db}\ =\ \frac{\tilde\mu_\al k_\db}{\langle\tilde\mu\tilde k\rangle}\eand
 \epsilon^-_{\al\db}\ =\ \frac{\tilde k_\al \mu_\db}{[k\mu]}~,
\end{equation}
where $\tilde\mu_\al$ and $\mu_\da$ are arbitrary co-spinors. Notice that the
$\epsilon^\pm_{\al\db}$ do not depend on the choice of these co-spinors---modulo
transformations of the form \eqref{eq:PolGT}. This can be seen as follows
\cite{Witten:2003nn}. Consider $\epsilon^+_{\al\db}$. Any change of $\tilde
\mu_\al$ is of the form
\begin{equation}
 \tilde\mu_\al\ \mapsto\ \tilde\mu_\al+a\tilde\mu_\al+b\tilde k_\al~,\efor a,b\ \in\ \IC\eand a\ \neq\ -1~,
\end{equation}
since the space of possible $\tilde \mu_\al$ is two-dimensional and $\tilde\mu_\al$ and $\tilde k_\al$ are linearly independent by assumption (if they were linearly dependent then $\langle\tilde\mu\tilde k\rangle=0$ and the above expression would not make sense). However, under such a change, $\epsilon^+_{\al\db}$ behaves as
\begin{equation}
 \epsilon^+_{\al\db}\ \mapsto\ \epsilon^+_{\al\db}+\frac{b}{1+a}\,\tilde k_\al k_\db\ =\ 
                                \epsilon^+_{\al\db}+\rho\,p_{\al\db}~,\ewith
\rho\ =\ \frac{b}{1+a}~,
\end{equation}
which is precisely \eqref{eq:PolGT}. Likewise, a similar argument goes through
for $\epsilon^-_{\al\db}$.
Furthermore, under the scaling \eqref{eq:MomSca}
we find that $\epsilon^\pm_{\al\db}\mapsto t^{\pm 2}\epsilon^\pm_{\al\db}$ and so we may conclude that the $\epsilon^\pm_{\al\db}$  indeed carry helicity $\pm1$.
Therefore, the helicity $h=\pm1$ plane waves are
\begin{equation}\label{eq:WFA}
 A^\pm_{\al\db}(x)\ =\ \epsilon^\pm_{\al\db}\de^{\di x^{\al\db}\tilde k_\al
k_\db}~.
\end{equation}

\Exercise{\label{exe:helicity} Consider the curvature
$F_{\al\da\be\db}=\partial_{\al\da}A_{\be\db}-\partial_{\be\db}A_{\al\da}=
\varepsilon_{\al\be}f_{\da\db}+\varepsilon_{\da\db}f_{\al\be}$. Show that for the choice $A^+_{\al\da}$ as given in \eqref{eq:WFA}, we have $F^+_{\al\da\be\db}=\di \varepsilon_{\al\be} k_\da k_\db 
\de^{\di x^{\al\db}\tilde k_\al k_\db}$ while
for $A^-_{\al\da}$ we have $F^-_{\al\da\be\db}=\di \varepsilon_{\da\db} \tilde k_\al \tilde k_\be 
\de^{\di x^{\al\db}\tilde k_\al k_\db}$.
}

\vspace*{5pt}

In summary, the different wavefunctions  
scale as $t^{2h}$ under \eqref{eq:MomSca}.
Therefore, instead of considering the scattering amplitude $\CA$ as a function
of the momenta and spins, we may equivalently regard it as as a function of the
two co-spinors specifying the momenta, and the helicity. Hence, we may write
\begin{equation}
 \CA\ =\ \CA(\{\tilde k_{r\al},k_{r\da},h_r\})~,\efor r\ =\ 1,\ldots,n~.
\end{equation}
In what follows, we shall also make use of the notation
$\CA=\CA(1^{h_1},\ldots,n^{h_n})$ where the
co-spinor dependence is understood.
When formulated in this way, the amplitude obeys an auxiliary condition for
each $r$ (i.e.~no summation over
$r$) \cite{Witten:2003nn}:
\begin{equation}\label{eq:AuxCon}
 \left( -\tilde k_{r\al}\frac{\partial}{\partial\tilde k_{r\al}}+
       k_{r\da}\frac{\partial}{\partial k_{r\da}}\right)\CA\ =\ 2h_r\CA~.
\end{equation}
This is easily checked by using \eqref{eq:SomeConventions}.

\vspace*{5pt}

\Remark{\label{rmk:WF} Above we have considered
wavefunctions in position space, but we may equivalently view them in momentum
space and in fact, in Sections \ref{sec:SMHV}~and \ref{sec:WHFT}, this will be
the more convenient point of view. We follow
\cite{Drummond:2008vq,Mason:2009sa}. Firstly, the
Klein--Gordon equation in
momentum space is $p^2\tilde\phi(p)=0$, where $p^2=4\det p_{\al\db}$; see
\eqref{eq:inner} and footnote \ref{foot:Metric}. Therefore, any solution is of
the form $\tilde\phi(p)=\delta(p^2)\tilde\phi_0(\tilde k,k)$ for some function
$\tilde\phi_0(\tilde k,k)$ defined on the null-cone in momentum space,
where $p_{\al\db}=\tilde k_\al k_\db$. Secondly, we can discuss the other
helicity fields. The momentum space versions of the field equations
\eqref{eq:EOMH>0}, \eqref{eq:EOMH<0} are given by
$p^{\al\da_1}\tilde\phi_{\da_1\cdots\da_{2h}}(p)=0$ and
$p^{\al_1\da}\tilde\phi_{\al_1\cdots\al_{2h}}(p)=0$. Away from the null-cone
$p^2=0$, the matrix $p^{\al\db}$ is invertible. Therefore, the general solutions
are given by 
$\tilde\phi_{\da_1\cdots\da_{2h}}(p)=\delta(p^2) k_{\da_1}\cdots k_{\da_{2h}}\tilde\phi_{-2h}(\tilde k,k)$ and 
$\tilde\phi_{\al_1\cdots\al_{2h}}(p)=\delta(p^2)\tilde k_{\al_1}\cdots\tilde
k_{\al_{2h}}\tilde\phi_{2h}(\tilde k,k)$, where again the $\tilde\phi_{\mp
2h}(\tilde k,k)$ are some functions defined on the null-cone (with $h>0$). Note
that the $\tilde\phi_{\mp2h}(\tilde k,k)$ for $h\geq0$ scale under
\eqref{eq:MomSca} like $\tilde\phi_{\mp2h}(t^{-1}\tilde k,t k)=t^{\mp2h}\tilde
\phi_{\mp2h}(\tilde k,k)$ in order to compensate the scaling of the co-spinor
pre-factors. We shall refer to $\tilde\phi_{\mp 2h}(\tilde k,k)$ as the on-shell
momentum space wavefunctions of helicity $\mp h$.}

\Exercise{Choose Minkowski signature. Show that the Fourier transform of
the solution \eqref{eq:H0PW} can be written as
$\tilde\phi(p)=\delta(p^2)\Theta(p_0)2|\vec{p}|\delta^{(3)}(\vec{p}-\vec{k})$,
where $p_0$ is the `time component' of $p_{\al\db}$ and $\vec{p}$ and $\vec{k}$
are the `spatial components' of $p_{\al\db}$ and
$k_{\al\db}:=\tilde k_\al k_\db$, respectively. Here, $\Theta(x)$
is the Heaviside step function with $\Theta(x)=1$ for $x>0$ and zero
otherwise. Note that on the support of the
delta function we have $p_{\al\db}=\tilde k_\al k_\db$. 
}

\vspace*{10pt}

\section{MHV amplitudes and twistor theory}
\subsection{Tree-level MHV amplitudes}\label{sec:TreeMHV}

In Section \ref{sec:MotAmp}, we already encountered some colour-stripped
tree-level $n$-gluon scattering amplitudes \eqref{eq:TreeAmp}, and with the
above discussion the notation should be clear. For the reader's
convenience, let us re-state them (in the complex setting):
\begin{subequations}\label{eq:MHVAmp}
\begin{eqnarray}
  A_{0,n}(1^\mp,\ldots,n^\mp)\! &=&\! 0~,\label{eq:MHVAmpa}\\
  \efor n\ \geq\ 4\quad A_{0,n}(1^\pm,2^\mp,\ldots,n^\mp)\! &=&\!
0~,\label{eq:MHVAmpb}\\
 A_{0,n}(\ldots,(r-1)^-,r^+,(r+1)^-,\ldots,(s-1)^-,s^+,(s+1)^-,\ldots)\! &=&\!
\notag \\
  &&\kern-8cm =\ g^{n-2}_{\rm YM}(2\pi)^4\delta^{(4)}\left(\sum_{r'=1}^n \tilde k_{r'\al}k_{r'\db}\right)
   \frac{[rs]^4}{[12][23]\cdots[n1]}~,\label{eq:MHVAmpd}\\
 A_{0,n}(\ldots,(r-1)^+,r^-,(r+1)^+,\ldots,(s-1)^+,s^-,(s+1)^+,\ldots)\! &=&\!
\notag \\ 
  &&\kern-8cm =\ g^{n-2}_{\rm YM}(2\pi)^4\delta^{(4)}\left(\sum_{r'=1}^n \tilde
k_{r'\al}k_{r'\db}\right)
   \frac{\langle rs\rangle^4}{\langle 12\rangle\langle 23\rangle\cdots\langle
n1\rangle}~.\label{eq:MHVAmpc}
 \end{eqnarray}
\end{subequations}
As already mentioned, for a Minkowski signature space-time and for $n=3$, the
amplitudes \eqref{eq:MHVAmpd} and \eqref{eq:MHVAmpc} actually vanish. 

The amplitude \eqref{eq:MHVAmpd} is referred to as the
maximally-helicity-violating amplitude or MHV amplitude for short. We shall
also write $A_{0,n}^{\rm MHV}(r^+,s^+)$. Furthermore, \eqref{eq:MHVAmpc} is
referred to as the $\overline{\rm MHV}$ amplitude and is also denoted by
$A_{0,n}^{\overline{\rm MHV}}(r^-,s^-)$. For a Minkowski signature, they are
the complex conjugates of each other. In addition, $n$-gluon scattering
amplitudes with three positive helicity gluons and $n-3$ negative helicity
gluons are called next-to-MHV amplitudes or NMHV amplitudes for short.
Analogously, one considers N$^k$MHV and $\overline{\mbox{N}^k\mbox{MHV}}$
amplitudes. Hence, an  N$^k$MHV has a `total helicity' of $-n+2(k+2)$. 

It is a straightforward exercise to show that the amplitudes \eqref{eq:MHVAmp}
indeed obey the condition \eqref{eq:AuxCon}.

\vspace*{5pt}

\Exercise{Consider four-gluon scattering. Show that we indeed have $A_{0,4}^{\rm
MHV}(1^+,2^+)=A_{0,4}^{\overline{\rm MHV}}(3^-,4^-)$ or more generally
$A_{0,4}^{\rm MHV}(r^+,s^+)=A_{0,4}^{\overline{\rm MHV}}({r'}^-,{s'}^-)$ for
$\{r,s\}\neq\{r',s'\}$. In order to demonstrate this, you will need to use
overall momentum conservation.}

\vspace*{5pt}

As a next step, it is instructive to verify
the invariance of the MHV amplitudes
under the action of the conformal group. Recall from Section \ref{sec:SCA}~that
the conformal group is generated by translations $P_{\al\da}$, Lorentz rotations
${L_\al}^\be$ and ${L_\da}^\db$, dilatations $D$ and special conformal
transformations $K^{\al\da}$. 
In terms of the co-spinors $\tilde k_{r\al}$ and $k_{r\da}$ for $r=1,\ldots,n$, these generators may be represented by \cite{Witten:2003nn}:
\begin{equation}\label{eq:CAMom}
 \begin{aligned}
   P_{\al\da}\ =\ -\di\sum_r\tilde k_{r\al} k_{r\da}~,\quad
   K^{\al\da}\ =\ -\di\sum_r\frac{\partial^2}{\partial\tilde k_{r\al}\partial k_{r\da}}~,\\
   D\ =\ \frac{\di}{2}\sum_r
    \left(\tilde k_{r\al}\frac{\partial}{\partial\tilde k_{r\al}}+
      k_{r\da}\frac{\partial}{\partial k_{r\da}}+2\right)~,\\
   {L_\al}^\be\ =\ \di\sum_r\left(\tilde k_{r\al}
      \frac{\partial}{\partial\tilde k_{r\be}}-\frac12\delta_\al^\be\tilde k_{r\ga}
                \frac{\partial}{\partial\tilde k_{r\ga}}\right),\quad
   {L_\da}^\db\ =\ \di\sum_r\left(k_{r\da}\frac{\partial}{\partial k_{r\db}}-\frac12\delta_\da^\db 
                k_{r\dc}\frac{\partial}{\partial k_{r\dc}}\right).
 \end{aligned}
\end{equation}
If not indicated otherwise, summations over $r,s,\ldots$ always run from $1$ to
$n$. Upon using \eqref{eq:SomeConventions}, 
one may straightforwardly check that
these generators obey the bosonic part of \eqref{eq:ComRelSuConAl}. Next we
observe that the amplitude is manifestly invariant under (Lorentz) rotations. It
is also invariant under translations because of the momentum conserving delta
function. 
What therefore remains is for us to verify dilatations and special conformal
transformations. Let us only check $D$ here; $K^{\al\da}$-invariance follows
from similar arguments and is left as an exercise. Let us write $D=\sum_r D_r=
\frac{\di}{2}\sum_r(\hat D_r+2)$, where $\hat D_r$ is given by
\begin{equation}
 \hat D_r\ :=\ \tilde k_{r\al}\frac{\partial}{\partial\tilde k_{r\al}}+
      k_{r\da}\frac{\partial}{\partial k_{r\da}}~.
\end{equation}
Then
\begin{subequations}
\begin{equation}\label{eq:eqA}
 \begin{aligned}
  & D\left(\delta^{(4)}\left(\sum_{s'} \tilde k_{s'\al}k_{s'\db}\right)
   \frac{[rs]^4}{[12][23]\cdots[n1]}\right)\ =\ \\
    & \kern1cm =\  \frac{\di}{2}\sum_{r'}
 \left[\hat D_{r'}\left(\delta^{(4)}\left(\sum_{s'} \tilde k_{s'\al}k_{s'\db}\right)[rs]^4\right)\frac{1}{[12][23]\cdots[n1]}
    \right.\\
  &\kern3cm +\, \left.\delta^{(4)}\left(\sum_{s'} \tilde k_{s'\al}k_{s'\db}\right)[rs]^4(\hat D_{r'}+2)\frac{1}{[12][23]\cdots[n1]} \right].
 \end{aligned}
\end{equation}
However,
\begin{equation}\label{eq:eqB}
\begin{aligned}
 \sum_{r'} \hat D_{r'}\left(\delta^{(4)}
    \left(\sum_{s'} \tilde k_{s'\al}k_{s'\db}\right)[rs]^4\right)\ =\ -8+8\ &=\ 0~,\\
(\hat D_{r'}+2)\frac{1}{[12][23]\cdots[n1]}\ &=\  0~,
\end{aligned}
\end{equation}
\end{subequations}
since under the rescaling $\delta^{(4)}(t \sum_{r} \tilde
k_{r\al}k_{r\db})=t^{-4}\delta^{(4)}(\sum_{r} \tilde k_{r\al}k_{r\db})$ for
$t\in\IC\setminus\{0\}$. Therefore, combining \eqref{eq:eqA} with \eqref{eq:eqB}
we find that $D A_{0,n}^{\rm MHV}=0$ as claimed.

\vspace*{5pt}

\Exercise{Show that $K^{\al\da}A_{0,n}^{\rm MHV}=0$.}

\vspace*{5pt}

The representation \eqref{eq:CAMom} of the generators of the conformal group is rather unusual in the sense of being a mix of first-order and second-order differential operators and of multiplication operators. The representation given in \eqref{eq:SCAP} on twistor space is more natural and indeed, following Witten \cite{Witten:2003nn}, we can bring \eqref{eq:CAMom} into the form of \eqref{eq:SCAP} by making the (formal) substitutions
\begin{equation}\label{eq:WittenFourier}
 \tilde k_{r\al}\ \mapsto\ \di\frac{\partial}{\partial z^\al_r}\eand 
 \frac{\partial}{\partial\tilde k_{r\al}}\ \mapsto\ \di z^\al_r
\end{equation}
and by re-labelling  $k_{r\da}$ by $\lambda_{r\da}$ for all $r=1,\ldots,n$. This
can also be understood by noting that after this substitution, the
transformation \eqref{eq:MomSca} will become the usual rescaling of projective
space, i.e.~$(z^\al_r,\lambda_{r\da})\sim (tz^\al_r,t\lambda_{r\da})$ for
$t\in\IC\setminus\{0\}$.
The substitution \eqref{eq:WittenFourier} will be referred to as Witten's half
Fourier transform to twistor space. The reason for this name becomes transparent
in Section \ref{sec:WHFT} For the sake of brevity, we shall also refer to it as
the Witten transform.\footnote{Notice that some people also use the
terminology twistor transform. However, the name twistor transform is reserved
for the transformation that acts between twistor space and dual twistor space.}

Before giving a more precise meaning to \eqref{eq:WittenFourier}, we can
already
make two observations about the Witten transform of scattering amplitudes.
Firstly, in making the substitution \eqref{eq:WittenFourier}, we have chosen to
transform $\tilde k_{r\al}$ instead of $k_{r\da}$. Naturally, the
symmetry between $\tilde k_{r\al}$ and $k_{r\da}$ is lost and
therefore parity symmetry is obscured. Henceforth, scattering amplitudes with,
say, $m$ positive helicity gluons and $n$
negative helicity gluons will be treated completely differently from those with
$m$ and $n$ interchanged. Secondly, let $\cW[\CA]$ be the Witten transform of an
$n$-particle scattering amplitude $\CA$. According to our above discussion, the
quantity $\cW[\CA]$ will live on 
the $n$-particle twistor space $\cP^{3n}$, which is defined as
\begin{equation}\label{eq:MPTS}
 \cP^{3n}\ :=\ P_1^3\times\cdots\times P_n^3~,
\end{equation}
together with the canonical projections $\mbox{pr}_r\,:\,\cP^{3n}\to P_r^3$. Each $P_r^3$ is equipped with
homogeneous coordinates $(z^\al_r,\lambda_{r\da})$.
The auxiliary condition \eqref{eq:AuxCon} then translates into
\begin{equation}\label{eq:ScaTwiAm}
 \left(z^\al_r\frac{\partial}{\partial z^\al_r}+\lambda_{r\da}\frac{\partial}{\partial\lambda_{r\da}}\right)
 \cW[\CA]\ =\ (2h_r-2)\cW[\CA]~.
\end{equation}
In other terms this says that $\cW[\CA]$ should be regarded as a section of 
\begin{equation}
 \CO(2h_1-2,\ldots,2h_n-2)\ :=\ \mbox{pr}_1^*\CO(2h_1-2)\otimes\cdots\otimes \mbox{pr}_n^*\CO(2h_n-2)~.
\end{equation}
Thus, like solutions to zero-rest-mass field equations, scattering
amplitudes can also be interpreted as holomorphic functions of a certain
homogeneity on twistor space (or more precisely on the multi-particle twistor
space
\eqref{eq:MPTS}). One may naturally wonder whether one can then also give
scattering amplitudes a sheaf cohomological interpretation as done for solutions
to the zero-rest-mass field equations (see Section \ref{sec:MFPT}). This would
in turn yield a precise definition of the Witten transform.
Unfortunately, this issue has not yet been completely settled and we will
therefore not concern ourselves with it any further in these notes.
When choosing a Kleinian signature space-time, on the other hand, one can
can by-pass sheaf cohomology and give the Witten
transform a precise meaning and this will be the subject
of Section
\ref{sec:WHFT}
Firstly, however, we would like to extend the above ideas to the $\CN=4$
supersymmetric theory.

\subsection{Tree-level MHV superamplitudes}\label{sec:SMHV}

In $\CN=4$ supersymmetric Yang--Mills theory one has different kind of particles, the gluons $f_{\al\be}$, $f_{\da\db}$, the gluinos $\chi^i_\al$, $\chi_{i\da}$ and the scalars $\phi^{ij}=\frac12\varepsilon^{ijkl}\phi_{kl}$.  According to Remark \ref{rmk:WF}, the corresponding momentum space wavefunctions may be taken as follows:
\begin{equation}\label{eq:N=4WF}
\begin{aligned}
\tilde  f_{\al\be}(p)\ =\ \delta(p^2)\tilde k_\al\tilde k_\be\tilde f^+(\tilde
k,k) ~,\quad
\tilde \chi^{i}_\al(p)\ =\  \delta(p^2)\tilde k_\al\tilde \chi^{i+}(\tilde
k,k)~,\\
\tilde\phi_{ij}(p)\ =\ \delta(p^2)\tilde\phi_{ij}(\tilde
k,k)~,\kern2.5cm\\
\tilde\chi_{i\da}(p)\ =\ \delta(p^2)k_\da\tilde\chi^-_i(\tilde
k,k)~,\quad
\tilde f_{\da\db}(p)\ =\ \delta(p^2)k_\da k_\db \tilde f^-(\tilde
k,k)~,
\end{aligned}
\end{equation} 
where 
$\tilde f^+$, $\tilde\chi^{i+}$, $\tilde\phi_{ij}$, $\tilde\chi^-_i$ and
$\tilde f^-$ are the on-shell momentum space wavefunctions of 
helicity $1$, $\frac12$, $0$, $-\frac12$ and $-1$, respectively. 
These particles can then scatter off each other in many different ways
eventually leading to a large variety of amplitudes. Despite this fact,
supersymmetry relates many of the amplitudes. It would therefore be desirable to
have a formulation of scattering amplitudes in which $\CN=4$ supersymmetry is
manifest. Fortunately, Nair \cite{Nair:1988bq} proposed a particular type of an
$\CN=4$ on-shell superspace\footnote{Notice that this superspace has a close
relationship
with the light-cone superspace of \cite{Mandelstam:1982cb,Brink:1982pd}.} twenty
years ago, used later by Witten \cite{Witten:2003nn}, that leads to a
manifestly supersymmetric formulation of the scattering amplitudes and in
addition forms the basis of the latter's twistor re-formulation. 

 The key idea of Nair's is to introduce additional spinless fermionic variables
$\psi^{ i}$, for $i=1,\ldots,4$, and to combine the different wavefunctions
\eqref{eq:N=4WF} into one single super wavefunction 
$\tilde\Phi=\tilde\Phi(\tilde k_{\al},\psi^{i},k_{\da})$:
\begin{equation}\label{eq:SuWaFu}
\tilde \Phi\ :=\ \tilde f^- + \psi^{i}\tilde\chi^{-}_i+\tfrac{1}{2!} \psi^{i} \psi^{j}\tilde\phi_{ij}
               +\tfrac{1}{3!} \psi^{i} \psi^{j}\psi^{k}\varepsilon_{ijkl}\tilde\chi^{l+}
              +\tfrac{1}{4!} \psi^{i} \psi^{j}\psi^{k}\psi^{l}\varepsilon_{ijkl}\tilde f^+~.
\end{equation}
For notational reasons, we shall not put the circles explicitly (as done in
Sections \ref{sec:SSDYM}~and \ref{sec:SYMTP}) to denote the component fields.
Furthermore, each $\psi^i$ is assumed to carry a helicity of $-1/2$ such that
the scaling \eqref{eq:MomSca} is augmented to
\begin{equation}\label{eq:MomScaSup}
 (\tilde k_\al,\psi^i,k_\da)\ \mapsto\ (t^{-1}\tilde k_\al,t^{-1}\psi^i,tk_\da)~,\efor t\ \in\ \IC\setminus\{0\}~.
\end{equation}
Consequently, the super wavefunction scales as $\tilde \Phi (t^{-1}\tilde
k_\al,t^{-1}\psi^i,tk_\da)= t^{-2}\tilde \Phi (\tilde k_\al,\psi^i,k_\da)$,
i.e.~it carries helicity $-1$. Notice that this may be re-expressed as a
differential constraint,
\begin{equation}\label{eq:MomScaSupDiff}
 \left( -\tilde k_{\al}\frac{\partial}{\partial\tilde
k_{\al}}-\psi^i\frac{\partial}{\partial \psi^i}+
       k_{\da}\frac{\partial}{\partial k_{\da}}+2\right)\tilde\Phi\ =\ 0~.
\end{equation}
Due to the scaling property of $\tilde\Phi$, we shall also write
$\tilde\Phi_{-2}$, in the following.
At this point it is worthwhile recalling the twistor space expression 
\eqref{eq:etaexp} and comparing it with the expression \eqref{eq:SuWaFu}.
Their resemblance is not coincidental as will be demonstrated in the next
section.
Moreover, in view of our twistor space application, it is useful to combine
$\tilde k_\al$ and $\psi^i$ into a super co-spinor $\tilde k_A=(\tilde k_\al,
\psi^i)$.  

In this superspace formulation, $n$-particle scattering amplitudes---also
referred to as superamplitudes or generating functions---are regarded as 
functions of $\tilde k_{rA}$ and $k_{r\da}$ and we will denote them by $\CF$,
\begin{equation}
 \CF\ =\ \CF(\tilde\Phi_1,\ldots,\tilde\Phi_n)\ =\ \CF(\{\tilde
k_{rA},k_{r\da}\})~,\efor r\ =\ 1,\ldots,n~,
\end{equation}
where $\tilde\Phi_r:=\tilde\Phi(\tilde k_{rA},k_{r\da})$. Likewise,
colour-stripped superamplitudes will be denoted by $F$. Thus, a term in a
superamplitude that is of $k$-th order in the $\psi^i_r$ for some $r$ describes
a scattering process in which the $r$-th particle has helicity $k/2-1$.
One may pick the appropriate terms by suitably differentiating the
superamplitude using the following set of operators:
\begin{equation}\label{eq:DOAmp}
\begin{aligned}
 \tilde f_r^-\ \leftrightarrow\ 1~,\qquad
 \tilde\chi^-_{ri}\ \leftrightarrow\ D_{ri}\ :=\
\frac{\partial}{\partial\psi^i_r}~,\qquad
\tilde\phi_{rij}\ \leftrightarrow\ D_{rij}\ :=\
\frac{\partial^2}{\partial\psi^i_r\partial\psi^j_r}~,\kern.8cm\\
\tilde\chi^{i+}_r\ \leftrightarrow\ D_r^i\ :=\ \frac{1}{3!}\,\varepsilon^{ijkl}
\frac{\partial^3}{\partial\psi^j_r\partial\psi^k_r\partial\psi^k_r}~,\qquad
\tilde f^{+}_r\ \leftrightarrow\ D_r\ :=\ \frac{1}{4!}\,\varepsilon^{ijkl}
\frac{\partial^4}{
\partial\psi^i_r\partial\psi^j_r\partial\psi^k_r\partial\psi^k_r}~.
\end{aligned}
\end{equation}
For instance, $\tilde f^+_r=D_r\tilde\Phi_r$, etc. In addition, when formulated
in this way, scattering amplitudes obey the auxiliary condition (again no
summation over $r$) \cite{Witten:2003nn}
\begin{equation}\label{eq:AuxConSuper}
 \left( -\tilde k_{rA}\frac{\partial}{\partial\tilde k_{rA}}+
       k_{r\da}\frac{\partial}{\partial k_{r\da}}+2\right)\CF\ =\ 0~,
\end{equation}
as follows directly from \eqref{eq:MomScaSupDiff}. This is the supersymmentric
analog of  \eqref{eq:AuxCon}.

Now we are in the position to write down our first superamplitude. If we let 
\begin{equation}\label{eq:OvDF}
 \delta^{(4|8)}\left(\sum_{r=1}^n \tilde k_{rA}k_{r\db}\right)\ :=\
 \delta^{(4|0)}\left(\sum_{r=1}^n \tilde k_{r\al}k_{r\db}\right)\delta^{(0|8)}\left(\sum_{r=1}^n \psi^i_rk_{r\db}\right)
\end{equation}
and recall that the fermionic delta function is $\delta(\psi)=\psi$ for one
fermionic coordinate $\psi$,\footnote{This definition is consistent with the
definition of the Berezin integral over fermionic coordinates:
$\int\dd\psi\,\delta(\psi)=\int\dd\psi\,\psi=1$. For more than one fermionic
coordinate one has $\int\dd^n\psi\,\delta^{(0|n)}(\psi)=1$ with
$\dd^n\psi=\dd\psi_1\cdots\dd\psi_n$ and
$\delta^{(0|n)}(\psi)=\psi_n\cdots\psi_1$.}
then the supersymmetric version of the MHV amplitude \eqref{eq:MHVAmpd} is given
by \cite{Nair:1988bq}
\begin{equation}\label{eq:MHVSuper}
 F_{0,n}^{\rm MHV}\ =\ g^{n-2}_{\rm YM}(2\pi)^4\delta^{(4|8)}\left(\sum_{r=1}^n
\tilde k_{r A}k_{r\db}\right)
   \frac{1}{[12][23]\cdots[n1]}~.
\end{equation}
Obviously, this expression obeys \eqref{eq:AuxConSuper}. Moreover, in order to
recover  \eqref{eq:MHVAmpd} from  \eqref{eq:MHVSuper}, we simply need to pick
the term that is of fourth order in both $\psi^i_r$ and $\psi^i_s$ corresponding
to the two positive helicity gluons; see \eqref{eq:SuWaFu}. This is most easily
accomplished by noting the identity
\begin{equation}\label{eq:DeltaIdentity}
 \delta^{(0|8)}\left(\sum_{r=1}^n \psi^i_rk_{r\da}\right)\ =\
 \frac{1}{16}\prod_{i=1}^4\sum_{r,s=1}^n[rs]\,\psi^i_r\psi^i_s
\end{equation}
and by using \eqref{eq:DOAmp}. Then, we find
\begin{equation}\label{eq:SomeSuMHVA}
 A_{0,n}^{\rm MHV}(r^+,s^+)\ =\ D_r D_s F_{0,n}^{\rm MHV}~.
\end{equation}
Note that since $\tilde\Phi$ carries helicity $-1$, the superamplitude carries a
total helicity of $-n$ and since $D_r$ carries helicity $2$ (remember that each
$\psi^i_r$ carries helicity $-1/2$), we recover the correct helicity of $-n+4$
for an MHV amplitude. 

Analogously, we may construct other MHV amplitudes. For instance, the
$n$-particle amplitude involving one positive helicity gluon, say particle $r$,
two gluinos of opposite helicity, say particles $s$ and $t$ and $n-3$ negative
helicity gluons is obtained by
\begin{equation}\label{eq:SomeSuMHVB}
\begin{aligned}
 A_{0,n}^{\rm MHV}(r^{+1},s^{-\frac12},t^{+\frac12})\ &=\ 
  D_r D_{si} D_t^j F_{0,n}^{\rm MHV}\\
 &=\ g^{n-2}_{\rm YM}(2\pi)^4\delta_i^j\delta^{(4)}\left(\sum_{r=1}^n
\tilde k_{r\al}k_{r\db}\right)
   \frac{[rs][rt]^3}{[12][23]\cdots[n1]}~.
\end{aligned}
\end{equation}
In general, all MHV amplitudes are of the type
$[r_1s_1][r_2s_2][r_3s_3][r_4s_4]/([12][23]\cdots[n1])$ and are obtained as
\begin{equation}\label{eq:AllMHV}
 A_{0,n}^{\rm MHV}\ =\ D^{(8)}F_{0,n}^{\rm MHV}~,
\end{equation}
where $D^{(8)}$ is an eighth-order differential operator made from
the operators \eqref{eq:DOAmp}.

\vspace*{5pt}

\Exercise{\label{exe:SupAmp} Verify \eqref{eq:SomeSuMHVA} and
\eqref{eq:SomeSuMHVB}.}

\vspace*{5pt}

In the preceeding section, we have shown that the MHV amplitudes are invariant
under the action of the conformal group. Likewise, the MHV superamplitudes are
invariant under the action of the $\CN=4$ superconformal group. Recall from
Section \ref{sec:SCA}~that the superconformal group for  $\CN=4$ is generated by
translations $P_{\al\da}$, Lorentz rotations ${L_\al}^\be$ and ${L_\da}^\db$,
dilatations $D$ and special conformal transformations $K^{\al\da}$ together with
the $R$-symmetry generators ${R_i}^j$ and the Poincar\'e supercharges
$Q_{i\al}$, $Q^i_\da$ and their superconformal partners $S^{i\al}$, $S^\da_i$.
In terms of $(\tilde k_{rA},k_\da)=(\tilde k_{r\al},\psi_r^i,k_{r\da})$ for $r=1,\ldots,n$, these generators may be represented by \eqref{eq:CAMom} together with \cite{Witten:2003nn}:
\begin{subequations}\label{eq:SCAMom}
\begin{equation}
 \begin{aligned}
   Q_{i\al}\ =\ \di\sum_r\tilde k_{r\al} \frac{\partial}{\partial\psi^i_r}~,\quad
   Q_{\da}^i\ =\ \sum_r\tilde k_{r\da} \psi_{r}^i~,\\
   S^{i\al}\ =\ -\di\sum_r \psi^i_r\frac{\partial}{\partial\tilde k_{r\al} }~,\quad
   S^{\da}_i\ =\ -\sum_r \frac{\partial^2}{\partial\psi^i_r\partial\tilde k_{r\da} }~,\\
   {R_i}^j\ =\ \frac{\di}{2}\sum_r
    \left(\psi^j_r\frac{\partial}{\partial\psi^i_r}-\frac14\delta_i^j\psi^k_r\frac{\partial}{\partial\psi^k_r}\right)~.
 \end{aligned}
\end{equation}
Furthermore, the central extension is given by
\begin{equation}
 Z\ =\   \frac{\di}{2}\sum_r\left( \tilde k_{r\al}\frac{\partial}{\partial\tilde k_{r\al}}+ \psi^i_{r}\frac{\partial}{\partial\psi^i_{r}}-k_{r\da}\frac{\partial}{\partial k_{r\da}}-2\right).
\end{equation}
\end{subequations}
Following the analysis presented in Section \ref{sec:TreeMHV}~in the bosonic
setting, one may demonstrate the invariance of \eqref{eq:MHVSuper} under
superconformal transformations. We leave this as an exercise.

\vspace*{5pt}

\Exercise{Verify that the MHV superamplitude \eqref{eq:MHVSuper} is invariant under the action of $\CN=4$ superconformal group.}

\vspace*{5pt}

In the supersymmetric setting we can again see that the representation
of the
generators of the superconformal group is rather unusual. However, we can bring
\eqref{eq:CAMom}, \eqref{eq:SCAMom} into the canonical form
\eqref{eq:SCAP} by making the (formal) substitutions \cite{Witten:2003nn}
\begin{equation}\label{eq:WittenFourierSuper}
 \tilde k_{rA}\ \mapsto\ \di\frac{\partial}{\partial z^A_r}\eand 
 \frac{\partial}{\partial\tilde k_{rA}}\ \mapsto\ \di (-)^{p_A} z^A_r
\end{equation}
and by re-labelling  $k_{r\da}$ by $\lambda_{r\da}$ for all $r=1,\ldots,n$.
After
this substitution, the transformation \eqref{eq:MomScaSup} becomes the usual
rescaling of projective superspace, i.e.~$(z^A_r,\lambda_{r\da})\sim
(tz^A_r,t\lambda_{r\da})$ for $t\in\IC\setminus\{0\}$. As before, we shall refer
to \eqref{eq:WittenFourierSuper} as  the Witten transform. Moreover, if we let
$\cW[\CF]$ be the Witten transform of an $n$-particle superamplitude $\CF$, then
$\cW[\CF]$ will live on 
the $n$-particle supertwistor space $\cP^{3n|4n}$, which is defined as
\begin{equation}\label{eq:MPTSsuper}
 \cP^{3n|4n}\ :=\ P_1^{3|4}\times\cdots\times P_n^{3|4}~,
\end{equation}
together with the canonical projections $\mbox{pr}_r\,:\,\cP^{3n|4n}\to P_r^{3|4}$. Each $P_r^{3|4}$ is equipped with
homogeneous coordinates $(z^A_r,\lambda_{r\da})$.
The auxiliary condition \eqref{eq:AuxConSuper} then translates into
\begin{equation}\label{eq:ScaTwiSuAm}
 \left(z^A_r\frac{\partial}{\partial z^A_r}+\lambda_{r\da}\frac{\partial}{\partial\lambda_{r\da}}\right)
 \cW[\CF]\ =\ 0~.
\end{equation}
In words this says that $\cW[\CF]$ should be regarded as a holomorphic function
of homogeneity zero on 
$\cP^{3n|4n}$. This is again analogous to what we encountered in Section
\ref{sec:HCS}, where one superfield \eqref{eq:etaexp} of homogeneity zero on
supertwistor space encodes all the wavefunctions. 

\vspace*{5pt}

\Remark{\label{rmk:HoloAno} In verifying the invariance of the MHV
(super)amplitudes under the
action of the (super)conformal group here and in Section \ref{sec:TreeMHV}, 
one should actually be a bit more careful. In the complex setting, where there
is
no relation between the co-spinors $\tilde k$ and $k$ or for Kleinian
signature space-times where $\tilde k$ and $k$ are real and independent of each
other (see also the subsequent section), the above procedure of verifying
(super)conformal invariance works as discussed. However, for a
Minkowski signature
space-time there is a subtlety. The Minkowski reality
conditions \eqref{eq:RC} imply that $\tilde k$ and $k$ are 
complex conjugates of each other: For a null co-vector
$p_{\al\db}=\tilde k_\al k_\db$, the reality condition $\bar
p_{\al\db}=-p_{\be\da}$ follows from $\tilde k_\al=\di \bar
k_\da$. This yields,
$$ \frac{\partial}{\partial \tilde k_\al}\,\frac{1}{[k k']}\ =\ -
2\pi\tilde k^\al\,\frac{\langle\tilde\xi\tilde k'\rangle}{\langle\tilde\xi\tilde
k\rangle}\,
\delta^{(2)}([kk'],\langle\tilde k\tilde k'\rangle)~,$$
for some arbitrary $\tilde\xi$; see also \eqref{eq:delta}. Note that on
the support  of the delta function, $k\propto k'$ and $\tilde k\propto \tilde
k'$ and so this expression is independent of $\tilde\xi$. Therefore, even though
the
functions multiplying the overall delta functions in the MHV (super)amplitudes
are independent of $\tilde k$, one produces delta functions whenever two
co-spinors become collinear, that is, whenever $[rs]=0$ for some $r$ and $s$.
This is known as the holomorphic anomaly \cite{Cachazo:2004by} and the problem
carries over to generic (super)amplitudes. One can resolve this
issue by modifying the generators of the (super)conformal group
\eqref{eq:CAMom}, \eqref{eq:SCAMom}, such that (super)conformal invariance 
holds also for Minkowski signature space-time. See
\cite{Bargheer:2009qu,Sever:2009aa,Korchemsky:2009hm}. At tree-level, the
holomorphic anomaly does not really matter as long as one sits at generic points
in momentum space. At loop-level, however, the anomaly becomes important as one
has to integrate over internal momenta.
}

\vspace*{10pt}

\subsection{Witten's half Fourier transform}\label{sec:WHFT}

The subject of this section is to give the Witten transform
\eqref{eq:WittenFourier}, \eqref{eq:WittenFourierSuper} a more precise
meaning.  We shall closely follow the treatment
of Mason \& Skinner \cite{Mason:2009sa}. 
As we have already indicated, we will focus on reality conditions that lead to a
Kleinian signature space-time. To jump ahead of our story a bit, the two main
reasons for doing this are as follows: Firstly, for Kleinian signature
(super)twistor space becomes a subset of real projective (super)space.
Secondly, solutions to zero-rest-mass field equations are represented in terms
of straightforward functions rather than representatives of sheaf cohomology
groups. This therefore simplifies the discussion. At this point it should be
noted that as far as perturbation theory is concerned, the choice of signature
is largely irrelevant as the scattering amplitudes are holomorphic functions of
the Mandelstam variables.

In Sections \ref{sec:PWtransform}~and \ref{sec:HCS}, we saw that Euclidean
reality conditions arise from a certain anti-holomorphic involution on the
(super)twistor space. We can discuss Kleinian reality conditions in an analogous
manner. To this end, we introduce the following anti-holomorphic involution
$\tau\,:\,P^{3|4}\to P^{3|4}$ on the supertwistor space: 
\begin{equation}\label{eq:RCKSS}
 \tau(z^A,\lambda_\da)\ :=\ (\hat z^A,\hat\lambda_\da)\ :=\ (\bar z^A ,\bar\lambda_\da)~.
\end{equation}
By virtue of the incidence relation $z^A=x^{A\da}\lambda_\da$, we find 
\begin{equation}\label{eq:xRelKlein}
 \tau(x^{A\da})\ =\ \bar x^{A\da}~
\end{equation}
such that the set of fixed points $\tau(x)=x$, given by $x=\bar x$ and denoted
by $M^{4|8}_\tau\subset M^{4|8}$, corresponds to Kleinian superspace
$\IR^{2,2|8}$ in $M^{4|8}$. Furthermore, it is important to emphasise that
unlike the Euclidean involution \eqref{eq:RCESS}, the involution 
\eqref{eq:RCKSS} has fixed points on $P^{3|4}$ since $\tau^2=1$ on $P^{3|4}$. In
particular, the set of fixed points $\tau(z,\lambda)=(z,\lambda)$, given by 
$(z,\lambda)=(\bar z,\bar \lambda)$ and denoted by $P^{3|4}_\tau\subset
P^{3|4}$,  is an open subset of real projective
superspace
$\IR P^{3|4}$ that is diffeomorphic to $\IR P^{3|4}\setminus \IR P^{1|4} $ and
fibred over $S^1\cong\IR P^1\subset\IC P^1$. Hence, the double fibration
\eqref{eq:DFsuper} becomes
\begin{equation}\label{eq:DFsuperKlein}
 \begin{picture}(50,40)
  \put(0.0,0.0){\makebox(0,0)[c]{$P^{3|4}_\tau$}}
  \put(64.0,0.0){\makebox(0,0)[c]{$M^{4|8}_\tau$}}
  \put(34.0,33.0){\makebox(0,0)[c]{$F^{5|8}_\tau$}}
  \put(7.0,18.0){\makebox(0,0)[c]{$\pi_1$}}
  \put(55.0,18.0){\makebox(0,0)[c]{$\pi_2$}}
  \put(25.0,25.0){\vector(-1,-1){18}}
  \put(37.0,25.0){\vector(1,-1){18}}
 \end{picture}
\end{equation}
where the correspondence space is $F^{5|8}_\tau\cong \IR^{2,2|8}\times S^1\cong \IR^{2,2|8}\times \IR P^1$.
The projection $\pi_2$ is again the trivial projection and
$\pi_1\,:\,(x^{A\da},\lambda_\da)\mapsto
(z^A,\lambda_\da)=(x^{A\da}\lambda_\da,\lambda_\da)$. Furthermore, the 
geometric correspondence in this real setting is as follows: A point $x\in
M^{4|8}_\tau\cong \IR^{2,2|8}$ corresponds to a real projective line
$\IR P^1\cong L_x=\pi_1(\pi_2^{-1}(x))\hookrightarrow P^{3|\CN}$, while a point
$p=(z,\lambda)\in P^{3|4}_\tau$ corresponds to a real $2|4$-plane in
$M^{4|8}_\tau$ that is parametrised by
$x^{A\da}=x^{A\da}_0+\mu^A\lambda^\da$, where $x_0^{A\da}$ is a particular
solution to $z^A=x^{A\da}\lambda_\da$. Notice that one has a similar double
fibration and geometric correspondence in the non-supersymmetric setting
(one simply drops the fermionic coordinates). For a discussion on the
subtleties of the Penrose--Ward transform in the Kleinian setting, see
e.g.~\cite{Popov:2004rb}. 

Now we are in a position to give the Witten transform
\eqref{eq:WittenFourier}, \eqref{eq:WittenFourierSuper} a
precise meaning. Let us start with \eqref{eq:WittenFourier} and consider Penrose's integral formula 
\begin{equation}
 \phi(x)\ =\ -\frac{1}{2\pi\di}\oint[\dd\lambda\lambda]\,f_{-2}(x^{\al\db}\lambda_\db,\lambda_\da)~,
\end{equation}
which was introduced in Section \ref{sec:IF} Note that in this real setting,
one also refers to this transformation as the X-ray
transform.\footnote{\label{foot:xray} The X-ray transform derives its name from
X-ray tomography
because the X-ray transform of some given function $f$ represents the scattering
data of a tomographic scan through an inhomogeneous medium whose density is
represented by $f$. It is closely related to the Radon transform; see footnote
\ref{foot:radon}.}
Recall that $\phi$ obeys the Klein--Gordon equation $\square\phi=0$. From Remark
\ref{rmk:WF}, we also know that the Fourier transform $\tilde\phi(p)$ of
$\phi(x)$ is of the form $\tilde\phi(p)=2\pi\di\,\delta(p^2)\tilde\phi_0(\tilde
k,k)$, where $p_{\al\db}=\tilde k_\al k_\db$. The pre-factor of $2\pi\di$ has
been chosen for later convenience. Then we may write
\begin{equation}
 2\pi\di\,\delta(p^2)\tilde\phi_0(\tilde k,k)\ =\ \int\dd^4x\, \de^{\di p\cdot x}\phi(x)
 \ =\  -\frac{1}{2\pi\di}\int\dd^4x\, \de^{\di p\cdot x}\oint[\dd\lambda\lambda]\,f_{-2}(x^{\al\db}\lambda_\db,\lambda_\da)~.
\end{equation}
Next let us choose some constant $\mu_\da$ with $\mu_\da=\bar\mu_\da$ and
decompose $x^{\al\db}$ as 
\begin{equation}
 x^{\al\db}\ =\ \frac{z^\al\mu^\db-w^\al\lambda^\db}{[\lambda\mu]}~.
\end{equation}
A short calculation shows that $z^\al=x^{\al\db}\lambda_\db$ and
$w^\al=x^{\al\db}\mu_\db$. Furthermore, $\bar x=x$ is satisfied. Note
that the choice of $\mu_\da$ does not really matter as will become clear
momentarily. The measure $\dd^4x$ is then given by
\begin{equation}
 \dd^4x\ =\ \frac{\dd^2z\,\dd^2w}{[\lambda\mu]^2}~.
\end{equation}
Making use of the short-hand notation $\langle z|p|\mu]=z^\al p_{\al\db}\mu^\db$, etc.,
we can integrate out $w^\al$ to obtain
\begin{eqnarray}
 2\pi\di\,\delta(p^2)\tilde\phi_0(\tilde k,k)\! &=&\! 
 -\frac{1}{2\pi\di}\int\dd^4x\, [\dd\lambda\lambda]\, \de^{\di p\cdot x}\,f(x^{\al\db}\lambda_\db,\lambda_\da)\notag\\
&=&\!  -\frac{1}{2\pi\di}\int\frac{\dd^2z\,\dd^2w\, [\dd\lambda\lambda]}{[\lambda\mu]^2}
             \exp\left(\di\frac{\langle z|p|\mu]-\langle w|p|\lambda]}{[\lambda\mu]}\right)f_{-2}(z,\lambda)\notag\\
 &=&\! 2\pi\di\int\frac{\dd^2z\, [\dd\lambda\lambda]}{[\lambda\mu]^2}\,\delta^{(2)}\left(\frac{p|\lambda]}{[\lambda\mu]}\right)
\exp\left(\di\frac{\langle z|p|\mu]}{[\lambda\mu]}\right)f_{-2}(z,\lambda)~.
\end{eqnarray}
The delta function may be converted into
\begin{equation}
 \delta^{(2)}\left(\frac{p|\lambda]}{[\lambda\mu]}\right)\ =\ \big| [\lambda\mu][k\mu] \big|\,\delta(p^2)\,\delta([k\lambda])~,
\end{equation}
where $p_{\al\db}=\tilde k_\al k_\db$ on the support of $\delta(p^2)$. Next
we may perform the integral over $\lambda_\da$,
\begin{eqnarray}
 2\pi\di\,\delta(p^2)\tilde\phi_0(\tilde k,k)\! &=&\! 
  2\pi\di\,\delta(p^2)\int\dd^2z\, [\dd\lambda\lambda]\left|\frac{[k\mu]}{[\lambda\mu]}\right|\delta([k\lambda])
 \exp\left(\di\frac{\langle z\tilde k\rangle[k\mu]}{[\lambda\mu]}\right)f_{-2}(z,\lambda)\notag\\
 &=&\! 2\pi\di\,\delta(p^2)\int\dd^2z\, \de^{\di\langle z\tilde k\rangle}
 f_{-2}(z,k)~.
\end{eqnarray}
Altogether, we arrive at
\begin{equation}
\begin{aligned}
 \tilde\phi_0(\tilde k,k)\ &=\ \int\dd^2z\, \de^{\di\langle z\tilde k\rangle}f_{-2}(z,\lambda)~,\\
 f_{-2}(z,\lambda)\ &=\  \int\frac{\dd^2\tilde k}{(2\pi)^2}\, \de^{-\di\langle z\tilde k\rangle}\tilde\phi_0(\tilde k,k)~.
\end{aligned}
\end{equation}
Here, we have re-labelled $k$ by $\lambda$ as before.

The integral formul{\ae} for other helicity fields introduced in Section
\ref{sec:IF}~may be treated similarly. Recall that they are given by ($h>0$)
\begin{equation}
\begin{aligned}
 \phi_{\da_1\cdots\da_{2h}}(x)\ &=\ -\frac{1}{2\pi\di}\oint [\dd
                 \lambda\lambda]\,\lambda_{\da_1}\cdots\lambda_{\da_{2h}}
                   f_{-2h-2}(x^{\al\db}\lambda_\db,\lambda_\da)~,\\
 \phi_{\al_1\cdots\al_{2h}}(x)\ &=\ -\frac{1}{2\pi\di}\oint[\dd
                 \lambda\lambda]\,\frac{\partial}{\partial z^{\al_1}}
                 \cdots \frac{\partial}{\partial z^{\al_{2h}}}
                   f_{2h-2}(x^{\al\db}\lambda_\db,\lambda_\da)~.
\end{aligned}
\end{equation}
According to Remark \ref{rmk:WF}, the corresponding Fourier transforms may be taken as
\begin{equation}
\begin{aligned}
 \tilde\phi_{\da_1\cdots\da_{2h}}(p)\ &=\ 2\pi\di\,\delta(p^2) k_{\da_1}\cdots k_{\da_{2h}}\tilde\phi_{-2h}(\tilde k,k)~,\\ 
 \tilde\phi_{\al_1\cdots\al_{2h}}(p)\ &=\ 2\pi(-\di)^{2h-1}\,\delta(p^2)\tilde k_{\al_1}\cdots\tilde k_{\al_{2h}}\tilde\phi_{2h}(\tilde k,k)~,
\end{aligned}
\end{equation}
where the additional factor of $(-\di)^{2h}$ has been inserted for convenience.
Repeating the above steps, one can show that
\begin{equation}
\begin{aligned}
 \tilde\phi_{\mp2 h}(\tilde k,k)\ &=\ \int\dd^2z\, \de^{\di\langle z\tilde k\rangle}f_{\mp2h-2}(z,\lambda)~,\\
 f_{\mp2h-2}(z,\lambda)\ &=\  \int\frac{\dd^2\tilde k}{(2\pi)^2}\, \de^{-\di\langle z\tilde k\rangle}\tilde\phi_{\mp2 h}(\tilde k,k)~.
\end{aligned}
\end{equation}

In summary, we may rephrase the Theorem \ref{thm:penrose} as follows
\cite{Mason:2009sa}:

 {\Thm\label{thm:KleinPen} Consider a
Kleinian signature space-time
$M_\tau^4\cong\IR^{2,2}$ with twistor space $P_\tau^3\cong\IR P^3\setminus\IR
P^1$.  Let $\tilde\phi_{\mp2h}(\tilde k,k)$ be a (sufficiently well-behaved)
on-shell momentum space wavefunction of helicity $\mp h$ ($h\geq 0$) with
$p_{\al\db}=\tilde k_\al k_\db$. Furthermore, let $(z^\al,\lambda_\da)$ be
homogenous coordinates on $P_\tau^3$. Then one can uniquely associate with
$\tilde\phi_{\mp 2h}(\tilde k,k)$ a function $f_{\mp 2h-2}(z,\lambda)$ of
homogeneity $\mp 2h-2$ on $P_\tau^3$  according to
$$
\begin{aligned}
 f_{\mp2h-2}(z,\lambda)\ &=\  \int\frac{\dd^2\tilde k}{(2\pi)^2}\,
\de^{-\di\langle z\tilde k\rangle}\tilde\phi_{\mp 2h}(\tilde k,k)~,\\
\tilde\phi_{\mp2 h}(\tilde k,k)\ &=\ \int\dd^2z\, \de^{\di\langle z\tilde
k\rangle}f_{\mp2h-2}(z,\lambda)~.
\end{aligned}
$$
}

\noindent
For the sake of brevity, we shall write
\begin{equation}\label{eq:WittenFourierIntegral}
 \begin{aligned}
 f_{2h-2}(z,\lambda)\ &=\  \int\frac{\dd^2\tilde k}{(2\pi)^2}\,
\de^{-\di\langle z\tilde k\rangle}\tilde\phi_{2h}(\tilde k,k)~,\\
\tilde\phi_{2 h}(\tilde k,k)\ &=\ \int\dd^2z\, \de^{\di\langle z\tilde
k\rangle}f_{2h-2}(z,\lambda)~
\end{aligned}
\end{equation}
and allow $h$ to run over $\frac12\IZ$.

Recall that by virtue of \eqref{eq:AuxCon}, the scattering amplitude
$\CA=\CA(\{\tilde k_{r\al},k_{r\da},h_r\})$ scales as
$t^{2h_r}$ under $(\tilde
k_{r\al},k_{r\da})\mapsto (t^{-1}\tilde k_{r\al},tk_{r\da})$. Also recall that
the Witten transform \eqref{eq:WittenFourier} 
of $\CA$, which we denoted by $\cW[\CA]$, is of homogeneity $2h_r-2$ in
$(z_r^\al,\lambda_{r\da})$ on the multi-particle twistor space $\cP^{3n}$; see
\eqref{eq:ScaTwiAm}. Therefore, $\cW[\CA]$ is obtained by performing the 
transformation \eqref{eq:WittenFourierIntegral} on each $\tilde k_{r\al}$, that
is 
\begin{subequations}
\begin{equation}\label{eq:WittenFourierIntegralAm}
 \cW[\CA](\{z_r^\al,\lambda_{r\da},h_r\})\ =\
\int\left(\prod_{r=1}^n\frac{\dd^2\tilde
k_r}{(2\pi)^2}\right)\de^{-\di\sum_{r=1}^n\langle z_r\tilde k_r\rangle}
 \CA(\{\tilde k_{r\al},k_{r\da},h_r\})~.
\end{equation}
This explains
the name `half Fourier transform' since we are essentially Fourier-transforming
only $\tilde k_r$ and not the `whole' of $p_r$. The inverse Witten transform is
just
\begin{equation}
 \CA(\{\tilde k_{r\al},k_{r\da},h_r\})\ =\    \int\left(\prod_{r=1}^n
\dd^2 z_r\right)\de^{\di\sum_{r=1}^n\langle z_r\tilde k_r\rangle}
 \cW[\CA](\{z_r^\al,\lambda_{r\da},h_r\})~.          
\end{equation}
\end{subequations}

Giving a precise meaning to \eqref{eq:WittenFourierSuper} is now a rather
small
step. As shown in the preceding section, the superamplitude $\CF=\CF(\{\tilde
k_{rA},k_{r\da}\})$ scales as $t^{-2}$ under $(\tilde k_{rA},k_{r\da})\mapsto
(t^{-1}\tilde k_{rA},tk_{r\da})$. Furthermore, the measure
$\dd^{2|4}\tilde k_r:=\dd^2\tilde k_r\dd^4\psi_r$ scales as $t^2$ because
of the definition of the Berezin integral over fermionic coordinates. If we
define
\begin{equation}
 \langle\!\langle z_r\tilde k_r\rangle\!\rangle\ :=\
 z^A_r\tilde k_{rA}\ =\ z^\al_r\tilde k_{r\al}+\eta_{ir}\psi^i_r
\end{equation}
then
\begin{equation}\label{eq:WittenFourierIntegralSuAm}
 \cW[\CF](\{z_r^A,\lambda_{r\da}\})\ =\
\int\left(\prod_{r=1}^n\frac{\dd^{2|4}\tilde
k_r}{(2\pi)^2}\right)\de^{-\di\sum_{r=1}^n\langle\!\langle z_r\tilde
k_r\rangle\!\rangle}
 \CF(\{\tilde k_{rA},k_{r\da}\})
\end{equation}
is of homogeneity zero on the multi-particle supertwistor space $\cP^{3n|4n}$.
This is essentially the implication of \eqref{eq:ScaTwiSuAm}. 

Finally, we would like to address the relationship between the twistor space
expression \eqref{eq:etaexp} and the on-shell super wavefunction
\eqref{eq:SuWaFu}. In Section \ref{sec:HCS}~we worked
in the Dolbeault picture where everything is encoded in a
differential $(0,1)$-form \eqref{eq:etaexp}. Upon imposing the linearised field
equations
$\dbar A^{0,1}=0$, one can use the Penrose transform to get the whole
$\CN=4$ multiplet on four-dimensional space-time. Now let 
$f_0=f_0(z,\lambda)$ be the \v Cech representative corresponding to 
$A^{0,1}$, that is, $f_0$ is a Lie algebra-valued element of
$H^1(P^{3|4},\CO)$. Like $A^{0,1}$, $f_0$ can also be expanded in powers
of $\eta_i$,
 \begin{equation}\label{eq:fetaexp}
 f_0\ =\ A+\eta_i\chi^i+\tfrac{1}{2!}\eta_i\eta_j\phi^{ij}+
 \tfrac{1}{3!}\eta_i\eta_j\eta_k\varepsilon^{ijkl}\chi_l+
 \tfrac{1}{4!}\eta_i\eta_j\eta_k\eta_l\varepsilon^{ijkl}G~,
\end{equation}
where the component fields are Lie algebra-valued elements of
$H^1(P^3,\CO(-2h-2))$ for
$h=-1,\ldots,1$. Upon imposing Kleinian reality conditions, the map between
$f_0$ and $\tilde\Phi_{-2}$ is simply the $\CN=4$ supersymmetric version of
Theorem \ref{thm:KleinPen} \cite{Mason:2009sa}:
\begin{equation}\label{eq:WittenFourierIntegralSu}
 \begin{aligned}
 f_0(z,\lambda)\ &=\  \int\frac{\dd^{2|4}\tilde k}{(2\pi)^2}\,
\de^{-\di\langle\!\langle z\tilde k\rangle\!\rangle}\tilde\Phi_{-2}(\tilde
k,k)~,\\
\tilde\Phi_{-2}(\tilde k,k)\ &=\ \int\dd^{2|4}z\, \de^{\di\langle\!\langle
z\tilde
k\rangle\!\rangle}f_0(z,\lambda)~.
\end{aligned}
\end{equation}

\subsection{MHV superamplitudes on supertwistor
space}\label{sec:MHVLocalisation}

Having given a precise meaning to the Witten transform, we
are now in a postion to explore the properties of the MHV amplitude
\eqref{eq:MHVAmpd} and its $\CN=4$ supersymmentric extension \eqref{eq:MHVSuper}
when re-interpreted in terms of twistors. Since
\eqref{eq:MHVAmpd} is just a special case of \eqref{eq:MHVSuper}, we shall
focus on the superamplitude only.

In order to transform \eqref{eq:MHVSuper} to $\cP^{3n|4n}$, let us re-write the
overall delta function as 
\begin{equation}
 \delta^{(4|8)}\left(\sum_{r=1}^n \tilde k_{r A}k_{r\db}\right)
 \ =\ \int\frac{\dd^{4|8}x}{(2\pi)^4}\, \exp\left(\di\sum_{n=1}^n x^{A\da}\tilde
k_{r A} k_{r\da}\right),
\end{equation}
where $x$ is some dummy variable to be interpreted.
Therefore, upon substituting the superamplitude \eqref{eq:MHVSuper} into 
\eqref{eq:WittenFourierIntegralAm}, we find
\begin{eqnarray}\label{eq:TwistorSMHV}
   \cW\big[F_{0,n}^{\rm MHV}\big]\! &=&\! g^{n-2}_{\rm YM}
 \int\dd^{4|8}x\int\left(\prod_{r=1}^n\frac{\dd^{2|4}\tilde
k_r}{(2\pi)^2}\right)\de^{-\di\sum_{r=1}^n(z_r^A-x^{A\da}k_{r\da})\tilde
k_{rA}}   \frac{1}{[12][23]\cdots[n1]}\notag \\
&=&\! g^{n-2}_{\rm
YM}\int\dd^{4|8}x\,
\frac{\prod_{r=1}^n\delta^{(2|4)}\big(z_r^A-x^{A\da}k_{r\da}\big)}{[12][23]
\cdots[n1]}~.
\end{eqnarray}

\begin{figure}[h]
\begin{center}
\begin{picture}(150,150)(0,0)
   \SetScale{1}
   \put(0,0){
   \Line(0,0)(120,120)
   \Vertex(20,20){2}
   \Vertex(40,40){2}
   \Vertex(60,60){2}
   \Vertex(80,80){2}
   \Vertex(100,100){2}
   \Text(30,15)[]{$+$}
   \Text(50,35)[]{$-$}
   \Text(70,55)[]{$+$}
   \Text(90,75)[]{$-$}
   \Text(110,95)[]{$+$}
   \Text(130,130)[]{$\IR P^1$}
   }
\end{picture}
\end{center}
\caption{\label{fig:Line}{\it\small The MHV (super)amplitudes localise on
projective lines
in (super)twistor space. Here, a five-gluon MHV amplitude is depicted as an
example.}}
\end{figure}
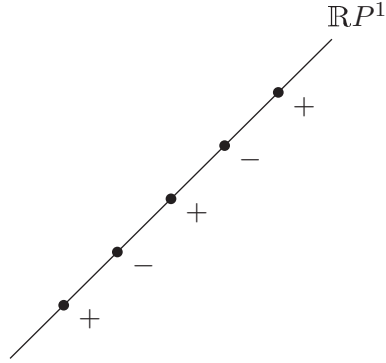

Let us interpret this result. The MHV superamplitude is supported on solutions
to the equations
\begin{equation}\label{eq:MHVlocalisation}
 z_r^A\ =\ x^{A\da}\lambda_{r\da}~,\efor r\ =\ 1,\ldots,n~
\end{equation}
on $\cP^{3n|4n}$, where we have re-labelled $k_{r\da}$ by $\lambda_{r\da}$. One
may rephrase this by saying that the MHV superamplitude `localises' on
solutions to these equations. We have encountered the relation
\eqref{eq:MHVlocalisation} on numerous occasions:
For each $x^{A\da}$, the incidence relation
$z^A=x^{A\da}\lambda_\da$ defines a real projective line inside supertwistor
space. Then \eqref{eq:TwistorSMHV} says that the MHV superamplitude
vanishes unless the equations \eqref{eq:MHVlocalisation} are satisfied, that
is, unless there is some real projective line inside supertwistor
space which is determined by $x^{A\da}$ via the incidence relation and which
contains all $n$ points $(z^A_r,\lambda_{r\da})$. Put differently, the
supertwistors $Z^I_r:=(z^A_r,\lambda_{r\da})$ are
collinear (see Figure \ref{fig:Line}). The $x$-integral in
\eqref{eq:TwistorSMHV}
is then to be understood
as the integral over the moduli space of this projective line. It is important
to stress that the $x^{A\da}$ should not be confused with the particle
coordinates
in superspace-time even though they have the same mass dimension as the
coordinates in superspace-time.

Finally, let us perform the moduli space integral \eqref{eq:TwistorSMHV} to
obtain the explicit expression of the Witten transform of the MHV
superamplitude. To this end, we use the formula
\begin{equation}
 \delta^{(m|n)}(f(x))\ =\
\sum_{\{x_\ell\,|\,f(x_\ell)=0\}}
\frac{\delta^{(m|n)}(x-x_\ell)}{|\,\mbox{sdet}\,f'(x_\ell)| } ~,
\end{equation}
where `sdet' is the superdeterminant \eqref{eq:sdet} and re-write 
two of the delta functions in \eqref{eq:TwistorSMHV}
according to
\begin{equation}
\delta^{(2|4)}\big(z_r^A-x^{A\da}k_{r\da}\big)\delta^{(2|4)}\big(z_s^A-x^{A\da}
k_{s\da}\big)\ =\ [rs]^2\,\delta^{(4|8)}\left(x^{A\da}-\frac{
z^A_r\lambda^\da_s-z^A_s\lambda^\da_r } { [ rs ] }\right).
\end{equation}
If we choose $r=1$ and $s=2$, for instance,  we end up with
\begin{equation}
   \cW\big[F_{0,n}^{\rm MHV}\big]\ =\  g^{n-2}_{\rm YM}
       \frac{[12]^{6-2n}}{[12][23]\cdots[n1]}
     \prod_{r=3}^n\delta^{(2|4)}\big(z_1^A[2r]+z_2^A[r1]+z_r^A[12]\big)~.
\end{equation}
Of particular interest will be the $n=3$ case, which we
record here:
\begin{equation}
   \cW\big[F_{0,3}^{\rm MHV}\big]\ =\  
       \frac{g_{\rm YM}}{[12][23][31]}\,
     \delta^{(2|4)}\big(z_1^A[23]+z_2^A[31]+z_3^A[12]\big)~.
\end{equation}

Summarising, we see that MHV superamplitudes localise on
projective lines in supertwistor space. These are special instances of algebraic
curves.
In Section \ref{sec:AmplitudeLocalisation}, we shall see that 
general superamplitudes also localise on
algebraic curves in supertwistor space. These localisation properties of
scattering amplitudes are among the key results of \cite{Witten:2003nn}.
This concludes our discussion about MHV amplitudes.

\Remark{Here, we have worked in a split signature space-time. However, one may
complexify and show that in a complex setting, the MHV (super)amplitudes also
localise on degree-one algebraic curves in (super)twistor space
\cite{Witten:2003nn}. Put differently, the MHV (super)amplitudes are supported
on
complex projective lines $\IC P^1$ in complex (super)twistor space.}

\vspace*{10pt}

\section{MHV formalism}

In the preceeding sections, we have extensively discussed
(colour-stripped) tree-level
maximally-helicity-violating amplitudes in both pure Yang--Mills
theory and $\CN=4$ supersymmetric Yang--Mills theory. In the remainder of
these lecture notes we would like to extend this discussion beyond the MHV case
and construct general tree-level scattering amplitudes. 

One of the striking developments that came out of Witten's twistor string theory
is the MHV formalism developed by Cachazo, Svr\v cek \& Witten
\cite{Cachazo:2004kj}. Recall that in the usual Feynman diagram expansion, the
interaction vertices are obtained from the (local) Lagrangian description of the
theory. In particular, interactions are supported on points in space-time. In
the first part of these lecture notes we have seen that points in space-time
correspond to projective lines in twistor space via the incidence relation
\eqref{eq:incidence}. Furthermore, as explained in Section
\ref{sec:MHVLocalisation}, MHV amplitudes localise on projective lines in
twistor space. So one can think of these amplitudes as representing, in some
sense, a generalisation of local interaction vertices. To take this analogy
further, one can try to build more complicated amplitudes from MHV amplitudes by
gluing together the latter in an appropriate way. As we shall see
momentarily, the gluing is done with the help of simple scalar
propagators. In order for this to work, however, one has to
continue the MHV amplitudes off-shell. The resulting diagrammatic expansion of
scattering amplitudes is referred to as the MHV diagram expansion.

We begin by studying the MHV formalism in the pure Yang--Mills setting and
then generalise it to the $\CN=4$ case. Before deriving this formalism from
twistor space, however, we first set up the basics and discuss some examples.

\subsection{Cachazo--Svr\v cek--Witten rules}

As indicated, we must continue the MHV amplitudes off-shell in order to use
them as vertices for general amplitudes. To this end, 
consider a generic momentum co-vector $p$ with $p^2\neq0$. On general grounds,
it can always be decomposed into the sum of two null-momenta. We shall adopt the
following decomposition \cite{Kosower:2004yz,Bena:2004ry}:
\begin{equation}
 p\ =\ q+t\ell~,
\end{equation}
where $q^2=0$, and $\ell$ is fixed but arbitrary with $\ell^2=0$.
Here, $t$ is some complex number which is determined as a function of $p$
according to
\begin{equation}
 t\ =\ \frac{p^2}{2(p\cdot\ell)}~.
\end{equation}
Since $q$ and $\ell$ are null, we may use co-spinors to represent them as
$q_{\al\db}=\tilde k_\al k_\db$ and $\ell_{\al\db}=\tilde\xi_\al\xi_\db$. It
then follows that
\begin{equation}
 \tilde k_\al\ =\ \frac{p_{\al\db}\xi^\db}{[k\xi]}\eand
 k_\da\ =\ \frac{\tilde\xi^\be p_{\be\da}}{\langle\tilde\xi\tilde k\rangle}
\end{equation}
or in our above short-hand notation,
\begin{equation}\label{eq:OffShell}
 \tilde k\ =\ \frac{p|\xi]}{[k\xi]}\eand
 k\ =\ \frac{\langle\tilde\xi|p}{\langle\tilde\xi\tilde k\rangle}
\end{equation}
These equations define our off-shell continuation (note that $p^2\neq
0$ in the above formul{\ae}): The co-spinors $k$ corresponding to the legs in
the MHV amplitudes that are going to be taken off-shell are represented
by \eqref{eq:OffShell}. Note that the denominators in
\eqref{eq:OffShell} turn out to be irrelevant for our applications since the
expressions we are dealing with are homogeneous in the $k$ that are continued
off-shell (see also below). Thus, we may discard the denominators and simply
write
\begin{equation}\label{eq:OffShellNew}
 \tilde k\ =\ p|\xi]\eand
 k\ =\ \langle\tilde\xi|p~.
\end{equation}
This is our off-shell formulation of MHV amplitudes. When continued
off-shell, we shall often refer to MHV amplitudes as MHV vertices. We
complete the definition of the MHV or Cachazo--Svr\v cek--Witten rules, by
taking $1/p^2$ for the propagator connecting the MHV vertices.

\vspace*{10pt}

\noindent
{\bf Cachazo--Svr\v cek--Witten rules.}
{\it The rules for computing colour-stripped tree-level
scattering amplitudes by gluing together MHV amplitudes are as follows:
\begin{itemize}\setlength{\itemsep}{-1mm}
 \item[(i)] Use the MHV amplitudes as vertices.
 \item[(ii)] For each leg of a vertex that joins with a propagator carrying
momentum $p$, define its corresponding co-spinor by  $k=\langle\tilde\xi|p$,
where $\tilde\xi$ is some fixed spinor.
 \item[(iii)] Glue the vertices together using scalar propagators $1/p^2$,
where $p$ is the momentum flowing between the vertices. The two ends of any
propagator must have opposite helicity labels, that is, plus at one end and
minus at the other. 
\end{itemize}
}

The reader may wonder whether the amplitudes constructed this way actually
depend on the choice reference spinor $\tilde\xi$. In fact, it can be shown
that this is not the case \cite{Cachazo:2004kj} and the resulting
overall amplitudes are indeed (Lorentz) covariant. This will also become 
transparent in Section \ref{sec:MHVTwistor}, where we deduce the
above rules from the twistor action \eqref{eq:full}:
$\tilde\xi$ will arise as an ingredient of a certain gauge condition.
BRST invariance in turn guarantees that the overall amplitudes are independent
of $\tilde\xi$.

Note that a direct proof of these rules was given by Risager in
\cite{Risager:2005vk}.

\begin{figure}[h]
\begin{center}
\vspace*{2cm}
\begin{picture}(250,150)(-50,0)
   \SetScale{1}
   \put(0,0){
   \Line(-20,20)(100,120)
   \Line(-20,20)(-60,-20)
   \Line(-20,20)(20,-20)
   \Line(100,120)(150,40)
   \Line(150,40)(100,0)
   \Line(150,40)(210,40)
   \Line(150,40)(170,-10)
   \Line(100,120)(150,170)
   \Line(100,120)(100,180)
   \Line(100,120)(50,170)
   \CCirc(100,120){15}{Black}{Gray}
   \CCirc(-20,20){15}{Black}{Gray}
   \CCirc(150,40){15}{Black}{Gray}   
   \Text(101,120)[]{{\small MHV}}
   \Text(-19,20)[]{{\small MHV}}
   \Text(151,40)[]{{\small MHV}}
   \Text(-65,-25)[]{$+$}
   \Text(25,-25)[]{$+$}
   \Text(45,175)[]{$+$}
   \Text(95,-5)[]{$+$}
   \Text(220,40)[]{$-$}
   \Text(172,-15)[]{$-$}
   \Text(155,175)[]{$-$}
   \Text(101,185)[]{$-$}
   \Text(18,35)[]{$-$}
   \Text(80,85)[]{$+$}
  \Text(135,98)[]{$-$}
   \Text(122,62)[]{$+$}
}
\end{picture}
\end{center}
\vspace*{15pt}
\caption{{\it\small A tree-level MHV diagram describing eight-gluon scattering
with four
gluons of positive helicity and four gluons of negative helicity. Amplitudes
with an equal number of positive and negative helicity gluons are referred to as
split helicity amplitudes. The amplitude displayed here is an N$^2$MHV
amplitude.}}
\end{figure}
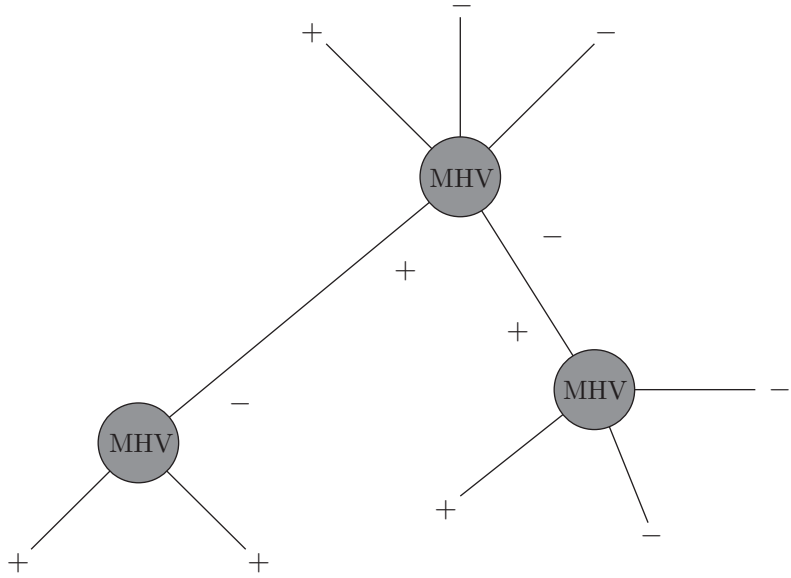

\subsection{Examples}\label{sec:CSWExamples}

The Cachazo--Svr\v cek--Witten rules for joining together the MHV amplitudes
are probably best illustrated with an example. 

A tree-level diagram with $n_V$ MHV vertices has $2n_V$ positive-helicity legs,
$n_V-1$ of which are connected by propagators. According to rule (iii),
each propagator must subsume precisely one positive-helicity leg and therefore,
we are left with $2n_V-(n_V-1)=n_V+1$ external positive helicities. Put
differently, if we wish to compute a scattering amplitude with $q$ positive
helicity gluons, we will need $n_V=q-1$ MHV vertices. Clearly, the MHV case is
$q=2$ and thus $n_V=1$. In additon, this construction 
makes the vanishing of the amplitudes \eqref{eq:MHVAmpa} and \eqref{eq:MHVAmpb}
(both for the upper sign choice) manifest: In the former case, $n_V=-1$ while in
the latter case
$n_V=0$.

The first non-trivial examples are  obtained for $q=3$ which are the NMHV
amplitudes. We shall denote them by $A^{\rm NMHV}_{0,n}(r^+,s^+,t^+)$ or simply
by $A^{\rm NMHV}_{0,n}$ if there is no confusion. Thus, we have schematically
\begin{equation}
\begin{aligned}
 A^{\rm NMHV}_{0,n}\ &=\ \sum_{\{\{\Gamma_L,\Gamma_R\}|n_L+n_R=n+2\}} A^{\rm
MHV}_{0,n_L}(\Gamma_L)\,\frac{1}{p_{L,R}^2}\,A^{\rm
MHV}_{0,n_R}(\Gamma_R)\\[10pt]
 &=\ \sum_{\{\{\Gamma_L,\Gamma_R\}|n_L+n_R=n+2\}}
\begin{picture}(150,40)(-30,0)
   \SetScale{1}
   \put(0,0){
    \Line(20,0)(100,0)
    \Line(20,0)(-10,30)
    \Line(20,0)(-10,-30)
    \Line(100,0)(130,30)
    \Line(100,0)(130,-30)
    \CCirc(20,0){15}{Black}{Gray}
    \CCirc(100,0){15}{Black}{Gray}
    \DashCArc(20,0)(30,160,200){2}
    \DashCArc(100,0)(30,340,20){2}
    \Text(-10,40)[]{$p_{n_L-1}$}
    \Text(-10,-40)[]{$p_1$}
     \Text(130,40)[]{$p_{n_L}$}
    \Text(130,-40)[]{$p_n$}
    \Text(21,0)[]{$\Gamma_L$}
    \Text(101,0)[]{$\Gamma_R$}
    \Text(60,10)[]{$1/p^2_{L,R}$}
}
\end{picture}\\[30pt]
\end{aligned}
\end{equation}
where the sum is taken over all possible MHV vertices $\Gamma_{L,R}$
contributing to the amplitude such that $n_L+n_R=n+2$ and $n_{L,R}\geq3$. One
may check that 
there are a total of $n(n-3)/2$ distinct diagrams contributing to this sum (see
e.g.~\cite{Elvang:2008na,Elvang:2008vz}).

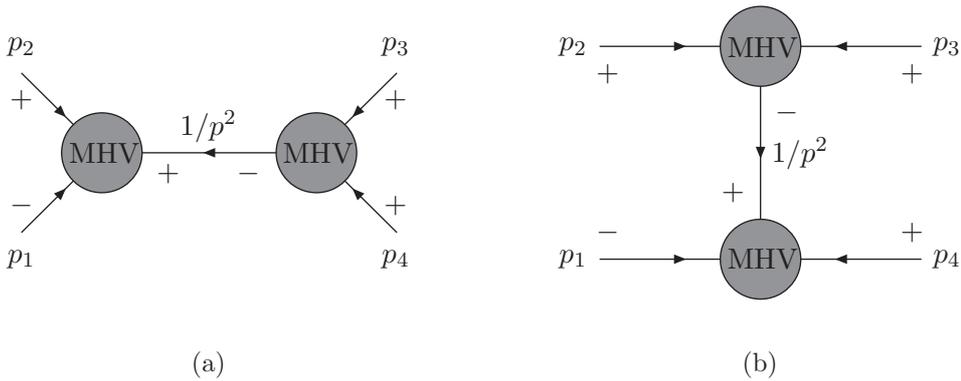
\begin{figure}[h]
\vspace*{-.5cm}
\begin{center}
\begin{picture}(150,100)(-180,0)
   \SetScale{1}
   \put(0,0){
    \Text(-10,-40)[]{$p_1$}
    \Text(-10,40)[]{$p_2$}
    \Text(130,40)[]{$p_3$}
    \Text(130,-40)[]{$p_4$}    
    \Text(3,-30)[]{$-$}
    \Text(3,30)[]{$+$}
    \Text(117,30)[]{$+$}
    \Text(117,-30)[]{$+$}
    \Text(50,-15)[]{$+$}
    \Text(70,15)[]{$-$}
    \Text(75,0)[]{$1/p^2$}
    \ArrowLine(60,40)(60,-40)
    \ArrowLine(0,40)(60,40)
    \ArrowLine(0,-40)(60,-40)
    \ArrowLine(120,40)(60,40)
    \ArrowLine(120,-40)(60,-40)
    \CCirc(60,40){15}{Black}{Gray}
    \CCirc(60,-40){15}{Black}{Gray}
    \Text(61,40)[]{MHV}
    \Text(61,-40)[]{MHV}   
    \Text(60,-80)[]{{\small(b)}}
}
\end{picture}
\begin{picture}(150,100)(180,0)
   \SetScale{1}
   \put(0,0){
  \ArrowLine(100,0)(20,0)
    \ArrowLine(-10,30)(20,0)
    \ArrowLine(-10,-30)(20,0)
    \ArrowLine(130,30)(100,0)
    \ArrowLine(130,-30)(100,0)
    \CCirc(20,0){15}{Black}{Gray}
    \CCirc(100,0){15}{Black}{Gray}
    \Text(21,0)[]{MHV}
    \Text(101,0)[]{MHV}    
   \Text(-10,40)[]{$p_2$}
    \Text(-10,-40)[]{$p_1$}
     \Text(130,40)[]{$p_3$}
    \Text(130,-40)[]{$p_4$}
    \Text(-10,20)[]{$+$}
    \Text(-10,-20)[]{$-$}
     \Text(130,20)[]{$+$}
    \Text(130,-20)[]{$+$}
    \Text(45,-8)[]{$+$}
     \Text(75,-8)[]{$-$}
    \Text(60,10)[]{$1/p^2$}
    \Text(60,-80)[]{{\small(a)}}
}
\end{picture}
\end{center}
\vspace*{2.5cm}
\caption{\label{fig:NMHVExample}{\it\small The two diagrams contributing to the
${-}{+}{+}{+}$ amplitude. All external momenta are taken to be incoming.}}
\end{figure}

The simplest example of an NMHV amplitude is the four-gluon amplitude
$A^{\rm NMHV}_{0,4}(2^+,3^+,4^+)$ $=A_{0,4}(1^-,2^+,3^+,4^+)$. This amplitude 
is of the type \eqref{eq:MHVAmpb} for the lower sign choice and hence, it
should vanish. Let us demonstrate this by using the above rules. The relevant
diagrams contributing to this amplitude are depicted in Figure
\ref{fig:NMHVExample}. For each diagram we should write down the MHV amplitudes
corresponding to each vertex and join them together with the relevant
propagator, remembering to use the above off-shell continuation to
deal with the co-spinors associated with the internal lines. Let us start with
diagram (a). We have $k=\langle\tilde\xi|p$ and therefore
\begin{equation}\label{eq:DiagramA}
 \mbox{(a)}\ =\
\frac{[2k]^4}{[k1][12][2k]}\,\frac{1}{p^2}\,\frac{[34]^4}{[4k][k3][34]}
\ = \
\frac{[2k]^3}{[k1][12]}\,\frac{1}{p^2}\,\frac{[34]^3}{[4k][k3]},
\end{equation}
where
\begin{equation}\label{eq:MomCon}
 p\ =\ -p_1-p_2\ =\ p_3+p_4
\end{equation}
due to momentum conservation. Note that $p_r^2=0$ and therefore,
$p_{r\al\db}=\tilde k_{r\al} k_{r\db}$ for $r=1,\ldots,4$. Next we define
$c_r:=\langle\tilde\xi\tilde k_r\rangle$. Hence,
\begin{equation}
 k\ =\ -c_1k_1-c_2k_2 \ =\ c_3 k_3+c_4 k_4~,
\end{equation}
which follows from \eqref{eq:MomCon} upon contracting with $\tilde\xi$. This
last equation can in turn be used to compute the quantities $[kr]$ which can
then be substituted into \eqref{eq:DiagramA}. Noting that $p^2=-2p_1\cdot
p_2$, we eventually arrive at
\begin{equation}
 \mbox{(a)}\ =\ -\frac{c_1^3}{4c_2c_3c_4}\,\frac{[34]}{\langle12\rangle}~.
\end{equation}

\vspace*{10pt}

Going through the same procedure for the second diagram in Figure
\ref{fig:NMHVExample} gives
\begin{equation}
 \mbox{(b)}\ =\ \frac{c_1^3}{4c_2c_3c_4}\,\frac{\langle23\rangle}{[14]}~
\end{equation}
and therefore,
\begin{equation}
 \mbox{(a)}+\mbox{(b)}\ =\
-\frac{c_1^3}{4c_2c_3c_4}\left(\frac{[34]}{\langle12\rangle}-
\frac{[23]}{\langle 14\rangle}\right).
\end{equation}
Momentum conservation is $\sum_r p_r=\sum_r\tilde k_r k_r=0$ and upon
contracting this equation with $k_3$, one realises that $[34]\langle
14\rangle=[23]\langle12\rangle$. Altogether, $\mbox{(a)}+\mbox{(b)}=0$ as
expected. For all the NMHV amplitudes, see
e.g.~\cite{Kosower:2004yz,Georgiou:2004by,Elvang:2008na,Elvang:2008vz}. 

\vspace*{5pt}

\Exercise{Derive the five-gluon NMHV amplitude $$A^{\rm
NMHV}_{0,5}(2^+,4^+,5^+)\ =\ A_{0,5}(1^-,2^+,3^-,4^+,5^+)$$
using the above rules. You should find $5(5-3)/2=5$ diagrams that contribute
to the amplitude. Notice that this amplitude is actually an
$\overline{\mbox{MHV}}$ amplitude and according to \eqref{eq:MHVAmpc}, you
should find $\langle
13\rangle^4/(\langle12\rangle\langle23\rangle\cdots\langle 51\rangle)$.}

\vspace*{5pt}

The MHV diagram expansion has been implemented both for amplitudes with more
external gluons and amplitudes with more positive helicities. In both cases the
complexity grows, but the number of diagrams one has to consider follows a power
growth for large $n$ (e.g.~$n^2$ for the NMHV diagrams \cite{Cachazo:2004kj})
which is a marked improvement on the factorial
growth of the number of Feynman diagrams needed to compute the same process;
see Table \ref{tab:gluons}.

\subsection{MHV diagrams from twistor space}\label{sec:MHVTwistor}

Let us now examine the Cachazo--Svr\v cek--Witten rules in the
context of the twistor action \eqref{eq:full}. The canonical way of deriving
a perturbative or diagrammatic expansion of some theory that is described by
some
action functional is to fix the gauge symmetries first and then to derive
the corresponding propagators and vertices. As we shall see momentarily, there
exists a gauge for the twistor space action \eqref{eq:full} that leads directly
to the Cachazo--Svr\v cek--Witten rules, that is, to the perturbative expansion
of Yang--Mills theory in  terms of MHV diagrams.
For concreteness, we
again assume
that Euclidean reality conditions have been imposed. The subsequent discussion
follows \cite{Mason:2005zm,Boels:2006ir} closely (see also 
\cite{Boels:2007gv,Mason:2008jy}). For a treatment of perturbative
Chern--Simons
theory, see e.g.~\cite{Axelrod:1991vq,Axelrod:1993wr}. 

As we wish to describe pure Yang--Mills
theory here, we assume that only $\underline{A}$ and $\underline{G}$ are
present in the superfield expansion \eqref{eq:etaexp}. Thus,
\begin{equation}
 A^{0,1}\ =\
\underline{A}+\tfrac{1}{4!}\eta_i\eta_j\eta_k\eta_l\varepsilon^{ijkl}\underline{
G }\ =\ \underline{A}+(\eta)^4\underline{G}~,
\end{equation}
where we have used the notation of \eqref{eq:Nairlemma}.
As before, we shall use the circles to denote the component fields. Upon
substituting the above expansion into the holomorphic Chern--Simons
part of the action \eqref{eq:full}, we arrive at
\begin{equation}
 S\ =\ \int\Omega_0\wedge\mbox{tr}\left\{\underline{G}
\wedge(\dbar\underline{A}+\underline{A}\wedge\underline{A})\right\}
\end{equation}
once the fermionic coordinates are integrated over. Notice that we have
re-scaled the measure $\Omega_0$ in order to get rid of certain numerical
pre-factors which are not essential for the discussion that follows. 

Furthermore, using
\begin{equation}\label{eq:AGexp}
 \underline{A}\ =\ \bar e^0\underline{A}_0+\bar e^\alpha\underline{A}_\alpha
 \eand
 \underline{G}\ =\ \bar e^0\underline{G}_0+\bar e^\alpha\underline{G}_\alpha~,
\end{equation}
we may perform an analysis similar to \eqref{eq:power} to find the contribution
coming
from \eqref{eq:HCSmod}:
\begin{eqnarray}\label{eq:ExpMHVTw}
 {\rm tr}\log\nabla^{0,1}|_{L_x}
   \!&=&\! {\rm tr}\log\left(\dbar+\underline{A}+
            (\eta)^4\underline{G}\right)|_{L_x} \notag\\
  && \kern-3cm =\ {\rm tr}\,\left\{\log\left(\dbar+\underline{A}\right)|_{L_x}+
           \log\left[1+(\eta)^4(\dbar+\underline{A})^{-1}
           \underline{G}\right]|_{L_x}\right\} \notag\\
  && \kern-3cm =\ {\rm tr}\left\{\log\left(\dbar+\underline{A}\right)|_{L_x}+
           \sum_{r=1}^\infty\frac{(-1)^r}{r}\int_{L_x}\prod_{s=1}^r
           K(\lambda_s)\,(\eta_s)^4\,G_A(\lambda_{s-1},\lambda_s)\,
           \underline{G}_0(\lambda_s)\right\}.
\end{eqnarray}
Here, $G_A(\lambda,\lambda')$ is the Green function of the
$(\dbar+\underline{A})|_{L_x}$-operator on the projective line $L_x$,
\begin{equation}\label{eq:GreensWithA}
 (\dbar+\underline{A})^{-1}\omega(\lambda)|_{L_x}\ =\ \int_{L_x} K(\lambda')\,
G_A(\lambda,\lambda')\,\omega_0(\lambda')~,\efor
 \omega\ =\ \bar e^0\omega_0+\bar e^\alpha\omega_\alpha~,
\end{equation}
and $K(\lambda)$ is the K\"ahler form on $L_x$ which was 
introduced in Remark \ref{rmk:fubini}  Now we may employ Nair's lemma
\eqref{eq:Nairlemma} and integrate over $\dd^8\eta$ in \eqref{eq:ExpMHVTw}
to obtain the following form of the action \eqref{eq:HCSmod}:
\begin{equation}
 S_\varepsilon\ =\ -\frac12\varepsilon
 \int\dd^4 x\int_{L_x} K(\lambda_1)K(\lambda_2)[12]^4
 \,{\rm tr}\left\{G_A(\lambda_2,\lambda_1)\,\underline{G}_0(\lambda_1)
 \,G_A(\lambda_1,\lambda_2)\,\underline{G}_0(\lambda_2)\right\}.
\end{equation}
Altogether, the twistor space action \eqref{eq:full} reduces to the Mason
twistor action \cite{Mason:2005zm}:
\begin{equation}\label{eq:MasonAction}
 \begin{aligned}
   &S_{{\rm tot}}\ =\ \int\Omega_0\wedge\mbox{tr}\left\{\underline{G}
\wedge(\dbar\underline{A}+\underline{A}\wedge\underline{A})\right\}\\
 &\kern1.5cm-\,\frac12\varepsilon
 \int\dd^4 x\int_{L_x} K(\lambda_1)K(\lambda_2)[12]^4
 \,{\rm tr}\left\{G_A(\lambda_2,\lambda_1)\,\underline{G}_0(\lambda_1)
 \,G_A(\lambda_1,\lambda_2)\,\underline{G}_0(\lambda_2)\right\}
 \end{aligned}
\end{equation}
of pure Yang--Mills theory.

Finally, we need to find an explicit expression for the Green function
$G_A(\lambda,\lambda')$. Obviously, for $\underline{A}_0=0$,
$G_A(\lambda,\lambda')$ coincides with the Green function
$G(\lambda,\lambda')$ for the $\dbar|_{L_x}$-operator as given in  
\eqref{eq:Greens}.
For $\underline{A}_0\neq0$, we have the expansion
\begin{equation}
 (\dbar+\underline{A})^{-1}|_{L_x}\ =\
\dbar^{-1}|_{L_x}-\dbar^{-1}\underline{A}\,\dbar^{-1}|_{L_x}+\dbar^{-1}
\underline{A}\,\dbar^{-1} \underline{A}\,\dbar^{-1}|_{L_x}+\ \cdots.
\end{equation}
With the help of \eqref{eq:Greens} and
\eqref{eq:GreensWithA}, this can be re-written according to
\begin{equation}
 G_A(\lambda,\lambda')\ =\ G(\lambda,\lambda')+
   \sum_{r=1}^\infty(-1)^r\int_{L_x}\left(\prod_{s=1}^r
           K(\lambda_s)\,G(\lambda_{s-1},\lambda_s)\,
           \underline{A}_0(\lambda_s)\right) G(\lambda_r,\lambda')~,
\end{equation}
with $\lambda_0\equiv\lambda$. This expression
may now be substituted into the Mason action \eqref{eq:MasonAction} to obtain
\begin{equation}\label{eq:MasonAction2}
 \begin{aligned}
   &S_{{\rm tot}}\ =\ \int\Omega_0\wedge\mbox{tr}\left\{\underline{G}
\wedge(\dbar\underline{A}+\underline{A}\wedge\underline{A})\right\}\\
 &\kern1.5cm-\,\frac12\varepsilon
 \int\dd^4 x\sum_{r=2}^\infty(-1)^r\int_{L_x} \left(\prod_{s=1}^r
           K(\lambda_s)\,G(\lambda_{s-1},\lambda_s)\right)\\
            &\kern2.5cm\times\ \sum_{t=2}^r[1t]^4
 \,{\rm
tr}\left\{\underline{G}_0(\lambda_1)\,\underline{A}_0(\lambda_2)\,\cdots\,
\underline{A}_0(\lambda_{t-1})\,\underline{G}_0(\lambda_t)\,
\underline{A}_0(\lambda_{t+1})\,\cdots\,\underline{A}_0(\lambda_r)
 \right\}.
 \end{aligned}
\end{equation}
Notice that the infinite sum strongly resembles a sum of MHV vertices
(recall that $\underline{G}$ carries helicity $+1$ while $\underline{A}$
carries helicity $-1$).
Also notice that \eqref{eq:MasonAction2} is invariant under the following gauge
transformations:
\begin{equation}\label{eq:MasActGS}
 \delta\underline{A}\ =\ \dbar\al+[\underline{A},\al]
 \eand
 \delta\underline{G}\ =\
\dbar\be+[\underline{A},\be]+[\underline{G},\al]~,
\end{equation}
where $\al$ and
$\be$ are smooth functions of homogeneities $0$ and $-4$, respectively. 
Notice that these transformations are an immediate consequence of
\eqref{eq:HCSgauge}.

Next we would like to restore the dependence on the Yang--Mills coupling
constant $g_{{\rm YM}}$ in $S_{\rm tot}$ that is compatible with the
perturbation theory we have discussed so far. Remember from our discussion in
Section
\ref{eq:MotN4SYM} that in this perturbative context, the parameter $\varepsilon$
is identified with $g_{{\rm YM}}^2$. Rescaling
$\underline{A}$ as $\underline{A}\mapsto g_{{\rm YM}}\underline{A}$ and
$\underline{G}$ as $\underline{G}\mapsto g_{{\rm YM}}^{-1}\underline{G}$,
respectively, we find
\begin{equation}\label{eq:MasonAction3}
 \begin{aligned}
   &S_{{\rm tot}}\ =\ \int\Omega_0\wedge\mbox{tr}\left\{\underline{G}
\wedge(\dbar\underline{A}+g_{{\rm YM}}
\underline{A}\wedge\underline{A})\right\}\\
 &\kern1.5cm-\,\frac12
 \int\dd^4 x\sum_{r=2}^\infty(-1)^r g_{{\rm YM}}^{r-2}\int_{L_x}
\left(\prod_{s=1}^r
           K(\lambda_s)\,G(\lambda_{s-1},\lambda_s)\right)\\
            &\kern2.5cm\times\ \sum_{t=2}^r[1t]^4
 \,{\rm
tr}\left\{\underline{G}_0(\lambda_1)\,\underline{A}_0(\lambda_2)\,\cdots\,
\underline{A}_0(\lambda_{t-1})\,\underline{G}_0(\lambda_t)\,
\underline{A}_0(\lambda_{t+1})\,\cdots\,\underline{A}_0(\lambda_r)
 \right\}.
 \end{aligned}
\end{equation}

Let us now decompose this expression as
\begin{subequations}
 \begin{equation}
   S_{\rm tot}\ =\ S_{\rm kin}+S_{\rm int}~,
 \end{equation}
where
\begin{equation}
 \begin{aligned}
  S_{\rm kin}\ &:=\ \int\Omega_0\wedge\mbox{tr}\left\{\underline{G}
\wedge\dbar\underline{A}\right\}\\
 S_{\rm int}\ &:=\ g_{{\rm YM}}\int\Omega_0\wedge\mbox{tr}\left\{\underline{G}
\wedge\underline{A}\wedge\underline{A}\right\}\\
  &\kern0.8cm-\,\frac12
 \int\dd^4 x\sum_{r=2}^\infty(-1)^r 
    g_{{\rm YM}}^{r-2}\int_{L_x} \left(\prod_{s=1}^r
           K(\lambda_s)\,G(\lambda_{s-1},\lambda_s)\right)\\
            &\kern1.4cm\times\ \sum_{t=2}^r[1t]^4
 \,{\rm
tr}\left\{\underline{G}_0(\lambda_1)\,\underline{A}_0(\lambda_2)\,\cdots\,
\underline{A}_0(\lambda_{t-1})\,\underline{G}_0(\lambda_t)\,
\underline{A}_0(\lambda_{t+1})\,\cdots\,\underline{A}_0(\lambda_r)
 \right\}.  
 \end{aligned}
\end{equation}
\end{subequations}
Note that the first term in the infinite sum of $S_{\rm int}$
contains a term that is quadratic in $\underline{G}$ and that involves no
$\underline{A}$ fields. We are always free to treat this term either as
a vertex or as part of
the kinetic energy, but as our re-writing already suggests, we shall do the
former.

After some algebraic manipulations, the interaction part 
$S_{\rm int}$ can be further re-written:
\begin{equation}\label{eq:SintMHV}
 \begin{aligned}
 S_{\rm int}\ &=\ g_{{\rm YM}}\int\Omega_0\wedge\mbox{tr}\left\{\underline{G}
\wedge\underline{A}\wedge\underline{A}\right\}\\
  &\kern0.8cm-\,
 \int\dd^4 x\sum_{r=2}^\infty\frac{(-1)^r}{r!} 
    g_{{\rm YM}}^{r-2}\sum_{\sigma\in S_r/Z_r}\int_{L_x} \left(\prod_{s=1}^r
      K(\lambda_{\sigma(s)})\,
     G(\lambda_{\sigma(s-1)},\lambda_{\sigma(s)})\right)\\
            &\kern1.4cm\times\ \sum_{s<t}[\sigma(s)\sigma(t)]^4
 \,{\rm
tr}\left\{\underline{A}_0(\lambda_{\sigma(1)})\,\cdots\,
\underline{A}_0(\lambda_{\sigma(s-1)})\,\underline{G}_0(\lambda_{\sigma(s)})\,
\underline{A}_0(\lambda_{\sigma(s+1)})\,\cdots\,\right.\\
&\kern3cm\left.\cdots\,
\underline{A}_0(\lambda_{\sigma(t-1)})\, \underline {G}_0(\lambda_{\sigma(t)})\,
\underline{A}_0(\lambda_{\sigma(t+1)})\,\cdots\,\underline{A}_0(\lambda_{
\sigma(r) })
 \right\}.  
 \end{aligned}
\end{equation}
Here, $S_r$ is the permutation group of degree $r$ and $Z_r$ is the group of
cyclic permutations of order $r$.

In order to proceed further, we need to fix the gauge symmetries; see
\eqref{eq:MasActGS}. In Sections
\ref{sec:HCS} and \ref{sec:SYMFromTwistor} we partially fixed the gauge
symmetries by imposing the space-time gauge 
\eqref{eq:BMSgauge}. This partial gauge-fixing turned out to be suitable for
deriving the
usual space-time action from the twistor action. However, we may certainly
impose other gauges and the one
that will eventually yield the Cachazo--Svr\v cek--Witten rules is the
following:
\begin{equation}\label{eq:CSWgauge}
 \tilde\xi_1^\al\bar V_\al\lrcorner\underline{A}\ =\ 0 \ =\ 
 \tilde\xi_2^\al\bar V_\al\lrcorner\underline{G} 
\end{equation}
for some fixed but arbitrary spinors $\tilde\xi_1$ and $\tilde\xi_2$. This is an
axial gauge, so the corresponding ghost terms will decouple. Notice that this
gauge choice differs somewhat from the one of
\cite{Mason:2005zm,Boels:2006ir}\footnote{See also
\cite{Jiang:2008xw} for a similar treatment.}
since here we are using two a priori different spinors $\tilde\xi_1$ and
$\tilde\xi_2$.
Any
solution to these gauge constraints
is of the form 
\begin{equation}\label{eq:SolGaugeConst}
 \underline{A}_\al\ =\ \tilde\xi_{1\al}A
 \eand 
 \underline{G}_\al\ =\ \tilde\xi_{2\al}G
\end{equation}
for some smooth functions $A$ and $G$ of
homogeneities $1$ and $-3$ (remember that $\underline{A}_\al$ and
$\underline{G}_\al$ are of these homogeneities).

Having chosen gauge fixing conditions, we are now in a position to
derive the propagator or the Green function of
the $\dbar$-operator on twistor space $P^3$. Upon
using \eqref{eq:AGexp}, $S_{\rm kin}$ is explicitly given by
\begin{equation}\label{eq:Skinexp}
 S_{\rm kin}\ =\ \int\frac{\Omega_0\wedge\bar\Omega_0}{[\lambda\hat\lambda]^4}
 \,\mbox{tr}\left\{\underline{G}^\al(\bar V_0\underline{A}_\al-\bar V_\al
 \underline{A}_0)+\underline{G}_0\bar V_\al\underline{A}^\al\right\}.
\end{equation}
On solutions \eqref{eq:SolGaugeConst} to the gauge constraints
\eqref{eq:CSWgauge}, this becomes
\begin{equation}\label{eq:Skingf1}
\begin{aligned}
 S_{\rm kin, gf}\ &=\ \int \dd^4x\,K(\lambda)
 \,\mbox{tr}\left\{
  \begin{pmatrix}
   \underline{G}_0~, & \!\!\!G
  \end{pmatrix}
 \begin{pmatrix}
  0 & \tilde\xi_1^\al\bar V_\al \\
  -\tilde\xi_2^\al\bar V_\al & \langle \tilde\xi_2 \tilde\xi_1\rangle\bar V_0
 \end{pmatrix}
 \begin{pmatrix}
  \underline{A}_0 \\ A
 \end{pmatrix}
\right\}\\
   &=\ \int \frac{\dd^4p}{(2\pi)^4}\,K(\lambda)
 \,\mbox{tr}\left\{
  \begin{pmatrix}
   \underline{\tilde{G}}_0~, & \!\!\!\tilde{G}
  \end{pmatrix}(p)
 \begin{pmatrix}
  0 & \di\langle\tilde\xi_1|p|\lambda] \\
  -\di\langle\tilde\xi_2|p|\lambda] & \langle \tilde\xi_2 \tilde\xi_1\rangle\bar
V_0
 \end{pmatrix}
 \begin{pmatrix}
  \underline{\tilde{A}}_0 \\ \tilde{A}
 \end{pmatrix}(-p)
\right\},
\end{aligned}
\end{equation}
where in the first step we have used \eqref{eq:measure} to re-write the
measure in \eqref{eq:Skinexp}
while in the second step we have performed a Fourier transform
on the $x$-coordinates. 

Next it is useful to introduce certain differential
$(0,1)$-form-valued weighted delta functions of spinor products
\cite{Cachazo:2004kj,Boels:2006ir},
\begin{equation}\label{eq:Wdelta}
 \bar\delta_{(m)}([\lambda_1\lambda_2])\ :=\
 \left(\frac{[\lambda_1\xi]}{[\lambda_2\xi]}\right)^{m+1}\dbar(\lambda_2)\,
 \frac{1}{[\lambda_1\lambda_2]}
\end{equation}
where $\xi$ is some constant spinor. Using \eqref{eq:delta}, this is explicitly 
\begin{equation}\label{eq:WdeltaCom}
\begin{aligned}
 \bar\delta_{(m)}([\lambda_1\lambda_2])\ &=\
\underbrace{\frac{[\dd\hat\lambda_2\hat
\lambda_2]}{[\lambda_2\hat\lambda_2]^2}}_{=\ \bar e^0(\lambda_2)}
 \underbrace{(-2\pi)[\lambda_2\hat\lambda_2]^2
\left(\frac{[\xi\lambda_1]}{[\xi\lambda_2]}\right)^{m+1}
\frac{[\xi\hat\lambda_1]}{[\xi\hat\lambda_2]}\,
\delta^{(2)}([\lambda_1\lambda_2],
[\hat\lambda_1\hat\lambda_2])}_{=:\ \bar\delta_{(m)}^0([\lambda_1\lambda_2])}\\
&=\ \bar e^0(\lambda_2)\, \bar\delta_{(m)}^0([\lambda_1\lambda_2])~.
\end{aligned}
\end{equation}
This expression is independent of $\xi$ since on support of the delta function
we have $\lambda_1\propto\lambda_2$. Note that under a re-scaling
$\lambda_{1,2}\mapsto t_{1,2}\lambda_{1,2}$ for $t_{1,2}\in\IC\setminus\{0\}$,
we have
\begin{equation}
 \bar\delta_{(m)}([\lambda_1\lambda_2])\ \mapsto\ 
 t_1^{m}t_2^{-m-2}\bar\delta_{(m)}([\lambda_1\lambda_2])\eand
 \bar\delta_{(m)}^0([\lambda_1\lambda_2])\ \mapsto\ 
 t_1^{m}t_2^{-m}\bar\delta_{(m)}^0([\lambda_1\lambda_2])~.
\end{equation}

Using \eqref{eq:Wdelta}, $S_{\rm kin,gf}$ given by \eqref{eq:Skingf1} becomes
\begin{subequations}
\begin{equation}\label{eq:Skingf2}
\begin{aligned}
 S_{\rm kin, gf}\ 
   &=\ \int \frac{\dd^4p}{(2\pi)^4}\,K(\lambda_1)K(\lambda_2)\\
&\kern1.5cm\times
 \,\mbox{tr}\left\{
  \begin{pmatrix}
   \underline{\tilde{G}}_0~, & \!\!\!\tilde{G}
  \end{pmatrix}(p,\lambda_1)
 \begin{pmatrix}
  0 & \CK_1(\lambda_1,\lambda_2) \\
  \CK_2(\lambda_1,\lambda_2) & \CK_3(\lambda_1,\lambda_2)
 \end{pmatrix}
 \begin{pmatrix}
  \underline{\tilde{A}}_0 \\ \tilde{A}
 \end{pmatrix}(-p,\lambda_2)
\right\},
\end{aligned}
\end{equation}
where
\begin{equation}
\begin{aligned}
 \CK(\lambda_1,\lambda_2)\ &:=\ 
\begin{pmatrix}
  0 & \CK_1(\lambda_1,\lambda_2) \\
  \CK_2(\lambda_1,\lambda_2) & \CK_3(\lambda_1,\lambda_2)
 \end{pmatrix}\\
  &:=\ -\frac{1}{2\pi}
 \begin{pmatrix}
   0 & 
   \di\bar\delta_{(2)}^0([\lambda_1\lambda_2])\langle\tilde\xi_1|p|\lambda_2]\\
   -\di\bar\delta_{(3)}^0([\lambda_1\lambda_2])\langle\tilde\xi_2|p|\lambda_2] &
\bar\delta_{(3)}^0([\lambda_1\lambda_2])
 \langle \tilde\xi_2 \tilde\xi_1\rangle\bar
V_0(\lambda_2)
 \end{pmatrix}
\end{aligned}
\end{equation}
\end{subequations}
The Green function or propagator $\CG(\lambda_1,\lambda_2)$,
\begin{equation}\label{eq:PropagatorDef}
\begin{aligned}
  \CG(\lambda_1,\lambda_2)\ &:=\ 
\begin{pmatrix}
  \CG_1(\lambda_1,\lambda_2) & \CG_2(\lambda_1,\lambda_2) \\
  \CG_3(\lambda_1,\lambda_2) & 0
 \end{pmatrix}\\
  &=\ 
 \begin{pmatrix}
  \langle\underline{\tilde{A}}_0(p,\lambda_1)
  \underline{\tilde{B}}_0(-p,\lambda_2)\rangle &
  \langle\underline{\tilde{A}}_0(p,\lambda_1)\tilde{B}(-p,\lambda_2)\rangle \\
  \langle \tilde{A}(p,\lambda_1)\underline{\tilde{B}}_0(-p,\lambda_2)\rangle & 0
 \end{pmatrix}
\end{aligned}
\end{equation}
is determined by
\begin{equation}
 \int K(\lambda_3)\,\CK(\lambda_1,\lambda_3)\,\CG(\lambda_3,\lambda_2)\ =\ 
\CI(\lambda_1,\lambda_2)~,
\end{equation}
where $\CI(\lambda_1,\lambda_2)$ is the identity
\begin{equation}
 \CI(\lambda_1,\lambda_2)\ :=\ -\frac{1}{2\pi}
 \begin{pmatrix}
   \bar\delta_{(2)}^0([\lambda_1\lambda_2]) & 0 \\
   0 & \bar\delta_{(-3)}^0([\lambda_2\lambda_1])
 \end{pmatrix}.
\end{equation}
A short calculation shows that
 \begin{equation}
  \begin{aligned}
    \CG_1(\lambda_1,\lambda_2)\ &=\ \frac{1}{2\pi}
    \frac{\langle\tilde\xi_2\tilde\xi_1\rangle}{\langle\tilde\xi_2|p|\lambda_1]}
   \left(\bar V_0(\lambda_1)\frac{1}{\langle\tilde\xi_1|p|\lambda_1]}\right)
    \bar\delta_{(2)}^0([\lambda_1\lambda_2])~,\\
   \CG_2(\lambda_1,\lambda_2)\ &=\ \frac{1}{2\pi\di}
 \frac{\bar\delta_{(-3)}^0([\lambda_2\lambda_1])}{\langle
\tilde\xi_2|p|\lambda_2]}~,\\
 \CG_3(\lambda_1,\lambda_2)\ &=\ -\frac{1}{2\pi\di}
 \frac{\bar\delta_{(2)}^0([\lambda_1\lambda_2])}{\langle
\tilde\xi_1|p|\lambda_1]}~.
  \end{aligned}
 \end{equation}
The propagator $\CG_1$ may be simplified further to give
\begin{equation}
 \begin{aligned}
  \CG_1(\lambda_1,\lambda_2)\ &=\ \frac{1}{2\pi}
  \frac{\langle\tilde\xi_2\tilde\xi_1\rangle}{\langle\tilde\xi_2|p|\lambda_1]}
  \frac{\langle\tilde\xi_1|p|\xi]}{[\lambda_1\xi]}\,
  \bar\delta_{(-2)}^0(\langle\tilde\xi_1|p|\lambda_1])\,
  \bar\delta_{(2)}^0([\lambda_1\lambda_2])\\
    &=\ -\frac{1}{2\pi}\frac{4}{p^2}\,
         \bar\delta_{(-2)}^0(\langle\tilde\xi_1|p|\lambda_1])\,
         \bar\delta_{(2)}^0(\langle\tilde\xi_1|p|\lambda_2])~,
 \end{aligned}
\end{equation}
where in the first step we have inserted the definition \eqref{eq:WdeltaCom}
while in the second step we have used the fact that 
$\lambda_1\propto \langle\tilde\xi_1| p$ on support of the first delta
function together with the identity  $p_{\al\da}p^{\be\da}=\frac14 p^2
\delta_\al^\be$.

Altogether, we arrive at
 \begin{equation}\label{eq:PropagatorDefRes}
  \CG(\lambda_1,\lambda_2)\ =\ -\frac{1}{2\pi}
  \begin{pmatrix}
   \frac{4}{p^2}\,
        \bar\delta_{(-2)}^0(\langle\tilde\xi_1|p|\lambda_1])\,
         \bar\delta_{(2)}^0(\langle\tilde\xi_1|p|\lambda_2]) & 
  \frac{\di}{\langle \tilde\xi_2|p|\lambda_2]} 
   \bar\delta_{(-3)}^0([\lambda_2\lambda_1])\\
  - \frac{\di}{\langle\tilde\xi_1|p|\lambda_1]} 
     \bar\delta_{(2)}^0([\lambda_1\lambda_2]) & 0
  \end{pmatrix}.
 \end{equation}
A particularly simple form of the propagator is achieved for the choice 
$\tilde\xi_1\to\tilde\xi$ and 
$\tilde\xi_2\to\tilde\xi$. We shall refer to this gauge as 
Cachazo--Svr\v cek--Witten gauge, and the reason for its name will become
apparent momentarily. Upon multiplying \eqref{eq:PropagatorDef},
\eqref{eq:PropagatorDefRes} by the appropriate differential forms $\bar e^0$
and $\bar e^\al$ and by $\tilde\xi^\al$, our final result for the propagator is
\cite{Boels:2006ir}
\begin{equation}\label{eq:CSWPropagator}
\begin{aligned}
\langle\underline{\tilde{A}}(p,\lambda_1)\wedge
 \underline{\tilde{B}}(-p,\lambda_2)\rangle\ &=\ 
 -\frac{1}{2\pi}\left[
   \frac{4}{p^2}\,
        \bar\delta_{(-2)}(\langle\tilde\xi|p|\lambda_1])\wedge
         \bar\delta_{(2)}(\langle\tilde\xi|p|\lambda_2])
 \phantom{\frac{\bar\delta_{(-3)}}{\di\langle\tilde\xi|p|\lambda_2]}}\right.
\\ 
 &\kern1.5cm\left.+\
\frac{\bar\delta_{(-3)}([\lambda_2\lambda_1])\wedge\langle\tilde\xi
        \bar e(\lambda_2)\rangle}{\di\langle\tilde\xi|p|\lambda_2]}
 -\frac{\langle\tilde\xi
        \bar e(\lambda_1)\rangle\wedge\bar\delta_{(2)}([\lambda_1\lambda_2])
}{\di\langle\tilde\xi|p|\lambda_1]}
     \right].
\end{aligned}
\end{equation}

Having constructed the propagator, we shall now take a closer look at the
vertices appearing in \eqref{eq:SintMHV}. One
pleasing feature of the
Cachazo--Svr\v cek--Witten gauge is that the interaction vertex
$\underline{G}\wedge\underline{A}\wedge\underline{A}$ does not contribute.
Therefore, the only remaining interactions come from the infinite sum in
\eqref{eq:SintMHV}. We shall now show that these are the MHV vertices, i.e.
the on-shell version of
\eqref{eq:SintMHV} does indeed give rise to the MHV amplitudes.

Recall from
Theorem \ref{thm:penrose}~that solutions to the zero-rest-mass field equations
of
helicity $h\in\frac12\IZ$ are given by elements of the \v Cech cohomology group
$H^1(P^3,\CO(-2h-2))$. By Remark \ref{rmk:HodgeDolbeault}~we know that
$H^1(P^3,\CO(-2h-2))\cong H^{0,1}_{\dbar}(P^3,\CO(-2h-2))$. Therefore,
solutions to free field equations may equivalently be described in terms of
differential $(0,1)$-forms $\omega_{-2h-2}$ of homogeneity $-2h-2$. In
particular, the plane wave solutions of momentum $p=\tilde k k$ which were
constructed in Section \ref{sec:Revisted}~arise from
\begin{equation}\label{eq:TwistorWF}
 \omega_{-2h-2}(z,\lambda)\ =\ \bar\delta_{(2h)}[k\lambda])
 \exp\left(\di\langle z\tilde k\rangle\frac{[k\xi]}{[\lambda\xi]}\right) 
\end{equation}
by suitably integrating over $\int_{L_x}[\dd\lambda\lambda]$. For more details,
see Exercise \ref{exe:PWD} As before, the parameter $\xi$ plays no role since
on support of the delta function $\lambda\propto k$.

\vspace*{5pt}

\Exercise{\label{exe:PWD} Show that the wavefunction \eqref{eq:H0PW} arises from
$\int_{L_x}[\dd\lambda\lambda]\wedge\omega_{-2}$. Show further that
the $h=\pm\frac12$ wavefunctions \eqref{eq:H1/2PW} are given by 
$\int_{L_x}[\dd\lambda\lambda]\wedge\lambda_\da\omega_{-3}$ and
$\int_{L_x}[\dd\lambda\lambda]\wedge\partial_\al\omega_{-1}$ where
$\partial_\al:=\partial/\partial z^\al$.}

\vspace*{5pt}

Notice that the twistor space wavefunctions \eqref{eq:TwistorWF} obey the
Cachazo--Svr\v cek--Witten gauge condition, since there are no components
along $\bar e^\al$. Therefore, we may take
\begin{equation}
 \begin{aligned}
  \underline{A}(x,\lambda_{\sigma(s)})\ &=\ t_{\sigma(s)}
   \bar\delta_{(-2)}[k_{\sigma(s)}\lambda_{\sigma(s)}])
 \exp\left(\di\langle z_{\sigma(s)}
 \tilde k_{\sigma(s)}\rangle
\frac{[k_{\sigma(s)}\xi]}{[\lambda_{\sigma(s)}\xi]}\right),\\
  \underline{G}(x,\lambda_{\sigma(s)})\ &=\ t_{\sigma(s)}
   \bar\delta_{(2)}[k_{\sigma(s)}\lambda_{\sigma(s)}])
 \exp\left(\di\langle z_{\sigma(s)}
 \tilde k_{\sigma(s)}\rangle
\frac{[k_{\sigma(s)}\xi]}{[\lambda_{\sigma(s)}\xi]}\right),\\
 \end{aligned}
\end{equation}
where $t_{\sigma(s)}$ is an arbitrary element of the Lie algebra of the gauge
group. Inserting these wavefunctions and the Green function \eqref{eq:Greens}
into \eqref{eq:SintMHV}, one may trivially
perform the $\lambda$-integrals by replacing $\lambda_{\sigma(s)}$ by
$k_{\sigma(s)}$. The final $x$-integral then gives an overall delta function of
momentum conservation. The result of these rather straightforward
manipulations is
\begin{equation}
\begin{aligned}
 S_{\rm int, gf}\ &=\ \sum_{r=0}^\infty  
(\di\, g_{\rm YM})^{r-2}(2\pi)^4\delta^{(4)}\left(
 \sum_{s=1}^r\tilde k_s k_s\right)\\
&\kern2cm\times\frac{1}{r!}\sum_{\sigma\in
S_r/Z_r}\frac{\sum_{s<t}[k_{\sigma(s)}k_{\sigma(t)}]^4}{
[k_{\sigma(r)}k_{\sigma(1)}][k_{\sigma(1)}k_{\sigma(2)}]\cdots
[k_{\sigma(r-1)}k_{\sigma(r)}]}\,\mbox{tr}\{t_{\sigma(1)}\cdots
t_{\sigma(r)}\}~.
\end{aligned}
\end{equation}
This, however, is a sum over all the MHV amplitudes \eqref{eq:MHVAmpd}
including the appropriate colour-trace factors (see also \eqref{eq:TreeAmp}).

Let us now summarise these results. We have shown\footnote{See also
\cite{Jiang:2008xw} for a more detailed exposition of this material.}
 that the twistor space 
action \eqref{eq:full} indeed gives the MHV diagram expansion once
the Cachazo--Svr\v cek--Witten gauge \eqref{eq:CSWgauge} with 
$\tilde\xi_{1,2}\to\tilde\xi$ has been imposed: Firstly, for this choice of
gauge, the interaction vertices come solely from the second part of
\eqref{eq:SintMHV} and as we have just seen, these vertices coincide with
the MHV amplitudes when put on-shell. This is the first of the Cachazo--Svr\v
cek--Witten rules. Secondly, the delta functions
in the propagator \eqref{eq:CSWPropagator} lead to the prescription for the
insertion of $k=\langle\tilde\xi| p$ as the co-spinor corresponding to the
off-shell momentum $p$. This is the second Cachazo--Svr\v cek--Witten rule.
Lastly, the third rule also follows from \eqref{eq:CSWPropagator} due to the
explicit appearance of $1/p^2$. Thus, we have obtained the 
Cachazo--Svr\v cek--Witten rules directly from twistor space. As
already indicated, since the spinor $\tilde\xi$ arises here as part of the gauge
condition, BRST invariance ensures that the overall amplitudes will not depend
on it.

Finally, we would like to emphasise that the above procedure makes manifest the
equivalence between the traditional Feynman diagram expansion and the MHV
diagram
expansion for Yang--Mills theory: The
space-time Yang--Mills action including the resulting Feynman diagram expansion
on the one side and the action or generating functional yielding the
MHV diagram expansion on the other side are merely consequences of
different gauge
choices for the twistor space action. This is possible since the gauge symmetry
of
the twistor space action, being defined on a real six-dimensional manifold, is
much larger than that of the Yang--Mills action on four-dimensional
space-time.
Note that the described equivalence can
only be valid at tree-level in perturbation theory, since in pure Yang--Mills
theory at loop-level there
are
diagrams that cannot be constructed from MHV vertices and propagators alone.
See e.g.~\cite{Brandhuber:2006bf} for a full discussion.
Therefore, even though the space-time gauge \eqref{eq:BMSgauge}
yields the fully fledged Yang--Mills action, the above
derivation of the MHV diagrams appears to work only at tree-level (and so do
the
Cachazo--Svr\v cek--Witten rules stated in the preceding section).\footnote{See
\cite{Brandhuber:2004yw,Quigley:2004pw,Bedford:2004py,Bedford:2004nh,
Glover:2008ffa} for an extension of the
Cachazo--Svr\v cek--Witten rules to one-loop level.} Presumably, the issue
with loops is due to the following two problems. The first problem, as argued
in \cite{Boels:2006ir}, lies in the change of the path integral measure: When
going from the space-time gauge to the Cachazo--Svr\v cek--Witten gauge 
one must perform a complex gauge transformation while the path
integral is only invariant under real gauge transformations. The second problem
has to do with certain regularisation issues as discussed in
\cite{Boels:2008ef}. As we
are only concerned with tree-level scattering amplitudes in these lecture notes,
we shall not worry about these issues any further.

\vspace*{5pt}

\Remark{\label{rmk:CSWMHV} Let us point out that there is an alternative
(but equivalent \cite{Boels:2008ef})
approach that maps the ordinary Yang--Mills action to a functional where the
vertices are explicitly the MHV vertices. This approach has been developed by
Mansfield in \cite{Mansfield:2005yd} (see also
\cite{Gorsky:2005sf,Ettle:2006bw,Feng:2006yy,Brandhuber:2006bf}).
For a recent review, see \cite{Ettle:2008ed}. This approach works directly
in four-dimensional space-time and involves formulating Yang--Mills theory in
light-cone gauge and performing a non-local field transformation. In similarity
with 
the twistor space approach above, the spinor $\tilde\xi$ arises in fixing the
light-cone gauge. Finally, let us mention in passing that the above procedure
has also been extended to gravity \cite{Mason:2008jy} 
(see also
\cite{Giombi:2004ix,Nair:2005iv,Bjerrum-Bohr:2005jr,Nasti:2007sr,Bianchi:2008pu}
)
using the gravity twistor
space action of \cite{Mason:2007ct} (see also \cite{Wolf:2007tx}).
}

\vspace*{10pt}

\subsection{Superamplitudes in the MHV formalism}\label{sec:SuAmpMHVForm}

So far, we have discussed the MHV formalism in pure Yang--Mills theory only.
Let us now briefly explain how the above discussion extends to the $\CN=4$
supersymmetric theory. 

In Section \ref{sec:SMHV}, we have seen that all the different scattering
amplitudes occuring in $\CN=4$ supersymmetric Yang--Mills theory are encoded
in terms of certain superamplitudes which we also called generating functions.
In particular,  all the different (colour-stripped) MHV amplitudes follow from
Nair's formula
\eqref{eq:MHVSuper} via \eqref{eq:AllMHV}. The
question then arises of: How do we generalise \eqref{eq:MHVSuper} to N$^k$MHV
scattering
amplitudes, that is, how can we encode all the N$^k$MHV amplitudes
in terms of N$^k$MHV superamplitudes with a relationship similar to
\eqref{eq:AllMHV}? 

To answer this question, let us first recall that the total helicity of an
$n$-particle N$^k$MHV scattering amplitude is $-n+2k+4$ while an $n$-particle
superamplitude has always a total helicity of $-n$.
Notice that $k=0,\ldots,n-4$ since 
$\CA^{{\rm N}^{0}{\rm MHV}}_n\equiv\CA^{\rm MHV}_n$ and 
$\CA^{{\rm N}^{n-4}{\rm MHV}}_n=\CA^{\overline{\rm MHV}}_n$. Furthermore,
as we have explained in Section \ref{sec:SMHV}, 
the differential operators \eqref{eq:DOAmp} will pick
from a superamplitude the appropriate particles  one
wishes to scatter (i.e.~the external states). This therefore suggests that an 
$n$-particle N$^k$MHV superamplitude $\CF^{{\rm N}^k{\rm MHV}}_n$ is a
homogeneous
polynomial of degree $2(2k+4)=4k+8$ in the fermionic coordinates $\psi^i_r$:
Each $\psi^i_r$ carries a helicity of $-1/2$ and therefore, a
differential operator $D^{(4k+8)}$ of order $4k+8$ that is made of the
operators \eqref{eq:DOAmp} will have a total helicity of $2k+4$.
Consequently, $D^{(4k+8)}\CF^{{\rm N}^k{\rm MHV}}_n$ will carry a
helicity of $-n+2k+4$, as desired.

Besides this, the superamplitudes are supersymmetric by construction and so
they are
annihilated by the Poincar\'e supercharges $Q_{i\al}$ and $Q^i_\da$ and by the
generator of space-time translations $P_{\al\db}$. Since $Q^i_\da$ and
$P_{\al\db}$ are represented by multiplication operators \eqref{eq:CAMom},
\eqref{eq:SCAMom} when acting on superamplitudes, the latter must be
proportional to the overall delta function 
\eqref{eq:OvDF} of (super)momentum conservation. Collecting what we have said so
far, we may write
\cite{Drummond:2008vq}
\begin{subequations}\label{eq:GenericSuAmp}
 \begin{equation}
  \CF_n\ =\ \CF^{{\rm MHV}}_n+\CF^{{\rm NMHV}}_n+\cdots
 + \CF^{{\rm N}^{n-4}{\rm MHV}}_n
 \end{equation}
for a generic superamplitude, together with
\begin{equation} 
\CF^{{\rm N}^k{\rm MHV}}_n\ =\ g_{\rm YM}^{n-2}(2\pi)^4
\delta^{(4|8)}\left(\sum_{r=1}^n \tilde k_{rA}k_{r\db}\right) \CP_n^{(4k)}~,
\end{equation}
\end{subequations}
where $\CP_n^{(4k)}$ is a polynomial 
of degree $4k$ in the fermionic coordinates $\psi^i_r$. Hence,
\begin{equation}
 \CA^{{\rm N}^k{\rm MHV}}_n\ =\ D^{(4k+8)}\CF_n|_{\psi=0}\ =\ 
 D^{(4k+8)}\CF^{{\rm N}^k{\rm MHV}}_n~.
\end{equation}
We should emphasise that we are suppressing $R$-symmetry indices here; see
e.g.~\eqref{eq:SomeSuMHVB}.
As before, we will now restrict our attention to the
colour-stripped amplitudes which  will be denoted by 
$F^{{\rm N}^k{\rm MHV}}_n$,
\begin{equation}
 F^{{\rm N}^k{\rm MHV}}_n\ =\ g_{\rm YM}^{n-2}(2\pi)^4
\delta^{(4|8)}\left(\sum_{r=1}^n \tilde k_{rA}k_{r\db}\right) P_n^{(4k)}~.
\end{equation}

According to the Cachazo--Svr\v cek--Witten rules,
N$^k$MHV amplitudes in pure Yang--Mills theory are obtained by gluing together
$k+1$ MHV vertices. While there is basically only one type of vertex in pure
Yang--Mills theory, in the $\CN=4$ supersymmetric extension we
have plenty of different
MHV vertices corresponding to the different MHV amplitudes \eqref{eq:AllMHV}.
Nevertheless,
the Cachazo--Svr\v cek--Witten rules can be extended to this case rather
straightforwardly as we simply need to glue together the different MHV vertices 
that follow from \eqref{eq:AllMHV} when continued off-shell appropriately.
The types of vertices we need to glue together in order to obtain a particular 
N$^k$MHV amplitude will depend on the $R$-symmetry index structure of the
external states for this amplitude. A generic
N$^{k}$MHV amplitude 
$A^{{\rm N}^k{\rm MHV}}_{0,n}$ is given by
\begin{equation}
 A^{{\rm N}^k{\rm MHV}}_{0,n}\ =\ \sum_{\{\{\Gamma_1,\ldots,\Gamma_{k+1}\}
 |n_1+\dots+n_k=n+2k\}}
  \frac{A^{\rm MHV}_{0,n_1}(\Gamma_1)
 \cdots A^{\rm
MHV}_{0,n_{k+1}}(\Gamma_{k+1})}{p^2_{\Gamma_1}\cdots p^2_{\Gamma_k}}~,
\end{equation}
where the different 
MHV vertices 
correspond to the different MHV amplitudes \eqref{eq:AllMHV}.
Here, $p_{\Gamma_r}$ represents an internal line flowing into the vertex
$\Gamma_r$ and is given by the sum over the remaining momenta entering
$\Gamma_r$ (see also below). As before, we are suppressing $R$-symmetry indices.
For instance, we have
\begin{subequations}
\begin{eqnarray}
 A^{\rm NMHV}_{0,n}\! &=& \! \sum_{\{\{\Gamma_1,\Gamma_2\}|n_1+n_2=n+2\}} A^{\rm
MHV}_{0,n_1}(\Gamma_1)\,\frac{1}{p_{\Gamma_1}^2}\,A^{\rm
MHV}_{0,n_2}(\Gamma_2)\notag\\[10pt]
 &=&\! \sum
\begin{picture}(150,40)(-20,0)
   \SetScale{1}
   \put(0,0){
    \Line(20,0)(100,0)
    \Line(20,0)(-10,30)
    \Line(20,0)(-10,-30)
    \Line(100,0)(130,30)
    \Line(100,0)(130,-30)
    \CCirc(20,0){15}{Black}{Gray}
    \CCirc(100,0){15}{Black}{Gray}
    \DashCArc(20,0)(30,160,200){2}
    \DashCArc(100,0)(30,340,20){2}
    \Text(21,0)[]{$\Gamma_1$}
    \Text(101,0)[]{$\Gamma_2$}
    \Text(60,10)[]{$1/p^2_{\Gamma_1}$}
}
\end{picture}\\[50pt]
 A^{{\rm N}^2{\rm MHV}}_{0,n}\! &=&\!
\sum_{\{\{\Gamma_1,\Gamma_2,\Gamma_3\}|n_1+n_2+n_3=n+4\}} 
\frac{A^{\rm MHV}_{0,n_1}(\Gamma_1)A^{\rm MHV}_{0,n_2}(\Gamma_2) 
A^{\rm MHV}_{0,n_3}(\Gamma_3)}{p_{\Gamma_1}^2p_{\Gamma_2}^2p_{\Gamma_3}^2}\notag
\\[10pt]
 &=&\! \sum
\begin{picture}(250,40)(-20,0)
   \SetScale{1}
   \put(0,0){
    \Line(20,0)(100,0)
    \Line(20,0)(-10,30)
    \Line(20,0)(-10,-30)
    \Line(100,0)(130,30)
    \Line(100,0)(130,-30)
    \Line(100,0)(70,30)
    \Line(100,0)(70,-30)
    \Line(100,0)(180,0)
    \Line(180,0)(210,30)
    \Line(180,0)(210,-30)
    \CCirc(20,0){15}{Black}{Gray}
    \CCirc(100,0){15}{Black}{Gray}
    \CCirc(180,0){15}{Black}{Gray}
    \DashCArc(20,0)(30,160,200){2}
    \DashCArc(180,0)(30,340,20){2}
    \DashCArc(100,0)(30,70,110){2}
    \DashCArc(100,0)(30,250,290){2}
    \Text(21,0)[]{$\Gamma_1$}
    \Text(101,0)[]{$\Gamma_3$}
    \Text(181,0)[]{$\Gamma_2$}
    \Text(60,10)[]{$1/p^2_{\Gamma_1}$}
    \Text(140,10)[]{$1/p^2_{\Gamma_2}$}
}
\end{picture}
\end{eqnarray}

\vspace*{25pt}

\noindent
for the NMHV and N$^2$MHV amplitudes, while for the 
N$^3$MHV amplitudes we have two different graph topologies:
\begin{eqnarray}
 A^{{\rm N}^3{\rm MHV}}_{0,n}\! &=&\!
\sum_{\{\{\Gamma_1,\ldots,\Gamma_4\}|n_1+\cdots+n_4=n+6\}} \frac{A^{\rm
MHV}_{0,n_1}(\Gamma_1)
 \cdots A^{\rm
MHV}_{0,n_{4}}(\Gamma_{4})}{p^2_{\Gamma_1}p^2_{\Gamma_2}p^2_{\Gamma_3}}
\notag
\\[10pt]
 &=&\! \sum
\begin{picture}(350,40)(-20,0)
   \SetScale{1}
   \put(0,0){
    \Line(20,0)(100,0)
    \Line(20,0)(-10,30)
    \Line(20,0)(-10,-30)
    \Line(100,0)(130,30)
    \Line(100,0)(130,-30)
    \Line(100,0)(70,30)
    \Line(100,0)(70,-30)
    \Line(100,0)(180,0)
    \Line(180,0)(210,30)
    \Line(180,0)(210,-30)
    \Line(180,0)(260,0)
    \Line(180,0)(150,30)
    \Line(180,0)(150,-30)
    \Line(260,0)(290,30)
    \Line(260,0)(290,-30)
    \CCirc(20,0){15}{Black}{Gray}
    \CCirc(100,0){15}{Black}{Gray}
    \CCirc(180,0){15}{Black}{Gray}
    \CCirc(260,0){15}{Black}{Gray}
    \DashCArc(20,0)(30,160,200){2}
    \DashCArc(260,0)(30,340,20){2}
    \DashCArc(100,0)(30,70,110){2}
    \DashCArc(100,0)(30,250,290){2}
    \DashCArc(180,0)(30,70,110){2}
    \DashCArc(180,0)(30,250,290){2}
    \Text(21,0)[]{$\Gamma_1$}
    \Text(101,0)[]{$\Gamma_2$}
    \Text(181,0)[]{$\Gamma_4$}
    \Text(261,0)[]{$\Gamma_3$}
    \Text(60,10)[]{$1/p^2_{\Gamma_1}$}
    \Text(140,10)[]{$1/p^2_{\Gamma_2}$}
    \Text(220,10)[]{$1/p^2_{\Gamma_3}$}
}
\end{picture}\notag\\[50pt]
&&\kern1cm+\ \sum
\begin{picture}(250,100)(-20,0)
   \SetScale{1}
   \put(0,0){
    \Line(20,0)(100,0)
    \Line(20,0)(-10,30)
    \Line(20,0)(-10,-30)
    \Line(100,0)(140,20)
    \Line(100,0)(120,35)
    \Line(100,0)(130,-30)
    \Line(100,0)(60,20)
    \Line(100,0)(80,35)
    \Line(100,0)(70,-30)
    \Line(100,0)(100,80)
    \Line(100,0)(180,0)
    \Line(180,0)(210,30)
    \Line(180,0)(210,-30)
    \Line(100,80)(70,110)
    \Line(100,80)(130,110)
    \CCirc(20,0){15}{Black}{Gray}
    \CCirc(100,0){15}{Black}{Gray}
    \CCirc(180,0){15}{Black}{Gray}
    \CCirc(100,80){15}{Black}{Gray}
    \DashCArc(20,0)(30,160,200){2}
    \DashCArc(180,0)(30,340,20){2}
    \DashCArc(100,0)(30,130,140){2}
    \DashCArc(100,0)(30,40,50){2}
    \DashCArc(100,0)(30,250,290){2}
    \DashCArc(100,80)(30,70,110){2}
    \Text(21,0)[]{$\Gamma_1$}
    \Text(101,0)[]{$\Gamma_4$}
    \Text(101,80)[]{$\Gamma_2$}
    \Text(181,0)[]{$\Gamma_3$}
    \Text(60,-10)[]{$1/p^2_{\Gamma_1}$}
    \Text(85,45)[]{$1/p^2_{\Gamma_2}$}
    \Text(140,-10)[]{$1/p^2_{\Gamma_3}$}
} 
\end{picture}
\end{eqnarray}
\end{subequations}
\vspace*{10pt}

\noindent
Similarly, one may draw the diagrams contributing to 
the N$^k$MHV amplitude for larger and larger $k$ but, of course, the complexity
in the graph topology will grow. In this notation, Figure
\ref{fig:NMHVExample} becomes:

\begin{figure}[h]
\vspace*{-.5cm}
\begin{center}
\begin{picture}(150,100)(-180,0)
   \SetScale{1}
   \put(0,0){
    \Text(-10,-40)[]{$p_1$}
    \Text(-10,40)[]{$p_2$}
    \Text(130,40)[]{$p_3$}
    \Text(130,-40)[]{$p_4$}    
    \Text(3,-30)[]{$-$}
    \Text(3,30)[]{$+$}
    \Text(117,30)[]{$+$}
    \Text(117,-30)[]{$+$}
    \Text(50,-15)[]{$+$}
    \Text(70,15)[]{$-$}
    \Text(75,0)[]{$1/p^2_{14}$}
    \ArrowLine(60,40)(60,-40)
    \ArrowLine(0,40)(60,40)
    \ArrowLine(0,-40)(60,-40)
    \ArrowLine(120,40)(60,40)
    \ArrowLine(120,-40)(60,-40)
    \CCirc(60,40){15}{Black}{Gray}
    \CCirc(60,-40){15}{Black}{Gray}
    \Text(61,40)[]{$\Gamma_2$}
    \Text(61,-40)[]{$\Gamma_1$}   
    \Text(60,-80)[]{{\small(b)}}
}
\end{picture}
\begin{picture}(150,100)(180,0)
   \SetScale{1}
   \put(0,0){
  \ArrowLine(100,0)(20,0)
    \ArrowLine(-10,30)(20,0)
    \ArrowLine(-10,-30)(20,0)
    \ArrowLine(130,30)(100,0)
    \ArrowLine(130,-30)(100,0)
    \CCirc(20,0){15}{Black}{Gray}
    \CCirc(100,0){15}{Black}{Gray}
    \Text(21,0)[]{$\Gamma_1$}
    \Text(101,0)[]{$\Gamma_2$}    
   \Text(-10,40)[]{$p_2$}
    \Text(-10,-40)[]{$p_1$}
     \Text(130,40)[]{$p_3$}
    \Text(130,-40)[]{$p_4$}
    \Text(-10,20)[]{$+$}
    \Text(-10,-20)[]{$-$}
     \Text(130,20)[]{$+$}
    \Text(130,-20)[]{$+$}
    \Text(45,-8)[]{$+$}
     \Text(75,-8)[]{$-$}
    \Text(60,10)[]{$1/p^2_{12}$}
    \Text(60,-80)[]{{\small(a)}}
}
\end{picture}
\end{center}
\vspace*{2.5cm}
\caption{\label{fig:NMHVExample2}{\it\small The two diagrams contributing to the
${-}{+}{+}{+}$ amplitude. In (a), vertex $\Gamma_1$ is
characterised by the momenta $p_1$ and $p_2$ and thus, $p_{\Gamma_1}\equiv
p_{12}$ and similarly for (b).}}
\end{figure}
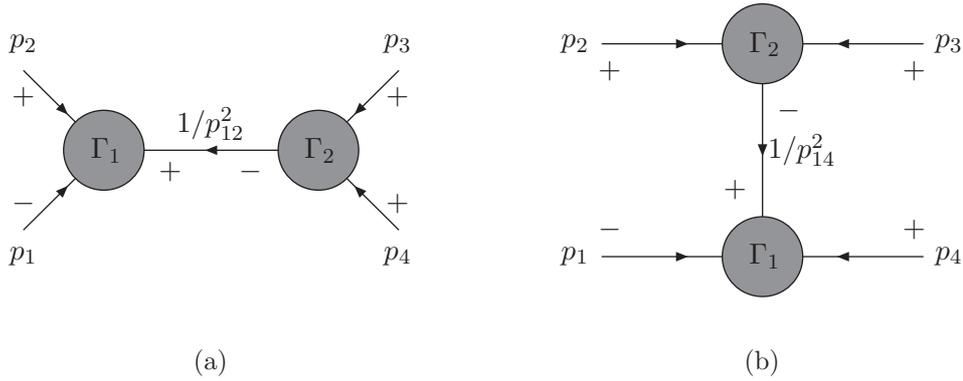

The above procedure of gluing MHV vertices together is not manifestly $\CN=4$
supersymmetric. Instead we would like to directly glue MHV
supervertices\footnote{i.e.~the vertices that follow from the MHV
superamplitudes after an appropriate off-shell continuation}
together as this would give a manifestly 
supersymmentric formulation of the N$^k$MHV amplitudes. In
order to do this, we need one additional ingredient since each MHV supervertex
is of eighth-order in the fermionic coordinates such that $k+1$ of such vertices
are of order $8k+8$. However, above we have argued that an 
N$^k$MHV superamplitude is of order $4k+8$ in the fermionic coordinates
$\psi_r^i$. Elvang, Freedman \& Kiermaier \cite{Elvang:2008vz} propose the
introduction of additional fermionic coordinates $\psi^i_{\Gamma_r}$ for each
internal
line $p_{\Gamma_r}$ and the definition of a  fourth-order differential operator
\begin{equation}
 D^{(4)}_{\Gamma_r}\ :=\ \frac{1}{4!}\varepsilon^{ijkl}
       \frac{\partial^4}{\partial\psi^i_{\Gamma_r}
       \partial\psi^j_{\Gamma_r}\partial\psi^k_{\Gamma_r}
       \partial\psi^l_{\Gamma_r}}~.
\end{equation}
All the N$^k$MHV amplitudes  $A^{{\rm N}^k{\rm MHV}}_{0,n}$ in $\CN=4$
supersymmetric Yang--Mills theory are then given by the following formul{\ae}
\cite{Elvang:2008vz}:
\begin{subequations}
\begin{equation}
 A^{{\rm N}^k{\rm MHV}}_{0,n}\ =\ D^{(4k+8)}  F^{{\rm N}^k{\rm MHV}}_{0,n}~,
\end{equation}
with
\begin{equation}\label{eq:EZK1}
F^{{\rm N}^k{\rm MHV}}_{0,n}\ =\ 
\sum_{\{\{\Gamma_1,\ldots,\Gamma_{k+1}\}
 |n_1+\dots+n_k=n+2k\}}
  \left(\prod_{r=1}^k D^{(4)}_{\Gamma_r}\right)
  \frac{F^{\rm MHV}_{0,n_1}(\Gamma_1)
 \cdots F^{\rm
MHV}_{0,n_{k+1}}(\Gamma_{k+1})}{p^2_{\Gamma_1}\cdots p^2_{\Gamma_k}}~.
\end{equation}
\end{subequations}
The justification of these formul{\ae} is tied to the action of the
differential operator $D^{(4k+8)}$ which is made from \eqref{eq:DOAmp} and 
which selects
the external states of the
amplitude: The $R$-symmetry index structure of the external states uniquely
determines the $R$-symmetry indices of the internal states (i.e.~the
internal lines). The operator $D^{(4)}_{\Gamma_r}$ then uniquely splits  into
two factors that correspond to the required split of the $R$-symmetry index
structure at each end of the internal line.

The expression \eqref{eq:EZK1} can be simplified further. We shall just state
the result here and refer the interested reader to \cite{Elvang:2008vz} for the
explicit derivation. We have
\begin{subequations}
\begin{equation}
F^{{\rm N}^k{\rm MHV}}_{0,n}\ =\ g_{\rm YM}^{n-2}(2\pi)^4
\delta^{(4|8)}\left(\sum_{r=1}^n \tilde k_{rA}k_{r\db}\right) P^{(4k)}_{0,n}~,
\end{equation}
with
\begin{equation}\label{eq:PolynomialNMHV}
P^{(4k)}_{0,n}\ =\ 
\sum_{\{\{\Gamma_1,\ldots,\Gamma_{k+1}\}
 |n_1+\dots+n_k=n+2k\}}
  \frac{\prod_{r=1}^k
  \frac{1}{p^2_{\Gamma_r}}\sum_{s\in\Gamma_r}\langle\tilde\xi |
  p_{\Gamma_r}|k_s](\psi_s)^4}{{\rm cyc}(\Gamma_1)\cdots{\rm cyc}
 (\Gamma_{k+1})}~.
\end{equation}
\end{subequations}
Here, we have used the notation \eqref{eq:Nairlemma} for $\psi^i_s$ and
`cyc' denotes the usual denominator of the spinor brackets appearing in the MHV
amplitudes, e.g.~${\rm cyc}(1,\ldots,n):=[12]\cdots[n1]$. In addition,
the external momenta are $p_r=\tilde k_r k_r$ for $r=1,\ldots,n$. 
Note that the result for the tree-level
NMHV superamplitude was first obtained by Georgiou, Glover \& Khoze in
\cite{Georgiou:2004by}. In \cite{Kiermaier:2009yu}, a refined version of
\eqref{eq:PolynomialNMHV} was given in which the number of diagrams
one needs to sum over is reduced substantially.

Next one should show that this expression is independent of the choice of the
spinor $\tilde\xi$ that defines the off-shell continuation. This was done
explicitly in \cite{Elvang:2008vz}. Notice, however, that 
analogously to our discussion presented in Section \ref{sec:MHVTwistor},
$\tilde\xi$ can
also be
shown to arise from a gauge fixing condition of the twistor space action
\eqref{eq:full}. This
time, of
course, one should include the full supermultiplet \eqref{eq:etaexp}.
Therefore, BRST invariance will guarantee the $\tilde\xi$-independence of the
overall
scattering amplitudes.

\vspace*{5pt}

\Remark{Note that there is an alternative method \cite{Drummond:2008cr} for
constructing the polynomials $\CP_n^{(4k)}$ occurring in
\eqref{eq:GenericSuAmp}.
Drummond \& Henn \cite{Drummond:2008cr} express the
$\CP_n^{(4k)}$ in terms of
invariants of the so-called dual superconformal symmetry group
\cite{Drummond:2006rz,Alday:2007hr,Ricci:2007eq,Drummond:2007aua,Alday:2007he,
Drummond:2007au,Drummond:2008vq,Berkovits:2008ic,Beisert:2008iq,
Brandhuber:2008pf,Drummond:2008bq,Drummond:2009fd,Beisert:2009cs,
Alday:2009zm,Henn:2010bk,Drummond:2010qh}
(see also
\cite{Hodges:2009hk,ArkaniHamed:2009dn,Mason:2009qx,ArkaniHamed:2009vw,
ArkaniHamed:2009dg} and \cite{Alday:2008yw,Henn:2009bd,Korchemsky:2009zz}
for recent reviews). The dual
superconformal symmetry is a recently discovered hidden (dynamical) symmetry,
planar scattering amplitudes in $\CN=4$ supersymmetric Yang--Mills theory appear
to exhibit. }

\vspace*{10pt}
 
\subsection{Localisation properties}\label{sec:AmplitudeLocalisation}

\begin{figure}[ht]
\begin{center}
\begin{picture}(150,150)(10,0)
   \SetScale{1}
   \put(0,0){
   \Line(0,0)(120,120)
   \Line(140,120)(180,0)
   \DashLine(158,66)(70,70){2}
   \Vertex(20,20){2}
   \Vertex(40,40){2}
   \Vertex(60,60){2}
   \Vertex(80,80){2}
   \Vertex(100,100){2}
   \Vertex(145,105){2}
   \Vertex(154,78){2}
   \Vertex(163,51){2}
   \Vertex(172,24){2}
   }
\end{picture}
\end{center}
\caption{\label{fig:disconnected}{\it\small An NMHV amplitude 
represented in twistor space: Each line represents an MHV amplitude (or vertex)
while the
dotted line
represents the propagator of holomorphic Chern--Simons theory; see 
\eqref{eq:CSWPropagator} for the case of pure Yang--Mills theory.}}
\end{figure}
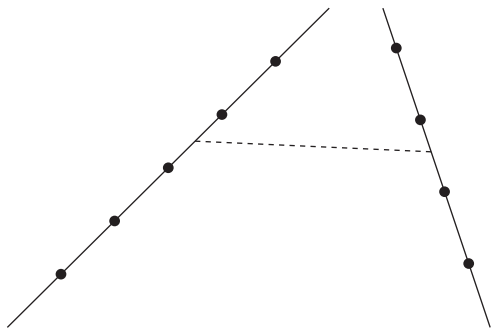

In this section, we would like to comment on the localisation properties of
generic scattering amplitudes in twistor space. In Section
\ref{sec:MHVLocalisation}, we saw that MHV superamplitudes localise on
genus-$0$ degree-$1$ curves $\Sigma$ in supertwistor space:
$\Sigma\cong\IC P^1$ in the complex setting or $\Sigma\cong\IR P^1$ in
the Kleinian setting. According to the MHV formalism, generic
tree-level superamplitudes are obtained by gluing the MHV
superamplitudes together (after a suitable off-shell continuation). This is
depicted in
Figure \ref{fig:disconnected} for the case
of NMHV amplitudes.

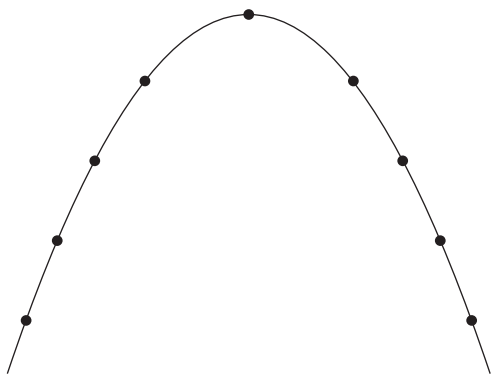
\begin{figure}[h]
\begin{center}
\begin{picture}(150,150)(10,0)
   \SetScale{1}
   \put(0,0){
   \Curve{(0,0)(120,120)(180,0)}
   \Vertex(7,20){2}
   \Vertex(18.6,50){2}
   \Vertex(32.6,80){2}
   \Vertex(51.3,110){2}
   \Vertex(90,135){2}
   \Vertex(129,110){2}
   \Vertex(147.4,80){2}
   \Vertex(161.4,50){2}
   \Vertex(173.1,20){2}
   }
\end{picture}
\end{center}
\caption{\label{fig:connected}{\it\small An NMHV amplitude described by a
connected degree-$2$ rational curve in supertwistor space.}}
\end{figure}

This therefore suggests that tree-level N$^k$MHV superamplitudes localise on
maximally disconnected rational curves $\Sigma$ of genus zero, that is,
each $\Sigma$ is the union of $k+1$ (projective) lines in supertwistor space.
Therefore, the total degree $d$ of each $\Sigma$ is
\begin{equation}\label{eq:degreek+1}
  d\ =\ k+1~.
\end{equation}
 This picture is
the so-called maximally disconnected description of scattering amplitudes.
As was observed in \cite{Roiban:2004vt,Roiban:2004ka,Roiban:2004yf},
scattering
amplitudes may also be described by completely connected rational curves. In
particular, tree-level NMHV superamplitudes may also be described by connected
genus-$0$ degree-$2$ rational curves as is depicted in Figure
\ref{fig:connected}. This is
referred to as the connected description of scattering amplitudes. It was
argued, however, by Gukov, Motl \& Neitzke in \cite{Gukov:2004ei} that the two
descriptions---that is, the disconnected and connected prescriptions---are in
fact equivalent.

In general, the conjecture
put forward by Witten \cite{Witten:2003nn} is that $\ell$-loop N$^k$MHV
superamplitudes localise on rational curves in supertwistor space which are of
genus $g\leq\ell$ and degree
\begin{equation}\label{eq:WittenConjecture}
 d\ =\ k+1+g~.
\end{equation}
Following the earlier work of
\cite{Hodges:2005bf,Hodges:2005aj,Mason:2009sa,ArkaniHamed:2009si}, Korchemsky
\& Sokatchev
\cite{Korchemsky:2009jv} (see also
\cite{Cachazo:2004zb,Cachazo:2004by,Bena:2004xu,Bern:2004ky,Bern:2004bt,
Britto:2004tx})
performed the Witten transform \eqref{eq:WittenFourierIntegralSuAm} of all
tree-level scattering amplitudes explicitly.
In fact, they transformed the Drummond \& Henn expressions
\cite{Drummond:2008cr} of the quantities $\CP^{(4k)}_n$ appearing in
\eqref{eq:GenericSuAmp} for the N$^k$MHV superamplitudes to supertwistor space.
They found that the
N$^k$MHV superamplitudes are supported on $2k+1$ intersecting lines in
supertwistor space in contra-distinction to what we have said
above
\eqref{eq:degreek+1}.
This puzzle was resolved very recently by Bullimore, Mason \& Skinner in
\cite{Bullimore:2009cb} (see also \cite{Kaplan:2009mh}) by proving that the
Drummond \& Henn expressions for
the tree-level amplitudes actually contain loop-level information: They showed
that the $2k+1$ intersecting lines form a curve of genus $g=k$ and
therefore, $d=k+1+k=k+1+g$. Their result generalises the fact 
\cite{Bern:2004bt,Drummond:2008bq} that the
tree-level NMHV superamplitudes can be written in terms of loop-level
information\footnote{More concretely, they can be written in terms of the
so-called leading singularities of the
one-loop amplitudes; see
also
\cite{Britto:2004nc,Bern:2005iz,Brandhuber:2005jw,Bern:2006ew,Bern:2007ct,
Buchbinder:2005wp,Cachazo:2008dx,Cachazo:2008vp,Cachazo:2008hp}
for a discussion on the
generalised unitarity and leading
singularity methods.}. Altogether, the result
of Korchemsky \& Sokatchev
\cite{Korchemsky:2009jv} is therefore in support of Witten's conjecture
\eqref{eq:WittenConjecture}. 

\section{Britto--Cachazo--Feng--Witten recursion relations}\label{sec:BCFW}

Central to this final section will be the dicussion of a powerful method for
constructing tree-level amplitudes first introduced by Britto, Cachazo \&
Feng
in
\cite{Britto:2004ap} (stemming from observations made in \cite{Roiban:2004ix})
and later proven by Britto, Cachazo, Feng \& Witten in
\cite{Britto:2005fq}.
We shall start with pure Yang--Mills theory and closely follow
the treatment of \cite{Britto:2005fq} before moving
on to the $\CN=4$ supersymmetric extension developed by 
\cite{Bianchi:2008pu,Brandhuber:2008pf,ArkaniHamed:2008gz}. 

\subsection{Recursion relations in pure Yang--Mills theory}

The basic philosophy of the method developed by Britto, Cachazo, Feng \& Witten
is to construct tree-level scattering amplitudes in terms of lower-valence
on-shell amplitudes and a scalar propagator.
Therefore, one also speaks of recursion relations since one constructs
higher-point amplitudes from lower-point ones.
 Recursion relations have been known for some time in field theory
since Berends \& Giele \cite{Berends:1987me} proposed them in terms of off-shell
currents. However, the recursion relations we are about to discuss are somewhat
more powerful as they directly apply to on-shell scattering amplitudes and in
addition are particularly apt when the amplitudes are written in the spinor
helicity formalism (see our above discussion).

\begin{figure}[t]
\begin{center}
\begin{picture}(150,150)(10,0)
   \SetScale{1}
   \put(0,0){
   \ArrowArc(75,75)(90,0,360)
   \ArrowArc(75,75)(20,0,360)
   \Vertex(75,75){2}
   \ArrowArc(90,20)(20,0,360)
   \Vertex(90,20){2}
   \ArrowArc(30,40)(20,0,360)
   \Vertex(30,40){2}
   \ArrowArc(50,120)(20,0,360)
   \Vertex(50,120){2}
   \ArrowArc(130,110)(20,0,360)
   \Vertex(130,110){2}
   \Line(170,150)(185,150)
   \Line(170,150)(170,168)
   \DashLine(95,114)(85,116){2}
   \Text(178,160)[]{$t$}
   \Text(75,68)[]{$t=0$}
   \Text(45,75)[]{$\cC_0$}
   \Text(92,10)[]{$t_1$}
   \Text(60,20)[]{$\cC_1$}
   \Text(32,30)[]{$t_m$}
   \Text(0,40)[]{$\cC_m$}
   \Text(52,110)[]{$t_{m-1}$}
   \Text(15,120)[]{$\cC_{m-1}$}
   \Text(132,100)[]{$t_2$}
   \Text(102,105)[]{$\cC_2$}
   \Text(-25,75)[]{$\cC_\infty$}
   }
\end{picture}
\end{center}
\caption{\label{fig:integration}{\it\small Integration in the complex $t$-plane
to obtain a recursion relation.}}
\end{figure}
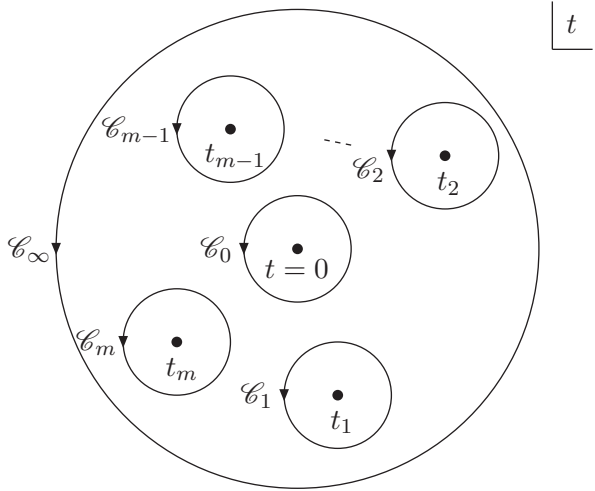

To derive a recursion relation for scattering amplitudes, let $A_{0,n}(t)$ be a
complex one-parameter family of $n$-particle colour-stripped scattering
amplitudes at tree-level with $t\in\IC$ such that
$A_{0,n}(t=0)$ is the amplitude we are interested in. As before, we shall work
in
the complex setting, i.e.~all momenta are taken to be complex. One can then
consider the contour integral \cite{Bern:2005hs}
\begin{equation}
 c_\infty\ :=\ \frac{1}{2\pi\di}\oint_{\cC_\infty}\!\!\!\! \dd t\,
\frac{A_{n,0}(t)}{t}~,
\end{equation}
where the integration is taken counter-clockwise around a circle $\cC_\infty$ at
infinity in the complex $t$-plane; see Figure \ref{fig:integration}. Using
the residue theorem, we may write this contour integral as
\begin{equation}
 c_\infty\ =\ A_{0,n}(t=0)+\sum_{k\geq1} \frac{1}{t_k}\,\mbox{Res}_{t=t_k}\,
A_{0,n}(t)~,
\end{equation}
where we have performed the contour integral around $\cC_0$ explicitly; see
Figure \ref{fig:integration}. Equivalently, we have
\begin{equation}\label{eq:residue1}
 A_{0,n}(t=0)\ =\ c_\infty-\sum_{k\geq1} \frac{1}{t_k}\,\mbox{Res}_{t=t_k}\,
A_{0,n}(t)~.
\end{equation}

So far, we have not given the definition of the one-parameter family
$A_{0,n}(t)$. There are some obvious requirements for $A_{0,n}(t)$. The main
point is to define the family $A_{0,n}(t)$ such that poles in $t$ correspond to
multi-particle poles in the scattering amplitude $A_{0,n}(t=0)$. If this is
done, the corresponding residues can be computed from factorisation properties
of the scattering amplitudes; see e.g.~\cite{Weinberg:1995mt,Dixon:1996wi}. 

In order to accomplish this, Britto, Cachazo, Feng \& Witten
\cite{Britto:2004ap,Britto:2005fq} defined $A_{0,n}(t)$ by shifting the momenta
of two external particles (gluons) in the original scattering amplitude.
Obviously, for this to make sense, one has to ensure that even with these
shifts, overall momentum conservation is preserved and that all particles remain
on-shell. Thus, let us shift the momenta of the particles $r$ and $s$ according
to
\begin{equation}
 p_r(t)\ :=\ p_r+t\ell\eand 
 p_s(t)\ :=\ p_s-t\ell~.
\end{equation}
Clearly, momentum conservation is maintained by performing these shifts. To
preserve the on-shell conditions, $p^2_r(t)=0=p^2_s(t)$, we choose
\begin{equation}
 \ell_{\al\db}\ =\ \tilde k_{s\al}k_{r\db}~,
\end{equation}
where $\tilde k_r$, $k_r$ and $\tilde k_s$, $k_s$ are the co-spinors for
the momenta $p_r$ and $p_s$, respectively.\footnote{Alternatively, one may take 
$\ell_{\al\db}=\tilde k_{r\al}k_{s\db}$.} This then corresponds to shifting
the co-spinors,
\begin{equation}\label{eq:BCFWcospinors}
 \tilde k_{r\al}(t)\ :=\ \tilde k_{r\al}+t\tilde k_{s\al}\eand
 k_{s\da}(t)\ :=\ k_{s\da}-tk_{r\da}~,
\end{equation}
with $k_{r\da}$ and $\tilde k_{s\al}$ unshifted. As a result, we can define
\begin{equation}
 A_{0,n}(t)\ :=\
A_{0,n}(p_1,\ldots,p_{r-1},p_r(t),p_{r+1},\ldots,p_{s-1},p_s(t),p_{s+1},
\ldots,p_n)~.
\end{equation}
The right-hand-side is a physical on-shell amplitude for all $t\in\IC$ (since
all the momenta are null).

The family $A_{0,n}(t)$ is a rational function of $t$: According to our above
discussion, the original tree-level amplitude is a rational function of the
spinor brackets and thus, performing the shifts \eqref{eq:BCFWcospinors}
clearly renders $A_{0,n}(t)$ a rational function in $t$. In addition,
$A_{0,n}(t)$ can only have simple poles as a function of $t$, since
singularities can only come from the poles of a propagator in a Feynman diagram:
In tree-level Yang--Mills theory, the momentum in a propagator is always a sum
of momenta of adjacent external particles, say $p=\sum_{r'\in R}p_{r'}$ where
$R$ denotes the set of these external particles. A propagator with this momentum
is $1/p^2$. Let now $s\in R$ but $r\not\in R$. Then, $p(t)=p-t\ell$ and 
$p^2(t)=p^2-2t(p\cdot\ell)$
vanishes at 
\begin{equation}
 t\ =\ t_{p}\ :=\ \frac{p^2}{2(p\cdot\ell)}~.
\end{equation}
These are the only poles of $A_{0,n}(t)$. Notice that there might be poles in
$t$
arising from the shifts \eqref{eq:BCFWcospinors} on the denominators of the
polarisation vectors \eqref{eq:PolarisationCoVectors}, but these may always be
removed by an appropriate choice of $\tilde \mu$, $\mu$ in
\eqref{eq:PolarisationCoVectors} and hence they are merely gauge
artifacts. Thus, $A_{0,n}(t)$ only has simple poles in $t$, as claimed.
Below
we shall also denote $p(t_p)$ by $\hat p$. Since $\hat p$ is null, we may
introduce the co-spinors $\tilde k_{\hat p}$, $k_{\hat p}$ and write $\hat
p=\tilde k_{\hat p}k_{\hat p}$. 

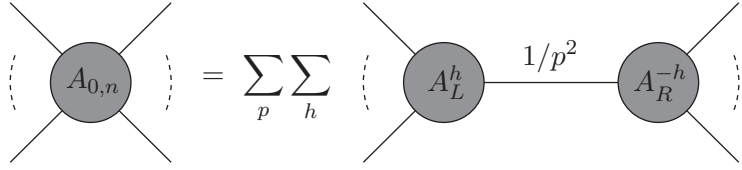
\begin{figure}[t]
 \begin{center}
 \begin{equation}\notag
\begin{picture}(55,40)(0,0)
   \SetScale{1}
   \put(0,0){
    \Line(20,0)(-10,30)
    \Line(20,0)(-10,-30)
    \Line(20,0)(50,30)
    \Line(20,0)(50,-30)
    \CCirc(20,0){15}{Black}{Gray}
    \DashCArc(20,0)(30,160,200){2}
    \DashCArc(20,0)(30,340,20){2}
    \Text(21,0)[]{$A_{0,n}$}
}
\end{picture}
  \ =\ \sum_p\sum_h
\begin{picture}(150,40)(-20,0)
   \SetScale{1}
   \put(0,0){
    \Line(20,0)(100,0)
    \Line(20,0)(-10,30)
    \Line(20,0)(-10,-30)
    \Line(100,0)(130,30)
    \Line(100,0)(130,-30)
    \CCirc(20,0){15}{Black}{Gray}
    \CCirc(100,0){15}{Black}{Gray}
    \DashCArc(20,0)(30,160,200){2}
    \DashCArc(100,0)(30,340,20){2}
    \Text(21,0)[]{$A_L^h$}
    \Text(101,0)[]{$A_R^{-h}$}
    \Text(60,10)[]{$1/p^2$}
}
\end{picture}
\end{equation}
 \end{center}

\vspace*{10pt}
\caption{\label{fig:BCFW}{\it\small Mnemonic
diagram illustrating the Britto--Cachazo--Feng--Witten recursion relation.}}
\end{figure}

Now we would like to evaluate the sum in \eqref{eq:residue1}. To get a pole at
$p^2(t)=0$, a tree-level diagram must contain a propagator that
divides it into a `left sub-amplitude' that contains all external
particles not in $R$ and a `right sub-amplitude'
that contains all external particles that are in $R$. The internal line
connecting the left and the right parts has momentum 
$p(t)$ and we need to sum over the helicity $h=\pm1$ at, say, the left
of this line. Therefore,
\begin{equation}
 \begin{aligned}
  \mbox{Res}_{t=t_{p}}\,A_{0,n}(t)
  \ &=\ \sum_h A^h_L(t_{p})\,
      \mbox{Res}_{t=t_{p}}\left\{
\frac{1}{p^2-2t(p\cdot\ell)}\right\} A^{-h}_R(t_{p})\\
  &=\ -t_{p}
 \sum_h A^h_L(t_{p})\,
\frac{1}{p^2}\, A^{-h}_R(t_{p})  
 \end{aligned}
\end{equation}
and \eqref{eq:residue1} becomes
\begin{equation}\label{eq:residue2}
 A_{0,n}(t=0)\ =\ c_\infty+\sum_p
\sum_h A^h_L(t_{p})\,
\frac{1}{p^2}\, A^{-h}_R(t_{p})~.
\end{equation}

Furthermore, in \cite{Britto:2005fq} it was
shown that $A_{0,n}(t)\to 0$ for $|t|\to\infty$ for pure Yang--Mills theory; see
also \cite{ArkaniHamed:2008yf} for a more detailed treatment. This results in
the remarkable property that $c_\infty=0$. Altogether, we obtain the following
recursion relation \cite{Britto:2004ap,Britto:2005fq}:
\begin{subequations}\label{eq:BCFW}
\begin{equation}
 A_{0,n}\ =\ \sum_p
\sum_h A^h_L(t_{p})\,
\frac{1}{p^2}\, A^{-h}_R(t_{p})~,  
\end{equation}
which is depicted in Figure \ref{fig:BCFW}. Here,
\begin{equation}
 \begin{aligned}
  A^h_L(t_{p})\ &=\ A^h_L(\underbrace{\ldots,p_{r-1},\hat
p_r,p_{r+1},\ldots}_{\not\in\,R},\hat p)~,\\
  A^{-h}_R(t_{p})\ &=\ A^{-h}_R(-\hat p,\underbrace{\ldots,p_{s-1},\hat
p_s,p_{s+1},\ldots}_{\in\, R})~,
 \end{aligned}
\end{equation}
\end{subequations}
where $\hat p_r:=p_r(t_p)$ and $\hat p_s:=p_s(t_p)$, respectively.
Note that ultimately \eqref{eq:BCFW} tells us that every scattering amplitude
reduces to a sum of products of MHV and $\overline{\rm MHV}$ amplitudes.
This is exemplified in the exercise below.

\vspace*{5pt}

\Exercise{Consider the MHV amplitude
$$ A_{n,0}^{\rm MHV}(1^+,r^+)\ =\ 
   g^{n-2}_{\rm YM}(2\pi)^4\delta^{(4)}\left(\sum_{r'} \tilde
   k_{r'}k_{r'}\right)\frac{[1r]^4}{[12]\cdots[n1]}~.$$
Assume that this formula (and the one for the $\overline{MHV}$ amplitude) is
valid up to $n$ gluons. Use the above recursion relation to prove its validity
for $n+1$ gluons by appropriately shifting two of the co-spinors. Note that you
will need to use both, MHV and $\overline{MHV}$ amplitudes as left and right
sub-amplitudes entering the recursion relation.}

\vspace*{10pt}

\subsection{Recursion relations in maximally supersymmetric Yang--Mills theory}

Finally, we would like to extend the above recursion relation to the $\CN=4$
supersymmetric setting. In Section \ref{sec:SuAmpMHVForm}, we saw that all the
different scattering amplitudes in $\CN=4$ supersymmetric Yang--Mills theory
are best understood when formulated in terms of superamplitudes or generating
functions $F$. In order to formulate a supersymmetric version of the recursion
relation for the tree-level superamplitudes $F_{0,n}$, we follow 
\cite{Bianchi:2008pu,Brandhuber:2008pf,ArkaniHamed:2008gz}\footnote{See also
Arkani-Hamed's talk at 
\htmladdnormallink{\sl ``Wonders of Gauge Theory and Supergravity'', Paris
2008}{http://ipht.cea.fr/Images/Pisp/pvanhove/Paris08/}.} and consider the
shifts
\begin{equation}\label{eq:BCFWcospinorsSu}
 \tilde k_{rA}(t)\ :=\ \tilde k_{rA}+t\tilde k_{sA}\eand
 k_{s\da}(t)\ :=\ k_{s\da}-tk_{r\da}~,
\end{equation}
with $k_{r\da}$ and $\tilde k_{sA}$ unshifted. Indeed, this is a direct
supersymmetric extension of \eqref{eq:BCFWcospinors}. 

The supersymmetric recursion relations then follow from arguments similar to
those which led to \eqref{eq:BCFW}. We therefore just record the result here
\cite{Brandhuber:2008pf}:
\begin{equation}\label{eq:BCFWSu}
 F_{0,n}\ =\ \sum_p\int\dd^4\psi_{\hat p}\, F_L(t_{p})\,
\frac{1}{p^2}\, F_R(t_{p})~.
\end{equation}
Here, the $\psi_{\hat p}^i$ are the fermionic variables associated with the
internal on-shell line with momentum $\hat p$. Notice that the sum over
helicities gets replaced by the integral over the fermionic coordinates: The
superamplitudes $F_L$ and $F_R$ carry total helicities of $-n_L-1$ and $-n_R-1$
where $n_L$ and $n_R$ are the number of external states on the left and right
sub-amplitudes; see also Section \ref{sec:TreeMHV} Since the measure
$\dd^4\psi_{\hat p}$ carries a helicity of $+2$ (remember that each $\psi_{\hat
p}^i$ carries a helicity of $-1/2$), the total helicity carried by the
right-hand-side of \eqref{eq:BCFWSu} is $-n_L-n_R$ and since $F_{0,n}$ has
helicity $-n$ we have $n=n_L+n_R$ as it should be.

In order to derive the recursion relations \eqref{eq:BCFWSu}, one needs to
verify that the one-parameter family $F_{0,n}(t)$ of superamplitudes induced by 
\eqref{eq:BCFWcospinorsSu} vanishes as $|t|\to\infty$. This was explicitly shown
in \cite{Brandhuber:2008pf} (see also \cite{Luo:2005rx,Bianchi:2008pu} for an
earlier account). In fact, as proved by Arkani-Hamed, Cachazo \& Kaplan in
\cite{ArkaniHamed:2008gz}, any $\CN=4$ superamplitude behaves as\footnote{As
also shown in \cite{ArkaniHamed:2008gz}, all superamplitudes in $\CN=8$
supergravity behave as $t^{-2}$ implying similar recursion relations for maximal
supergravity. See
\cite{Bedford:2005yy,Cachazo:2005ca,Benincasa:2007qj,Bianchi:2008pu} for an
earlier account of gravity recursion relations.}
\begin{equation}
F_{n,0}(t)\ \to\ t^{-1}\efor |t|\ \to\ \infty~.
\end{equation}

In the pure Yang--Mills setting, we have seen that the recursion relations imply
that all amplitudes are given in terms of MHV and $\overline{\rm MHV}$
amplitudes. Likewise, \eqref{eq:BCFWSu} implies that every superamplitude is
given in terms of  MHV and $\overline{\rm MHV}$ superamplitudes. So far, we have
only given the MHV superamplitudes \eqref{eq:MHVSuper} explicitly. Therefore,
in order to use \eqref{eq:BCFWSu} we also need the  $\overline{\rm MHV}$
superamplitudes. Working---for a moment---in Minkowski signature, the 
$\overline{\rm MHV}$
superamplitudes are obtained from \eqref{eq:MHVSuper} via complex conjugation;
see also Remark \ref{rmk:HoloAno} Thus, we may write 
\begin{equation}\label{eq:BarMHVSuper}
 \tilde F_{0,n}^{\overline{\rm MHV}}\ =\ g^{n-2}_{\rm
YM}(2\pi)^4\delta^{(4|0)}\left(\sum_{r=1}^n
\tilde k_{r \al}k_{r\db}\right)\delta^{(0|8)}\left(\sum_{r=1}^n
\bar \psi_{r i}\tilde k_{r\al}\right)
   \frac{1}{\langle 12\rangle \langle 23\rangle \cdots\langle n1\rangle}~,
\end{equation}
where we have used $\tilde k_\al=\di\bar k_\da$. This, however, is not yet quite
the form of the $\overline{\rm MHV}$ superamplitudes that is suitable for the
recursion relations as the left and right sub-amplitudes entering
\eqref{eq:BCFWSu} are holomorphic in the fermionic coordinates. To get an
expression which depends holomorphically on the fermionic coordinates, we simply
perform a Fourier transform on all the $\bar\psi_{ri}$ in \eqref{eq:BarMHVSuper}
(see \cite{Roiban:2004yf,Drummond:2008vq,Drummond:2008bq,Elvang:2008na} for
general amplitudes), that is,
\begin{equation}
 F_{0,n}^{\overline{\rm MHV}}\ =\
\int\left(\prod_{r=1}^n\dd^4\bar\psi_r\right)\de^{-\di\sum_{r=1}^n\bar\psi_{ri}
\psi^i_r}\, \tilde F_{0,n}^{\overline{\rm MHV}}~.
\end{equation}
Using the identity \eqref{eq:DeltaIdentity}, we straightforwardly find (see also
the exercise below)~\cite{Elvang:2008na}
\begin{equation}\label{eq:BarMHVSuperFin}
 F_{0,n}^{\overline{\rm MHV}}\ =\
\frac{g^{n-2}_{\rm
YM}}{16}(2\pi)^4\delta^{(4|0)}\left(\sum_{r=1}^n
\tilde k_{r \al}k_{r\db}\right)
\left(\prod_{i=1}^4\sum_{r,s=1}^n\langle rs\rangle
\frac{\partial^2}{\partial\psi^i_r\partial\psi^i_s}\prod_{r'=1}^n
\psi^i_{r'}\right)
   \frac{1}{\langle 12\rangle \cdots\langle n1\rangle}~.
\end{equation}

We may now return to the complex setting by analytically continuing this
expression, i.e.~all the $\tilde k_r$, $k_r$ and $\psi^i_r$ are to be regarded
as complex. We have thus found a formula for the $\overline{\rm MHV}$
superamplitudes which depends holomorphically on the fermionic coordinates and
which can therefore be used in the supersymmetric recursion relations
\eqref{eq:BCFWSu}. A special case of \eqref{eq:BarMHVSuperFin} is the
three-particle $\overline{\rm MHV}$ superamplitude \cite{Brandhuber:2008pf}
\begin{equation}\label{eq:BarMHVSuper3}
 F_{0,3}^{\overline{\rm MHV}}\ =\
g_{\rm YM}(2\pi)^4
   \frac{\delta^{(4|0)}(
p_1+p_2+p_3)
\delta^{(0|4)}\left(\psi_1\langle 23\rangle+\psi_2\langle 31\rangle+
\psi_3\langle 12\rangle\right)}{\langle 12\rangle\langle 23\rangle\langle
31\rangle}~.
\end{equation}

\vspace*{5pt}

\Exercise{Verify \eqref{eq:BarMHVSuperFin} and \eqref{eq:BarMHVSuper3}.}

\vspace*{5pt}

We have now provided all the necessary ingredients to construct general
superamplitudes from the MHV and $\overline{\rm MHV}$ superamplitudes.
In fact, the recursion relation generates all the tree-level amplitudes from
the three-particle amplitudes $F_{0,3}^{\rm MHV}$ and $F_{0,3}^{\overline{\rm
MHV}}$. The interested reader might wish to consult the cited references (see
e.g.~\cite{Brandhuber:2008pf,ArkaniHamed:2008gz}) for explicit examples.
Notice that the three-particle amplitudes just follow from helicity
information and Lorentz invariance and without actually refering to a
Lagrangian \cite{Benincasa:2007xk}.

\vspace*{5pt}

\Remark{The Britto--Cachazo--Feng--Witten recursion relations have also been
investigated from the point of view of twistor theory. The first twistor
formulation of these recursion relations was given in terms of twistor diagrams
by Hodges in \cite{Hodges:2005bf,Hodges:2005aj,Hodges:2006tw}. The Hodges
construction is an ambidextrous approach as it uses both twistors and dual
twistors. This approach has been re-considered by Arkani-Hamed, Cachazo, Cheung
\& Kaplan in \cite{ArkaniHamed:2009si}. Mason \& Skinner \cite{Mason:2009sa}
discuss the recursion relations in terms of twistors only. In particular, if we
let
$\cW[F_{0,n}](Z^I_1,\ldots,Z^I_n)$ be the Witten transform
\eqref{eq:WittenFourierIntegralSuAm} of any tree-level $n$-particle
superamplitude $F_{0,n}$ (in Kleinian signature) where
$Z^I_r:=(z^A_r,\lambda_{r\da})$ is the supertwistor corresponding to particle
$r$, then the shift \eqref{eq:BCFWcospinorsSu} with $r=1$ and $s=n$ corresponds
to the simple shift
$\cW[F_{0,n}](Z^I_1,\ldots,Z^I_n)\to \cW[F_{0,n}](Z^I_1,\ldots,Z^I_n-t Z^I_1)$.
See\cite{Mason:2009sa} for more details.  }

\newpage

\thispagestyle{empty}
\vspace*{5cm}
\addtocontents{toc}{\hspace{-0.61cm}{\bf \centerline{Appendices}\\}} 
\appendix
\begin{center}
 {\bf\Large Appendices}
\end{center}
\renewcommand{\thesection}{\Alph{section}.}
\setcounter{subsection}{0} \setcounter{equation}{0}
\renewcommand{\theequation}{\thesection\arabic{equation}}


\newpage

\vspace*{10pt}

\noindent
The main purpose of the subsequent appendices is to give an overview of some of
the mathematical concepts underlying these lecture notes that the reader might
not
be so familiar with. For obvious reasons, it is impossible to give full-length
explanations here. For a detailed account of the following material, we refer
to the literature cited at the beginning of these notes.

\section{Vector bundles}\label{app:VB}

Let $X$ be a manifold (either smooth or complex). A complex rank-$r$ vector bundle over $X$ is defined to be a manifold $E$ together with a mapping $\pi$,
\begin{equation}
 \pi\,:\,E\ \to\ X~,
\end{equation}
such that the following is fulfilled:
\begin{itemize}\setlength{\itemsep}{-1mm}
 \item[(V1)] $\pi$ is surjective,
 \item[(V2)] for all $x\in X$, $E_x:=\pi^{-1}(x)$ is a complex vector space of dimension $r$, i.e.~there
exists an isomorphism $h_x\,:\,E_x\to\IC^r$
 \item[(V3)] and for all $x\in X$ there exists a neighbourhood $U\ni x$ and a
diffeomorphism $h_U\,:\,U\times\IC^r\to\pi^{-1}(U)$  such that $\pi\circ
h_U\,:\,U\times\IC^r\to U$ with $(x,v)\mapsto x$ for any $v\in \IC^r$.
\end{itemize}
The manifold $E$ is called total space and $E_x$ is the fibre over $x\in X$. 
Because of (V3), $E$ is said to be locally trivial and $h_U$ is called a
trivialisation of $E$ over $U\subset X$. 
A trivial vector bundle $E$ is globally of this form, i.e.~$E\cong X\times\IC^r$.
A map $s\,:\,X\to E$ satisfying $\pi\circ s=\mbox{id}_X$ is called a section section of $E$. Notice that we could replace $\IC^r$ in the above definition by $\IR^r$ in which case one speaks of real vector bundles of real rank $r$.

A useful description of vector bundles is in terms of transition functions. Let $E\to X$ be a complex vector bundle and $\{U_i\}$ be a covering of $X$ with trivialisations $h_i\,:\,U_i\times\IC^r\to\pi^{-1}(U_i)$ of $E$ over $U_i$. If $U_i\cap U_j\neq\emptyset$ then the functions $f_{ij}:=h_i^{-1}\circ h_j\,:\,U_i\cap U_j\times\IC^r\to U_i\cap U_j\times\IC^r$ are called transition functions with respect to the covering. By definition, they are diffeomorphisms. Furthermore, they define maps $\tilde f_{ij}\,:\,U_i\cap U_j\to\sGL(r,\IC)$ via $f_{ij}(x,v)=(x,\tilde f_{ij}v)$.
In the following, we shall not make a notational distinction between $f_{ij}$
and $\tilde f_{ij}$ and simply write $f_{ij}$. By construction, the transition
functions obey
\begin{equation}\label{eq:defvb}
 f_{ii}\ =\ 1~~\mbox{on}~~U_i~,\quad
 f_{ij}\ =\ f^{-1}_{ji}~~\mbox{on}~~U_i\cap U_j\ \neq\ \emptyset~,\quad
 f_{ij}\circ f_{jk}\ =\ f_{ik}~~\mbox{on}~~U_i\cap U_j\cap U_k\ \neq\ \emptyset~.
\end{equation}
The last of these equations is called the cocycle condition. Conversely, given a
collection of functions $f_{ij}$ which obey \eqref{eq:defvb}, then one can
construct a vector bundle such that the $f_{ij}$ are its transition functions.
See e.g.~\cite{Ward:1990vs} for a proof. So far, we have talked about complex
vector bundles. Likewise, one has holomorphic vector bundles $E\to X$ where $X$
is a complex manifold and the transition functions of $E$ are assumed to be
bi-holomorphic. 

Let $E$ and $E'$ be two vector bundles over $X$. A morphism (bundle map)
$\phi\,:\,E\to E'$ is a mapping such that $\phi$ restricted to the fibre $E_x$
is a linear mapping to the fibre $E'_x$ for all $x\in X$. We call $\phi$ a
monomorphism if it is one-to-one on the fibres, an epimorphism if it is
surjective on the fibres and an isomorphism if it is one-to-one and surjective
on the fibres. The vector bundle $E$ is a subbundle of $E'$ if $E$ is a
submanifold of $E'$ and $E_x$ is a linear subspace of $E'_x$ for all $x\in X$.

An important concept we use throughout these notes is that of a pull-back. Let $E\to X$ be a complex vector bundle and $g\,:\,Y\to X$ be a smooth mapping, then the pull-back bundle $g^*E$ is a complex vector bundle over $Y$ of the same rank as $E$ such that
\begin{equation}
  \begin{CD}
   g^*E@>{} >> E\\
   @V{} VV               @VV{} V\\
   Y@>> {} >X
  \end{CD}
 \end{equation}
is commutative. Put differently, the fibre of $g^*E$ over $y\in Y$ is just a copy of the fibre of $E$ over $g(y)\in X$. Furthermore, if $\{U_i\}$ is a covering of $X$ and $f_{ij}$ are the transition functions of $E$ then $\{g^{-1}(U_i)\}$ defines a covering of $Y$ such that $g^*E$ is locally trivial. The transition functions $g^*f_{ij}$ of the pull-back bundle $g^*E$ are then given by $g^*f_{ij}=f_{ij}\circ g$.

Moreover, given two complex vector bundles $E$ and $E'$ over $X$, we can form
new vector bundles. For instance, we can form the dual vector bundles, the
direct sum of $E$ and $E'$, the 
tensor product of $E$ and $E'$, the symmetric tensor product of $E$ and $E'$,
which we respectively denote by
\begin{equation}
 E^*~,\qquad E\oplus E'~,\qquad E\otimes E'~,\qquad E\odot E'~,
\end{equation}
or if $E$ is a subbundle of $E'$, one can form the quotient bundle $E'/E$. We
can also form the exterior product of a vector bundle
\begin{equation}
 \Lambda^kE~,\efor k\ =\ 0,\ldots,\mbox{rank}\,E~.
\end{equation}
In each case, one defines the derived bundles fibrewise by the linear algebra
operation indicated. For example, the fibres of $E\oplus E'$ are defined by
$(E\oplus E')_x:=E_x\oplus E'_x$ for all $x\in X$. If $r=\mbox{rank}\,E$ then
$\Lambda^rE$ is given a special symbol $\det E$ and called the determinant line
bundle since the transition functions of $\Lambda^rE$ are given by the
determinants of the transition functions of $E$. In the case  where $E$ is the
cotangent bundle $T^*X$ for some manifold $X$, $\det T^*X$ is called the
canonical bundle and denoted by $K$ or $K_X$, respectively.

Another important concept is that of exact sequences. Let $E_1$, $E_2$ and $E_3$
be three complex vector bundles over $X$. Then the sequence 
\begin{equation}
 E_1\ \overset{\phi}{\longrightarrow}\ E_2\ \overset{\psi}{\longrightarrow}\ E_3
\end{equation}
is exact at $E_2$ if $\ker\psi=\mbox{im}\,\phi$. A short exact sequence is a
sequence of the form
\begin{equation}\label{eq:SESV}
 0\ \longrightarrow\ E_1\ \longrightarrow\ E_2\ \longrightarrow\ E_3\
\longrightarrow\ 0
\end{equation}
which is exact at $E_1$, $E_2$ and $E_3$. We say that the sequence splits if
$E_2\cong E_1\oplus E_3$. Hence, one can understand $E_2$ in \eqref{eq:SESV} as
a deformation of the direct sum $E_1\oplus E_3$.

Let us give an example. Let $X=\IC P^m$. Then
\begin{equation}\label{eq:Euler}
 0\ \longrightarrow\ \IC\ \overset{\phi}{\longrightarrow}\ \CO(1)\otimes\IC^{m+1}\ \overset{\psi}{\longrightarrow}\ T\IC P^m\ \longrightarrow\ 0
\end{equation}
is a short exact sequence which is called the Euler sequence (see also Remark
\ref{rmk:jet}). This can be understood as follows. Let
$\pi\,:\,\IC^{m+1}\setminus\{0\}\to\IC P^m$ be the canonical projection and let
$z^a$, for $a=0,\ldots,m$, be linear coordinates on $\IC^{m+1}$ (or
equivalently, homogeneous coordinates on $\IC P^m$). In Remark
\ref{rmk:pojection}~we saw that $\pi_*(z^a\partial_a)=0$, where
$\partial_a:=\partial/\partial z^a$.
If we let $s=(s^0,\ldots,s^m)$ be a section of $\CO(1)\otimes\IC^{m+1}$, then
the mapping $\psi$ in \eqref{eq:Euler} is given by
$\psi(s)=\pi_*(s^a(z)\partial_a)$. Clearly, $\psi$ is surjective and moreover,
its kernel is the trivial line bundle spanned by the section
$s_0=(z^0,\ldots,z^m)$, i.e.~$\psi(s_0)=0$. Thus, \eqref{eq:Euler} is indeed a
short exact sequence as claimed.

The next concept we need is that of connections and curvature. Let $E\to X$ be a rank-$r$ complex vector bundle and let $\Omega^p(X,E)$ be the differential $p$-forms on $X$ with values in $E$. A connection $\nabla$ on $E$ is a differential operator 
\begin{equation}
 \nabla\,:\,\Omega^p(X,E)\ \to\ \Omega^{p+1}(X,E)
\end{equation}
which obeys the Leibniz rule: For $f$ some function on $X$ and $s$ a section of $T^pX\otimes E$, we have
\begin{equation}
 \nabla(fs)\ =\ \dd f\wedge s+f\nabla s~.
\end{equation}
Suppose now that $e=\{e_1,\ldots,e_r\}$ is a local frame field of $E$ over
$U\subset X$, i.e.~the $e_a$ are sections of $E$ over $U$, for $a=1,\ldots,r$,
and for all $x\in U$, $\{e_1(x),\ldots,e_r(x)\}$ is a basis for $E_x$. Then we
may introduce the connection one-form (or also referred to as the gauge
potential) $A$ according to
\begin{equation}\label{eq:defCOF}
  \nabla e_a\ =\ {A_a}^b e_b~,
\end{equation}
i.e.~$A$ is a differential one-form with values in the endomorphism bundle $\mbox{End}\,E\cong E^*\otimes E$
of $E$. Thus, $\nabla$ is locally of the form $\nabla=\dd+A$. Let $\{U_i\}$ be a covering of $X$ which trivialises $E$ and let $e_i:=e|_{U_i}=\{e_1|_{U_i},\ldots,e_r|_{U_i}\}$ be the frame field with respect to $\{U_i\}$. On $U_i\cap U_j\neq\emptyset$ we have $e_i=f_{ij}e_j$, where the $f_{ij}$ are the transition functions of $E$. Upon substituting this into \eqref{eq:defCOF}, we find
\begin{equation}
 A_j\ =\ f_{ij}^{-1}\dd f_{ij}+f_{ij}^{-1} A_i f_{ij}~,\ewith A_i\ :=\ A|_{U_i}~.
\end{equation}
Hence, patching together the $A_i$ according to this formula, we obtain a globally defined connection one-form $A$.
The curvature (also referred to as the field strength) of $\nabla$ is defined
as
\begin{equation}
 F\ :=\ \nabla^2
\end{equation}
and is a section of $\Lambda^2 T^*X\otimes\mbox{End}\,E$, i.e.~under a change of trivialisation, $F$ behaves as
\begin{equation}
 F_j\ =\ f^{-1}_{ij} F_i f_{ij}~,\ewith F_i\ :=\ F|_{U_i}~.
\end{equation}
Furthermore, the curvature is locally of the form
\begin{equation}
 F\ =\ \dd A+A\wedge A~.
\end{equation}
Notice that $\nabla F=0$, which is the so-called Bianchi identity.

\section{Characteristic classes}\label{app:CC}

In this section, we shall define certain characteristic classes. However, we
shall avoid the general definition in terms of invariant polynomials and
cohomology classes and instead focus on the example of Chern classes
and characters.

Let $E\to X$ be a complex vector bundle and $F$ be the curvature two-form of a conncection $\nabla=\dd+A$ on $E$. The total Chern class $c(E)$ of $E$ is defined by
\begin{equation}
 c(E)\ :=\ \det\left(1+\frac{\di}{2\pi}F\right).
\end{equation}
Since $F$ is a differential two-form, $c(E)$ is a sum of forms of even degrees. The Chern classes $c_k(E)$ are defined by the expansion
\begin{equation}
 c(E)\ =\ 1+c_1(E)+c_2(E)+\cdots~.
\end{equation}
Notice that one should be more precise here, since the $c_k(E)$ written
here are actually the Chern forms: They are even differential forms which are
closed. The Chern classes are defined as the cohomology classes of the Chern
forms and so the Chern forms are representatives of the Chern classes. However,
we shall loosely refer to the $c_k(E)$ as Chern classes, i.e.~$c_k(E)\in
H^{2k}(X,\IZ)$ (see also Section \ref{sec:Cech} for the definition of
the cohomology groups). Notice also that the above definitions do not depend on
the choice of connection.
If $E=TX$ then one usually writes $c_k(X)$. 

If $2k>\dim_\IR X$ then $c_k(E)=0$. Furthermore, if $k>\mbox{rank}\,E$ then $c_k(E)=0$, as well. Hence, if $E$ is a line bundle then $c(E)=1+c_1(E)$ and if in addition $c_1(E)=0$ then $E$ is a trivial bundle.
A few explicit Chern classes are
\begin{equation}
\begin{aligned}
 c_0(E)\ &=\ 1~,\\
 c_1(E)\ &=\ \frac{\di}{2\pi}\,\mbox{tr}\,F~,\\
 c_2(E)\ &=\ \frac12\left(\frac{\di}{2\pi}\right)^2[\mbox{tr}\,F\wedge\mbox{tr}\,F-\mbox{tr}\,(F\wedge F)]~,\\
         &\ \, \vdots\\
 c_r(E)\ &=\ \left(\frac{\di}{2\pi}\right)^r\det F~,
\end{aligned}
\end{equation}
where $r=\mbox{rank}\,E$. 

If we consider the direct sum $E_1\oplus E_2$ then $c(E_1\oplus
E_2)=c(E_1)\wedge c(E_2)$ as
follows from the properties of the determinant.  Furthermore, this property is
deformation
independent, i.e.~it is also valid for the short exact sequence \eqref{eq:SESV}.
This is known as the splitting principle. An immediate consequence is then that
$c_1(E_1\oplus E_2)=c_1(E_1)+c_1(E_2)$. This is an important fact that we used
in
Section
\ref{sec:HCS} when talking about Calabi--Yau spaces. Finally, if $g^*E$ denotes
the pull-back bundle of $E$ via some map $g$ then $c(g^*E)=g^*c(E)$. This is
called naturality or functoriality (see below).

A related quantity is the total Chern character. It is defined by
\begin{equation}
 \mbox{ch}(E)\ :=\ \mbox{tr}\,\exp\left(\frac{\di}{2\pi}F\right)\ =\ 
      \mbox{ch}_0(E)+\mbox{ch}_1(E)+\mbox{ch}_2(E)+\cdots~.
\end{equation}
where the $k$-th Chern character is 
\begin{equation}
 \mbox{ch}_k(E)\ =\ \frac{1}{k!}\mbox{tr}\,\left(\frac{\di}{2\pi}F\right)^k.
\end{equation}
The Chern characters can be expressed in terms of the Chern classes according to
\begin{equation}
 \begin{aligned}
  \mbox{ch}_0(E)\ &=\ \mbox{rank}\,E~,\\
  \mbox{ch}_1(E)\ &=\ c_1(E)~,\\ 
  \mbox{ch}_2(E)\ &=\ \frac12[c_1(E)\wedge c_1(E)-2c_2(E)]~,\\
   &\ \,\vdots
 \end{aligned}
\end{equation}
The total Chern character has the following properties. Let $E_1$ and $E_2$ be
two complex vector bundles. Then $\mbox{ch}(E_1\oplus
E_2)=\mbox{ch}(E_1)+\mbox{ch}(E_2)$ and
$\mbox{ch}(E_1\otimes E_2)=\mbox{ch}(E_1)\wedge\mbox{ch}(E_2)$. If $g^*E$
denotes the pull-back bundle of $E$ via $g$ then
$\mbox{ch}(g^*E)=g^*\mbox{ch}(E)$ which is also called naturality or
functoriality (see below).

Likewise, we may introduce Chern classes and characters for supervector bundles
over supermanifolds. Given a rank-$r|s$ complex vector bundle $E$ over a
complex supermanifold $X$, we define the $k$-th Chern class of $E$ to be
\begin{equation}
  c_k(E)\ :=\ \frac{1}{k!}\left.\frac{\dd^k}{\dd t^k}\right|_{t=0}
            \mbox{sdet}\left(1+t\,\frac{\di}{2\pi}F\right)
            \efor k\ \leq\ r+s~, 
\end{equation}
where $F$ is again the curvature two-form of a connection $\nabla$ on $E$ and
`sdet' is the superdeterminant defined in \eqref{eq:sdet}.
The first few Chern classes are given by:
\begin{equation}
 \begin{aligned}
 c_0(E)\ &=\ 1~,\\
        c_1(E)\ &=\ \frac{\di}{2\pi}\,\mbox{str}\,F~,\\
 c_2(E)\ &=\
\frac12\left(\frac{\di}{2\pi}\right)^2[\mbox{str}\,F\wedge\mbox{str}\,F-
\mbox{str} \,(F\wedge F)]~,\\
         &\ \, \vdots\\
\end{aligned}
\end{equation}
Here, `str' is the supertrace defined in \eqref{eq:str}.
The total Chern class is then $c(E):=\sum_{k=0}^{r+s}c_k(E)$.
In a similar fashion, we may also introduce the $k$-th 
Chern character according to
\begin{equation}
 \mbox{ch}_k(E)\ :=\
\left.\frac{\dd^k}{\dd t^k}\right|_{t=0}\mbox{str}\,
 \exp\left(t\,\frac{\di}{2\pi}F\right)
            \efor k\ \leq\ r+s~.
\end{equation}
More details can be found, e.g.~in the book by Bartocci et al.
\cite{Bartocci:1991}.

\pagebreak[4]

\section{Categories}\label{app:Ca}
 
A category $\cC$ consists of the following data:
\begin{itemize}\setlength{\itemsep}{-1mm}
 \item[(C1)] a collection ${\rm Ob}(\cC)$ of objects,
 \item[(C2)] sets ${\rm Mor}(X,Y)$ of morphisms for each pair 
 $X,Y\in{\rm Ob}(\cC)$, including a distinguished identity morphism 
 ${\rm id}={\rm id}_X\in{\rm Mor}(X,X)$ for each $X$,
 \item[(C3)] a composition of morphisms function $\circ\,:\,{\rm Mor}(X,Y)
 \times{\rm Mor}(Y,Z)\rightarrow{\rm Mor}(X,Z)$ for each triple
 $X,Y,Z\in{\rm Ob}(\cC)$ satisfying $f\circ{\rm id}=f={\rm id}\circ f$ and 
 $(f\circ g)\circ h=f\circ(g\circ h)$.
\end{itemize}
There are many examples of categories, such as:
\begin{itemize}\setlength{\itemsep}{-1mm}
\item[(i)] The category of vector spaces over $\IR$ or $\IC$ consists of all 
vector spaces over $\IR$ or $\IC$ (= objects). The morphisms are linear maps.
\item[(ii)] The category of topological spaces consists of all topological 
spaces (= objects). The morphisms are continuous maps. 
\item[(iii)] The category of $C^k$-manifolds consists of all $C^k$-manifolds 
(= objects). The morphisms are $C^k$-maps. Notice that $k$ can also be infinity 
 in which case one speaks of smooth manifolds.
\item[(iv)] The category of complex manifolds consists of all complex manifolds 
(= objects). The morphisms are  holomorphic maps. 
\item[(v)] The categories of complex/holomorphic vector bundles
consist of all complex/holomorphic vector bundles 
(= objects). The morphisms are the smooth/holomorphic bundle maps.
\item[(vi)] The category of Lie algebras over $\IR$ or $\IC$ consists of all 
Lie algebras over $\IR$ or $\IC$ (= objects) and the morphisms are those linear maps 
respecting the Lie bracket.
\item[(vii)] The category of Lie groups consists of all Lie groups (= objects). 
The morphisms are the Lie morphisms, which are smooth group morphisms.
\end{itemize}

Besides the notion of categories, we need so-called functors which relate 
different categories. 
A functor $F$ from a category $\cC$ to another category $\cD$ takes each object
$X$ in ${\rm Ob}(\cC)$ and assigns an object $F(X)$ in ${\rm Ob}(\cD)$ to
it. Similarly, it takes each morphism $f$ in ${\rm Mor}(X,Y)$ and assigns a
morphism $F(f)$ in ${\rm Mor}(F(X),F(Y))$ to it such that $F({\rm id})={\rm
id}$ and 
$F(f\circ g)=F(f)\circ F(g)$. Pictorially, we have
\begin{equation}
  \begin{CD}
   X@>f >> Y\\
   @VF VV               @VVF V\\
   F(X)@>>F(f)>F(Y)
  \end{CD}
 \end{equation}
What we have just defined is a covariant functor. A contravariant 
functor differs from the covariant functor 
by taking $f$ in ${\rm Mor}(X,Y)$ and assigning the morphism $F(f)$ in
${\rm Mor}(F(Y),F(X))$ to it with $F({\rm id})={\rm id}$ and $F(f\circ
g)=F(g)\circ F(f)$. 

The standard example 
is the so-called dual vector space functor. 
This functor takes a vector space V over $\IR$ or $\IC$ and assigns a dual
vector space $F(V)=V^*$ to it and to each linear map $f\,:\,V\to W$ it assigns
the dual map
$F(f)=f^*\,:\,W^*\rightarrow V^*$, with $\omega\mapsto\omega\circ f$ and
$\omega\in W^*$, 
in the reverse direction. Hence, it is a 
contravariant functor. 

Another functor we have already encountered is the
parity map $\Pi$ defined in \eqref{eq:paritymap}, which is a functor
from the category of $R$-modules to the category of $R$-modules for $R$ being a supercommutative ring.
To understand how it acts, let us take a closer look to morphisms between
$R$-modules.
An additive
mapping of $R$-modules, $f\,:\,M\to N$, is called an 
even morphism if it preserves the grading and is $R$-linear. We
denote the group of such morphisms by ${\rm Hom}_0(M,N)$. On the
other hand, we call an additive mapping of $R$-modules odd if
it reverses the grading, $p_{f (m)}=p_m+1$, and is
$R$-linear, that is, $f (rm)=(-)^{p_r}rf (m)$ and
$f (mr)=f (m)r$. The group of such morphisms is denoted by 
${\rm Hom}_1(M,N)$. Then we set
\begin{equation}
 {\rm Hom}(M,N)\ :=\ {\rm Hom}_0(M,N)\oplus{\rm Hom}_1(M,N)
\end{equation}
and it can be given an $R$-module structure.
Then $\Pi$ is defined in \eqref{eq:paritymap}, where we implicitly assumed 
that i) addition in $\Pi M$ is the same as in $M$,
ii) right multiplication by $R$ is the same as in $M$ and iii) left
multiplication differs by a sign, i.e.~$r\Pi(m)=(-)^{p_r}\Pi(rm)$
for $r\in R$, $m\in M$ and $\Pi(m)\in\Pi M$. Corresponding to the
morphism $f\,:\,M\to N$, we let $f^\Pi\,:\,\Pi M\to\Pi N$
be the morphism which agrees with $f$ as a mapping of 
sets. Moreover, corresponding to the morphism $f\,:\,M\to N$, we
can find morphisms
\begin{equation}
 \begin{aligned}
  \Pi f\,:\,M\ \to\ \Pi N~,\ewith
        (\Pi f)(m)\ :=\ \Pi(f(m))~,\\
        f\Pi\,:\,\Pi M\ \to\ N~,\ewith
        (f\Pi)(\Pi m)\ :=\ f(m)~,\kern.2cm
 \end{aligned}
\end{equation}
and hence $f^\Pi=\Pi f\Pi$. Therefore, $\Pi$ is a covariant functor.

\section{Sheaves}\label{app:Sh}

Let $X$ be a topological space. A pre-sheaf $\CS$ of Abelian groups on $X$
consists of the following data:
\begin{itemize}\setlength{\itemsep}{-1mm}
\item[(P1)] for all open subsets $U\subset X$, an Abelian group $\CS(U)$ and
\item[(P2)] for all inclusions $V\subset U$ of open subsets of $X$, a
morphism of Abelian groups $r^U_V\,:\,\CS(U)\to\CS(V)$
\end{itemize}
subject to the conditions:
\begin{itemize}\setlength{\itemsep}{-1mm}
\item[(P3)] $r(\emptyset)=0$,
\item[(P4)] $r^U_U={\rm id}_U\,:\,\CS(U)\to\CS(U)$ and
\item[(P5)] $W\subset V\subset U$, then $r^U_V\circ r^V_W=r^U_W$.
\end{itemize}
Put differently, a 
pre-sheaf is just a contravariant functor from the category
$\mathfrak{Top}(X)$ (objects = open sets of $X$, morphisms = inclusion maps)
to the category $\mathfrak{Ab}$ of Abelian groups. In fact, we can replace
$\mathfrak{Ab}$ by any other fixed category $\cC$.
Let $\CS$ be a pre-sheaf  on $X$, then $\CS(U)$ are the sections of 
$\CS$ over $U$. The $r^U_V$ are called restriction maps.

A pre-sheaf $\CS$ on a topological space is a sheaf if it satisfies:
\begin{itemize}\setlength{\itemsep}{-1mm}
\item[(S1)] $U$ open, $\{V_a\}$ open covering of $U$, $s\in\CS(U)$ such that
           $r^U_{V_a}(s)=0$ for all $a$, then $s=0$,
\item[(S2)] $U$ open, $\{V_a\}$ open covering of $U$, $s_a\in\CS(V_a)$, with
           $r^U_{V_a\cap V_b}(s_a)=r^U_{V_a\cap V_b}(s_b)$, then there exists
           an $s\in\CS(U)$ such that $r^U_{V_a}(s)=s_a$ for all $a$.
\end{itemize}
Point (S1) says that any section is determined by its local behaviour while
(S2) means that local sections can be pieced together to give global sections.

Examples are:
\begin{itemize}\setlength{\itemsep}{-1mm}
\item[(i)] Let $X$ be a real manifold and $U\subset X$. Then $\CS(U):=\{\mbox{smooth functions on }U\}$,
$\Omega^p(U):=\{\mbox{smooth }p\mbox{-forms on }U\}$, etc. are pre-sheaves,
where the restriction mappings are
the usual restriction mappings. They are also sheaves.
\item[(ii)] Let $X$ be a complex manifold and $U\subset X$. Then $\CO(U):=\{\mbox{holomorphic functions on }U\}$,
$\Omega^{p,q}(U):=\{\mbox{smooth }(p,q)\mbox{-forms on }U\}$, etc. are
pre-sheaves, where the restriction mappings are
the usual restriction mappings. They are also sheaves.
\item[(iii)] Let $E\to X$ be a complex vector bundle over some manifold $X$ and
$U\subset X$. Then $\CS(E)(U):=\{\mbox{smooth sections of $E$ on }U\}$ is
a pre-sheaf, where the restriction mappings are
the usual restriction mappings. It is also a sheaf.
\item[(iv)] Let $E\to X$ be a holomorphic vector bundle over some complex
manifold $X$ and $U\subset X$. Then $\CO(E)(U):=\{\mbox{holomorphic sections of
$E$ on }U\}$ is a pre-sheaf, where the restriction mappings are
the usual restriction mappings. It is also a sheaf.
\item[(v)] Let $R$ be a ring, $X$ a topological space and  $U\subset X$. 
Then $R(U)$ is the ring of
locally constant continuous functions on $U$. This determines a pre-sheaf that
is a sheaf. We call this the constant sheaf $R$ on $X$, e.g., 
$R=\IR,\IC,\ldots$.
\item[(vi)] Consider $\CB(U):=\{\mbox{bounded holomorphic functions on }
U\subset\IC\}$. Then $U\to \CB(U)$ is a pre-sheaf but not a sheaf.
\end{itemize}

The remainder of this appendix collects some basic notions regarding
(pre-)sheaves. Firstly, we need the notion of morphisms between (pre-)sheaves.
A morphism of (pre-)sheaves $\phi\,:\,\CS\to\CS'$ consists of a morphism
of Abelian groups $\phi_U\,:\,\CS(U)\to\CS'(U)$ for all open subsets $U$ such 
that whenever $V\hookrightarrow U$ is an inclusion, the diagram
\begin{equation}
  \begin{CD}
   \CS(U)@>\phi_U >> \CS'(U)\\
   @V{r^U_V} VV               @VV{{r'}^U_V} V\\
   \CS(V)@>>\phi_V>\CS'(V)
  \end{CD}
 \end{equation}
is commutative. An isomorphism is a morphism that has a two-sided 
inverse. A typical example is the de Rham complex on a real manifold, where 
the sheaf morphism is the usual exterior derivative.

Let $\CS$ be a pre-sheaf and
\begin{equation}
 \tilde{\CS}_x\ :=\ \dot{\bigcup_{U\ni x}}\,\CS(U)~. 
\end{equation}
Then we say that two elements of $\tilde{\CS}_x$, $f\in\CS(U)$, $U\ni x$ and
$g\in\CS(V)$, $V\ni x$, are equivalent if there exists an open set
$W\subset(U\cap V)$, with $x\in W$ such that
\begin{equation}
 r^U_W(f)\ =\ r^V_W(g)~.
\end{equation}
We define the stalk $\CS_x$ to be the set of equivalence classes
induced by this equivalence relation. Of course, $\CS_x$ inherits the 
algebraic structure of the pre-sheaf $\CS$, i.e.~we can add elements in $\CS_x$
by adding representatives of equivalence classes. We shall let 
\begin{equation}
\pi\ :=\ r^U_x\,:\,\CS(U)\ \to\ \CS_x 
\end{equation}
be the natural restriction mapping to stalks.

Let $\phi\,:\,\CS\to\CS'$ be a morphism of sheaves on a topological space
$X$. Then $\phi$ is an isomorphism if and only if the induced map on the stalks
$\phi_x\,:\,\CS_x\to\CS'_x$ is an isomorphism for all $x\in X$. 
Note that this is not true for pre-sheaves.

Secondly, let us say a few words about exact sequences. We say that a sequence
of morphisms of sheaves
\begin{equation}
 \CS_1\ \overset{\phi}{\longrightarrow}\ \CS_2\ \overset{\psi}{\longrightarrow}\
\CS_3
\end{equation}
on a topological space $X$ is exact at $\CS_2$ if the induced sequence
\begin{equation}
 \CS_{1\,x}\ \overset{\phi_x}{\longrightarrow}\ \CS_{2\,x}\
\overset{\psi_x}{\longrightarrow}\ \CS_{3\,x}
\end{equation}
is exact, i.e.~$\ker\psi_x=\mbox{im}\,\phi_x$ for all $x\in X$. A short exact sequence of sheaves
is a sequence of the form
\begin{equation}\label{eq:SES}
 0\ \longrightarrow\ \CS_1\ \longrightarrow\ \CS_2\ \longrightarrow\ \CS_3\
\longrightarrow\ 0
\end{equation}
which is exact at $\CS_1$, $\CS_2$ and $\CS_3$. 

Before moving on, let us pause and give an example. Let $\CO^*$ be the sheaf of
non-vanishing holomorphic functions on a complex manifold $X$. Then 
 \begin{equation}
 0\ \longrightarrow\ \IZ\ \longrightarrow\ \CO\ \overset{\exp}{\longrightarrow}\ \CO^*\ \longrightarrow\ 0
\end{equation}
is a short exact sequence, where $\exp(f):=\de^{2\pi\di f}\in\CO^*(U)$ for $f\in\CO(U)$ with $U\subset X$.
This sequence is called the exponential sheaf sequence.

In Section \ref{sec:Cech}, we introduced the notion of sheaf cohomology, so let
us
state a basic fact about the sheaf cohomology for short exact sequences. Let
$X$ be a topological space together with a short exact sequence of the form
\eqref{eq:SES}. Then \eqref{eq:SES} always induces a long exact sequence of \v
Cech cohomology groups
according to:
\begin{equation}
\begin{aligned}
 &0\ \longrightarrow\ H^0(X,\CS_1)\ \longrightarrow\ H^0(X,\CS_2)\
\longrightarrow \ H^0(X,\CS_3)\\
 &\kern1cm\longrightarrow \ H^1(X,\CS_1)\ \longrightarrow\ H^1(X,\CS_2)\
\longrightarrow \ H^1(X,\CS_3)\\
 &\kern2cm\longrightarrow \ H^2(X,\CS_1)\ \longrightarrow\ H^2(X,\CS_2)\
\longrightarrow \ H^2(X,\CS_3)\ \longrightarrow\ \cdots
\end{aligned}
\end{equation}
For a proof, see e.g.~\cite{Ward:1990vs,Griffiths:1978}.

Let us come back to the exponential sheaf sequence. We have
\begin{equation}
 \cdots\ \longrightarrow\ H^1(X,\IZ)\ \longrightarrow\ H^1(X,\CO)\ \overset{\exp}{\longrightarrow}\ H^1(X,\CO^*)\
 \overset{c_1}{\longrightarrow}\ H^2(X,\IZ)\ \longrightarrow\ \cdots
\end{equation}
By virtue of our discussion in Section \ref{sec:HCS}, $H^1(X,\CO^*)$ parametrises the holomorphic line bundles $E\to X$. The image $c_1(E)\in H^2(X,\IZ)$ of a line bundle $E\in H^1(X,\CO^*)$ is the first Chern class.
If $X=\IC P^m$ then
$H^1(\IC P^m,\CO)=0=H^2(\IC P^m,\CO)$ and the above long exact cohomology sequence yields
$H^1(\IC P^m,\CO^*)\cong H^2(\IC P^m,\IZ)\cong\IZ$, i.e.~we have a classification of all holomorphic line bundles on complex projective space $\IC P^m$. Another related example we already encountered is given for the choice
$X=P^3$ and $E\to P^3$ a holomorphic line bundle over $P^3$. Then
$H^1(P^3,\IZ)=0$ since $P^3$ is simply connected and if in addition $c_1(E)=0$,
i.e.~$E|_{L_x}$ is holomorphically trivial on any $L_x\hookrightarrow P^3$, then
we conclude from the above sequence that
\begin{equation}
 H^1(P^3,\CO)\ \cong\ \{E\in H^1(P^3,\CO^*)\,|\,c_1(E)=0\}~.
\end{equation}
This is basically the content of Exercise \ref{exe:PWT} 


\end{document}